\documentclass[11pt,headings=big,tightenlines,numbers=noenddot,DIV=14,a4paper]{article}%

\pdfoutput=1

\usepackage[margin=1 in]{geometry}
\usepackage{comment}
\usepackage{graphicx}  
\usepackage{dcolumn} 
\usepackage{bm,relsize}   
\usepackage{amsfonts,amsmath,amssymb,amsthm,mathtools}
\usepackage{physics}
\usepackage{slashed}
\usepackage{placeins}
\usepackage{adjustbox}
\usepackage[normalem]{ulem}
\usepackage{lscape}
\usepackage{multirow}
\usepackage{lipsum, color}
\usepackage{tabularray}
\usepackage{chngpage}
\usepackage[usenames,dvipsnames,svgnames,table]{xcolor}
\usepackage[linktoc=page,bookmarks=false,colorlinks=true,linkbordercolor=RoyalBlue,citebordercolor=ForestGreen,urlbordercolor=CornflowerBlue]{hyperref}
\hypersetup{
    colorlinks=true,
    linkcolor=ForestGreen,
    filecolor=ForestGreen,      
    urlcolor=ForestGreen,
    citecolor=ForestGreen,
}

\usepackage[sort&compress,numbers,merge]{natbib}
\usepackage{breakcites}

\newcommand{\X}{\rm X}

\def\mX		{m_{\mathsmaller{\X}}}
\def\pX		{p_{\mathsmaller{\X}}}
\def\EX		{E_{\mathsmaller{\X}}}

\newcommand{\px}{p_{\mathsmaller{X}}}
\newcommand{\pxw}{p_{\mathsmaller{X,\mathrm{w}}}}

\def\thetacm		{\theta_{\mathsmaller{\rm CM}}}

\def\qty			{}
\def\abs			{}

\def\pcm		{p_{\mathsmaller{\rm CM}}}
\def\MPl		{M_{\mathsmaller{\rm Pl}}}

\def\gD		{g_{\mathsmaller{\rm D}}}
\def\alphaD		{\alpha_{\mathsmaller{\rm D}}}

\newcommand{\SM}{\mathsmaller{\rm SM}}

\newcommand{\LL}{\mathsmaller{\rm LL}}

\newcommand{\Tnuc}{T_{n}}

\newcommand{\Tn}{T_{n}}
\newcommand{\Teq}{T_{\rm eq}}

\usepackage{tikz-feynman}

\usepackage{empheq}
\usepackage[most]{tcolorbox}
\newtcbox{\mymath}[1][]{%
    nobeforeafter, math upper, tcbox raise base,
    enhanced, colframe=blue!20!black,
    colback=blue!15, boxrule=1pt,
    #1}

\begin{document}

\begin{flushright}
\footnotesize
\end{flushright}
\color{black}

\begin{center}

{\huge \bf Particle shells from relativistic bubble walls}

\medskip
\bigskip\color{black}\vspace{0.5cm}

{
{\large Iason Baldes}$^{a}$,
{\large Maximilian Dichtl}$^{b}$,
{\large Yann Gouttenoire}$^{c}$,
{\large Filippo Sala}$^{d}$
}
\\[7mm]

{\it \small $^a$ Laboratoire de Physique de l'\'Ecole Normale Sup\'erieure, ENS, \\ Universit\'e PSL, CNRS, Sorbonne Universit\'e, Universit\'e Paris Cit\'e, F-75005 Paris, France}\\
{\it \small $^b$ Laboratoire de Physique Th\'eorique et Hautes \'Energies, \\ CNRS, Sorbonne Universit\'e, F-75005 Paris, France}\\
{\it \small $^c$ School of Physics and Astronomy, Tel-Aviv University, Tel-Aviv 69978, Israel}\\
{\it \small $^c$ Dipartimento di Fisica e Astronomia, Universit\`a di Bologna and INFN sezione di Bologna, via Irnerio 46, 40126 Bologna, Italy}\\
\end{center}

\bigskip

\centerline{\bf Abstract}
\begin{quote}

Relativistic bubble walls from cosmological phase transitions (PT) necessarily accumulate expanding shells of particles. We systematically characterize shell properties, and identify and calculate the processes that prevent them from free streaming: phase-space saturation effects, out-of-equilibrium $2\to2$ and $3\to2$ shell-shell and shell-bath interactions, and shell interactions with bubble walls.
We find that shells do not free stream in scenarios widely studied in the literature, where standard predictions will need to be reevaluated, including those of bubble wall velocities, gravitational waves (GW) and particle production. Our results support the use of bulk-flow GW predictions in all regions where shells free stream, irrespectively of whether or not the latent heat is mostly converted in the scalar field gradient.

\end{quote}

\clearpage
\noindent\makebox[\linewidth]{\rule{\textwidth}{1pt}} 
\setcounter{tocdepth}{2}
\tableofcontents
\noindent\makebox[\linewidth]{\rule{\textwidth}{1pt}}

\section{Introduction}

Cosmological phase transitions (PTs) constitute some of the most prominent changes that our universe underwent in its early age. 
The PTs predicted by the standard model (SM) of particle physics, associated with QCD confinement and electroweak (EW) symmetry breaking, are crossovers~\cite{Aoki:2006we,Kajantie:1996mn}. Physics beyond the SM (BSM) can make the EW PT first order, a possibility that has been extensively studied to realise electroweak baryogenesis~\cite{Bodeker:2020ghk}.
Other first-order PTs are predicted in extensions of the SM that explain, for example, the origin of the weak scale~\cite{Creminelli:2001th,Nardini:2007me,Konstandin:2011dr,Craig:2020jfv}, neutrino oscillations~\cite{Jinno:2016knw}, the strong CP~\cite{DelleRose:2019pgi,VonHarling:2019rgb} and flavour~\cite{Greljo:2019xan} problems. 
First-order PTs have recently attracted an enormous amount of interest because they can source a spectrum of gravitational waves~\cite{Witten:1984rs,Hogan:1986qda} that is observable in foreseen experiments, such as LISA~\cite{LISACosmologyWorkingGroup:2022jok}.
Speculatively, such a PT could already have been observed by pulsar timing arrays~\cite{NANOGrav:2023hvm,Antoniadis:2023zhi,Gouttenoire:2023bqy}.

First-order PTs happen via the nucleation of bubbles of the broken phase of some symmetry, into the cosmological bath which is still sitting in the unbroken phase, for pedagogical introductions see~\cite{Hindmarsh:2020hop,Gouttenoire:2022gwi}.
Phase transitions with relativistic bubble walls, in particular, have the brightest detection prospects in GWs and offer unique phenomenological possibilities, for example,
as ultra-high energy colliders in the early universe~\cite{Baldes:2023fsp}, in the production of dark matter~\cite{Falkowski:2012fb,Hambye:2018qjv,Baldes:2020kam,Baldes:2021aph,Azatov:2021ifm,Baldes:2022oev,Giudice:2024tcp}, the baryon asymmetry of the universe~\cite{Katz:2016adq,Azatov:2021irb,Baldes:2021vyz,Chun:2023ezg,Dichtl:2023xqd} and primordial black holes~\cite{Kodama:1982sf,Liu:2021svg,Hashino:2021qoq,Kawana:2022lba,Lewicki:2023ioy,Gouttenoire:2023naa,Baldes:2023rqv,Gouttenoire:2023bqy,Gouttenoire:2023pxh,Lewicki:2024ghw,Flores:2024lng}.

The evolution of the relativistic bubble walls, and of the surrounding bath, has crucial implications in all the PT properties mentioned above. 
The evolution of relativistic walls has been the object of several studies, which identified and computed different sources of pressure, which in turn determine the wall velocities~\cite{Bodeker:2009qy,Bodeker:2017cim,Azatov:2020ufh,Gouttenoire:2021kjv,Azatov:2023xem,GarciaGarcia:2022yqb,Ai:2024shx}.
Much less attention has so far been given to the evolution of the shells of energetic particles, that necessarily accumulate around relativistic bubble walls due to their interactions with the cosmological bath, for example due to splitting radiation at the bubble walls~\cite{Gouttenoire:2021kjv}.
To the best of our knowledge, both the origin and the evolution of such shells have been studied so far only partially, and in a few specific cases~\cite{Baldes:2020kam,Gouttenoire:2021kjv,Baldes:2022oev,GarciaGarcia:2022yqb}. Progress in this direction is a necessary ingredient to reliably compute many observational consequences of PTs, including GWs~\cite{Jinno:2022fom}, and could lead to spectacular phenomenological consequences, such as heavy particle production~\cite{Baldes:2023fsp}.

\begin{figure}[p]
\raisebox{0cm}{\makebox{\includegraphics[width=0.49\textwidth, scale=1]{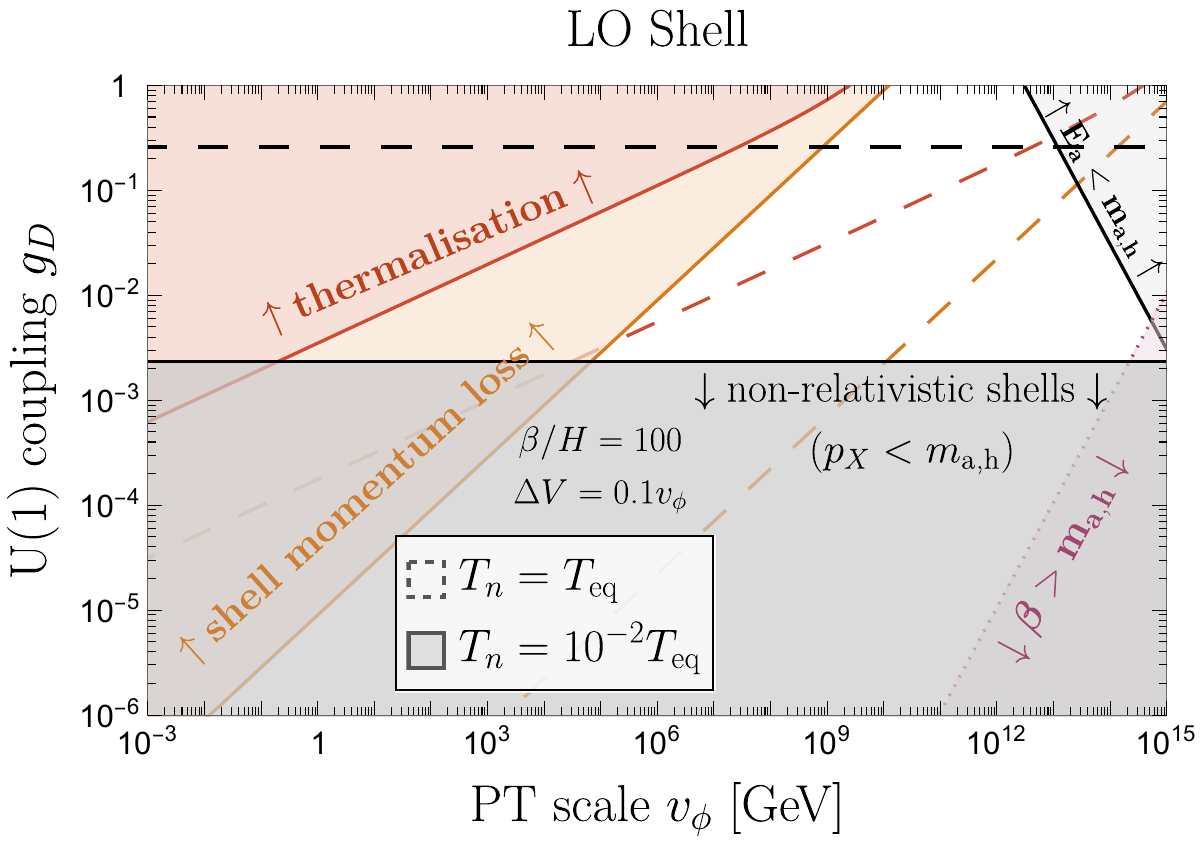}}}
{\makebox{\includegraphics[width=0.49\textwidth, scale=1]{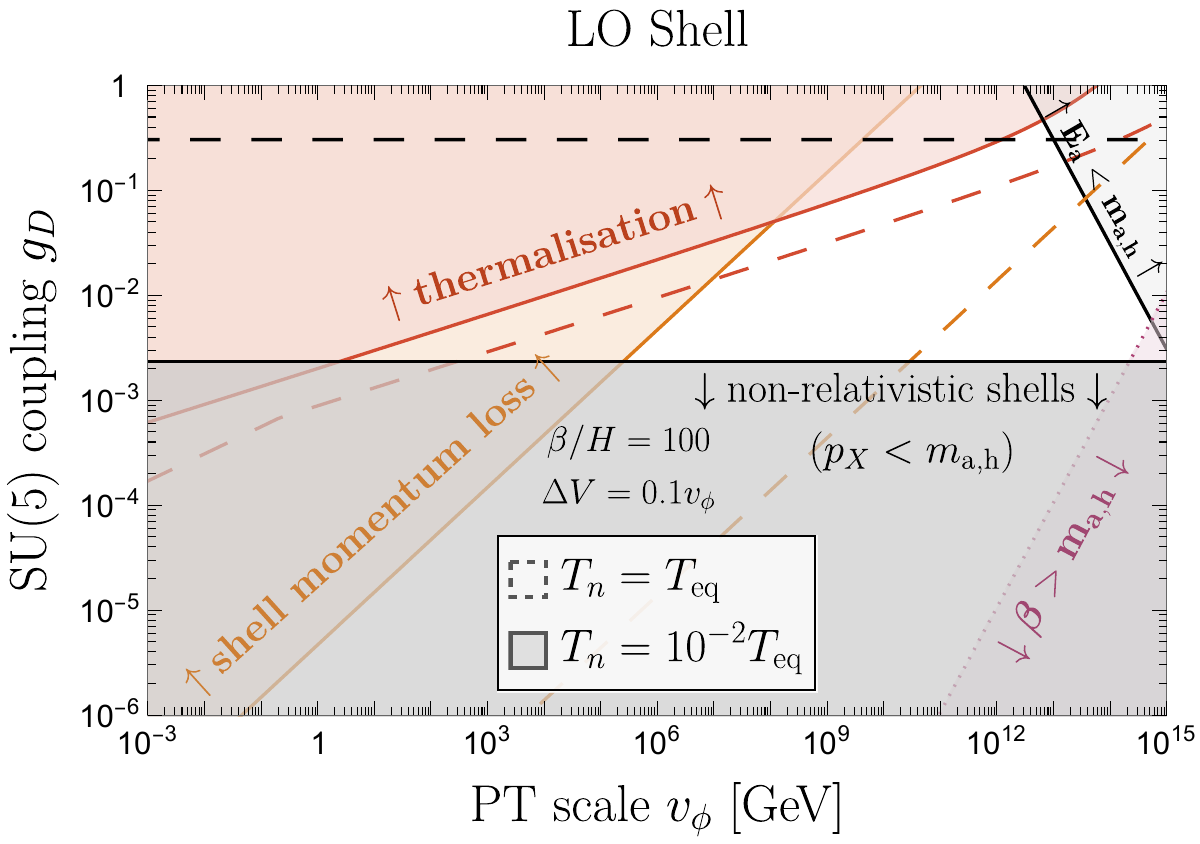}}}
{\makebox{\includegraphics[width=0.49\textwidth, scale=1]{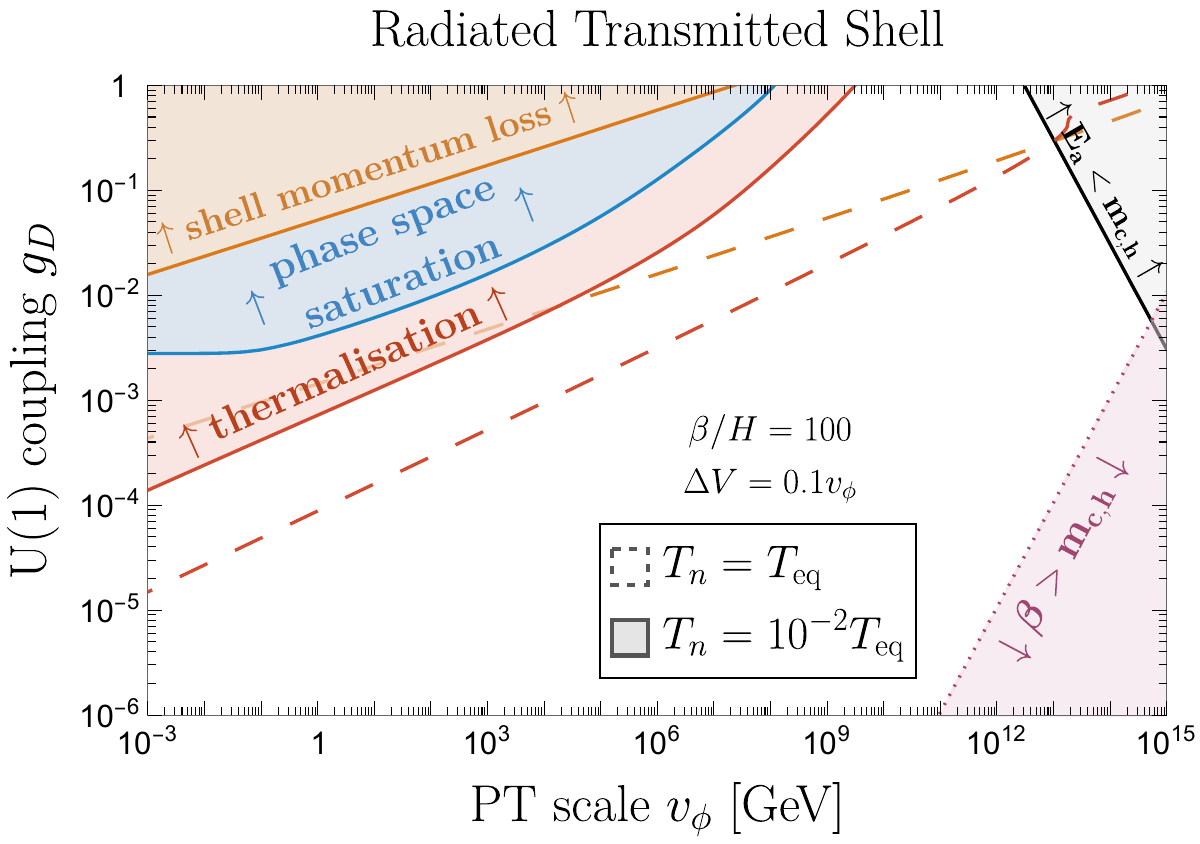}}}
{\makebox{\includegraphics[width=0.49\textwidth, scale=1]{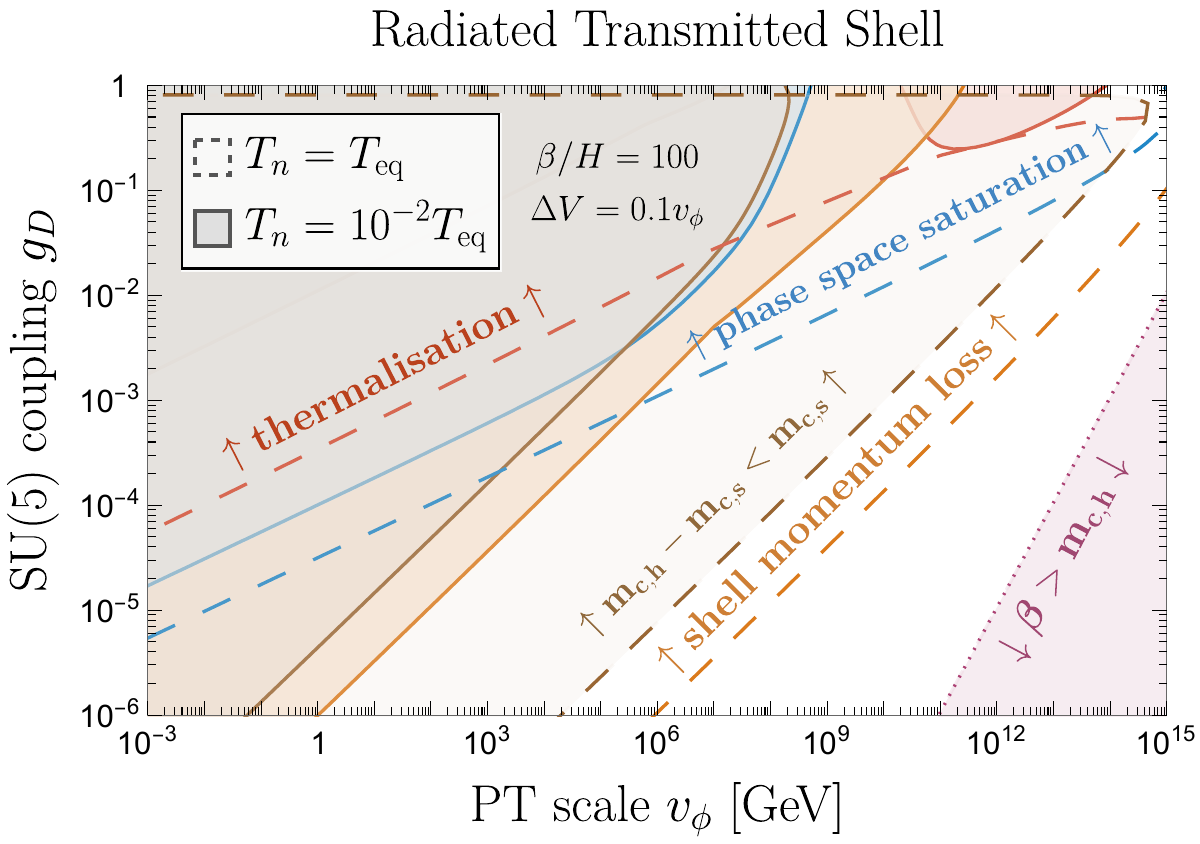}}}
{\makebox{\includegraphics[width=0.49\textwidth, scale=1]{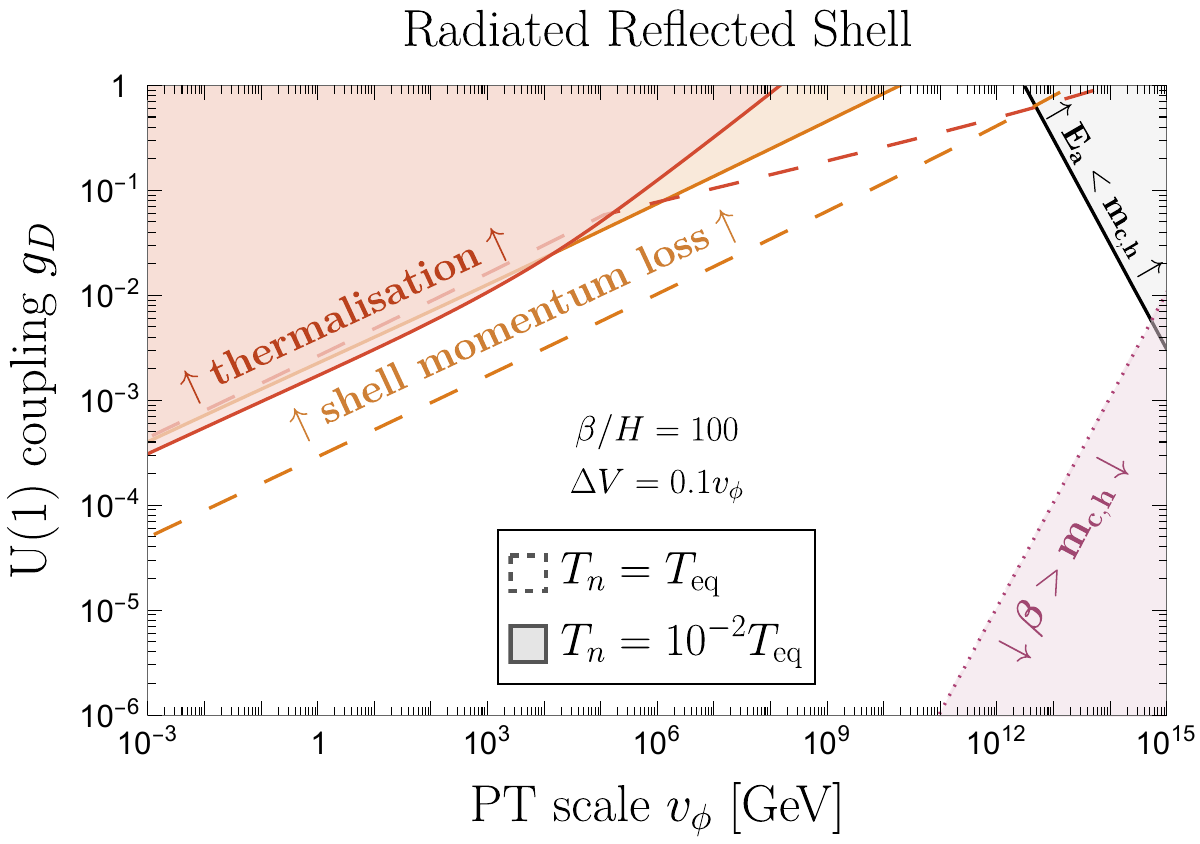}}}
{\makebox{\includegraphics[width=0.49\textwidth, scale=1]{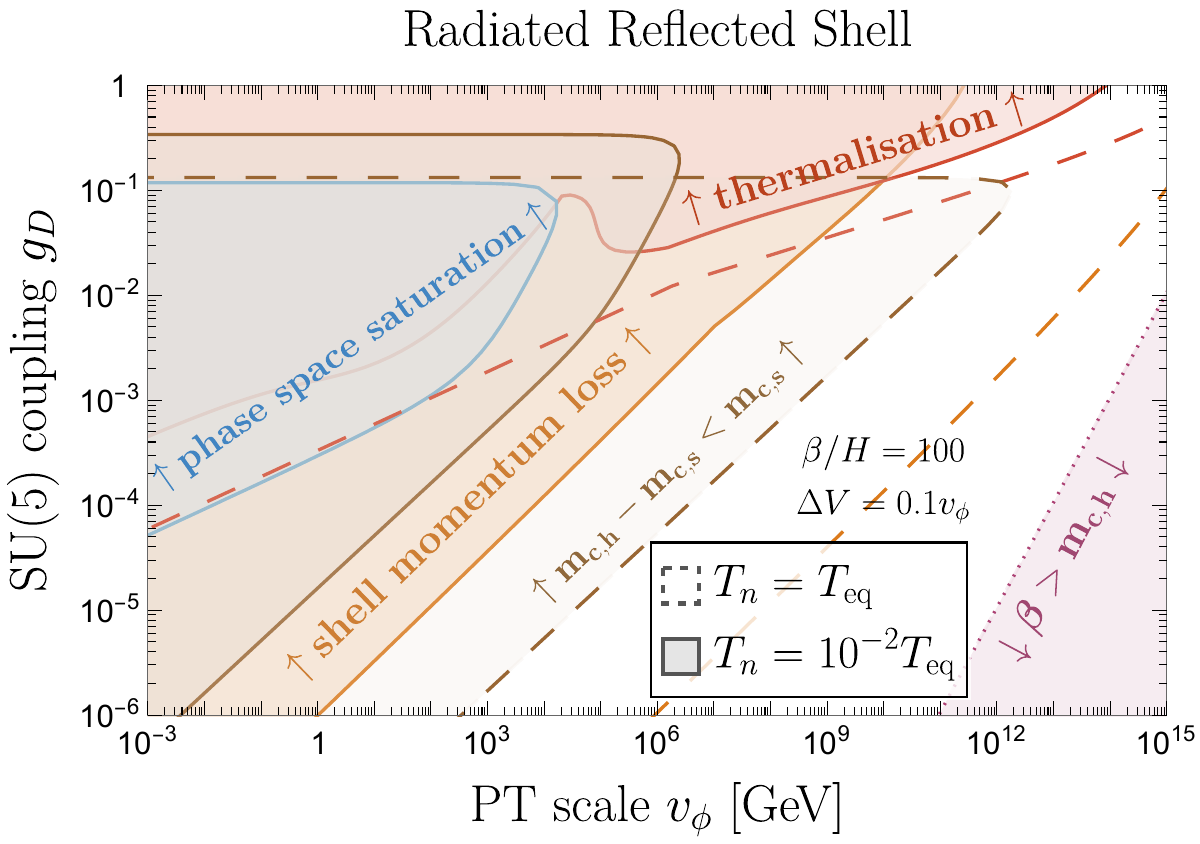}}}
\caption{\it \small  \textbf{Summary plots:}
In the shaded colored areas the particle shells around relativistic bubble walls cannot be considered as free-streaming until wall collision, and thus their evolution will affect standard PT predictions like wall velocities, GW and particle production.
We consider shell particles produced from $LO$ interaction (\textbf{top}), Bremsstrahlung radiation, either transmitted (\textbf{middle}) or reflected (\textbf{bottom}), assuming an abelian (\textbf{left}) or non-abelian (\textbf{right}) gauge interaction, see the first two rows of Table~\ref{tab:production_mechanism}, for small ($\Tn/T_{\rm eq} = 1$ \textbf{dashed}) and large ($\Tn/T_{\rm eq} = 10^{-2}$, \textbf{solid}) supercooling.
In the \textbf{blue} regions finite-density corrections become large and shells can not be described by perturbative theory anymore, see Sec.~\ref{sec:phase_space_sat} (Eqs.~\eqref{eq:PSF_non-abelian} and \eqref{eq:PSF_abelian}). 
In the \textbf{orange} region, momentum exchange between the shell and the thermal bath significantly affects the momentum of the shell particles in either the wall frame (Eqs.~\eqref{eq:Moller_lplasma_refersal} and \eqref{eq:Compton_lplasma_refersal}, relevant only for reflected shells) or the bath frame ((Eqs.~\eqref{eq:l_meta_moller} and \eqref{eq:l_meta_compton})), see Sec.~\ref{sec:mom_loss}.
In the \textbf{red} region, 3-to-2 interactions within the shells become efficient and shells are expected to thermalise, see Sec.~\ref{sec:3to2} (Eqs.~\eqref{eq:P3to2} and \eqref{eq:kperpsq}).
For the readers interested in free streaming up until shell-shell collision (as opposed to up until wall-wall collision) see Sec.~\ref{sec:wall_int} and the additional shaded regions in Fig.~\ref{fig:wall_shell_collision}.
In the \textbf{brown} region, relevant only to radiated shells in the non-abelian case, finite-density corrections drive the vector boson mass $m_{c,s}=\mu$ in the symmetric phase, cf. Eq.~\eqref{eq:IRcutoff}, above its vacuum value $m_{c,h}^\infty=\gD v_{\phi}/\sqrt{2}$ in the Higgs (i.e. broken) phase. This cuts off vector boson production, see Sec.~\ref{sec:phase_space_sat} and App.~\ref{app:large_IR_cutoff}, and our treatment breaks down. 
Similarly, our treatment breaks down in the \textbf{pink} region, where the IR cutoff due to the bubble size, $\beta$, is larger than $m_{c,h}$. In the \textbf{grey} region, the incoming bath particle energy, $E_{a}$, is insufficient to create the shell particle.
In the \textbf{black} region relativistic LO shells do not exist (see Sec.~\ref{sec:shell_boost_factor}).}
\label{fig:free_stream_all_cases}
\end{figure}

In this paper we perform the first systematic study of the evolution of particle shells around relativistic walls of first-order PTs. We then identify the possible interactions that shells undergo until they collide with those from other bubbles, and compute the conditions that allow shells to free stream without interacting nor affecting the mechanisms that source them, thus conserving both their momenta and number densities.  Our main results are summarised in Fig.~\ref{fig:free_stream_all_cases}.

The structure of the paper is as follows. In Sec.~\ref{sec:PTs} we provide a short review of first-order PTs, with a focus on wall velocities, and we list the different kinds of shells at bubble walls and we derive their properties. We then study the processes that could prevent shells from free streaming:
\begin{itemize}
\item[$\diamond$]
In Sec.~\ref{sec:phase_space_sat} we study phase space saturation of the shells. This is associated with large finite-density corrections, that backreact on the shells' production mechanism and in turn affect the computation of all other processes considered in this paper;
\item[$\diamond$]
In Sec.~\ref{sec:mom_loss} we study momentum changes of the shell and bath particles due to $2 \to 2$ scatterings. They could for example push the shell inside the bubble and/or significantly change its properties. We also provide a general argument why such elastic scatterings cannot prevent bath particles from entering the bubbles;
\item[$\diamond$]
In Sec.~\ref{sec:3to2} we study thermalisation (i.e.~efficient $3\to2$ number-changing interactions) of shells within themselves and with the bath, which would also affect the properties of both;
\item[$\diamond$] In Sec.~\ref{sec:wall_int} we study the shell interactions with the collided bubble walls.
\end{itemize}
In Sec.~\ref{sec:GW} we estimate the GW spectrum in the free-streaming regime and in Sec.~\ref{sec:conclusions} we conclude.

\section{Shell production}
\label{sec:PTs}

\subsection{Phase transition parameters}
We consider a cosmological first-order phase transition taking place in the early universe with latent heat
\begin{equation}
\alpha \equiv \frac{\Delta V}{\rho_{\rm rad}}\Big|_{\Tn} \equiv \left( \frac{T_{\rm eq}}{\Tn} \right)^4,
\end{equation}
where $T_n$ is the nucleation temperature and $T_{\rm eq}$ is the temperature when the radiation energy density, $\rho_{\rm rad}$, would drop below the vacuum energy difference $\Delta V$ of the potential at zero-temperature. That is to say,
\begin{equation}
\Delta V~\equiv~\frac{\pi^2}{30}g_* T_{\rm eq}^4,
\end{equation}
where $g_*$ is the effective number of degrees of freedom (dofs).
Denoting $\Delta V = c_{\rm vac}v_\phi^4$, where $v_\phi$ is the phase transition scale (\emph{i.e.}~the vacuum expectation value gained by the field driving the transition), we can write
\begin{equation}
T_{\rm eq}~=~ \left(\frac{30c_{\rm vac}}{g_* \pi^2}\right)^{1/4} v_\phi.
\end{equation}
Because of its connection to the latent heat parameter $\alpha$, we show all our plots as function of $T_{\rm eq}/\Tn$. It can be useful to note its relation with another PT strength parameter, $v_\phi/\Tn$, used in the literature
\begin{equation}
     \frac{v_\phi}{\Tn} = 4.3\left( \frac{g_*}{106.75}\right)^{\!1/4}\left( \frac{0.1}{c_{\rm vac}}\right)^{\!1/4} \frac{T_{\rm eq}}{\Tn} .
\end{equation}
Scenarios with $\alpha > 1$ are accompanied  with a period of supercooling during which the universe inflates for a number of e-folds, 
    \begin{equation}
    N_e \equiv \log \left( \frac{ T_{\rm eq} } { T_n } \right)=\frac{\log(\alpha)}{4}.
    \end{equation}
The size of bubbles at collision, $R_{c}$, is related to the time derivative of the nucleation rate per unit volume, $\Gamma_V$, through~\cite{Enqvist:1991xw}\footnote{The average distance between nucleation sites is $2R_{c}$. }
\begin{equation}
R_{c}~\simeq~\frac{ \pi^{1/3} }{ \beta }, \qquad \mathrm{where} \qquad
\beta \equiv \frac{1}{\Gamma_V}\frac{d\Gamma_V}{dt}.
\end{equation}

\subsection{Context: relativistic bubble walls}
Bubble walls reach relativistic velocities, $v_w\simeq 1$, if the driving pressure $\Delta V$ is larger than the leading order pressure, $\mathcal{P}_{\rm LO}$. This pressure comes from particles gaining a mass across the bubble wall, and reads~\cite{Bodeker:2009qy}
\begin{equation}
\mathcal{P}_{\rm LO} =  \frac{g_*}{24}\Delta m^2 T_n^2, 
\end{equation}
where, for convenience, we have introduced the mass difference $\Delta m$ averaged over the population of the thermal bath,
    \begin{equation}
    \Delta m^2 \equiv \sum_i c_i \frac{g_i}{g_*} \Delta m_i^2, \qquad \mathrm{with} \qquad g_* \equiv\sum_i g_i,
    \end{equation}
and where $\Delta m_i$ is the mass acquired by species $i$ inside the bubble, $g_i$ is its number of relativistic dofs, and $c_i = 1 \; (1/2)$ for bosons (fermions). 
In this work, we suppose bubble walls to be relativistic, $\gamma_{\rm w} \gg 1$, which implies that the following condition is satisfied
\begin{equation}
\label{eq:alpha_LO}
\alpha ~\gtrsim~\alpha_{\rm LO} \equiv 0.1\left(\frac{\Delta m}{T_n}\right)^2.
\end{equation}
In fact, the Bodeker and Moore criterion \cite{Bodeker:2009qy} for bubble walls to be relativistic, defined by Eq.~\eqref{eq:alpha_LO}, has a caveat. Friction pressure is non-monotonic and features a peak at the Jouguet velocity ($\sim$ speed of sound) \cite{Konstandin:2010dm,Cline:2021iff,Laurent:2022jrs,DeCurtis:2023hil,Ai:2024shx}.
This is due to the presence of a compression wave heating the thermal bath in front of the wall, called hydrodynamic obstruction~\cite{Konstandin:2010dm}.\footnote{
A peak in the pressure as a function of velocity can also arise from a large reflection probability of the longitudinal component of vector bosons, in the particular case that most of the vector boson mass is already different from zero in the unbroken phase~\cite{GarciaGarcia:2022yqb}.
If, in addition, the parameters of the PT are such that this specific pressure is the one stopping the wall acceleration, then those reflected vector bosons also lead to a shell propagating ahead of the wall, which we leave for future study.
} In principle, this peaked pressure can stop the wall from accelerating around the Jouguet velocity \cite{Espinosa:2010hh}
\begin{equation}
    \xi_J = \frac{\sqrt{\frac{2}{3}\alpha+\alpha^2}+\sqrt{1/3}}{1+\alpha},
\end{equation}
even if Eq.~\eqref{eq:alpha_LO} is satisfied.  In models that minimally extend the Standard Model, such as by adding a single singlet scalar, the enhancement of pressure has been found effective in halting wall acceleration only for deflagration-type PTs with strength parameter $\alpha \lesssim 0.01$ \cite{Laurent:2022jrs}. 
So in PTs with $\alpha \gtrsim 0.01$ walls would start running away, until possible effects beyond the hydrodynamic regime (discussed later) become important.
When instead the drop in number of relativistic dofs from the symmetric to broken phase is large (to be precise larger than about $15\%$, as it can happen e.g. in confining PTs), the recent study~\cite{Ai:2024shx} suggests that an hydrodynamic obstruction to wall acceleration could arise also for $\alpha$ up to order 1, so reducing the parameter space where walls become relativistic.
However, the results from \cite{Ai:2024shx}  could be affected when going beyond the stationary regime which they presuppose. 
Indeed, upon accounting for transitory hydrodynamic effects, the even more recent Ref.~\cite{Krajewski:2024gma} finds that bubbles start to run away in much larger regions of parameter space than predicted in the stationary regime, i.e. at $\alpha$ way smaller than 1, thus raising a doubt about the solidity of the conclusions that one can draw in the stationary regime~\cite{Ai:2024shx}.
In this study, we assume that the pressure is monotonic, which essentially means applying the Bodeker and Moore criterion of Eq.~\eqref{eq:alpha_LO}, while keeping in mind these interesting ongoing developments.

\begin{figure}[!ht]
\centering
\raisebox{0cm}{\makebox{\includegraphics[width=0.49\textwidth, scale=1]{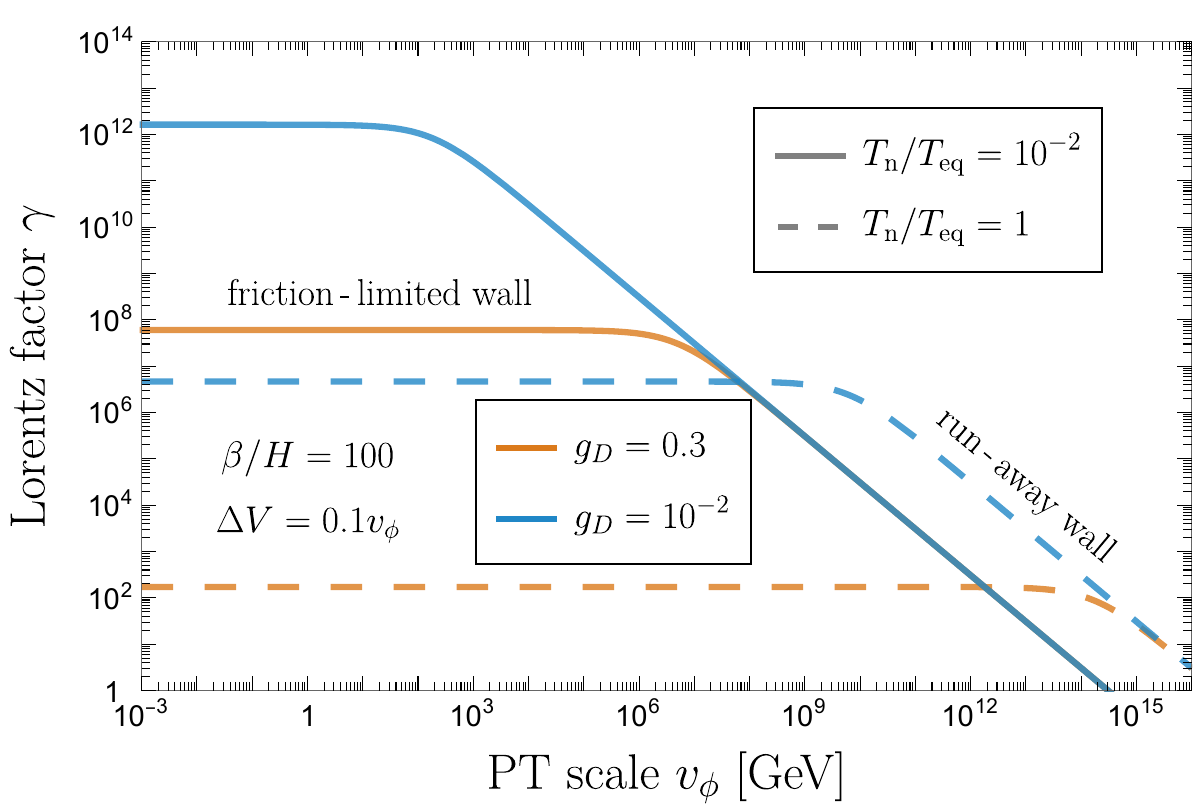}}}
{\makebox{\includegraphics[width=0.49\textwidth, scale=1]{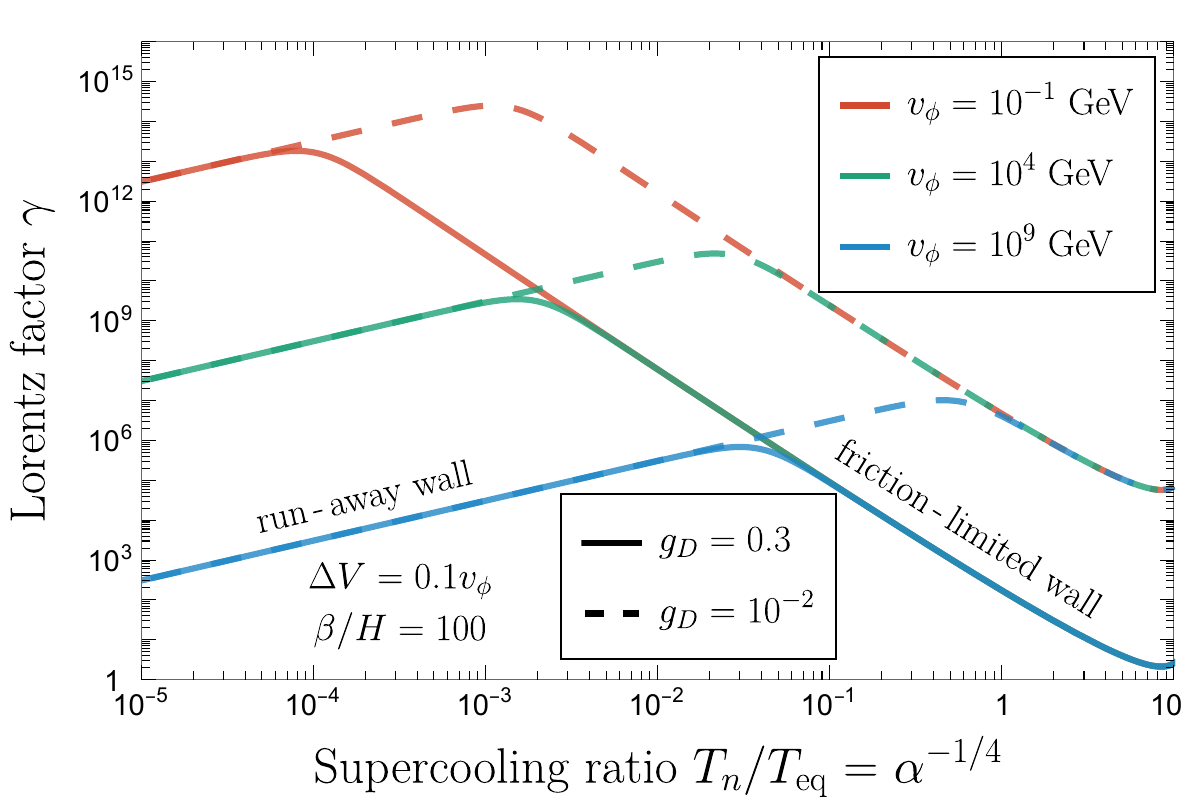}}}
\caption{\it \small Bubble wall Lorentz factor $\gamma_{\rm w}$ in Eq.~\eqref{eq:gamma_final} as a function of the scalar field vacuum expectation value $v_\phi$ or confining scale for strongly-coupled theories (\textbf{left}) and the ratio of the nucleation temperature $T_n$ over the temperature $T_{\rm eq}$ when the universe would become vacuum dominated (\textbf{right}). The supercooling ratio can be linked to the so-called latent heat fraction $\alpha = (T_{\rm eq}/\Tn)^4$, up to changes in dofs. In both panels we assume $\alpha \gtrsim \alpha_{\rm LO}$ in Eq.~\eqref{eq:alpha_LO} such that walls reach relativistic velocities. The gauge coupling constant $\gD$ dictates the magnitude of quantum corrections to the frictional pressure, while the phase transition completion rate $\beta/H$ determines the bubble size upon collision.}
\label{fig:gamma} 
\end{figure}

\begin{figure}[!ht]
\centering
\raisebox{0cm}{\makebox{\includegraphics[width=0.49\textwidth, scale=1]{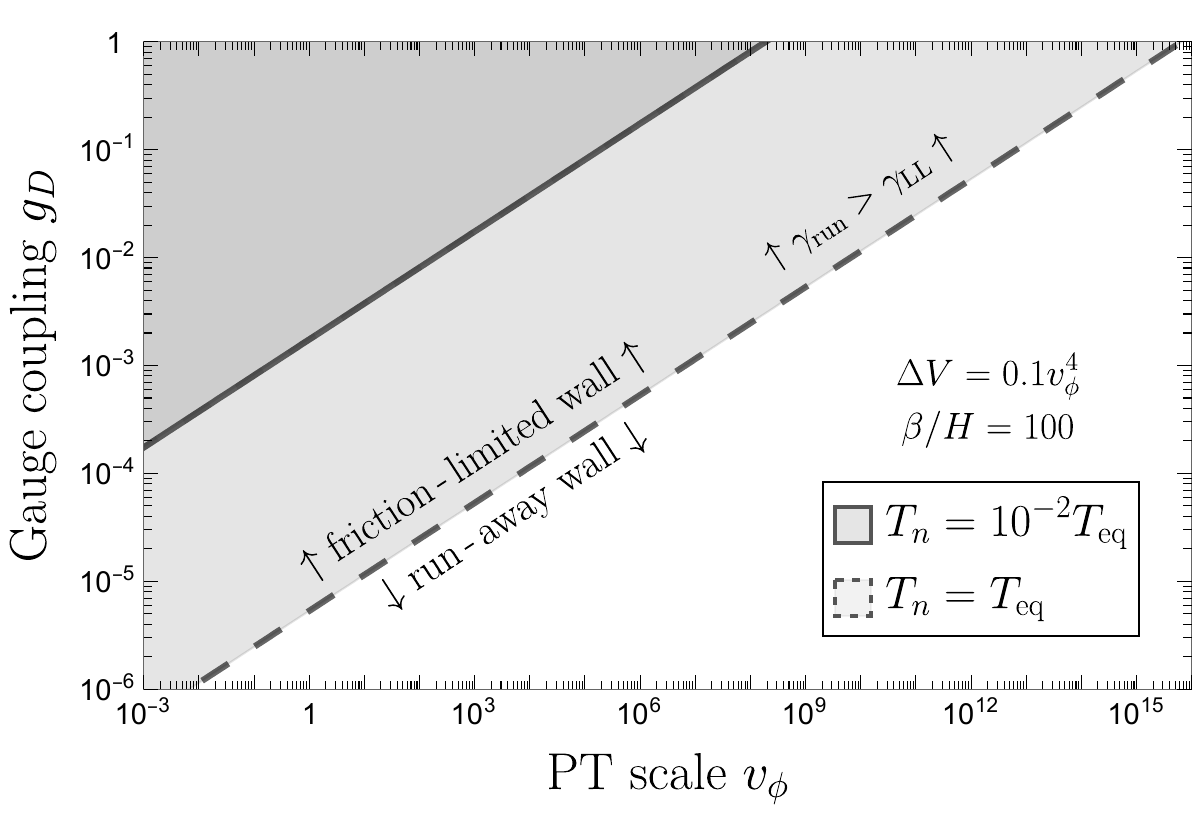}}}
{\makebox{\includegraphics[width=0.49\textwidth, scale=1]{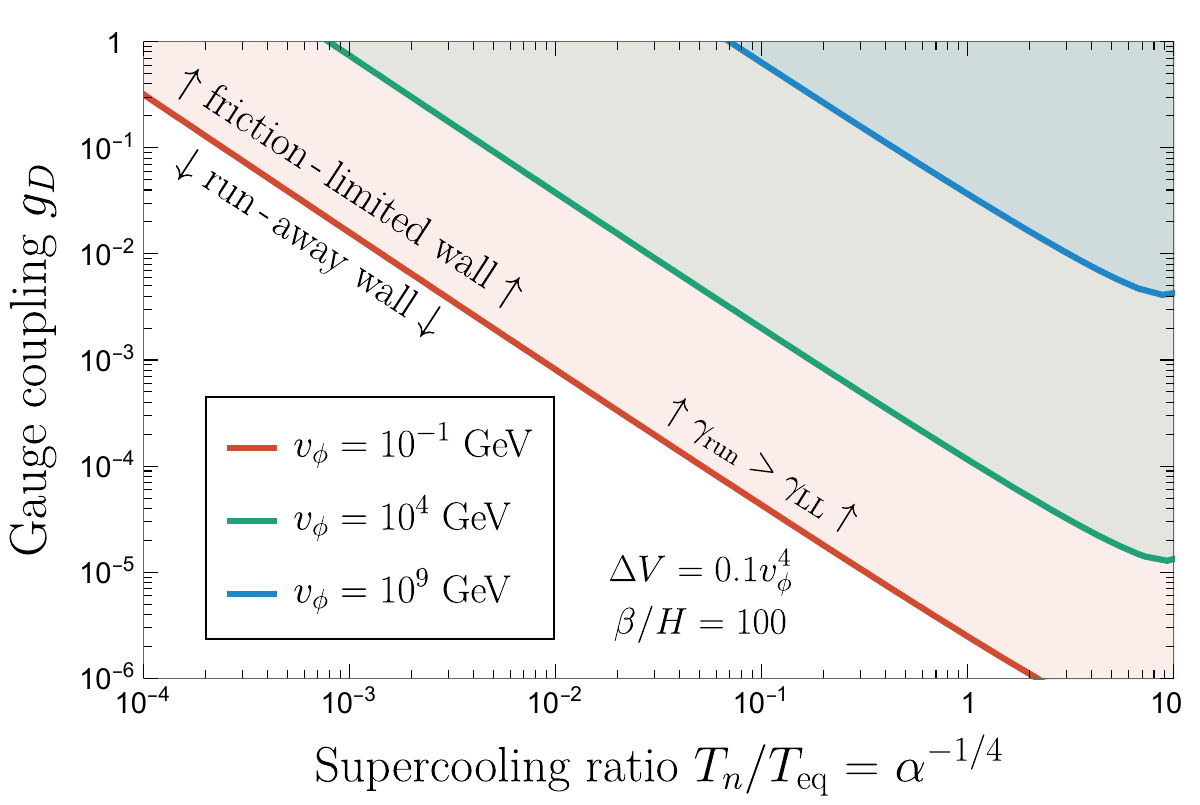}}}
\caption{\it \small In shaded regions, the friction pressure causes bubble walls to reach a constant Lorentz factor before collision, for different values of the supercooling ratio $T_n/T_{\rm eq}$ (\textbf{left}) and the PT scale $v_{\phi}$ (\textbf{right}). $\gamma_{\rm run}$ and $\gamma_\LL$ are the Lorentz factors at collision in the run-away and friction-dominated regime respectively, cf. Eqs.~\eqref{eq:gamma_run} and \eqref{eq:gamma_LL}.}
\label{fig:runaway} 
\end{figure}

Assuming the Bodeker and Moore criterion is satisfied, $\alpha\gtrsim \alpha_{\rm LO}$, bubble walls reach ultra-relativistic velocities $v_{w}\simeq 1$ and two distinct possibilities present themselves:
\begin{itemize}
\item[$\diamond$]
Either the Lorentz factor $\gamma_{\rm w}$ of the wall grows linearly with time until bubbles collide, with~\cite{Bodeker:2017cim,Gouttenoire:2021kjv}
\begin{align}
\gamma_{\rm run} 
\simeq \frac{R_{c}}{3\,R_n}
\simeq \frac{\pi^{\!\frac{1}{3}}}{3 c_w} \frac{\Tn }{\beta } 
\simeq 3 \times 10^{10} \!\!\;
\left(\frac{0.5}{c_{w}}\right) \!\left(\frac{100}{g_*}\right)^{\!\frac{1}{4}}\!\!\left(\frac{\Tn}{10^{-4}\,T_{\rm eq}}\right)
\left(\frac{10}{\beta/H_*}\right)\!\left(\frac{\rm TeV}{v_\phi}\right)\!\left(\frac{0.1v_\phi^4}{\Delta V}\right)^{\!\frac{1}{4}}\!\!,
\label{eq:gamma_run}
\end{align}
where $R_c \simeq \pi^{1/3}/\beta$, $R_{n} = c_w/\Tn$ and $c_w$ is a model-dependent order-one factor, see~\cite{Baldes:2020kam} for examples of its calculation.
\item[$\diamond$]
Alternatively, the driving pressure on the wall, $\Delta V$, can become balanced by quantum corrections to the friction pressure~\cite{Bodeker:2017cim}. Resummed at leading-log order, the corrections read~\cite{Gouttenoire:2021kjv,Azatov:2023xem}
    \begin{equation}
    \label{eq:P_LL}
    \mathcal{P}_\LL
    \simeq  \gamma_{\rm w} \alphaD \Delta m_V T_n^3 \log\left( \frac{v_\phi}{T_n}\right),
    \end{equation}
where $\Delta m_V$ is the mass difference of a vector boson across the bubble wall, and  $ \alphaD \equiv \gD^2/4\pi$ is the associated fine structure constant. The balance occurs if $\gamma_{\rm w} $ reaches
\begin{align}
\gamma_\LL &\simeq \frac{3.3 \times 10^{13}}{\log\left( {v_\phi/T_n}\right)} \times \left(\frac{g_*}{100}\right)^{\!3/4} \left(\frac{1/30}{ \alphaD } \right)^{\!3/2} \left(\frac{10^{-4}\,T_{\rm start}}{\Tn}\right)^{\!3}\left( \frac{\sqrt{2\pi \alphaD} v_\phi}{\Delta m_V } \right)\left(\frac{\Delta V}{0.1v_\phi^4}\right)^{\!1/4},
\label{eq:gamma_LL}
\end{align}
after which the Lorentz factor no longer increases. 
\end{itemize}
The Lorentz factor at collision is therefore the minimum of Eq.~\eqref{eq:gamma_run} and Eq.~\eqref{eq:gamma_LL},
\begin{equation}
	\boxed{
\gamma_{\rm coll}  \simeq \frac{\gamma_\LL \gamma_{\rm run}}{\gamma_\LL + \gamma_{\rm run}}
}
\label{eq:gamma_final}
\end{equation}
The bubble wall Lorentz factor $\gamma_{\rm w}$ is shown in Fig.~\ref{fig:gamma} as a function of the PT scale $v_{\phi}$ and the supercooling ratio $\Tn/T_{\rm eq}$. The regions in which bubble walls reach a terminal velocity are shaded in Fig.~\ref{fig:runaway}. In those regions, most of the latent heat $\Delta V$ of the PT is converted into the particle shells. Hence, they would become the main source for GW emission.

\begin{table}[htp!]
\begin{adjustwidth}{-.5in}{-.5in}  
\begin{center}
\centering
\begin{tabular}{| c | c | c | c |c| }
 \hline
 \multicolumn{2}{|c|}{Channel} & 
 \begin{tabular}[c]{@{}c@{}} 
 $\textrm{Multiplicity}$  $\mathcal{N}$ per \\
 $\textrm{incoming particle}$ 
\end{tabular}
  &   \begin{tabular}[c]{@{}c@{}}
  \\[-0.9em]
 $\textrm{Momentum of}$\\
  $\textrm{shell particles}$\\
 ($p_c$ or $\pX$)\\[0.3em]
\end{tabular} &  \begin{tabular}[c]{@{}c@{}} 
$\bar{L}_b = (L_b^2 -\frac{1}{p_X^2})^{\!\frac{1}{2}} $\\
($L_b =$ \textrm{effective}\\
\textrm{shell thickness)}
\end{tabular} \\
 \hline\hline 
 \multicolumn{2}{|c|}{\begin{tabular}[c]{@{}c@{}} \\[-0.8em]
 $\textrm{Leading-order interaction (LO):}$\\
 $a\to a$ \\
 $\textrm{Particles acquiring a mass \cite{Bodeker:2009qy,Jinno:2022fom}}$\\[0.2em]
 \end{tabular} }
  & $ 1 $ &  $ \Delta m^2/T_n$ & $  \dfrac{R_c}{2 (\Delta m/T_n)^2} $\\[1em]
 \hline\hline \rule{0pt}{20pt}
 \multirow{4}{*}{
 \begin{tabular}[c]{@{}c@{}} 
 $\textrm{Gauge interaction}$ $\alphaD \ll 4\pi:$\\
 $\textrm{Bremsstrahlung radiation}$ \\
 $a\to bc$\\
\cite{Bodeker:2017cim,Azatov:2020ufh,Gouttenoire:2021kjv,Azatov:2023xem} $\textrm{ and App.~\ref{app:weak}}$ \\
\end{tabular} }& \multirow{1}{*}{transmitted}
  & \multirow{1}{*}{$2\dfrac{\alphaD}{\pi} L_m L_E$} & \multirow{4}{*}{$ \gamma_{\rm w} \,m_{c,h}$ } & \multirow{4}{*}{
  $\dfrac{R_c}{2\gamma_w^2}$}  \\[1em]  \cline{2-2}
   & \multirow{3}{*}{reflected }&  \multirow{2}{*}{$\dfrac{\alphaD}{\pi} L_m^2$} & & \\[1em]
   &&&&\\
 \hline\hline \rule{0pt}{20pt}
 \multirow{2}{*}{\begin{tabular}[c]{@{}c@{}} 
 $\textrm{Gauge interaction $\alphaD \simeq 4\pi$:}$\\
 $\textrm{Hadronization}$ \\
\cite{Baldes:2020kam}
\end{tabular} }& \begin{tabular}[c]{@{}c@{}} string\\ fragmentation
\end{tabular}
  & \multirow{3}{*}{$\dfrac{\alphaD}{\pi} L_E$} & \multirow{3}{*}{$ \gamma_{\rm w} \,v_\phi$ } & \multirow{3}{*}{ $\dfrac{R_c}{2\gamma_w^2}$}  \\ \cline{2-2} 
   & \begin{tabular}[c]{@{}c@{}}  ejected \\
   quarks \end{tabular} & & & \\[1em]
   \hline\hline \rule{0pt}{20pt} 
   \multirow{4}{*}{\begin{tabular}[c]{@{}c@{}} 
 $\textrm{Scalar interaction}$ $\lambda \phi^4/4!:$\\
  $\textrm{Scalar Bremsstrahlung}$ \\
    $a\to bc$\\
 $\textrm{App.~\ref{app:scalar}}$ \\
\end{tabular} }& \multirow{1}{*}{transmitted}
  & \multirow{1}{*}{$\lambda^2v_\phi^2/192\pi^2m_{c,h}^2$} &\multirow{1}{*}{$ \gamma_{\rm w} \,m_{c,h}^2/E_a $}  & \multirow{4}{*}{ $\dfrac{R_c}{2\gamma_w^2}$}  \\[1em]  \cline{2-2}
   &  \multirow{3}{*}{reflected } &  \multirow{3}{*}{$\lambda^2v_\phi^2/32\pi^2E_a^2$} & \multirow{3}{*}{$\gamma_{\rm w} m_{c,h}$}  & \\[1em]
   &&&&\\
  \hline\hline 
  \multicolumn{2}{|c|}{\begin{tabular}[c]{@{}c@{}} 
  \\[-0.8em]
 $\textrm{Heavier particle production}$ $\lambda \phi^2X^2/4$ \\
 (Azatov-Vanvlasselaer mechanism $\phi\to XX$) \\
 $M_X\gg v_\phi$ \cite{Azatov:2020ufh}\\[0.2em]
\end{tabular} }
&    \begin{tabular}[c]{@{}c@{}}
\\[-0.8em]
 $\lambda^2 v_\phi^2/192\pi^2M_X^2\times $\\
  $ \Theta\left(\gamma_{\rm w} -M_X^2/\Tn v_\phi \right)$\\[0.5em]
\end{tabular}
&  \begin{tabular}[c]{@{}c@{}} 
$M_X^2/\Tn$
\end{tabular} 
&   \begin{tabular}[c]{@{}c@{}}
$\dfrac{R_c}{2 (M_X/T_n)^2}$
\end{tabular}
 \\[1em]
 \hline  
\end{tabular}
\end{center}
\end{adjustwidth}
\caption{\it \small Estimates of quantities $\mathcal{N}$, $\pX$ and $\bar{L}_b$ that characterize the particle shells surrounding relativistic bubble walls, generated by a variety of selected channels among those described in Sec.~\ref{sec:shells_general}.
All quantities are expressed in the bath frame and are valid for free-streaming shells.  These quantities are necessary ingredients to determine whether shells free stream or not, e.g.~the number density of shells is determined as a function of $\mathcal{N}$, $\pX$ and $L_b = \sqrt{\bar{L}_b^2 + 1/p_{X}^{2}}$ via Eq.~(\ref{eq:shelldensity1}).
 The bubble radius at collision is $R_c$, and we define $L_m\equiv \log\left(m_{c,h}/m_{c,s}\right)$, and $L_E\equiv\textrm{log} \left(E_a/m_{c,h}\right)$, where $E_a \simeq 3 \gamma_{\rm w} \Tn$ is the energy of the incoming bath particle in the wall frame, and $m_{c,s}$, $m_{c,h}$, are the masses of the radiated particle in the symmetric and Higgs phases respectively. The order parameter of the PT is $v_\phi$, either the vacuum expectation value of the scalar field driving the PT or the confining scale.
The effective shell thickness $\bar{L}_b$ and the momentum of the shell particles, denoted $p_c$ or $\pX$ depending on the shell under investigation, are given in the frame of the thermal bath, i.e. the one of the CMB.
For derivation of these quantities see our discussion in Sec~\ref{sec:shells_general} plus the references/appendices reported in the Table for each channel.
}
\label{tab:production_mechanism} 
\end{table}

\begin{figure}[!ht]
\centering
\raisebox{0cm}{\makebox{\includegraphics[width=1\textwidth, scale=1]{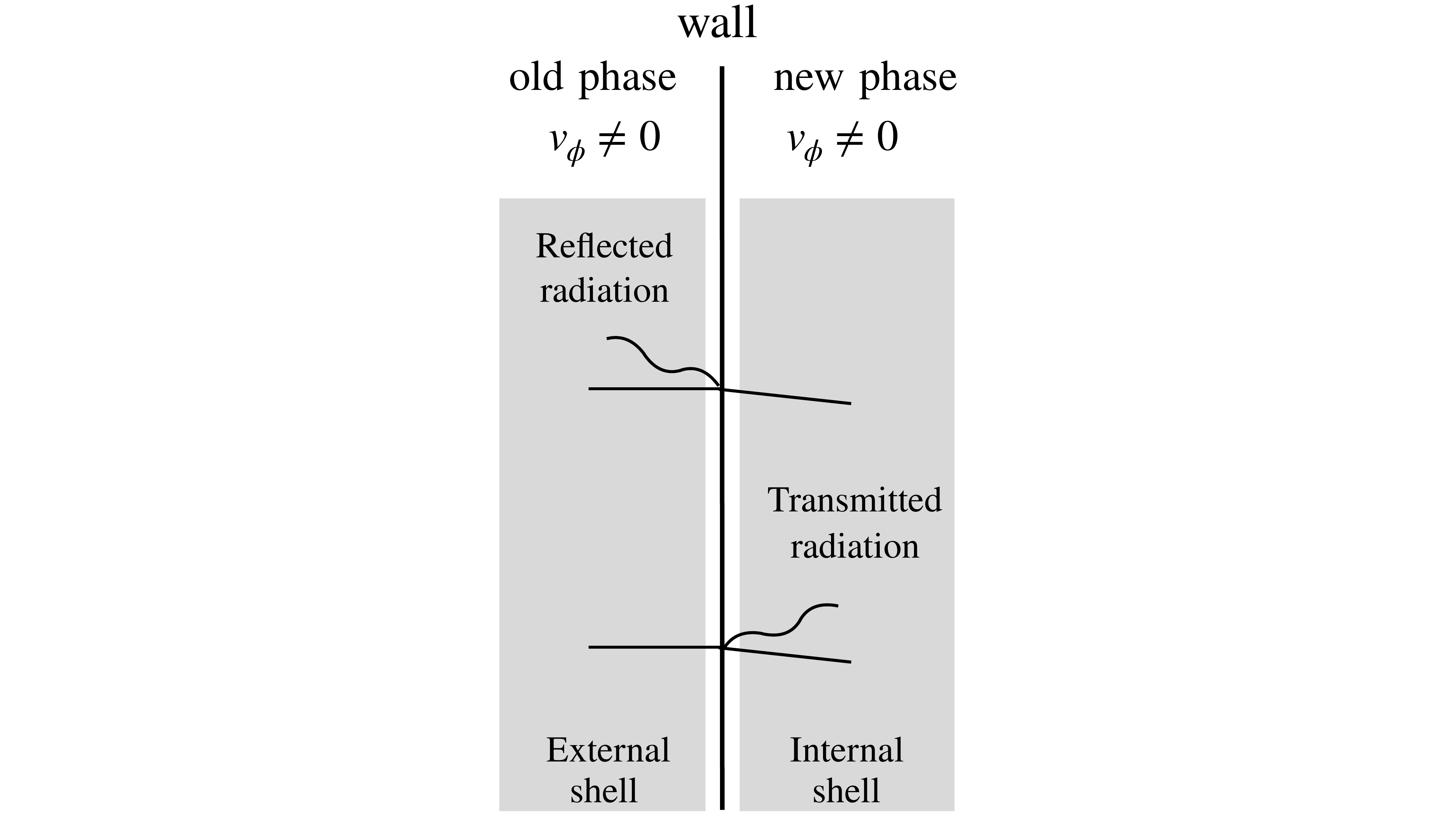}}}
\caption{\it \small \textbf{Shell productions:} Particles from the bath interact with bubble walls and radiate other particles which accumulate in shells, preceding and following the walls. See Table~\ref{tab:production_mechanism} for the different shell production mechanisms at bubble walls.}
\label{fig:shelL_production} 
\end{figure}

\subsection{Shell production mechanisms and their properties}
\label{sec:shells_general}

Diverse mechanisms can produce shells of primary particles $\X$, or $c$, propagating either in front or behind bubble walls
(to avoid clutter, in what follows we will dub kinematic quantities related to shell particles with either only $X$ or only $c$).
In Sec.~\ref{sec:shells_origin} we discuss the origin of the shells and report from the literature the typical values of the momenta and of the number of particles that constitute them.
We then derive their number density in Sec.~\ref{sec:shell_densities} and their thickness in Sec.~\ref{sec:shell_thickness}.
These will constitute necessary ingredients to determine the rates of the various interactions that govern the evolution of shells. These of course further depend on the interactions under consideration, that we will discuss in the rest of this paper.

\noindent
\subsubsection{Shells: origin, momenta, number of particles}
\label{sec:shells_origin}
We describe the shell-production mechanisms, or channels, that are known to us below. For a subset of them we provide, in Table~\ref{tab:production_mechanism}, quantitative estimates for properties of the associated shells that will be useful in the rest of the discussion.

\begin{itemize}
    \item[$\diamond$]
    \textbf{LO shell}.
   Particles that acquire a mass when passing through the bubble wall also lose part of their momentum in the wall frame, because of energy conservation. This implies that, in the bath frame, they accumulate in a shell following the wall. The mass gain happens via tree-level interactions, like a scalar potential for spin-0 bosons, Yukawa interactions for fermions, and the covariant derivative for spin-1 bosons.
    We dub these shells `leading-order' (LO) in relation to the associated pressure on the wall~\cite{Bodeker:2009qy}.

    \item[$\diamond$]
    \textbf{Transmitted and reflected shells}.
    Particles that obtain a mass when passing through the wall, $c$, can also be emitted as quantum radiation by other particles that couple to them, $a$, via a process $a \to b c$ (more than two final state particles are possible depending on the interaction).
    The breaking of Lorentz boosts by the wall background makes these emissions possible even if they would be kinematically forbidden in vacuum, so that e.g. they happen also for $a = b$.
    Quantum radiation results in a shell of particles $c$ (and/or $b$) following the wall and in one preceding the wall: the first is generated by the particles radiated with enough energy to penetrate the wall, the second by those that do not have enough energy,  $E_{c}\lesssim  m_{c,h}$, and so are reflected.
    If the PT is associated to the breaking of a gauge group then the radiation of gauge bosons $c$, by particles $a$ charged under that group, is enhanced for small energies of the gauge bosons. The multiplicity and momentum of particles in these shells depend on the particular interaction under study.
    For concreteness we report in Table~\ref{tab:production_mechanism} those generated at a PT associated with 
    \begin{enumerate}
        \item
        the breaking of an Abelian gauge group, with Lagrangian
    \begin{equation}
          \mathcal{L} = (D_{\mu} \phi)^{\dagger} D^{\mu} \phi + \bar{\Psi} i \slashed{D} \Psi,
    \end{equation}
    where $\phi$ is a scalar, $\Psi$ a fermion, and  $D_{\mu} = \partial_{\mu} - i \gD Q_{D} A_{\mu}$ is the covariant derivative with $\gD$ and $Q_{D}$ the respective gauge coupling and charge.
        \item
        the breaking of a non-Abelian gauge group, with Lagrangian
         \begin{equation}
          \mathcal{L} = (D_{\mu} \phi)^{\dagger} D^{\mu} \phi + \bar{\Psi} i \slashed{D} \Psi - \frac{1}{4} F_{\mu \nu}^a F^{a \, \mu \nu},
        \end{equation}
        where $F^{a}_{\mu \nu} = \partial_{\mu}A_{\nu}^{a} - \partial_{\nu}A_{\mu}^{a} +\gD f^{abc}A_{\mu}^{b}A_{\nu}^{c}$ with structure constants $f^{abc}$ now emphasises the possibility of radiating gauge bosons off gauge bosons upon wall crossing. Here the covariant derivative is given by the usual expression $D_{\mu} = \partial_{\mu} - i \gD A^{a}_{\mu} t^{a} $ where $t^{a}$ are the generators of the representation.
        \item
        the breaking of a symmetry that gives mass to a scalar $\phi$, with self-coupling
        \begin{equation}
        \mathcal{L} = \frac{\lambda}{4!} \phi^4.
        \end{equation}
    \end{enumerate} 
    For both the Abelian and non-Abelian cases, in the table we use $\alphaD = \gD^2/4\pi$, and normalize to $Q_{D}=1$, but in general we could have species of different charges in the cosmological thermal bath.
    The gauge-radiation shells are associated with the radiation pressure first evaluated in~\cite{Bodeker:2017cim} and first resummed in~\cite{Gouttenoire:2021kjv} (except for longitudinal gauge bosons, first included in~\cite{Azatov:2023xem}), reviewed in App.~\ref{app:weak}.
    Neither the $\phi^4$ interaction,  studied in App.~\ref{app:scalar}, nor a Yukawa-like coupling between fermions and scalars (not shown here), give rise to an IR-enhanced radiation. Therefore the induced pressure is subleading with respect to that from radiated gauge bosons and, importantly for our paper, the associated radiated reflected shells are way less dense than those from the radiation of gauge bosons.

    \item[$\diamond$]
    \textbf{Shells from a confining PT}.
    Particle production associated to confining PTs has first been modeled in~\cite{Baldes:2020kam}. The key aspect is that, if the bubble walls are fast enough, bath particles charged under the confining gauge group attach to the wall via a fluxtube when they are swiped by it. The fluxtube then breaks forming hadrons inside the bubble, that constitute a shell following the wall, and ejecting charged particles in the deconfined phase to conserve charge, that constitute a shell preceding the wall. We refer the reader to~\cite{Baldes:2020kam,Baldes:2021aph} for more information and for the quantitative estimates about these shells, and to~\cite{Bachmaier:2023wzz} for a later study supporting the picture proposed in~\cite{Baldes:2020kam}.

    \item[$\diamond$]
    \textbf{Shells of heavy particles}.
    Particles $X$ much heavier than the scale of the PT, $M_X \gg v_\phi$, can be produced from other particles when they enter the wall, if their interaction with $X$ `feels' the PT (i.e. if it contains the field $\phi$). The heavy $X$ particles then constitute a shell following bubble walls.
    According to the Heisenberg uncertainty principle, the maximal momentum exchange $\Delta p\simeq M_X^2/\gamma_{\rm w} \Tn$ in the wall frame is given by the inverse of the wall thickness $L_{\rm wall}\simeq v_{\phi}^{-1}$, which leads to the Heaviside function shown in Table~\ref{tab:production_mechanism}.
  Their emission and the associated pressure has first been computed in~\cite{Azatov:2020ufh}, for the computation of some of their other properties and their implications for dark matter or baryogenesis see e.g.~\cite{Azatov:2021ifm, Baldes:2021vyz,Azatov:2022tii,Baldes:2022oev}. For concreteness, in Table~\ref{tab:production_mechanism} we report estimates for shell quantities associated to the interaction
  \begin{equation}
    \mathcal{L} = \frac{\lambda}{4} \phi^2 X^2\,.
    \end{equation}

      \item[$\diamond$]
      \textbf{Shells from vector bosons acquiring a small part of their mass}. These shells arise in the particular case where the mass of a gauge boson is already non-zero in the unbroken phase, and it gets an extra subleading contribution upon the PT. The pressure on walls in this specific scenario has first been evaluated in~\cite{GarciaGarcia:2022yqb}, together with some of the quantities that characterize the associated shell.

     \item[$\diamond$]
     \textbf{Shells from decay of the wall}. The background field $\phi$ constituting the wall undergoes several oscillations inside the bubble, before relaxing to $v_\phi$ deep inside it. These oscillations experience friction both by Hubble and by emissions of the particles coupled to the background field. These particles are then produced by $\phi$ oscillations and constitute a shell following the wall.

     \item[$\diamond$]
     \textbf{Unruh and Casimir shells}. The motion of bubble walls acts as a time-dependent boundary condition for quantum fields. Hence, energy is not conserved and particles should be produced, analogously to what happens in quantum electrodynamics~\cite{Davies:1976hi,Davies:1977yv,Ford:1982ct}. The production of those particles can be viewed as equivalent to Unruh radiation \cite{Unruh:1980cg} or dynamical Casimir effects \cite{1970JMP....11.2679M} according to whether the bubble wall is accelerating or expanding at constant velocity.
\end{itemize}
To our knowledge, the existence of the last two kinds of shells mentioned in this list have never been pointed out in the context of cosmological PTs and we leave them for future work.

\subsubsection{Shell's boost factor}
\label{sec:shell_boost_factor}
We will be interested in shells with significant bulk motion distinguishable from the thermal bath. That is, shells in which the particles move through the bath in a direction outward from the bubble nucleation site, and with significant Lorentz factors associated with the averaged velocity vector.

Consider leading-order (LO) shells, coming from the transition of bath particles from the symmetric to the Higgs phase, $a \to a$, which gives a corresponding change in mass $m_{a,s} \to m_{a,h}$. Take the wall moving in the $z$-direction. Then the change in momentum in the $z$-direction is given
$   \pX
    \simeq \Delta m_{a}^{2}/\Tn = (m_{a,h}^{2} - m_{a,s}^{2})/\Tn .
$  
The averaged momentum of the shell particles is therefore relativistic in the bath frame provided 
\begin{equation}
\pX \, \gtrsim \,m_{a,h}   \,.
\label{eq:pXgreatma}
\end{equation}
Assuming this is the case, this will also be the dominant momentum component of our shell particles. 
If $\pX \lesssim m_{a,h} $ the particle's momentum will still be dominated by the initial value from the thermal bath, $|\vec{p} \,| \sim T_n$, rather than the ``kick" received by the passing wall. Taking the simple example with $m_{a,h} \gg m_{a,s}$, and a gauge boson shell with $m_{a,h} = \gD v_{\phi}$, Eq.~(\ref{eq:pXgreatma}) translates into
    \begin{equation}
      \boxed{ \gD \gtrsim \frac{\Tn}{v_{\phi}} \qquad \text{(LO shell)}. }
      \label{eq:LOshell}
    \end{equation}
In all figures for the LO cases we will shade the regions where Eq.~(\ref{eq:LOshell}) is not satisfied.

Going on to other types of shells, we read their momentum, $\px$, off Tab.~\ref{tab:production_mechanism} and see that for perturbative and non-perturbative gauge interactions we automatically have relativistic shells, provided the wall is relativistic $\gamma_{\rm w} \gg 1$. This also holds for the scalar bremsstrahlung in case of reflected scalars. For transmitted scalars we again require the final state mass to be large compared to the temperature $m_{c,h}/\Tn \gtrsim 1$ for the shell to be relativistic (similarly to the LO shell). Taking the mass to be $m_{c,h} \simeq \sqrt{\lambda}v_{\phi}$, one finds the condition
    \begin{equation}
      \boxed{ \lambda_{\phi} \gtrsim \left(\frac{\Tn}{v_{\phi}} \right)^{2} \qquad \text{(Transmitted scalar bremsstrahlung shell)}. }
    \end{equation}
Finally for the Azatov-Vanvlasselaer mechanism we also need $M_X/\Tn \gtrsim 1$. But this is automatically satisfied even in the weakly supercooled limit in this scenario, $v_{\phi} \gtrsim \Tn$, as the scenario operates in the limit $M_X \gg v_{\phi}$.

\subsubsection{Shell particles' number density}
\label{sec:shell_densities}

The number density of the shell particles $X$ in the wall frame, produced at a given time $t_X$ when the bubble radius is $r_X$, is given by
\begin{equation}
n_{X,{\rm w}}
= \mathcal{N}^i_X n_{i,{\rm w}}(t_X)
=\gamma_{\rm w}(t_X)\,n_{i,{\rm b}}\,,
\label{eq:nshell_wall}
\end{equation}
where $\mathcal{N}^i_X$ is the number of shell particles $X$ produced per each particle of population $i$ upon being swept by the bubble wall (see Table~\ref{tab:production_mechanism}), and where $n_{i,{\rm b}} = g_i \zeta(3) T_n^3/\pi^2$ is their number density in the bath frame. 
The number density of shell particles in the bath frame, $n_{X,{\rm b}}$, then depends on the time elapsed since their production at $t_X$, or equivalently on how much the bubble has expanded after their production.
Assuming that shells free stream until the bubble radius reaches the value $R > r_X$, their number density reads
\begin{equation}
n_{X,{\rm b}}
=2 \gamma_{\rm w}(R) \Big(\frac{r_X}{R}\Big)^2 n_{X,{\rm w}}
= 2 \gamma_{\rm w}(R) \gamma_{\rm w}(r_X)\, \Big(\frac{r_X}{R}\Big)^2 \mathcal{N}^i_X n_{i,{\rm b}}\,,
\label{eq:nX_bath}
\end{equation}
where the factor $2 \gamma_{\rm w}(R)$ arises from boosting the shell's density current from the wall to the bath frame, and the factor $(r_X/R)^2$ accounts for the spatial dilution of the number density due to the expansion of the bubble from $r_X$ to $R$.\footnote{
Our derivation of the shells number density and effective thickness generalises the one for ejected quanta at confining PTs of~\cite{Baldes:2020kam} in two directions: i) it is valid for any shell, the shell dependence being encoded in $\mathcal{N}^i_X$; ii) it is valid irrespective of whether walls run away or reach a terminal velocity, while~\cite{Baldes:2020kam} assumed runaway walls.
}

As we will prove in Sec.~\ref{sec:shell_thickness}, an effective shell's thickness $L_b$ exists such that, for any shell, $n_{X,{\rm b}}$ is approximately constant at distances from the wall shorter than $L_b$, and goes to zero at larger distances.
For simplicity, in this paper we will then use a constant value for the shells number density, obtained by dividing the total number of particles in that shell, $n_{i} \mathcal{N}^i_X 4 \pi R_c^3/3$, by the effective volume of the shell, $4 \pi R_c^2 L_b$, where $R_c$ is the bubble radius at collision:
\begin{equation}
	\boxed{
n_{X,\rm b}~= ~ \sum n_{i} \mathcal{N}^i_X \frac{R_c}{3L_b},
}
\label{eq:shelldensity1}
\end{equation}
 where all quantities are evaluated in the bath frame. We find it useful to also report the total number of particles $X$ produced per bubble
\begin{equation}
\label{eq:N_X_multiplicity}
N_X ~=~\sum n_{i} \mathcal{N}^i_X \frac{ 4\pi R_c^3}{3}.
\end{equation}

\subsubsection{Shell's thickness}
\label{sec:shell_thickness}

We start by noticing that the distance of shells' particles from the bubble wall cannot be localized to shorter values than their de Broglie wavelength $\sim 1/\pX$. This will constitute a lower bound to the effective thickness of any shell, that we write as
\begin{equation}
\boxed{
L_b^X
= \sqrt{(\bar{L}_b^X)^2 + \frac{1}{\pX^2}}\,,
}
\label{eq:LpX}
\end{equation}
where $\bar{L}_b^X$ is the contribution to the shell's thickness that depends on the shell's production dynamics.
We will find that, throughout the parameter space considered in our study, the de Broglie contribution will dominate over $\bar{L}_b^X$ at very small $\gD$ and very large $v_\phi$.
We now turn to the computation of $\bar{L}_b^X$ for the various shells.

Let us define $x$ as the radial distance of a given layer of the shell from the bubble wall, which is limited between $x=0$ (i.e. the shell's layer produced last before collision), and the position of the layer emitted first, which corresponds to $x=L_b^\text{max}$. $x$ can be positive or negative depending on whether the shell of interest precedes or follows the walls.
All the $x$ dependence in the number density Eq.~(\ref{eq:nX_bath}) is encoded in $r_X(x)$.
While $L_b^\text{max}$ defines the maximal extension of a shell, we anticipate that there always exists a value of $x$, which we define as $\bar{L}_b$, below which $n_{X,{\rm b}}$ goes to a constant value and above which it gets very suppressed. We now derive $r_X(x)$ and $\bar{L}_b$.

The distance $x$ between a shell particle ``$X$'' and the wall is found by integrating the difference between the world lines of ``$X$'' and the bubble wall ``$w$'' from the moment of particle production $t_X$ until the time of bubble collision $t_c$,
\begin{equation}
x
= \int_{t_X}^{t_c}\!dt \left[ v_{\rm X}(t_X) - v_{\rm w}(t)\right]
\simeq  \int_{t_X}^{t_c}\!dt \left( \frac{1}{2\gamma_{\rm w}^2(t)} - \frac{1}{2\gamma_X^2(t_X)}\right)
=  - \frac{t_c-t_X}{2\gamma_X^2(t_X)} + \int_{t_X}^{t_c} \!dt \frac{1}{2\gamma_{\rm w}^2(t)}\,.
\label{eq:x}
\end{equation}
where we have used  $v\simeq 1-1/(2\gamma^2)$ in the second equality, i.e. we have assumed $\gamma_{X,{\rm w}} \gg 1$. 

\medskip

Let us start by considering shells that follow the wall, like LO shells, shells of heavy particles, and radiated transmitted shells. For them one has $\gamma_X \ll \gamma_{\rm w}$, so that the second term in Eq.~(\ref{eq:x}) for $x$ is negligible and
\begin{equation}
    x \simeq \frac{r_X - R_c}{2 \gamma_X^2(r_X)} \qquad \text{(shells~behind~wall)}
    \label{eq:x_shell_behind_wall}
\end{equation}
irrespective of whether walls runaway until collision or not.
By inverting Eq.~(\ref{eq:x_shell_behind_wall}) one obtains $r_X(x)$ and, via Eq.~(\ref{eq:nX_bath}), the dependence of $n_{X,\rm{b}}$ in $x$. If $\gamma_X$ is constant in $r_X$, as it is for LO shells ($\gamma_X \simeq \Delta m^2/(T_n m)$, $\simeq \Delta m/T_n$ if $m \simeq \Delta m$) heavy particle shells ($\gamma_X \simeq M_X/T_n$) and radiated transmitted ones for terminal velocity walls ($\gamma_X \lesssim \gamma_{\rm w}^\LL$), one finds
\begin{align}
n_{X,\rm{b}}(x)
&=2\gamma_{w}(R_c) \gamma_{\rm w}\big(r_X(x)\big)\, \Big(\frac{r_X(x)}{R_c}\Big)^2\mathcal{N}_X^i n_{i,b} \\
&= 2 \mathcal{N}_X^i n_{i,b}\Big(1 + \frac{x}{\bar{L}_b^X}\Big)^2 \times
\begin{cases}
(\gamma_{\rm w}^\LL)^2,&\quad \text{terminal velocity},\\
(\gamma_{\rm w}^{\rm run}(R_c))^2 \big(1+\dfrac{x}{\bar{L}_b^X}\big) , & \quad \text{runaway}\,,
\end{cases}
\end{align}
where we used $\gamma^{\rm run}_{w}(r_X)=\gamma^{\rm run}_{w}(R_c)(r_X/R_c)$ given by Eq.~\eqref{eq:gamma_run} and we have defined
\begin{equation}
\boxed{
\bar{L}_b^X = \frac{R_c}{2 \gamma_X^2}\,, \qquad X = \text{shells~behind~wall}\,.
}
\label{eq:LpX_behindwall}
\end{equation}
$\bar{L}_b^X$ contributes to what we defined as the effective thickness of shell $X$ because for values of $x >- \bar{L}_b^X$ the shells number density is approximately constant, while it goes to zero as $x$ approaches $-\bar{L}_b^X$ (we remind that $x < 0$ for shells following the wall).

For shells made of radiated transmitted particles, this time in the case that bubbles collide while walls run away, we assume for simplicity that all radiated transmitted particles are produced in the regime of runaway walls, so that $\gamma_X(r_X) \propto r_X$.\footnote{\label{foot:produced_in_runaway}
Most of the shell is produced in the last stages of bubble expansion, which are the stages where walls swipe through most of the volume available to a bubble before collision. Therefore this approximation is accurate up to an $O(1)$ factor in the regions of parameter space where $\gamma_{\rm w}^\LL(R_c) \simeq \gamma_{\rm w}^{\rm run}(R_c)$, and to a much better precision as soon as $\gamma_{\rm w}^\LL(R_c) > \gamma_{\rm w}^{\rm run}(R_c)$.
}
Eq.~(\ref{eq:x_shell_behind_wall}) then becomes $x = (R_c^2/r_X^2) (r_X/R_c - 1) R_c/(2 \gamma_{\rm w}^{\rm run}(R_c)^2)$, its solution $r_X(x)$ satisfies $r_X(x) \to R_c$ for $x \ll R_c/(2 \gamma_{\rm w}^{\rm run}(R_c)^2)$ and $r_X(x) \to R_c\sqrt{R_c/(2 \gamma_{\rm w}^{\rm run}(R_c)^2\,x)}$ for $x \ll R_c/(2 \gamma_{\rm w}^{\rm run}(R_c)^2$, so that
\begin{equation}
\label{eq:n_trans}
n_{\text{transm.},\rm{b}}^{\rm run}(x)
= 2 \gamma_{\rm w}^2(R_c) \mathcal{N}^i_\text{transm.} n_{i{\rm b}} \times\begin{cases}
1, &\qquad x \ll \bar{L}_b^X\\
\Big(\dfrac{\bar{L}_b^X}{x}\Big)^{\frac{3}{2}}, &\qquad x \gg \bar{L}_b^X\\
\end{cases}
\end{equation}
where $\bar{L}_b^X$ has again the expression of Eq.~(\ref{eq:LpX_behindwall}), this time of course using $\gamma_X = \gamma_{\rm w}^{\rm run}(R_c)$.

\medskip

Let us now consider shells that precede the wall, like those formed by radiated reflected particles, or by ejected techniquanta in a confining PT. In this case $\gamma_X  = \pX/m_X \sim \gamma_{\rm w} v_\phi/\mu \gg \gamma_{\rm w}$, where we have anticipated from Sec.~(\ref{sec:IR_cutoff}) that in the unbroken phase $m_X \sim \mu$ where $\mu$ is the IR cutoff. Eq.~(\ref{eq:x}) for $x$ is then dominated by the second term and it becomes
\begin{equation}
 x \simeq \frac{(R_c - r_X)}{2 \gamma_{\rm w, coll}^2} \times
 \begin{cases}
 1 &\quad \text{terminal~vel.}\\
 \frac{\gamma_{\rm w, coll}}{\gamma_{\rm w}^{\rm run}(r_X)} &\quad \text{runaway}
 \end{cases}
 \qquad \text{(shells~ahead~of~wall)},
\label{eq:x_shell_aheadof_wall}
\end{equation}
where $\gamma_{\rm w, coll}$ is given in Eq.~(\ref{eq:gamma_final}). In the runaway case, we have again assumed for simplicity that all shell's particles are produced in the regime of runaway walls, so that  $\gamma_{\rm w}(r_X) \propto r_X$ (see again footnote~\ref{foot:produced_in_runaway}).
Using Eq.~\eqref{eq:nX_bath}, we then find
\begin{equation}
n_{X,\rm{b}}(x)
= 2 \gamma_{\rm w}^2(R_c) \mathcal{N}^i_X n_{i{\rm b}} \times\begin{cases}
\Big(1-\frac{x}{\bar{L}_b^X}\Big)^2, &\qquad \text{terminal~vel.}\\
\Big(1+\frac{x}{\bar{L}_b^X}\Big)^{-3}, &\qquad \text{runaway}\,,
\end{cases}
\end{equation}
where we have defined, analogously to above,
\begin{equation}
\boxed{
\bar{L}_b^X = \frac{R_c}{2 \gamma_{\rm w,coll}^2}\,, \qquad X=\text{shells~ahead~of~wall}\,.
}
\label{eq:LpX_aheadofwall}
\end{equation}

Finally, we find it useful to report the expressions for the contributions to the effective thickness of shells, $\bar{L}_b^X$ of Eqs.~(\ref{eq:LpX_behindwall}) and (\ref{eq:LpX_aheadofwall}), in the wall frame. They read
\begin{equation}
\bar{L}_{\rm w}^X = R_c \times\begin{cases}
\frac{1}{2 \gamma_{\rm w, coll}},& \quad X = \text{shells~ahead~of~wall},\\
\frac{\gamma_{\rm w, coll}}{\gamma_X^2 + \gamma_{\rm w, coll}^2},& \quad X = \text{shells~behind~wall},
\end{cases}
\end{equation}
where the non-trivial factor, in the case of shells behind the wall, comes from the transformation law of velocities $v_X\big|_{\rm wall}=(v_X-v_{\rm w})/(1-v_w v_X)\simeq\gamma_{\rm w}^2/(\gamma_X^2+\gamma_{\rm w}^2)$.  If  $\gamma_X \lesssim \gamma_{\rm w}$, then we have $\bar{L}_w \simeq R_c/(2\gamma_{\rm w, coll})$.
The expressions for the shells thickness is reported in Table~\ref{tab:production_mechanism}, for convenience for specific shells.

\subsection{IR cut-off}
\label{sec:IR_cutoff}

We anticipate that most squared amplitudes of the processes that we will study have IR  singularities in vacuum QFT, once integrated over the final phase space. This is unphysical however, since interactions with the plasma (be it thermal, i.e. the bath, or out of equilibrium, i.e. the shell) screen long-range forces. This is equivalent to providing an IR cutoff, which takes the form of a Debye mass appearing in the propagators, which reads~\cite{Weldon_1999,Rebhan_1993,Gorda_2023}
\begin{align}
    m_{\rm D}^2 &\simeq 2 g_{\rm eff} \gD^2 \int \dfrac{\mathrm{d}^3 p}{(2\pi)^3} \dfrac{f(p,T)}{p^{0}} \approx 2 \gD^2 \frac{n}{\langle E \rangle} \,,
    \label{eq:Debye_mass}
\end{align}
where $\gD$ is the coupling strength of the interaction of interest, $f$ is the phase-space distribution of the plasma particles that are charged under that interaction, and $g_{\rm eff}$ is their number of degrees of freedom weighted by their charge. 
For example, in the case of an equilibrium thermal distribution $f(p,T)$ of gluons and fermions charged under a gauged $SU(N)$, this reproduces the well-known result $m_{\rm D}^2 \simeq \gD^2 T^2 (N/3 + N_{f}/6)$~\cite{Rebhan:1993az}, where $N_{f}$ is the number of Dirac fermions in the fundamental of $SU(N)$.
In case particles in the shells are charged under the gauge interaction of interest, e.g. as in the cases of LO shells and of shells formed by reflected or transmitted non-abelian gauge bosons, then Eq.~(\ref{eq:Debye_mass}) implies that $m_{\rm D}^2$ receives a contribution both from bath and from shell particles, and the latter can dominate over the former~\cite{Baldes:2020kam}. 
For PTs occuring in a gauge sector, once bubbles
are nucleated and shells are created, the IR cutoff reads
\begin{equation}
\mu^2 \simeq\,
    \begin{cases} 
        \frac{N_f}{6} \gD^2 \Tnuc^2  & \qquad U(1)\textrm{ gauge bosons}\,,
        \vspace{.2cm}\\ 
        \big(\frac{N_f}{6}+\frac{N}{3}\big) \gD^2 \Tnuc^2 + 2 \gD^2  \dfrac{n_s}{\pX} &  \qquad SU(N)\textrm{ gauge bosons} \,,
        \vspace{.2cm}\\ 
        \frac{N_fC_f}{8}\gD^2 \Tnuc^2 + d_f \gD^2  \dfrac{n_s}{\pX} &
        \textrm{fermions or scalars charged under $U(1)$ or $SU(N)$,}
    \end{cases}
    \label{eq:IRcutoff}
\end{equation}
where the shell densities $n_s$ depend on the shell of interest and can be obtained from Sec.~\ref{sec:shell_densities} plus Table~\ref{tab:production_mechanism}. The coefficient $C_f$ is the fermion color factor \cite{Bellac:2011kqa}. It is equal to $C_f=1$ for $U(1)$ fermion and $C_f=4/3$ for $SU(N)$ fermion in the fundamental representation. The factor $d_f$ is an order one factor which we take equal to $d_f=2$ for definiteness.
In the Abelian case, the IR cutoff as given in Eq.~(\ref{eq:IRcutoff}) is already a function of the fundamental parameters of the PT. This is not the case for non-Abelian gauge theories, for which we give below more explicit expressions for $\mu^2$.
    For LO shells, using $n_s$ and $\pX$ from Table~\ref{tab:production_mechanism}, we find that the non-Abelian contribution scales as the Abelian one,
    \begin{equation}
    	\boxed{
        \text{Non-abelian, LO shells:} \qquad \mu^2 \sim \gD^2 \Tnuc^2.
        }
    \end{equation}
    For shells coming from radiated gauge bosons, either transmitted or reflected, particles in the shells are squeezed and the number density $n_X$ defined in Eq.~\eqref{eq:shelldensity1} reads
\begin{equation}
  n_X \simeq   \mathcal{N}\,\frac{\zeta(3)}{\pi^2}g_*\Tn^3\frac{R_c}{3L_b} \simeq \gD\gamma_{\rm w}^2\Tn^3 \left(\frac{g_*}{12\pi^4}\right)\left(\frac{\gamma_{\rm run}}{\gamma_{\rm w}}\right)\left( \frac{\mathcal{N}}{\alphaD/\pi}\right)\left( \frac{\overline{L}_b}{L_b}\right),
  \label{eq:nX_IRcutoff}
\end{equation}
where $\overline{L}_b$ and $L_b$ are related by Eq.~\eqref{eq:LpX}.
For non-abelian gauge sector, it leads to the IR cut-off, cf. Eq.~\eqref{eq:IRcutoff}
\begin{equation}
	\boxed{
     \text{Non-abelian, radiated shells:} \qquad  \mu^2 \simeq \gD^4  R_c \frac{\Tn^4}{m_{c,h}}\left(\frac{g_*}{6\pi^4}\right)\left( \frac{\mathcal{N}}{\alphaD/\pi}\right)\left( \frac{\gamma_{\rm w} m_{c,h}}{\pX}\right)\left( \frac{\overline{L}_b}{L_b}\right).
     }
\end{equation}

\begin{figure}[!ht]
\centering
\raisebox{0cm}{\makebox{\includegraphics[width=1\textwidth, scale=1]{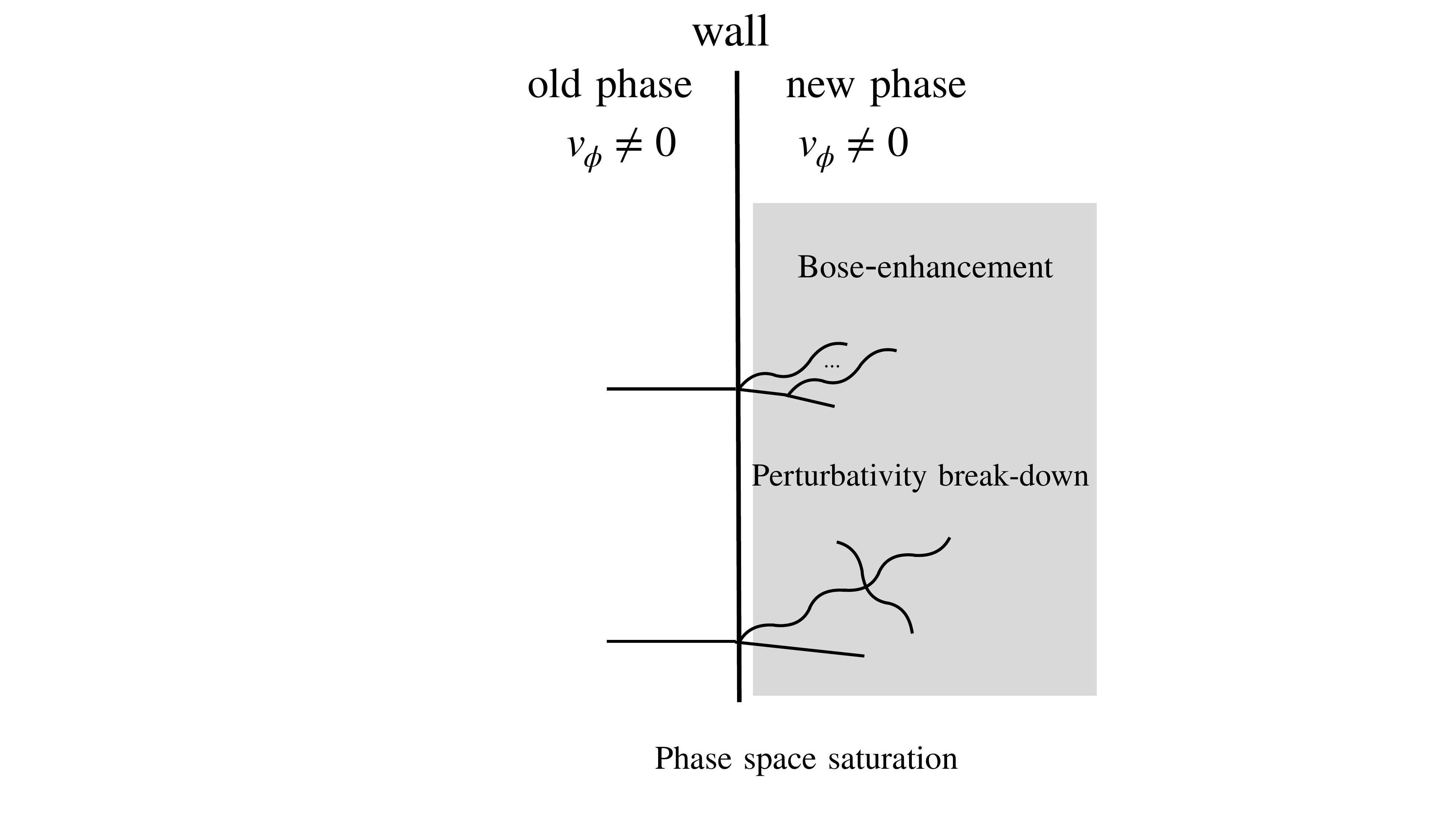}}}
\caption{\it \small \textbf{Phase space saturation:} A large occupation number of particles in the shell can lead to Bose-enhancement of the splitting radiation process (\textbf{top}) or to a break-down of perturbativity (\textbf{bottom}). See Sec.~\ref{sec:phase_space_sat} for the details.}
\label{fig:phase_space_saturation} 
\end{figure}

Another IR cutoff is provided by the size of the bubbles at collision,
\begin{equation}
	\boxed{
     \text{All shells:} \qquad  \mu^2 \simeq \beta^2\,.
     }
     \label{eq:IRcutoff_bubblesize}
\end{equation}
If the interaction of a shell particle is mediated by frequencies larger than the inverse of the collision time, $\beta$,
then that interaction cannot be effective before bubbles collide, and so it cannot affect the propagation of shells. In other words, wavelengths larger than the spatial dimension of bubbles at collision should affect the physics of bubbles nor vice-versa. 
The IR cutoff due to bubble size, Eq.~(\ref{eq:IRcutoff_bubblesize}), is subleading with respect to the IR cutoffs due to plasma effects, Eq.~(\ref{eq:IRcutoff}), in most of the regions of our parameter space. For example it is smaller than $\gD \Tn$ if 
\begin{equation}
    \gD \frac{\MPl}{v_\phi}\frac{\Tn}{v_\phi} \frac{1}{\beta/H} \gtrsim 1\,.
    \label{eq:cutoff_from_plasma}
\end{equation}
We anticipate that this condition will turn out to be satisfied at the border of all free-streaming regions, so that using the cutoff from the plasma-mass to determine that border, as we will do, is self-consistent. Deep in the regions of free streaming, far from that border, there will be parts of parameter space where the inequality of Eq.~(\ref{eq:cutoff_from_plasma}) is violated: this will only mean that amplitudes have a larger cutoff than the one we employ, so they are smaller, so a fortiori free-streaming in those regions is confirmed. 
The IR cutoff in Eq.~(\ref{eq:IRcutoff_bubblesize}) will however play a role in identifying a region, deep in the free-streaming area, where our predictions cannot be trusted due to $\mu \simeq m_{c,h}$, see Sec.~\ref{sec:large_plasma_mass}  and the regions shaded in pink in Figs.~\ref{fig:free_stream_all_cases} and \ref{fig:phase_space_sat}.

\subsection{Approach of the following calculations}
In the calculation of all effects that could prevent shells from free streaming (i.e. in Secs.~\ref{sec:phase_space_sat},~\ref{sec:mom_loss},~\ref{sec:3to2} and~\ref{sec:wall_int}), we use the determinations of the PT and shell properties within the free-streaming assumption. This allows one to self-consistently determine the regions of parameter space of a PT where shells are guaranteed to free-stream, which is the purpose of this paper.
Also, to our knowledge, no one ever determined (outside of the hydrodynamic limit) the PT and shell properties in regions where shells are relativistic and do not free stream. Our study is then not only self-consistent, but also a needed first step to determine the behaviour of walls and shells in regions of no free-streaming, which we leave for future work.
For example, in those regions, the Lorentz factor $\gamma_{\rm coll}$ in Eq.~\eqref{eq:gamma_final}, the shell number density and typical momentum, $n_X$ and $\pX$, and the shell thickness $L_w$ (see Table~\ref{tab:production_mechanism}) will be affected. These will in turn induce a backreaction on all calculations that depend on them.

\section{Phase space saturation}
\label{sec:phase_space_sat}

In this section we determine the regions where finite density corrections become large, either to the plasma mass discussed in Sec.~\ref{sec:IR_cutoff} or to the particle occupation number in the shells. These regions are shaded individually in Fig.~\ref{fig:phase_space_saturation}, and their envelope is shaded in Fig.~\ref{fig:free_stream_all_cases}.

\begin{figure}[!ht]
\raisebox{0cm}{\makebox{\includegraphics[width=0.49\textwidth, scale=1]{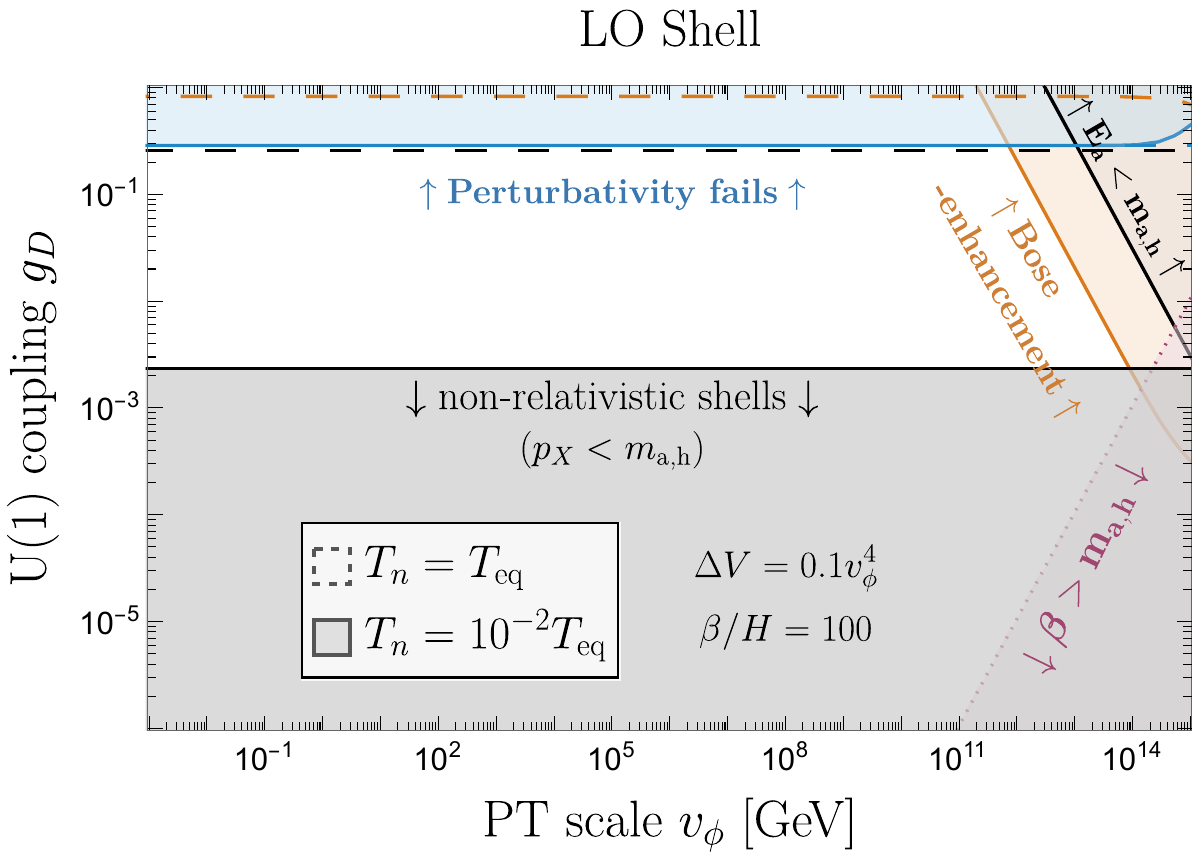}}}
{\makebox{\includegraphics[width=0.49\textwidth, scale=1]{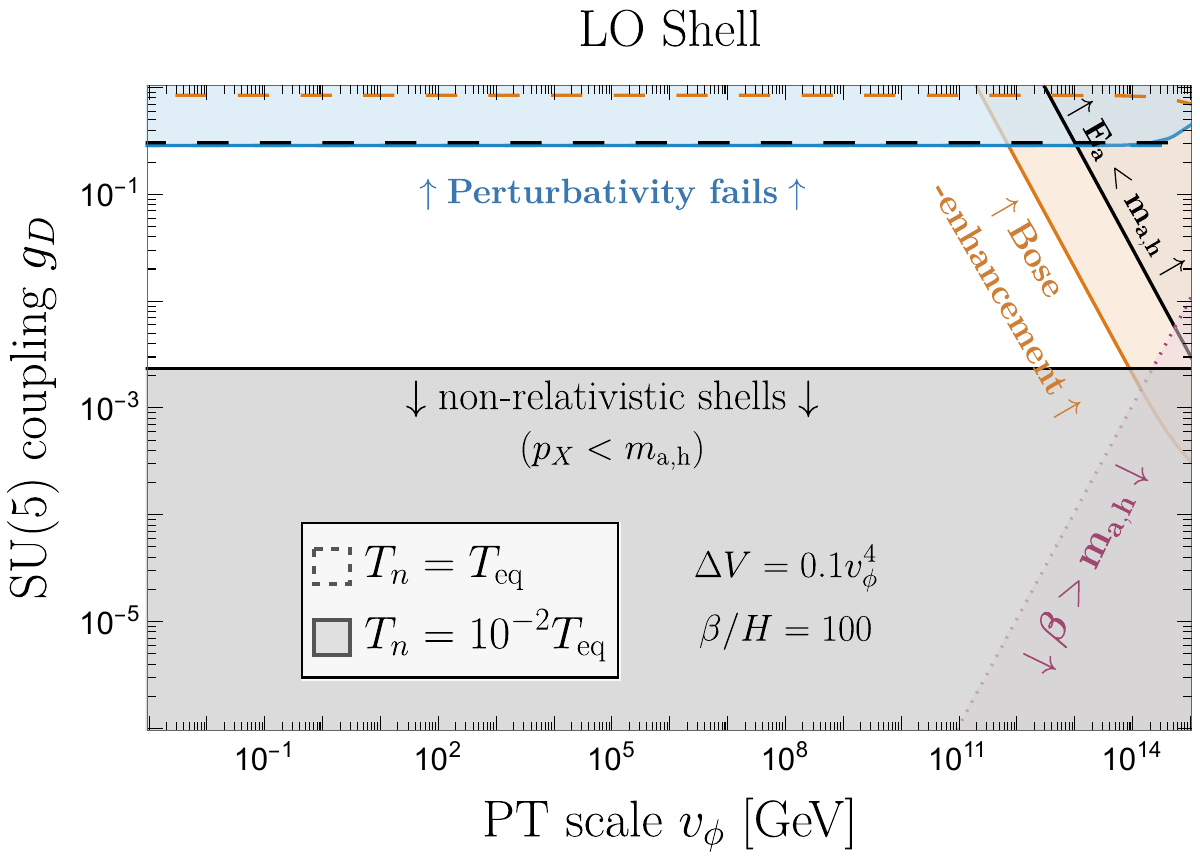}}}
{\makebox{\includegraphics[width=0.49\textwidth, scale=1]{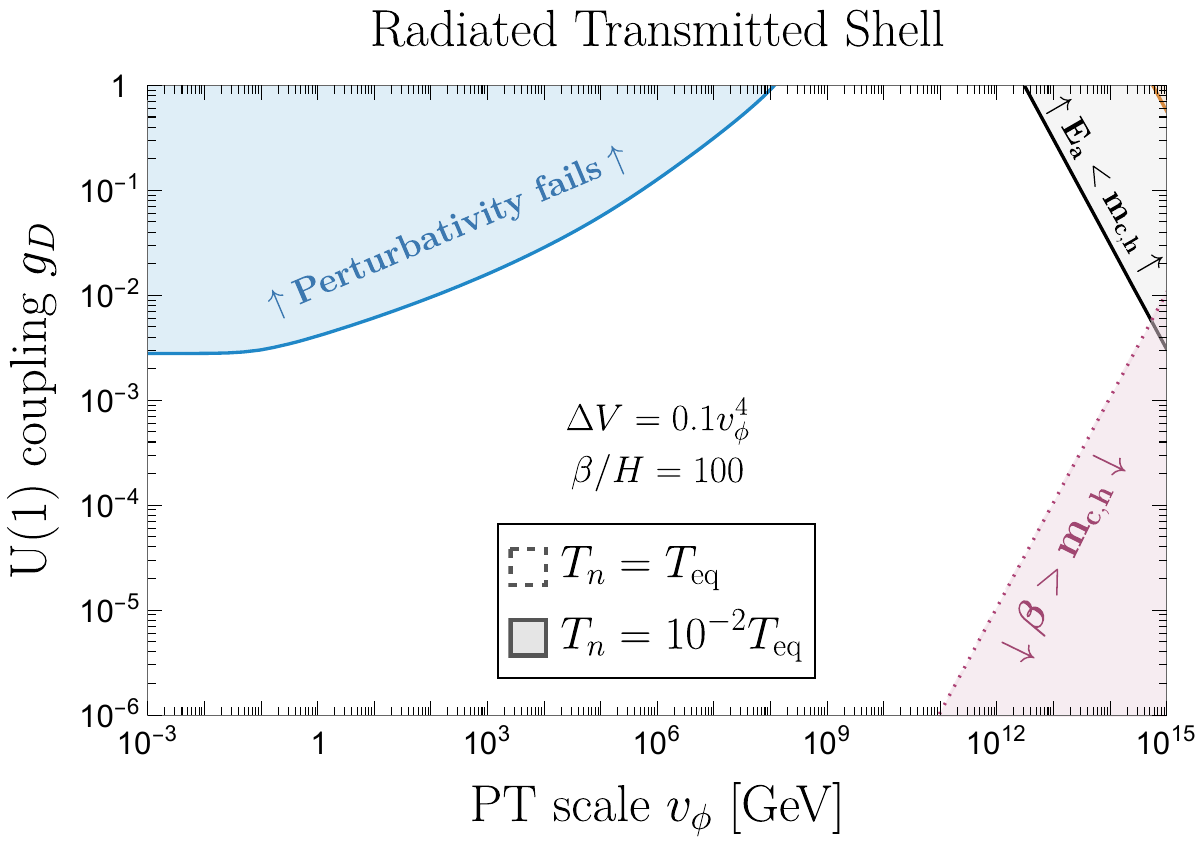}}}
{\makebox{\includegraphics[width=0.49\textwidth, scale=1]{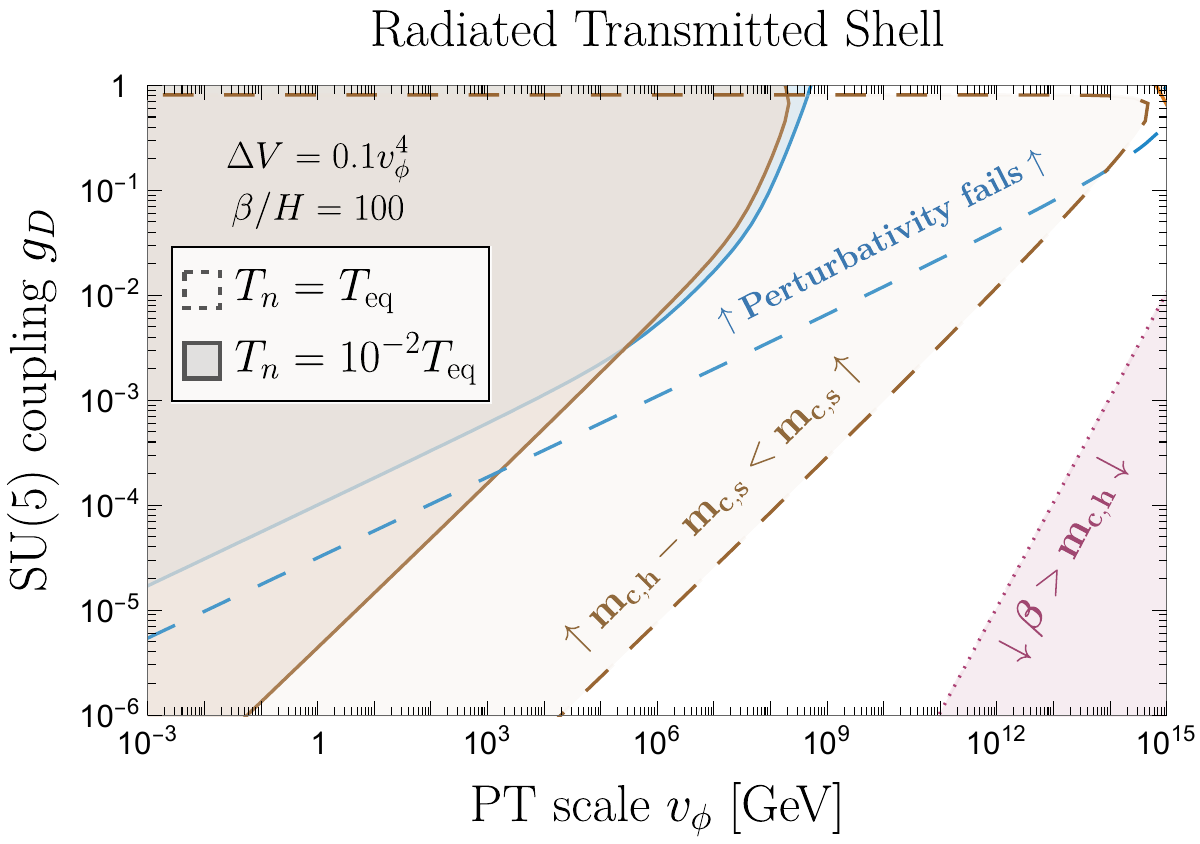}}}
{\makebox{\includegraphics[width=0.49\textwidth, scale=1]{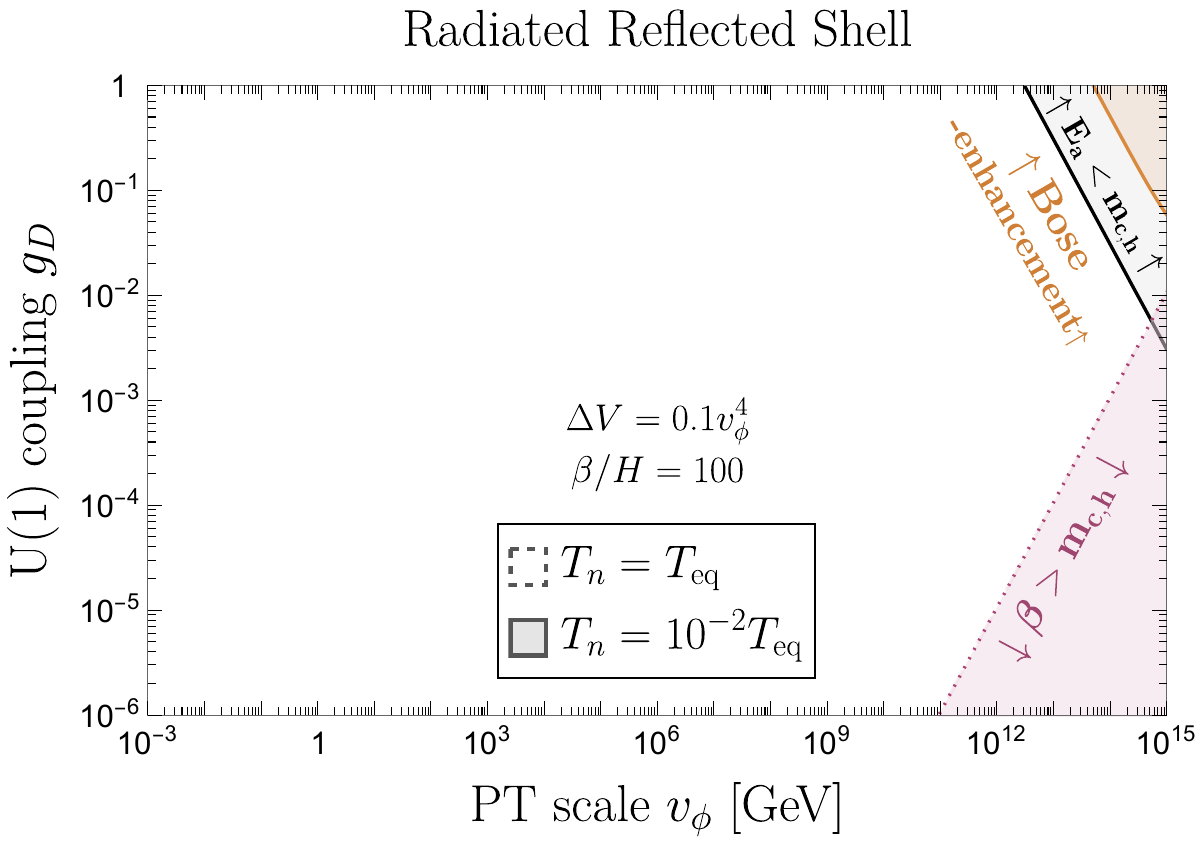}}}
{\makebox{\includegraphics[width=0.49\textwidth, scale=1]{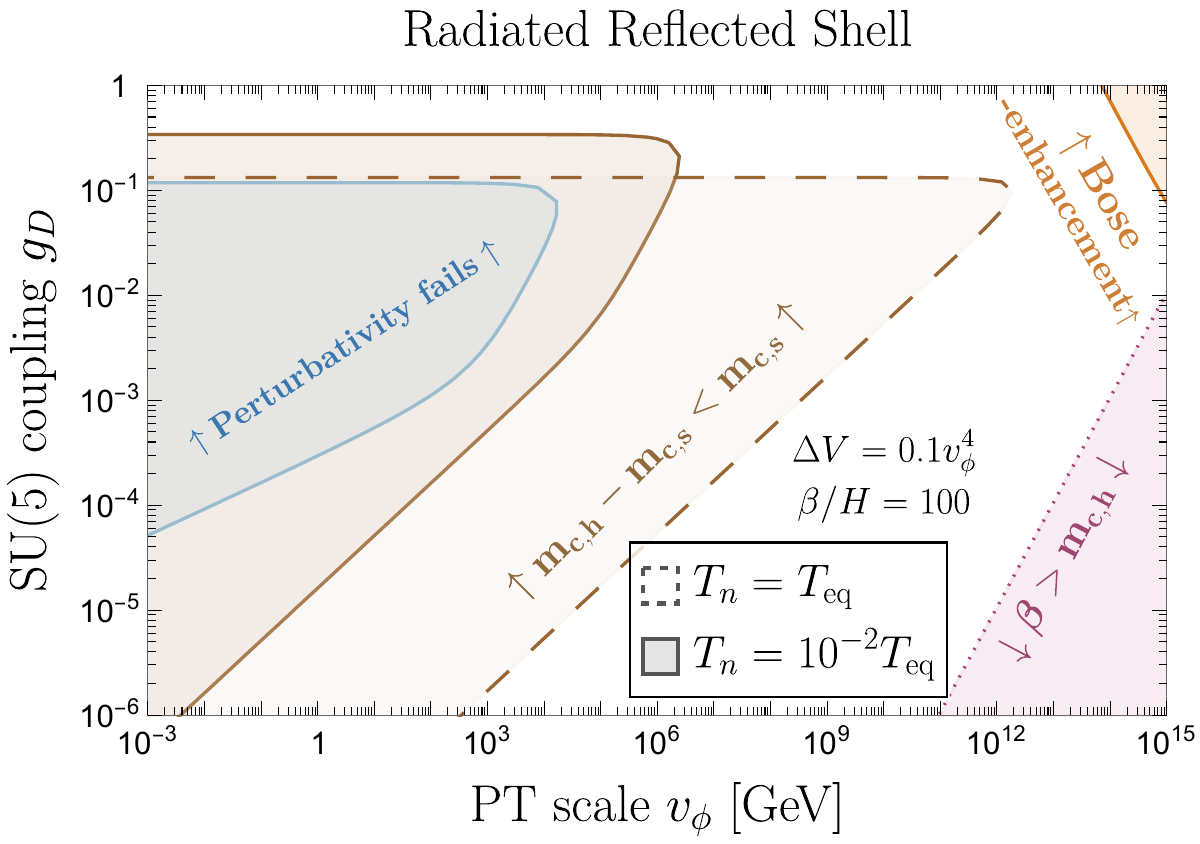}}}
\caption{\it \small  \textbf{Phase space saturation:} 
in the \textbf{blue} regions, the phase space is  saturated and the shells - their production and propagation - can not be described by perturbative theory, see conditions in Eq.~\eqref{eq:PSF_non-abelian} and \eqref{eq:PSF_abelian}. In the \textbf{orange} region, the splitting radiation probability producing the shell should be corrected by Bose-enhancement factors, see condition in Eq.~\eqref{eq:fc_fa_-_fb}. In the \textbf{brown} region, finite-density corrections drive the vector boson mass $m_{c,s}=\mu$, cf. Eq.~\eqref{eq:IRcutoff}, above its broken vacuum value $m_{c,h}^\infty=\gD v_{\phi}/\sqrt{2}$, cutting off vector boson production, see App.~\ref{app:large_IR_cutoff}. Our treatment based on $m_{c,s}\ll m_{c,h}$, can not be trusted in the brown region and is left to future works.  
Similarly, our treatment breaks down in the \textbf{pink} region, where the IR cutoff to shell processes due to the bubble size, $\beta$, is larger than $m_{c,h}$.
We consider shell particles produced from $LO$ interaction (\textbf{top}), Bremsstrahlung radiation, either transmitted (\textbf{middle}) or reflected (\textbf{bottom}), assuming abelian (\textbf{left}) or non-abelian (\textbf{right}) gauge interaction, see the first two rows of Table~\ref{tab:production_mechanism}, assuming the amounts of supercooling $\Tn/T_{\rm eq} = 1$ (\textbf{dashed}) and $\Tn/T_{\rm eq} = 10^{-2}$ (\textbf{solid}). We used $g_b = g_\SM + g_\text{emit}$ with $g_\SM = 106.75$ and $g_\text{emit} = 10 = N_F$. 
\textbf{Pink}, \textbf{grey} and \textbf{black} shaded areas as in Fig.~\ref{fig:free_stream_all_cases}.}
\label{fig:phase_space_sat} 
\end{figure}

\subsection{Large plasma mass}
\label{sec:large_plasma_mass}

Due to the high density of shells, the plasma mass $m_{c,s}\simeq \mu$ in the symmetric phase can become larger than the vector boson mass $m_{c,h}^{\infty}\simeq \gD v_\phi/\sqrt{2}$ in the broken phase far away from the wall
\begin{equation}
    \mu^2 \gtrsim (m_{c,h}^{\infty})^2 \quad \implies \quad \gD \Tn R_c \left(\frac{\Tn}{v_\phi}\right)^3 \left(\frac{g_*}{200}\right)\left( \frac{\mathcal{N}}{\alphaD/\pi}\right)\left( \frac{\gamma_{\rm w} m_{c,h}}{\pX}\right) 
    \frac{\bar{L}_b}{L_b}
    \gtrsim 1.
\end{equation}
Note that the vector boson mass in the broken phase close to the wall, within the shell, is also supplemented by the plasma correction
\begin{equation}
    m_{c,h}^2 \simeq  m_{c,s}^2+ (\gD v_\phi/\sqrt{2})^2,\qquad \textrm{with}\quad m_{c,s}\simeq \mu.
\end{equation}
As a result, for non-abelian gauge group in which the vector boson mass receives finite-density correction and in the ultra-relativistic limit in which this correction  is large, the mass difference $\Delta m_c \equiv m_{c,h}-m_{c,s}$ becomes small $\Delta m_c \ll m_{c,h}$.
In App.~\ref{app:large_IR_cutoff}, we show that in this regime the number of particles radiated becomes suppressed by $\mathcal{N}_R \propto (\Delta m_c/m_{c,h})^3$ and $\mathcal{N}_T\propto (\Delta m_c/m_{c,h})^2$, according to whether they are reflected on or transmitted though the wall. This suggests that the number of particles in the shell saturates. At the same time, the momentum exchange between an incoming particle and the wall is suppressed by $\left<\Delta p_R\right> \propto (\Delta m_c/m_{c,h})^4$ and $\left<\Delta p_T\right> \propto (\Delta m_c/m_{c,h})^{5/2}$, see App.~\ref{app:large_IR_cutoff} again, suggesting that the friction pressure drops, leading to an enhancement of the bubble wall Lorentz factor $\gamma_{\rm w}$. We shade the regions where this occurs in brown in Figs.~\ref{fig:free_stream_all_cases}, \ref{fig:phase_space_sat}, \ref{fig:shell_reversal}, \ref{fig:shell_meta} and \ref{fig:3to2}. Our results can not be trusted inside those regions since the number of particles in the shells and their kinematics should be modified accordingly. We leave its exploration for future works.
In an entirely analogous manner, we shade in pink the regions where the IR cutoff due to finite bubble-size, Eq.~(\ref{eq:IRcutoff_bubblesize}) is larger than $m_{c,h}$, where all the limitations mentioned above also apply. Note that this region is independent of $T_n/\Teq$.

\subsection{Bose enhancement}
In the wall frame, consider particle $a$ crossing the wall, splitting to produce particle $b$ of roughly similar momentum as $a$, and soft boson $x$ which will form the shell. The interaction Hamiltonian for the splitting process can be expressed as
\begin{equation}
H_{\rm int} = \mathcal{M}_0 a_c^\dagger a_b^\dagger a_a + \rm h.c.,
\end{equation}
where $a_x$ are the creation operators in Fock space. Then the transition amplitudes for emission and absorption read, respectively
\begin{align}
&\mathcal{M}_{a\to bc}=\left<f_a-1, f_b +1, f_c+1| H_{\rm int} | f_a, f_b, f_c \right> = \mathcal{M}_0 \sqrt{f_a}\sqrt{1\pm f_b} \sqrt{1 + f_c}, \label{eq:Matobc} \\
&\mathcal{M}_{bc\to a}=\left<f_a+1, f_b -1, f_c-1| H_{\rm int} | f_a, f_b, f_c \right> = \mathcal{M}_0 \sqrt{1\pm f_a}\sqrt{ f_b} \sqrt{f_c}, \label{eq:Mbctoa}
\end{align}
where $+/-$ refers to boson/fermion statistic.
We deduce the interaction rate accounting for both emission and absorption
\begin{equation}
|\mathcal{M}_{a\to bc}|^2 - |\mathcal{M}_{bc\to a}|^2 = |\mathcal{M}_0|^2 \left[ f_a(1 \pm f_b) + f_c(f_a -f_b) \right],
\end{equation}
where we have used that particle $c$ is bosonic so $a$ and $b$ necessarily are of the same statistical type\footnote{The phase space factors, $f_a$, $f_b$, $f_c$ which appear in the forward, Eq.~(\ref{eq:Matobc}), and reverse processes, Eq.~(\ref{eq:Mbctoa}), are the same ones. There is no switch in the momentum direction.
}. 
We can see that Bose-enhancement can be neglected as long as $(f_a - f_b) f_c \ll f_a$, in which case we get
\begin{equation}
|\mathcal{M}_{a\to bc}|^2 - |\mathcal{M}_{bc\to a}|^2 \simeq |\mathcal{M}_0|^2 f_a.
\end{equation}
We now investigate the condition for neglecting Bose-enhancement
\begin{equation}
\label{eq:fc_fa_-_fb}
	\boxed{
    f_c(p_{c})(f_{a}(p_{a})-f_b(p_{b})) = f_{c}(p_{c})(p_{a}-p_{b})\frac{\partial f_a}{\partial p_{a}}~\lesssim~f_{a}(p_{a}).
    }
\end{equation}
We now add a subscript `$p$' to indicate quantities evaluated in the bath frame. In the absence of `$p$', quantities are evaluated per default in the wall frame.
 Assuming a Maxwell-Boltzmann momentum distribution in the bath frame, $f_a(p_{a,p}) \propto \exp\left(-p_{a,p}/T_n \right)$, we can write 
\begin{equation}
\label{eq:df_a_dp_aw}
    \frac{\partial f_a}{\partial p_{a}}=\frac{\partial p_{a,p}}{\partial p_{a}}\frac{\partial f_a}{\partial p_{a,p}}= - \frac{f_a(p_{a})}{\gamma_{\rm w} T_n},
\end{equation}
where we used that $p_{a,p} = \gamma_{\rm w} p_{a} - \beta \gamma_{\rm w} \sqrt{m_a^2 + p_{a}^2} \simeq p_{a}/\gamma_{\rm w}$. 
Plugging Eq.~\eqref{eq:df_a_dp_aw} into Eq.~\eqref{eq:fc_fa_-_fb} leads to
\begin{equation}
\label{eq:f_c_f_a_-_f_b_2}
     f_c(p_c)(f_a(p_a)-f_b(p_b)) = f_c(p_c)f_a(p_a)\frac{\Delta m_a^2}{2p_a\gamma_{\rm w} T_n},
\end{equation}
Replacing $p_a \simeq 3 \gamma_{\rm w} T_n$ and $f_a(p_a)\simeq 1$, and using Eq.~\eqref{eq:gamma_run} we calculate
\begin{equation}
\label{eq:f_c_f_a_-_f_b_2_bis}
     f_c(p_c)(f_a(p_a)-f_b(p_b)) \simeq\frac{1}{6} f_c(p_c)\left(\frac{\Delta m_a}{R_c T_n^2} \right)^2\left( \frac{\gamma_{\rm run}}{\gamma_{\rm w}} \right)^2
\end{equation}
Note that Eq.~(\ref{eq:f_c_f_a_-_f_b_2_bis}) is general, in the sense that it applies to particles $c$ belonging to any shell considered in this paper.
We show the parameter space where Bose-enhancement matters in orange in Fig.~\ref{fig:phase_space_sat}.
In most of the parameter space, we find the condition for neglecting Bose-enhancement enhancement  $f_c(p_c)(f_a(p_a)-f_b(p_b))\ll 1$ less stringent than the condition for non-perturbativity $\gD^2 f_c(p_c) \ll 1$ or $\gD^4f_c(p_c) \ll (4\pi)^2$, shown in blue in the same figure and discussed in the next section.

\subsection{Perturbativity break-down}

Let us consider first shells produced by splitting radiation $a\to bc$. The occupation number $ f_c(p_c) $ of radiated vector bosons  is related to  the vector boson wave function $A_\mu$ by
\begin{equation}
(\partial A)^2 \sim
p^2 A^2
\sim \int_{p_c}\! d^3p\,p\, f_c(p)
\sim
p_c^4  \, f_c(p_c)\,,
\end{equation}
where, in the second wiggle, we have used that the boson kinetic term `corresponds' to the energy of the free gas of bosons.
For a non-abelian theory, the hierarchy between the Lagrangian terms
\begin{equation}
\mathcal{L} \supset \partial A \partial A + g A A \partial A + \gD^2 A A A A\,,
\label{eq:lagragian_vector_boson}
\end{equation}
which is essential for perturbation theory to apply, breaks down as soon as
\begin{equation}
\label{eq:PSF_non-abelian}
	\boxed{
\textrm{Non-abelian:} \qquad \gD^2 f_c(p_c) > 1.
}
\end{equation}
Instead, for an abelian theory, gauge bosons self-interaction terms are loop-suppressed so that perturbativity breaks down for 
\begin{equation}
\label{eq:PSF_abelian}
	\boxed{
\textrm{Abelian:} \qquad \frac{\gD^4}{(4\pi)^2} f_c(p_c) > 1.  
}
\end{equation}
Similarly, for emission of self-interacting scalars $\phi$ with $\mathcal{L}\supset \lambda \phi^4$, perturbativity breaks down as soon as
\begin{equation}
\label{eq:PSF_scalar}
	\boxed{
\textrm{Self-interacting scalar:} \qquad \lambda \,f_c (p_c) > 1. 
}
\end{equation}
We now proceed in calculating the phase space occupation number of radiated particles in the shells.
The $c$ particles  get accumulated within a thin shell whose thickness $L_b$ is computed using kinematics arguments in Sec.~\ref{sec:shells_general}.
The associated number density of particles in the bath frame is 
\begin{equation}
\label{eq:nXp}
n_{c,p}~ \simeq~ \mathcal{N}\,\frac{\zeta(3)}{\pi^2}g_*\Tn^3\frac{R_c}{3L_b},
\end{equation}
 We deduce the occupation number $f_{c}$ (number of particles per de Broglie wavelength cell)
\begin{equation}
\label{eq:occupation_number}
f_{c}~ \simeq ~ \frac{n_{c,p}}{k_\perp^2 p_{c,b}}~ \simeq ~ \frac{\zeta(3)}{3\pi^2}g_*\,\mathcal{N}\frac{R_c\Tn^3}{k_\perp^2 p_{c,b} L_b} 
\end{equation}
In Fig.~\ref{fig:phase_space_sat}, we show in blue the regions where perturbativity breaks-down, using Eqs.~\eqref{eq:PSF_non-abelian} and \eqref{eq:PSF_abelian}. In this plot we accounted for the full expressions for $\mathcal{N}$, $p_{c,b}$ and $k_\perp$ given in App.~\ref{app:weak}. For each quantity $X$, we included both the mean $\left<X\right>$ and width $\sigma_X$ by performing the simple quadratic sum $\sqrt{\left<X\right>^2 +\sigma_X^2}$.

We now provide analytical estimates of the occupation numbers in the simplified cases where $L_b \simeq \bar{L}_b$.
For shells of particles generated at leading order (LO), we can use $L_b\simeq {\color{red} \bar{L}_b} \simeq R_c(T_n/m_{c,h})^2$, $\mathcal{N}=1$, $p_{c,b}  \simeq m_{c,h}^2/T_n$ and $k_\perp \simeq T_n$ as shown in  the first row of Table~\ref{tab:production_mechanism} (with $X\equiv c$), leading to
\begin{equation}
    f_{c}(p_c) = \zeta(3)g_s/\pi^2,\qquad (\rm LO~ Shell).
\end{equation}
Replacing the quantities $L_b\simeq  \bar{L}_b \simeq R_c/(2\gamma_{\rm w}^2)$, $\mathcal{N} \simeq (\alphaD/\pi)L_m^2$, $p_{c,b}  \simeq 1.6\gamma_{\rm w} m_{c,h} /L_m^{\!1/2}$ and $k_\perp \simeq \Tn$ in Eqs.~\eqref{eq:LpX_behindwall} and~\eqref{eq:LpX_aheadofwall}, \eqref{eq:N_R}, \eqref{eq:sigma_E} and \eqref{eq:k_perp_sigma} for shells of reflected radiated particles and $\mathcal{N} \simeq (2\alphaD/\pi)L_mL_E$, $p_{c,b}  \simeq 0.4\gamma_{\rm w} m_{c,h} /L_E^{\!1/2}$ and $k_\perp \simeq \Tn$ for transmitted radiated particles, cf. second row of Table~\ref{tab:production_mechanism}, the occupation number of particles $c$ reads
\begin{equation}
\label{eq:f_c_p_c}
    f_c(p_c) = 
         \gD^2\left(\frac{g_*}{100} \right) \left(\frac{\gamma_{\rm w} T_n}{m_{c,h}} \right) \times  \begin{cases} 
         L_m^{5/2}/8,\qquad \qquad  \quad  \rm(Reflected~Shell),\\
        L_E^{3/2},\hspace{2cm} \rm(Transmitted~Shell),
    \end{cases}
\end{equation}
with $L_m\equiv \log\left(m_{c,h}/m_{c,s}\right)$, and $L_E\equiv\textrm{log} \left(E_a/m_{c,h}\right)$.
For shells of particles produced by Azatov-Vanvlasselaer mechanism \cite{Azatov:2021ifm,Baldes:2021vyz,Azatov:2022tii,Baldes:2022oev}, in Eq.~\eqref{eq:occupation_number} we plug $L_b \simeq \bar{L}_b\simeq R_c(T_n/M_X)^2$ to be conservative, $\mathcal{N}=\lambda^2v_\phi^2/192\pi^2M_X^2$, $p_{c,b}  \simeq M_X^2/T_n$ and $k_\perp \simeq T_n$ as shown in the last row of Table~\ref{tab:production_mechanism}, leading to
\begin{equation}
    f_c(p_c) = \frac{\zeta(3)}{\pi^2}g_* \left(y \frac{v_\phi}{M_X}\right)^2,\qquad (\text{Azatov-Vanvlasselaer mechanism}),
\end{equation}
which is much smaller than 1, implying that phase-space saturation effects (both the perturbativity breakdown given by Eqs.~\eqref{eq:PSF_scalar} and the Bose enhancement given by Eq.~\eqref{eq:fc_fa_-_fb}) are negligible.

\section{Shell momentum loss}
\label{sec:mom_loss}

In this section we consider elastic $2 \to 2$ interactions between shell and bath particles and determine the conditions necessary for shells to free stream without the momentum losses induced by these interactions.
We identify the possible impediments listed in Sec.~\ref{sec:list_momentum_losses}, and shade the associated regions of no-free streaming individually in Figs.~\ref{fig:shell_reversal}, \ref{fig:shell_meta} and their envelope in Fig.~\ref{fig:free_stream_all_cases}.

\subsection{Possible processes inducing momentum losses}
\label{sec:list_momentum_losses}

Consider a shell produced on either side of the wall and propagating outwards from the nucleation site in the same direction of the wall. Particles in the shell will see an incoming flux of particles from the thermal bath. Interactions with these bath particles may change the momentum of the shell particles and therefore modify the overall properties or propagation of the shell. Also the bath particles may be affected by the shell and perhaps the flux of particles reaching the wall be suppressed or otherwise modified.

\begin{itemize}
    \item[$\diamond$] 
    \textbf{Reversal of the shell}: Consider a shell travelling in front of the wall. In the bath frame, both wall and shell travel at close to the speed of light, but the shell is slightly faster than the wall. Interactions with the bath particles may lead to a change in momentum of the shell particles, insufficient for dissipation, but sufficient to slow the shell particles so that the latter are caught by the wall. The picture in the wall frame is the following: shell particles travel outward with momentum $\pxw$. Typically one has $\pxw \approx \mX$ in the wall frame for Bremstrahlung type production. Incoming bath particles have momentum $\approx \gamma_{\rm w} \Tn$. These interact with the shell particles and if the change of momentum of the latter in the wall frame is $\Delta \pxw \approx \pxw$ before the shells collide, then the shell particles are typically caught by the wall. More concretely, the condition to avoid the shell reversal is given by 
    \begin{equation}
    	\boxed{
    \frac{1}{\pxw} \frac{d \pxw}{dt}\Big|_{\rm wall} < \frac{\gamma_{\rm w}}{R_c},
    }
    \end{equation}
    where the $\gamma_{\rm w}$ factor takes into account the shorter propagation distance before wall collision in the wall frame. Note in the wall frame the density of bath particles is also Lorentz boosted $\sim \gamma_{\rm w} \Tn^3$.
    We calculate the conditions for shell reversal in Sec.~\ref{sec:shell_reversal}.
        \item[$\diamond$]  
    \textbf{Reversal of the bath}: If a relativistic shell is travelling in front of the wall, its interactions with the bath could in principle reverse the bath particles, and prevent them from reaching the wall.
    This can occur only if the shell carries more energy than the bath in the wall frame.
    We provide a general argument in Sec.~\ref{sec:shell_or_bath}, however, that this is energetically impossible because the shell itself comes from bath particles. 
        \item[$\diamond$] 
    \textbf{Shell dissipation/metamorphosis}: In the extreme case the shell may be partially or fully dissipated. Consider particles travelling outward with momentum $\px$ in the bath frame. Depending on the nature of the shell, we can have $\px \approx \gamma_{\rm w} v_\phi$ where $\gamma_{\rm w}$ is the Lorentz factor of the wall, or $\px \approx \mX^{2}/\Tn$, see Table~\ref{tab:production_mechanism}. Then if interactions with the bath particles leads to changes in the shell particle momentum $\Delta \px \approx \px$ before the shells collide, then the shell will become dissipated back into the thermal bath, provided the bath carries sufficient energy to do so. To ensure free streaming, we thus require the momentum exchange to small over the distance of propagation, here taken to be the bubble size at collision,  
        \begin{equation}
        	\boxed{
        \frac{1}{\px} \frac{d \px}{dt}\Big|_{\rm bath} < \frac{1}{R_c}.
        }
        \end{equation}
   However, if the total energy in the shell particles is larger than the total energy in the bath particles, $E_{\rm shell \, total} > E_{\rm bath \, total}$, the latter cannot fully dissipate the former. Then a short path length for $\Delta \px \approx \px$ interactions, indicates the shell will transfer an $\mathcal{O}(1)$ fraction of its momentum to bath particles, which themselves form part of an outward propagating shell, but now with modified properties. The number density, species, and momenta of shell particle is changed, so we dub this metamorphosis.
   The energy condition to have either dissipation or metamorphosis is determined in Sec.~\ref{sec:diss_or_meta}, and the associated free-streaming conditions are calculated in Sec.~\ref{sec:shell_diss_meta}.

\end{itemize}

\begin{figure}[!ht]
\begin{center}
\centering
\raisebox{0cm}{\makebox{\hspace{-1.5 cm}\includegraphics[width=1.25\textwidth, scale=1]{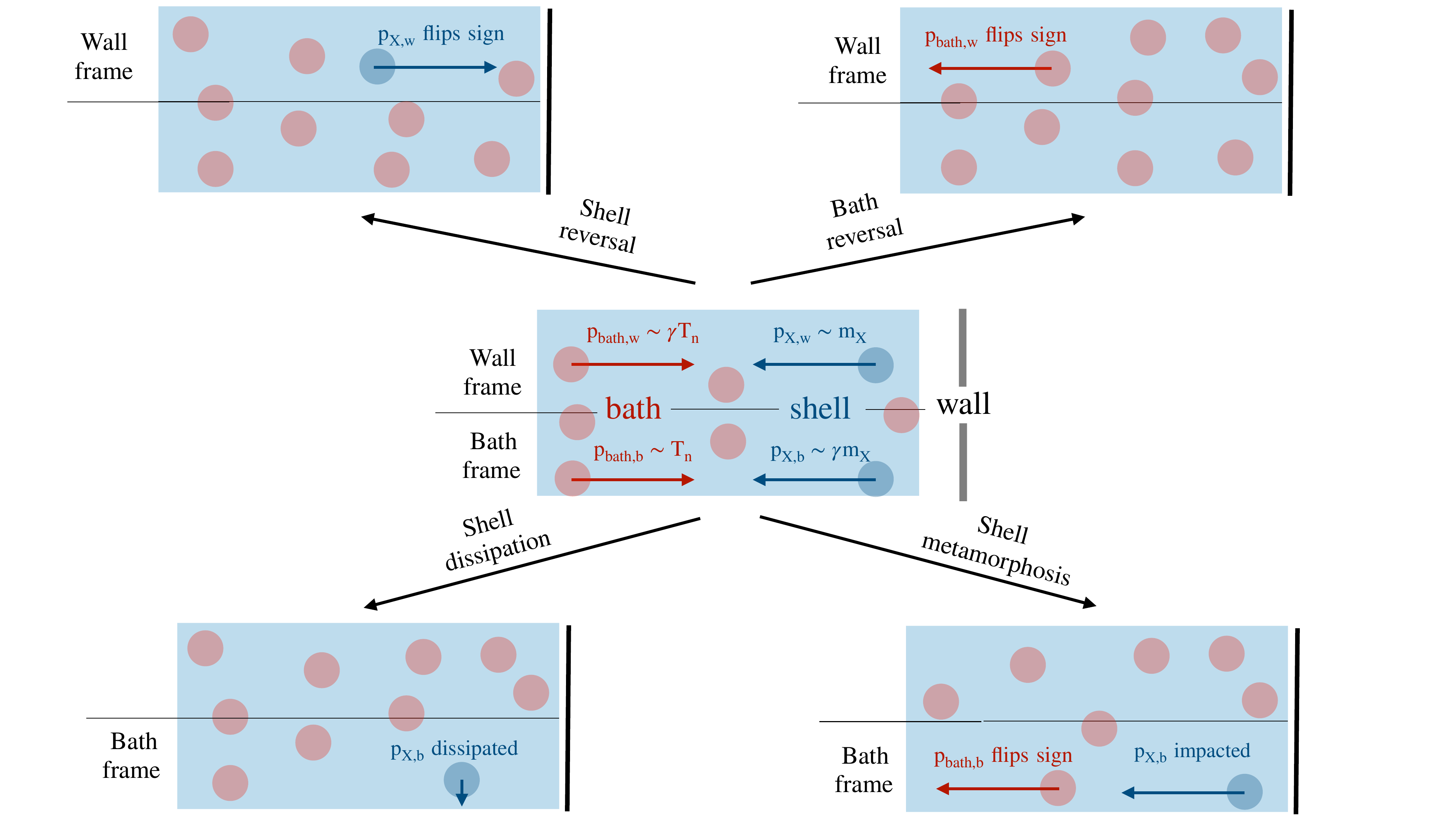}}}
\caption{\it \small  \textbf{Shell-bath momentum loss mechanisms:} Interactions between particles of the shells and particles of the bath generate momentum transfer between the two systems. Clockwise starting from top-left, different scenarios occur according to whether in the wall frame the shell typical momentum is reversed  (\textbf{shell reversal}), the bath momentum is reversed (\textbf{bath reversal}), in the bath frame the shell momentum is changed by $\mathcal{O}(1)$ (\textbf{shell metamorphosis}), or is fully dissipated (\textbf{shell dissipation}), see Sec.~\ref{sec:mom_loss} for the details. The last condition, shell dissipation, requires the bath to have more energy than the shell in the bath frame, see Eq.~\eqref{eq:Rshelldiss}.  The parameter space where shell reversal occurs is shown in Fig.~\ref{fig:shell_reversal}.  The one where shell metamorphosis/dissipation occurs is shown in Fig.~\ref{fig:shell_meta}. Bath reversal in the wall frame is never energetically feasible, see Eq.~\eqref{eq:R_shellObath_wall}.
}
\label{fig:shell_reversal} 
\end{center}
\end{figure}

The rest of this section is structured as follows. To set the stage, we first derive the required energy conditions, for shell dissipation/metamorphosis and for shell/bath reversal (showing the bath cannot be reversed in the wall frame). We then go on to evaluate the reversal and dissipation/metamorphosis path lengths for the shell, for different choices of underlying interactions, in greater detail.

\subsection{Total Energy considerations}

To better understand the end state of each the shell/bath interactions, it is instructive to consider the total energy/momentum of the shell and bath. The free energy difference between the true and false vacuaa leads to a pressure gradient which works on the bubble wall. The bath carries an associated pressure but no net momentum in the bath frame. Through the microphysical processes of bath-wall interactions, momentum of the wall is transferred to the resulting shell. This gives a net outward going momentum to the shell which in turn encounters the bath particles (either inside or outside the bubble, depending on the nature of the shell). Interactions between the shell and bath lead to further transfers of energy and momentum between the two when the above free streaming conditions are violated.

\subsubsection{Shell dissipation or metamorphosis?}
\label{sec:diss_or_meta}

As previously mentioned, our shell dissipation/metamorphosis condition takes as a threshold an $\mathcal{O}(1)$ change in the momenta of the shell particles, but this does not mean the complete erasure of the shell, only a significant change from its character in the limit of free streaming. Here we derive the required total energy of the bath in order for it to dissipate the shell completely back into a thermal state. We work in the bath frame. The total energy of bath particles in a Hubble volume, as seen in the bath frame, is given by
	\begin{equation}
	E_{\rm bath \; total,b} = \rho_{\rm bath} V_{\rm Hubble} = \frac{ g_{\ast} \pi^{2} \Tn^{4} }{ 30 } \frac{ 4\pi }{ 3 H^{3} }. 
	\end{equation}
Similarly, the total energy of relativistic shell particles in a Hubble volume, in the bath frame, is given by 
	\begin{equation}
	E_{\rm shell \; total,b} = n_{\rm bath} \mathcal{N} \px  V_{\rm Hubble} = \frac{ g_{\ast} \zeta(3) \Tn^{3} }{ \pi^{2} } \frac{ 4\pi \mathcal{N} \px }{ 3 H^{3} }.
	\end{equation}
The ratio of the two is thus
	\begin{equation}
	\boxed{
	R_{E \, \mathrm{shell \, diss.}} \equiv \frac{ E_{\rm shell \; total,b} }{ E_{\rm bath \; total,b} } \approx  \frac{ \mathcal{N}  \px }{ 3\Tn }.
	}
    \label{eq:Rshelldiss}
	\end{equation}
The simplified prefactor $3$ in the denominator of the above expression, in reality ranges between $2.7$ and $3.15$, where then former corresponds to a bath made purely of bosons and the latter to a bath made purely of fermions. (This subtlety is not captured by our simplified assignment of the same $g_{\ast}$ for $\rho_{\rm bath}$ and $n_{\rm bath}$.) Clearly, if $R_{E \, \mathrm{shell \, diss.}} > 1$, then the shell cannot be completely dissipated by the bath back into a purely thermal state before shell/bubble collision. The above condition should therefore be checked if the path length for dissipation/metamorphosis is short compared to $R_{c}$, in order to gain a better handle on the end state of the process.

 As a simple example, we can take the reflected radiated vector bosons, for which $\mathcal{N} \simeq \alphaD L_{m}^2/4\pi$, $\px \simeq \gamma_{\rm w} \sqrt{\alphaD} v_{\phi}$, and for $\gamma_{\rm w}$ assume the vacuum dominated runaway regime. Then we find 
	\begin{equation}
	R_{E \, \mathrm{shell \, diss.}} \simeq \frac{ \alphaD^{3/2} \MPl \log^2(v_{\phi}/\Tn) }{ \sqrt{48 c_{\rm vac}} \pi^{2/3} v_{\phi} \beta/H },
	\end{equation}
which for typical values implies $R_{E \, \mathrm{shell \, diss.}} \gg 1$. Thus if the shell dissipation free streaming condition is violated in this scenario, the shell will not be completely dissipated back into a thermal state, but rather be metamorphosised into a m\'elange of original shell and accelerated bath particles, before colliding with opposing shells/bubbles. Other scenarios of interest can easily be checked using Table~\ref{tab:production_mechanism} and the relevant $\gamma_{\rm w}$ factor. 
The more precise energy condition, Eq.~\eqref{eq:Rshelldiss}, will  be plotted together with the path lengths in our summary plots below, distinguishing regimes of dissipation/metamorphosis regarding the bath frame momentum of the shell.

\subsubsection{Shell or bath reversal?}
\label{sec:shell_or_bath}

We can also ask a similar question for shell or bath reversal in the wall frame. We are interested in whether the shell or bath particles change direction in the wall frame and so perform the calculation in said frame for convenience.\footnote{Working in the bath frame leads to additional complications, \emph{e.g.}~because the bath particles gain a net outward momentum during the reversal process, so it is not as easy as simply taking into account the energy change required for shell reversal in the bath frame, $\delta \EX = -\hat{t}/(4T) = \px^{2}/(\gamma_{\rm w}^{2}T)$.} The total energy of bath particles in the wall frame is
	\begin{equation}
	E_{\rm bath \; total \; w} = \gamma_{\rm w} E_{\rm bath \; total,b}.
	\end{equation}
The total energy of the shell particles in the wall frame is
	\begin{equation}
 \label{eq:R_shellObath_wall}
	E_{\rm shell \; total \; w} = \frac{E_{\rm shell \; total,b}}{\gamma_{\rm w}}.
	\end{equation}
If the  ratio of shell to bath energies in the wall frame is smaller than one, then the shell particles do not have carry enough energy to reverse the bath particles in the wall frame. In the wall frame the bath particles therefore reach the bubble without significant changes in momentum. Moreover, it is energetically allowed for the same bath particles to reverse the shell particles so the latter are pushed into the bubble. Actually, for consistency, we always have 
	\begin{equation}
	\boxed{
	R_{E \, \mathrm{shell/bath \, rev.}} \equiv \frac{ E_{\rm shell \; total. \; w} }{ E_{\rm bath \; total \; w} } = \frac{ \mathcal{N}  \px }{ 3 \gamma_{\rm w}^2 \Tn } < 1.
	}
    \label{eq:Eshellbathrev}
	\end{equation}
 This can be understood as follows. The wall can be treated as an inertial frame over the timescale of the particle transitions through it. Thus we can boost into the effectively time independent wall frame, for which energy conservation holds between the incoming particles (the bath), and their products, of which the shell is a subset. Indeed, shell particles produced earlier in the bubble expansion are produced at smaller $\gamma_{\rm w}$, so their energy contribution is actually smaller than this estimate. Thus, in the wall frame, the total energy of the shell is smaller than the total energy of the bath. In other words $ R_{E \, \mathrm{shell/bath \, rev.}} < 1$: out of the bath and the shell, it is only the latter which can be reversed in the wall frame.
 In other words, Eq.~\eqref{eq:Eshellbathrev} is satisfied in the entire parameter space of our interest.

Having derived the energy requirements, we now turn to calculating the relevant path lengths for shell reversal and shell dissipation/metamorphosis.

\subsection{Shell reversal}
\label{sec:shell_reversal}

\subsubsection{Basic picture}

We consider the particles created at the wall which are reflected back into the unbroken phase and calculate their reversal path length. Gauge bosons are of particular interest because they obtain an enhanced production rate compared to fermions and scalars. The gauge bosons have a typical momentum in the wall frame  $\pxw \approx m_{c,h} \approx  \gD v_\phi$ and zero vacuum mass (thus their momentum in the bath frame is $\px = \gamma_{\rm w} \pxw \approx \gamma_{\rm w} \gD v_\phi$, as in Table~\ref{tab:production_mechanism}).

We now derive the conditions for shell particles (either gauge bosons or others) to be sent back to the wall via interactions with the bath particles.
We also consider Compton scattering, which is of relevance when $U(1)$ gauge bosons interact with charged fermions or scalars in the bath, and  which gives a longer path length. The limits from Compton scattering are therefore the relevant ones in the absence of $t-$channel gauge boson exchange processes. Finally we consider a non-gauged PT in which a scalar boson is reflected back into the symmetric phase.

\subsubsection{Simple estimates}
\label{sec:reversal_simple}

\begin{figure}[t]
\centering
\includegraphics[width=200pt]{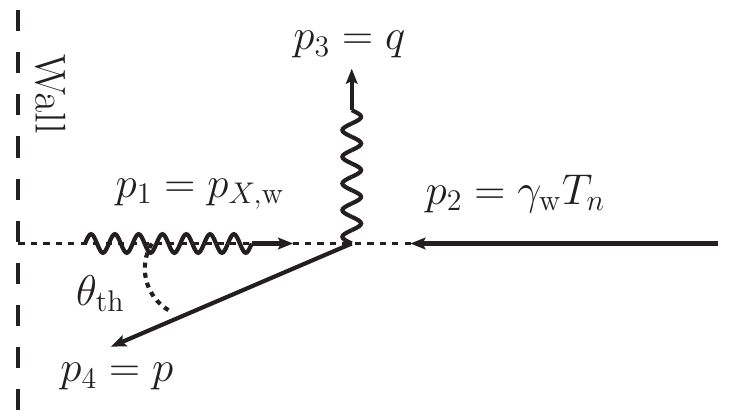}
\caption{\it \small
The scattering --- as seen in the wall frame --- leading to the threshold at which point the reflected particle will head back to the wall.   
}
\label{fig:reverse}
\end{figure}

We work in the wall frame. The basic picture is illustrated in Fig.~\ref{fig:reverse}. (The same situation, but limited to M\o{}ller scattering, has also been previously been considered in ~\cite{Gouttenoire:2021kjv}.) The shell particle momentum before and after scattering is denoted $p_1$ and $p_3$ respectively. The bath particle momentum is denoted $p_2$ or $p_4$. The four momenta are 
	\begin{subequations}
	\begin{align}
	p_{1} & = ( \pxw,        \quad 0,  \quad 0, \quad \pxw)  \\
	p_{2} & = ( \gamma_{\rm w} \Tn, \quad 0,  \quad 0, \quad -\gamma_{\rm w} \Tn)  \\
	p_{3} & = ( q,        \quad 0,  \quad q, \quad 0) \\
	p_{4} & = ( p,        \quad 0,  \quad -p \sin{\theta_{\rm th}}, \quad -p \cos{\theta_{\rm th}}),	
	\end{align}
	\end{subequations}
 where $\theta_{\rm th}$ is the minimal deflection angle of bath particles which lead to the reversal of the momentum of reflected shell particles, i.e. the deflection angle for which $\vec{p}_3$ becomes parallel to the wall.
Here we are ignoring particle masses and $\gamma_{\rm w}$ is the Lorentz factor of the wall in the bath frame. There are three unknowns, $q, p, \theta_{\rm th}$, and luckily for us three equations from energy and momentum conservation:
	\begin{subequations}
	\begin{align}
	 \gamma_{\rm w} \Tn + \pxw & = q + p \\
	\gamma_{\rm w} \Tn - \pxw  & = p \cos{\theta_{\rm th}}	\\
	q  	      & = p \sin{\theta_{\rm th}}. 
	\end{align}
	\end{subequations}
The solution to these equations is
	\begin{subequations}
	\begin{align}
	q & = \frac{ 2 \pxw \gamma_{\rm w} \Tn  }{ \pxw + \gamma_{\rm w} \Tn } \simeq 2\pxw \\
	p & = \frac{ \pxw^{2} + (\gamma_{\rm w} \Tn )^2 }{ \pxw + \gamma_{\rm w} \Tn } \simeq \gamma_{\rm w} \Tn \\
	\sin{\theta_{\rm th} } & = \frac{ 2 \pxw \gamma_{\rm w} \Tn  }{  \pxw^{2} + (\gamma_{\rm w} \Tn )^2 } \simeq \frac{ 2\pxw }{ \gamma_{\rm w} \Tn },
	\end{align}
	\end{subequations}
where we have also given the approximate solution in the limit $\gamma_{\rm w} \Tn \gg \pxw$. Clearly then a small deflection of the incoming bath particle is sufficient to reverse the reflected particle. The minimum momentum transfer squared needed to reverse the momentum of reflected shell particles is then
	\begin{equation}
 \label{eq:t_reversal}
	-\hat{t} = -(p_{1} - p_{3} )^2 = 2 \, p_{1} \cdot p_{3} = 4 \pxw^{2}.
	\end{equation}
Note this is far below the COM energy squared $\hat{s} = 2 p_{1} \cdot p_{2} = 4 \gamma_{\rm w} \Tn \pxw$. The COM momentum is likewise found in the massless limit $\pcm^{2} = \hat{s}/4$.

\begin{figure*}[h!]
\centering
\includegraphics[width=450pt]{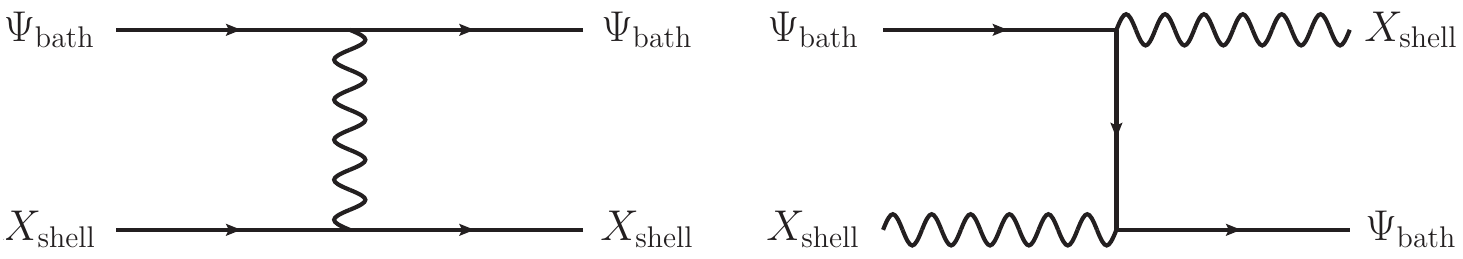}
\caption{ \it \small
M\o{}ller (\textbf{left}) and Compton  (\textbf{right}) scattering between shell and bath particles. 
}
\label{fig:feynmann_diagram}
\end{figure*}

\paragraph{M\o{}ller scattering.} 
We assume a spin-averaged matrix element inspired by $t-$channel gauge boson exchange, as in M\o{}ller scattering in regular fermionic (or scalar) QED, see Fig.~\ref{fig:feynmann_diagram}-left, and consider the leading contribution in the ($\hat{t} \to 0$, $\hat{u}\to -\hat{s}$) limit 
 	\begin{equation}
	|\mathcal{M}|^2 \simeq  \frac{ 128 \pi^2\alphaD^{2} \hat{s}^2}{\hat{t}^2},
	\end{equation}
where we have added a factor $2$ to effectively also account for the ($\hat{u} \to 0$, $\hat{t}\to -\hat{s}$) limit.
This gives us
	\begin{equation}
 \label{eq:Msq_Moller_fermion}
	\frac{d \sigma }{ d \hat{t} } =  \frac{|\mathcal{M}|^2}{16 \pi  \hat{s}^2 } \approx \frac{ 8 \pi \alphaD^{2} }{ \hat{t}^2 }. 
	\end{equation}
For other examples, such as Bhaba scattering, or gluon scattering in an $SU(3)$, we find $\mathcal{O}(1)$ differences, so for practical purposes the same as above.
Other choices would not give an IR enhancement, so the eventual path length before reversal is minimized by an assumption of $t-$channel gauge boson exchange.
All relevant cross sections for the $2\to2$ momentum-loss processes, computed in this Sec.~\ref{sec:mom_loss}, are reported synthetically in Table~\ref{tab:2-2_cross-section}.
The effective cross section to have one scattering impart the necessary momentum exchange is then
	\begin{equation}
	\label{eq:effmol}
	\sigma_{\rm eff} = \int_{-4 \pcm^{2} }^{ - 4\pxw^{2} } d \hat{t} \frac{d \sigma }{ d \hat{t} }   \simeq \frac{2 \pi \alphaD^{2} }{ \pxw^2 }.
	\end{equation}
 The upper boundary is the minimal momentum exchange in order to reverse the momentum of reflected shell particles, Eq.~\eqref{eq:t_reversal}, and the lower boundary is the maximal momentum exchange available.
We use this to find the rate of such scatterings in the wall frame,
	\begin{equation}
	\Gamma_{\rm wall} = n_{\rm bath,w} \sigma_{\rm eff} v_{\text{M\o{}l}} \simeq \frac{4 \gamma_{\rm w} \alphaD^{2} g_{\ast} \zeta(3) \Tn^{3} }{ \pi \pxw^{2} },
	\end{equation}
where we have used the bath particle density in the wall frame 
	\begin{equation}
	n_{\rm bath,w} \simeq  \frac{\gamma_{\rm w} g_{\ast}\zeta(3)\Tn^{3} }{ \pi^{2} },
	\end{equation}
and $v_{\text{M\o{}l}} \simeq 2$. This result leads to an effective path length before reversal
	\begin{equation}
	l_{\rm wall} \simeq \frac{ \pi \pxw^{2} }{ 4 \gamma_{\rm w} \alphaD^{2} g_{\ast} \zeta(3) \Tn^{3} }.
 \label{eq:revmolsimple}
	\end{equation}
The path length in the bath frame is larger by a factor of $l_{\rm bath}=\gamma_{\rm w}l_{\rm wall}$.	
We will eventually also study the effects of multiple soft scatters not taken into account in the above. But we first consider some other possibilities for the matrix element.

\paragraph{Compton scattering in fermion QED.}
In the massless limit, see Fig.~\ref{fig:feynmann_diagram}-right, the polarization and spin averaged matrix element squared reads
	\begin{equation}
 \label{eq:M_compton_QED_fermion}
	|\mathcal{M}|^2  \simeq  -32 \pi^{2} \alphaD^{2} \left( \frac{\hat{s}}{\hat{u}} + \frac{\hat{u}}{\hat{s}} \right).
	\end{equation}
Note the divergence in the limit $\hat{u} = (p_{1} - p_{4})^2 \to 0$. In terms of the scattering angle, this divergence occurs when the gauge boson scatters back directly toward the wall. The divergence is cut off by the finite fermion mass in the propagator. Using $\hat{u} = 2m_{f}^2 - \hat{s} - \hat{t}$, we write the effective cross section as
	\begin{align}
	\sigma_{\rm eff}   \simeq \frac{1}{16 \pi \hat{s}^2 }\int_{-\hat{s} }^{ - m_{f}^2 } d \hat{u} |\mathcal{M}|^2   \simeq  \frac{ 2\pi \alphaD^{2} }{ \hat{s} } \log \left( \frac{ \hat{s} }{ m_{f}^2 } \right) 
	 \simeq \frac{ \pi \alphaD^{2} }{ 2\gamma_{\rm w} \Tn \pxw }  \log \left( \frac{ 4 \gamma_{\rm w} \pxw \Tn}{ \mu^2} \right),
   \label{eq:sigmaefffermionqed}
	\end{align}
where we used $\hat{s} = 4 \gamma_{\rm w} \Tn \pxw$. The quantity $m_{f}^{2} = \mu^2$ is the IR cut-off of the fermion propagator due to thermal and finite-density correction, see last line of Eq.~\eqref{eq:IRcutoff}.
Note the suppression compared with Eq.~\eqref{eq:effmol}. The effective path length in the wall frame is then
	\begin{equation}
 \label{eq:revcompfermsimple}
	l_{\rm wall} \simeq \frac{ \pi \pxw }{ \alphaD^{2} g_{\ast} \zeta(3) \Tn^{2} \log \left( \frac{ 4 \gamma_{\rm w} \pxw \Tn }{ \mu^2 } \right)}.
	\end{equation}
The path length in the bath frame is therefore $l_{\rm bath} =\gamma_{\rm w}l_{\rm wall}  $ which is larger by a factor $\sim \gamma_{\rm w} \Tn/\pxw$ than the equivalent estimate for M\o{}ller scattering in Eq.~\eqref{eq:revmolsimple}.

\paragraph{Compton scattering in scalar QED.}
In the massless limit, the polarization averaged matrix element squared reads 
	\begin{equation}
	|\mathcal{M}|^2  \simeq  64 \pi^{2} \alphaD^{2},
	\end{equation}
and therefore 
	\begin{equation}
	\sigma_{\rm eff} =  \frac{|\mathcal{M}|^2}{16 \pi  \hat{s} } \simeq \frac{ 4 \pi \alphaD^{2} }{ \hat{s} }.
    \label{eq:sigmaeffscalarqed}
	\end{equation}
The effective path length in the wall frame is then
	\begin{equation}
 \label{eq:revcompscalsimple}
	l_{\rm wall} \simeq \frac{ \pi \pxw }{ 2 \alphaD^{2} g_{\ast} \zeta(3) \Tn^{2} }.
	\end{equation}
The path length in the bath frame is therefore $l_{\rm bath} =\gamma_{\rm w}l_{\rm bath}$.

\paragraph{Non-gauged Phase Transition.}

We now consider a case with no gauge bosons present. It is instructive to consider renormalizable scalar self interactions and a yukawa interaction between the scalar and some fermion
    \begin{equation}
    \mathcal{L} \supset \frac{ \mu_{\phi}^2 }{2} \phi^2 + \frac{ \mu_3 }{3!} \phi^{3} + \frac{ \lambda }{ 4!} \phi^{4} + \frac{y}{\sqrt{2}} \phi \bar{f}f + m_{f} \bar{f}f +  \mathrm{H.c.} 
    \label{eq:nongaugePTmodel}
    \end{equation}
In principle, the scalar can also interact with other scalars in the bath, but these effects can easily be taken into account in a similar way as what we do here for the self interactions. The momentum of the reflected radiated scalar is set by $\pxw \approx m_{\phi,t}$, its mass in the broken phase. 

For the $\phi \phi \to \phi \phi$ scattering, and after suppressing $\mathcal{O}(1)$ factors, we have
    \begin{equation}
    |\mathcal{M}|^2 \simeq  \lambda^2 + \lambda \mu_3^2
    \left[\frac{ 1 }{ (\hat{t} - m_{\phi,f}^2) } + \frac{ 1 }{ (\hat{s} - m_{\phi,f}^2)} \right] +  \mu_3^4 \left[\frac{ 1 }{ (\hat{t} - m_{\phi,f}^2) } + \frac{ 1 }{ (\hat{s} - m_{\phi,f}^2)} \right]^2.
        \label{eq:nongaugePTM2scalar}
    \end{equation}
Here the first term comes from the quartic scalar interaction, the third term from the $t-$channel and $s-$channel scalar exchange amplitude squared, and the middle term is the interference between the first and third. The effective scalar mass in the false vacuum, including thermal effects, is denoted $m_{\phi,f}$.
The leading three terms of the effective cross section reads
    \begin{equation}
    \sigma_{\rm eff} \approx \frac{ \lambda^2 }{32 \pi \hat{s} } + \frac{ \lambda \mu_{3}^2 }{ 32 \pi \hat{s}^2 } \log \left( \frac{ \hat{s} }{ 4 \pxw^2 } \right) + \frac{ \mu_{3}^4 }{32  \pi \hat{s}^2 \pxw^2 },
    \label{eq:sigmaeffnongaugescalar}
    \end{equation}
where we have assumed $\pxw \gtrsim m_{\phi,f}$ (otherwise the final two terms are further suppressed by the finite propagator mass).
A natural expectation is $\mu_{3} \sim m_{\phi,t}$ (or smaller in case of an approximate $Z_{2}$ symmetry), from which it immediately follows that the $\lambda^{2}$ term typically dominates, unless $\lambda \lesssim m_{\phi,t}/(\gamma_{\rm w} \Tn)$. As the quartic coupling is also unavoidably present in front of the bubble (the cubic can be suppressed by symmetry reasons in the unbroken phase), we focus on this coupling for simplicity. Then the shell reversal path length in the wall frame is
        \begin{equation}
    l_{\rm wall} \simeq \frac{ 64 \pi^3 \pxw }{  \zeta(3) g_{\ast} \lambda^{2} \Tn^2},
    \end{equation}
which in the bath frame is $l_{\rm bath}=\gamma_{\rm w}l_{\rm wall}$.

For $\phi f \to \phi f$ scattering we have $s-$channel $f$ mediation and $t-$channel $\phi$ mediation, which after dropping $\mathcal{O}(1)$ factors, gives
    \begin{equation}
    |\mathcal{M}|^2 \simeq y^{4} +  \frac{ \mu_{3} y^3 m_f  }{ (\hat{t} - m_{\phi,f}^2) } + \frac{ \mu_{3}^2 y^2 \hat{t} }{ (\hat{t} - m_{\phi,f}^2)^2}.
       \label{eq:nongaugePTM2fermion}
    \end{equation}
Here the first term comes from the $s-$channel amplitude squared, the second from the interference, and the third from the $t-$channel squared.
We thus find
    \begin{equation}
    \sigma_{\rm eff} \simeq \frac{ y^{4} }{ 16 \pi \hat{s} } + \frac{ \mu_{3} m_{f} y^3 }{ 16  \pi \hat{s}^2 } \log \left( \frac{ \hat{s} }{ 4 \pxw^2 } \right) + \frac{ \mu_{3}^2 y^2 }{ 16 \pi \hat{s}^2 } \log \left( \frac{ \hat{s} }{ 4 \pxw^2 } \right),
    \label{eq:sigmaeffnongaugefermion}
    \end{equation}
where $m_{f}$ is the fermion mass in the false vacuum. The first term will typically dominate, unless the Yukawa coupling is very small, $y < \mu^{3} m_{f}/(\gamma_{\rm w} \Tn m_{\phi,t})$, or $y \lesssim \mu_3/\sqrt{\gamma_{\rm w} \Tn m_{\phi, t}}$, in which case the second and/or third terms become important. For brevity, focusing on the large $y$ case, we have a reversal path length in the wall frame
        \begin{equation}
    l_{\rm wall} \simeq \frac{32 \pi^3 \pxw }{  \zeta(3) g_{\ast} y^{4} \Tn^2}.
        \end{equation}   
Translated to the bath frame the length is instead $l_{\rm bath} =\gamma_{\rm w}l_{\rm wall}$, which is of course qualitatively similar to the scalar self interaction and the scalar Compton scattering path lengths.

\subsubsection{Integral method}

In Sec.~\ref{sec:reversal_simple}, we have only considered scattering processes which are ``hard'' enough to reverse of momentum of reflected shells. We now consider the possibility for the momentum of shell particles to be reversed under the effect of a large number of ``soft'' scattering processes.
We again begin by working in the wall frame. From the above discussion, for reversal, we need to change the initial momentum (and energy) $p_{1}$ by an $\mathcal{O}(1)$ factor in the wall frame. Using energy conservation, the change in the gauge boson momentum magnitude is
	\begin{equation}
	\delta \pxw = \delta \EX = - (E_{2} - E_{4}).
	\end{equation}
Our first task is to find $\delta \pxw$ as a function of $\hat{t}$. In the COM frame the momenta are
	\begin{subequations}
	\begin{align}
	p_{1}' &  = ( \pcm, \quad 0, \quad 0, \quad \; \; \; \pcm), \\
	p_{2}' &  = ( \pcm, \quad 0, \quad 0, \quad -\pcm), \\
	p_{3}' &  = ( \pcm, \quad 0, \quad \; \; \; \pcm \sin{\thetacm}, \quad \; \; \; \pcm \cos{\thetacm}) \\
	p_{4}' &  = ( \pcm, \quad 0, \quad -\pcm \sin{\thetacm}, \quad -\pcm \cos{\thetacm}).
	\end{align} 
	\end{subequations}
Here $\hat{s} = 4 \pcm^{2} =   4  \gamma_{\rm w} \Tn \pxw$. 
To go from the wall frame to the COM frame requires a relativistic boost, $v_{\rm boost} \simeq 1$, in the direction of the wall with Lorentz factor
	\begin{equation}
	\gamma_{\rm boost} = \frac{ \pcm }{ 2 E_{1} } = \frac{ E_{2} }{ 2 \pcm } = \frac{1}{2} \sqrt{ \frac{ \gamma_{\rm w} \Tn }{ \pxw } }. 
	\end{equation}
Note we have the relation between the scattering angle in the COM frame and the momentum exchange
	\begin{equation}
	\cos{\thetacm} =  1 + \frac{ \hat{t} }{ 2 \pcm^{2} }
	\end{equation}
Boosting from the COM frame back into the wall frame we have
	\begin{subequations}
	\begin{align}
	E_{2}  & = \gamma_{\rm boost} ( E_{2}' - v_{\rm boost} p_{2z}') =  2 \gamma_{\rm boost} \pcm, \\
	E_{4}  & = \gamma_{\rm boost} ( E_{4}' - v_{\rm boost} p_{4z}')  =  \gamma_{\rm boost} \pcm (1 +  \cos{\thetacm} )  = 2 \gamma_{\rm boost} \pcm \left( 1 + \frac{ \hat{t} }{ 4 \pcm^{2} } \right).
	\end{align}
	\end{subequations}
Using these we find
	\begin{equation}
	\delta \pxw \simeq - \frac{ \hat{t} }{ 4 \pxw }.
	\label{eq:momchangewallframe}
	\end{equation}	
	We have to be careful with the above formula, as it captures the change in the magnitude of the momentum, but we can only reverse a particle once. Thus we should cut-off the weighting for $-\hat{t} \gtrsim 4\pxw^2$, by making the replacement
	\begin{equation}
	\delta \pxw \to \mathrm{Min} \left[ - \frac{ \hat{t} }{ 4 \pxw }, \quad \pxw \right],
	\end{equation}	
in regions of phase space where such hard scatterings occur.
Then the rate to lose an $\mathcal{O}(1)$ fraction of momentum in the wall frame is given by
	\begin{align}
	\frac{1}{\pxw} \frac{ d \pxw }{ dt } & \simeq \frac{ n_{\rm bath,w} v_{\text{M\o{}l}} }{ \pxw } \int_{-4\pcm^{2}}^{-\mu^{2}}  d \hat{t} \frac{d \sigma }{ d \hat{t} } \delta \pxw \label{eq:IBmomloss}.
	\end{align}

\paragraph{M\o{}ller scattering.}
We apply this formula to our M\o{}ller scattering cross section to find
    \begin{subequations}
	\begin{align}
	\frac{1}{\pxw} \frac{ d \pxw }{ dt } & \simeq n_{\rm bath,w} v_{\text{M\o{}l}}   \left [ \int_{-4\pxw^{2}}^{-\mu^{2}} d \hat{t} \frac{ - 2 \pi \alphaD^{2} }{ \pxw^{2}\hat{t} } + \int_{-4\pcm^2}^{-4\pxw^{2}} \frac{8  \pi \alphaD^2 }{\hat{t}^2} d \hat{t} \right] \\
    & \simeq \frac{4 \gamma_{\rm w} \alphaD^{2} \zeta(3) g_{\ast} \Tn^{3} }{ \pi \pxw^{2} } \left[ 1+  \log \left( \frac{  4  \pxw^2 }{ \mu^2 } \right) \right] . 
	\end{align}
    \end{subequations}
The streaming length before reversal in the wall frame is therefore
	\begin{equation}
	l_{\rm wall} \simeq \frac{ \pi \pxw^{2} }{ 4  \gamma_{\rm w} \alphaD^{2} g_{\ast} \zeta(3) \Tn^{3} \left[ 1+  \log \left( \frac{  4  \pxw^2 }{ \mu^2 } \right) \right]}.
 	\label{eq:Molsimplediss}
	\end{equation}
This finally leads us to the streaming length in the bath frame $l_{\rm bath}=\gamma_{\rm w}l_{\rm wall}$.
Note up to the logarithmic suppression factor appearing in the denominator, this is the same estimate as using the simple $\sigma_{\rm eff}$ method. Assuming a thermal mass cut-off  $\mu^2$ as in Eq.~\eqref{eq:IRcutoff}, the suppression is $\sim \log(v_{\phi}^2/\Tn^2)$ in the Abelian case.

\begin{figure}[ht!]
\centering
\raisebox{0cm}
{\makebox{\includegraphics[width=0.49\textwidth, scale=1]{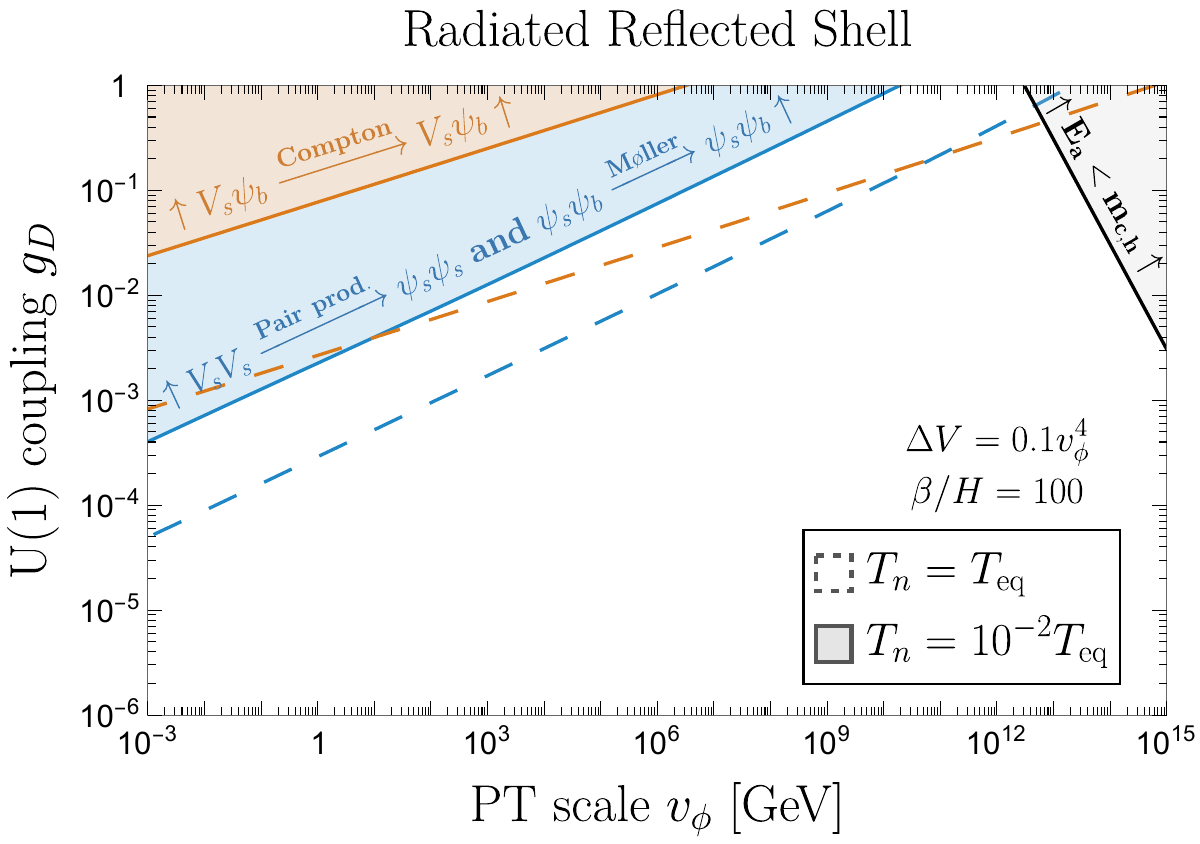}}}
{\makebox{\includegraphics[width=0.49\textwidth, scale=1]{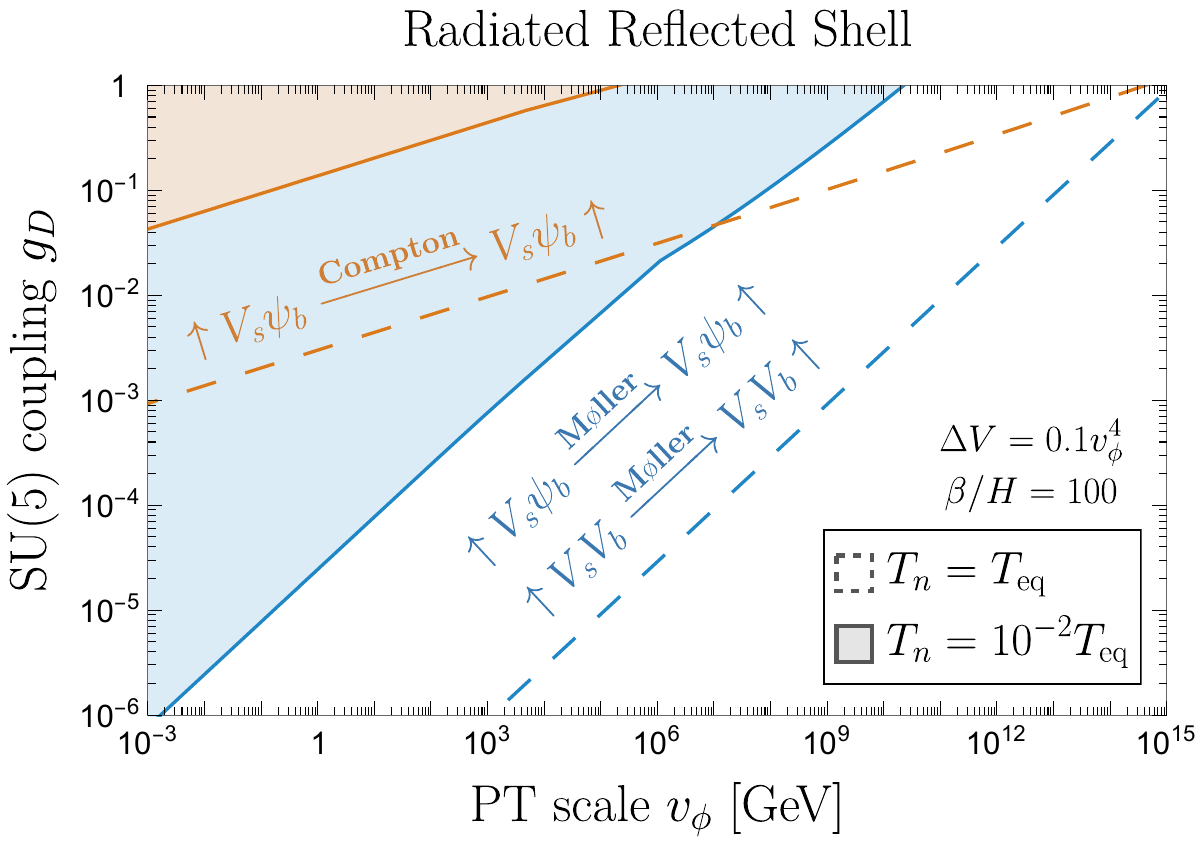}}}
\caption{\it \small   \textbf{Shell reversal:}  regions in which the momentum of shells of reflected radiated bosons is reversed in the wall frame, due to Compton-type (\textbf{orange}, Eq.~(\ref{eq:Compton_lplasma_refersal})) or M\o{}ller-type  (\textbf{blue}, Eq.~(\ref{eq:Moller_lplasma_refersal}) scatterings with bath particles. The symbol $``V"$ denotes vector bosons while $``\psi"$ denotes fermions and scalars. The subscripts $``s"$ and $``b"$ denote particles from the shell and from the bath, respectively. In the Abelian case, M\o{}ller-type scattering can only happen if there are fermions or scalars in the shells, which itself depends on whether $2 \to 2$ pair production processes are active or not, see Eq.~\eqref{eq:2-to-2_scattering}. The region for the scalar self interaction is similar to the Compton scattering region for $\lambda = \gD^2$. We consider reflected shell particles produced from Bremsstrahlung radiation, assuming abelian (\textbf{left}) or non-abelian (\textbf{right}) gauge interaction, see  first two rows  of Tab.~\ref{tab:production_mechanism}, assuming the amounts of supercooling $\Tn/T_{\rm eq} = 1$ (\textbf{dashed}) and $\Tn/T_{\rm eq} = 10^{-2}$ (\textbf{solid}).  We used $g_b = g_\SM + g_\text{emit}$ with $g_\SM = 106.75$ and $g_\text{emit} = 10 = N_F$. \textbf{Grey} shaded areas as in Fig.~\ref{fig:free_stream_all_cases}.
}
\label{fig:shell_reversal} 
\end{figure}

\paragraph{Compton scattering in fermion and scalar QED.}

Here the scatterings are dominantly hard, so the use of the integral method will not change the estimates of reversal path length compared to the effective cross section approach.

\paragraph{Non-gauged PT.} Here the leading terms are also hard scattering dominated, unless $\lambda$ and/or $y$ are sufficiently small to suppress the first terms in the effective cross sections of Eqs.~\eqref{eq:sigmaeffnongaugescalar} and \eqref{eq:sigmaeffnongaugefermion}. So the integral method will not change the results derived above for the reversal path length. The (typically subdominant) $\propto 1/\hat{t}^2$ term in the $\phi \phi \to \phi \phi$ scattering receives a modest $\sim \log( m_{\phi,t}/m_{\phi,f})$ correction. 

\subsubsection{Summary of reversal path lengths}

For M\o{}ller type scattering, \emph{i.e.}~$t-$channel gauge boson exchange, we can use the path length for reversal in the bath frame
	\begin{equation}
 \label{eq:Moller_lplasma_refersal}
	\boxed{ l_{\rm bath} \simeq \frac{ \pi \pxw^{2} }{ 4  \alphaD^{2} g_{\ast} \zeta(3) \Tn^{3} \log \left( \frac{  4  \pxw^2 }{ \mu^2 } \right) } =  \frac{  \pi \px^{2} }{ 4 \gamma_{\rm w}^{2} \alphaD^{2} g_{\ast} \zeta(3) \Tn^{3}  \log \left( \frac{  4  \px^2 }{ \gamma_{\rm w}^2 \mu^2 } \right)}. } 
	\end{equation}
For Compton scattering involving fermions 
	\begin{equation}
  \label{eq:Compton_lplasma_refersal}
 	\boxed{	l_{\rm bath} \simeq \frac{ \gamma_{\rm w} \pi \pxw }{ \alphaD^{2} g_{\ast} \zeta(3) \Tn^{2} \log \left( \frac{ 4 \gamma_{\rm w} \pxw \Tn }{ \mu^2 } \right)} = \frac{  \pi \px }{ \alphaD^{2} g_{\ast} \zeta(3) \Tn^{2} \log \left( \frac{ 4  \px \Tn }{ \mu^2 } \right)}. }
	\end{equation}
And for Compton scattering involving scalars
	\begin{equation}
 	\boxed{	l_{\rm bath} \simeq \frac{ \gamma_{\rm w} \pi \pxw }{2  \alphaD^{2} g_{\ast} \zeta(3) \Tn^{2} } = \frac{  \pi \px }{ 2  \alphaD^{2} g_{\ast} \zeta(3) \Tn^{2} } . }
	\end{equation}
For non-gauged PTs the quartic interaction typically gives the shortest reversal path length although some care must be taken in this case with the precise parameter space and field content, using the methods outlined above. For fermionic interactions with sizable Yukawa couplings, the path length is similar to the quartic interaction, with the replacement $\lambda^2 \to y^4$. We thus write 
    \begin{equation}
    \boxed{ l_{\rm bath} \simeq \frac{ 64  \gamma_{\rm w} \pi^3 \pxw }{  \zeta(3) g_{\ast} (\lambda^{2} + 2  y^{4}) \Tn^2} = \frac{ 64   \pi^3 \px }{  \zeta(3) g_{\ast} (\lambda^{2} + 2 y^{4}) \Tn^2}, }
    \end{equation}
In the above, $\mu^2$ is the gauge boson thermal mass from Eq.~\eqref{eq:IRcutoff}, $\pxw$ is the initial gauge boson momentum in the wall frame, and we reintroduce the momentum in the bath frame, $\px = \gamma_{\rm w} \pxw$, to aid later comparison using Table~\ref{tab:production_mechanism}. Finally in order, say, to check whether the shells meet before particle reversal, one simply compares the above path lengths to the required propagation distances, typically $\sim R_c$ in the bath frame. In Fig.~\ref{fig:shell_reversal} we compare path lengths, for selected shells, to the bubble radius for shell reversal for a shell of radiated and reflected gauge bosons.

\begin{table}[htp!]
\begin{adjustwidth}{-.5in}{-.5in}  
\begin{center}
\centering
\begin{tabular}{ | c | c|c |c|c | c| }
 \hline
 \multicolumn{3}{|c|}{Channel}&
 $d\sigma/d\hat{t}$&\multicolumn{2}{c|}{Application}
\\
 \hline\hline 
 \multicolumn{3}{|c|}{\begin{tabular}[c]{@{}c@{}} \\[-0.8em]
 M\o{}ller scattering\\
 $X_s X_b\to X_s X_b$ \\[0.2em]
 \end{tabular} }
  & \begin{tabular}[c]{@{}c@{}} \\[-0.8em]
 $ 8\pi \alphaD^2/\hat{t}^2 $ \\
 cf. Eq.~\eqref{eq:Msq_Moller_fermion} \\[0.2em]
 \end{tabular} 
   & 
  \multirow{6}{*}{
   \begin{tabular}[|c|]{@{}c@{}}
   \\[1.8em]
   Shell reversal 
  \end{tabular}}& \multirow{6}{*}{
   \begin{tabular}[|c|]{@{}c@{}}
   \\[1.8em]
   Shell dissipation\\
   and metamorphosis
  \end{tabular}}
   \\[1em]
  \cline{1-4}
   \multirow{4}{*}{
   \begin{tabular}[c]{@{}c@{}}
   \\[0.3em]
 Compton \\ scattering 
  \end{tabular}}&   \multicolumn{2}{c|}{\multirow{2}{*}{\begin{tabular}[c]{@{}c@{}} \\[-0.8em]
 Fermion QED\\
 $V_s\psi_b\to V_s \psi_b$ \\[0.5em]
 \end{tabular} }}
  & \multirow{2}{*}{\begin{tabular}[c]{@{}c@{}} \\[-0.8em]
 $ 2\pi \alphaD^2/\hat{s}\hat{u} $ \\
 cf. Eq.~\eqref{eq:M_compton_QED_fermion} \\[0.2em]
 \end{tabular} }
  & & \\[2em]
  \cline{2-4}
  & \multicolumn{2}{c|}{\multirow{2}{*}{\begin{tabular}[c]{@{}c@{}} \\[-0.8em]
  Scalar QED\\
 $V_s\phi_b\to V_s \phi_b$ \\
 \end{tabular} }}
  & \multirow{2}{*}{\begin{tabular}[c]{@{}c@{}} \\[-0.8em]
 $ 4\pi \alphaD^2/\hat{s}^2 $ \\
 cf. Eq.~\eqref{eq:sigmaeffscalarqed} \\[0.2em]
 \end{tabular} }
 &&\\[2em]
    \cline{1-4}
   \multicolumn{3}{|c|}{\begin{tabular}[c]{@{}c@{}} \\[-0.8em]
 Non-gauged scalar\\
 $\phi_s\phi_b\to \phi_s \phi_b$ \\[0.2em]
 \end{tabular} }
  & \begin{tabular}[c]{@{}c@{}} \\[-0.8em]
 $ \lambda^2/32\pi\hat{s}^2  $ \\
 cf. Eq.~\eqref{eq:sigmaeffnongaugescalar} \\[0.2em]
 \end{tabular}
  && \\[1em]
    \hline\hline 
   \multirow{4}{*}{
   \begin{tabular}[c]{@{}c@{}}
   \\[-0.4em]
 Pair \\ production
  \end{tabular}}&  \multirow{2}{*}{\begin{tabular}[c]{@{}c@{}} \\[-0.8em]
 Fermion QED \\[0.5em]
 \end{tabular} }&$V_sV_s\to \psi_s \psi_s$
  & \multirow{2}{*}{
  \begin{tabular}[c]{@{}c@{}} \\[-1.1em]
  $ 8\pi \alphaD^2/\hat{t}\hat{s} $ \\[-0.1em]
 cf. Eq.~\eqref{eq:fermion_QED_annihilating} \\[1em]
 \end{tabular}
 }  & \multicolumn{2}{c|}{\multirow{4}{*}{
   \begin{tabular}[c]{@{}c@{}}
   Thermalization 
  \end{tabular}}}  \\
  \cline{3-3}
  &&$V_sV_b\to \psi_b \psi_s$&&\multicolumn{2}{c|}{}\\
  \cline{2-4}
  &\multirow{2}{*}{\begin{tabular}[c]{@{}c@{}}  \\[-0.8em]
  Scalar QED \\
 \end{tabular} }&$V_sV_s\to \phi_s \phi_s$
  & \multirow{2}{*}{
  \begin{tabular}[c]{@{}c@{}} \\[-1.1em]
 $  2\pi\alphaD^2/\hat{s}^2 $ \\[-0.1em]
 cf. Eq.~\eqref{eq:scalar_QED_annihilating} \\[0.2em]
 \end{tabular}
  }    &\multicolumn{2}{c|}{}\\
  \cline{3-3}
  &&$V_sV_b\to \phi_s \phi_b$&&\multicolumn{2}{c|}{}\\
   \hline
  \end{tabular} 
\end{center}
\end{adjustwidth}
\caption{\it \small List of $ 2\to 2$ processes accounted in this work for the estimation of the momentum exchange between the shell and the bath (shell reversal in Sec.~\ref{sec:shell_reversal} and dissipation/metamorphosis in Sec.~\ref{sec:shell_diss_meta}) and the $2\to 2$ inelastic between shell-shell and shell-bath entering the thermalization conditions in Sec.~\ref{sec:2to2}. For $d\sigma/d\hat{t}$, we give the leading divergent behaviour, in the limit of massless particles. The greater the magnitude of the divergence as $\hat{t} \to 0$ or $\hat{u} \to 0$, the higher the efficiency of the corresponding process (depending also on the required momentum exchange).  The symbol $``V"$ denotes vector bosons, $``\psi"$ denotes fermions, $``\phi"$ denotes scalars, and $``X"$ denotes any charged particles. The subscripts $``s"$ and $``b"$ denote particles from the shell and from the bath, respectively.
}
\label{tab:2-2_cross-section} 
\end{table}

\subsection{Shell dissipation/metamorphosis}
\label{sec:shell_diss_meta}
\subsubsection{Basic picture}

We now consider the shell dissipation/metamorphosis path length. By this we mean any process which changes the momentum of shell particles in the bath frame by an $\mathcal{O}(1)$ factor. Note this could still leave an expanding shell with significant, albeit altered, mean momenta and particle types and number densities.  We shall make additional comments, in the context of specific examples, clarifying the two possibilities when relevant below. (Similar calculations to those below, in the case of non-gauged PTs, have been used in~\cite{Baldes:2022oev}.)

In our calculation of the shell reversal, we were interested whether a shell propagating in front of the bubble wall would remain there, or be sent back into the bubble. Thus the primary interest was for shells of the reflected radiated bosons of transition radiation. The shell dissipation/metamorphosis, however, is relevant not only for the shells considered in the reversal, but also for shells formed behind the bubble wall through the processes summarized in Sec.~\ref{sec:shells_general}. 

\subsubsection{Simple method}
We work in the bath frame and consider $2 \to 2$ scattering between a shell and bath particle. We assume the shell particles are relativistic in the bath frame so that $E_{1} \equiv \EX \simeq \px$. The initial momenta are approximately
	\begin{subequations}
	\begin{align}
	p_{1} &  = ( \px, \quad 0, \quad 0, \quad \px) \\
	p_{2} &  = ( \Tn, \quad 0, \quad 0, \quad -\Tn).
	\end{align} 
	\end{subequations}
The COM energy squared is $\hat{s} \simeq \mX^2 + 4 \px \Tn$ and the COM momentum squared is $\pcm^{2} \simeq 4 \px^2 \Tn^2/\hat{s}$.
To bring the bath particle energy from the bath to the COM frame requires a relativistic boost in the positive $z-$direction with $v_{\rm boost} \simeq 1$ and Lorentz factor $\gamma_{\rm boost} \simeq \pcm/2\Tn \gg 1$. Consider now a scattering between the two particles in the COM frame with a scattering angle $\thetacm$, corresponding to Mandelstam variable $\hat{t} = -2\pcm^{2}(1-\cos \thetacm)$. In the COM frame the four-momenta are
	\begin{align}
	p_{1}' &= ( \pcm,  \quad 0,  \quad  0 ,  \quad \pcm) 	\\
	p_{2}' &= ( \pcm, 		     \quad 0, \quad  0 ,  \quad  -\pcm ) 	\\
	p_{3}' &= ( \pcm,  \quad 0,  \quad  \pcm s_{\thetacm}, \quad \pcm c_{\thetacm} ), \\
	p_{4}' &= ( \pcm,	             \quad 0,  \quad  -\pcm s_{\thetacm},  \quad -\pcm c_{\thetacm} ),
	\end{align}
The change in momentum for the shell particle can be found by boosting back into the bath frame, and is given by
	\begin{equation}
	\delta \px = E_{3} - E_{1} = -\gamma_{\rm boost} v_{\rm boost} \pcm (1-\cos \thetacm) = \frac{ \hat{t} }{4\Tn}.
	\label{eq:dissmomchange}
	\end{equation}
Thus to achieve $\delta \px \approx -\px$ requires rather hard scattering $\hat{t} \approx -4\Tn\px$, of the order of $\hat{s}$. 

\paragraph{M\o{}ller scattering.}
For $t-$channel gauge boson exchange processes, such as M\o{}ller scattering, we have as above, 
 	\begin{equation}
	|\mathcal{M}|^2 \simeq 128 \pi^2\alphaD^{2} \frac{\hat{s}^2}{\hat{t}^2}.
	\end{equation}
and thus
	\begin{equation}
	\frac{d\sigma}{d\hat{t}}=  \frac{|\mathcal{M}|^2}{16 \pi  \hat{s}^2 } \simeq \frac{ 8 \pi \alphaD^{2} }{ \hat{t}^2 }.
    \label{eq:moldissipation}
	\end{equation}
For the hard scattering processes we are interested in for shell dissipation, we use $\hat{t} \approx -4\Tn\px$, so we take an effective cross section
	\begin{equation}
	\sigma_{\rm eff} \simeq \frac{ 2 \pi \alphaD^{2} }{ \px \Tn}.
	\end{equation}
Note the above ignores a suppression factor because the required momentum exchange is close to $\hat{s}$, nevertheless, it should give a suitable estimate up to $\mathcal{O}(1)$ factors. Remembering that we are working in the bath frame, so that the bath number density is
	\begin{equation}
	n_{\rm bath} \simeq \frac{ g_{\ast} \zeta(3) \Tn^{3} } { \pi^{2} }
	\end{equation}
we find the scattering rate 
	\begin{equation}
	\Gamma = n_{\rm bath} \sigma_{\rm eff} v_{\text{M\o{}l}} = \frac{ 4 \alphaD^{2} g_{\ast} \zeta(3) \Tn^{2} }{ \pi \px }.
	\end{equation}
Hence, the dissipation length in the bath frame is 
	\begin{equation}
	l_{\rm bath} = \frac{ \pi \px }{ 4 \alphaD^{2} g_{\ast} \zeta(3) \Tn^{2} }.
	\end{equation} 
It is instructive to compare this to the equivalent calculation for the shell reversal path length, Eq.~\eqref{eq:revmolsimple}, in the case of reflected radiated gauge bosons. Noting $\px = \gamma_{\rm w} \pxw \approx \sqrt{\alphaD} \gamma_{\rm w} v_{\phi}$, we find the path length for dissipation is a factor $\gamma_{\rm w} \Tn /(\sqrt{\alphaD}v_{\phi})$ larger than for reversal. Note the path length is indeed longer as kinematics requires $\gamma_{\rm w} > \sqrt{\alphaD}v_{\phi}/\Tn$ for the gauge bosons to be produced. As the large momentum change required for dissipation/metamorphosis, $\delta \px \approx \px$ given in Eq.~\eqref{eq:dissmomchange}, exceeds the momentum change required for reversal, $\delta \px \approx \px^2/(\gamma_{\rm w}^2 \Tn)$ given Eqs.~\eqref{eq:momchangewallframe} and \eqref{eq:t_reversal}, the former would necessarily imply also the latter. So it is self consistent to have a dissipation/metamorphosis path length longer than or approximately coinciding with reversal. 

\paragraph{Compton scattering in fermion QED.}
The matrix element is given by
	\begin{equation}
	|\mathcal{M}|^{2}  \simeq  -32 \pi^{2} \alphaD^{2} \left( \frac{\hat{s}}{\hat{u}} + \frac{ \hat{u} }{ \hat{s} } \right)	\end{equation}
Thus we have an effective cross-section
	\begin{align}
	\sigma_{\rm eff}  \simeq  \frac{1}{64 \pi \hat{s} \pcm^{2} } \int_{-4\pcm^2}^{-m_f^{2}} d\hat{u} |\mathcal{M}|^{2} \simeq \frac{ \pi \alphaD^{2} }{ 2 \pcm^{2} } \log \left( \frac{ 4 \pcm^{2} }{ m_{f}^{2} } \right)  \simeq  \frac{ \pi \alphaD^{2} \hat{s} }{ 8 \px^{2}  \Tn^{2} } \log \left( \frac{ 16 \px^{2}\Tn^2 }{ \mu^2 \hat{s} } \right), 
    \label{eq:shelldissfermqedsigmaeff}
	\end{align}
where we have assumed an IR cutoff from the fermion thermal mass $m_{f}^{2} \approx \mu^2$ with $\mu$ given by the second line of Eq.~\eqref{eq:IRcutoff}, and used $\pcm^{2} \simeq 4 \px^2 \Tn^2/\hat{s}$.  Note that scattering precisely at the $u-$channel singularity would correspond to replacing a shell particle of one type (say a gauge boson) with a shell particle of another type (say a fermion). This changes the nature of the shell, but one may hesitate to label it as dissipation. Nevertheless, at somewhat more moderate $\hat{t}$, the size of the final state momenta are also significantly altered, so this remains a valid estimate up to logarithmic factors. The above leads to a dissipation length in the bath frame of 
	\begin{equation}
	l_{\rm bath}  \simeq  \frac{  \pi \px  }{ \alphaD^{2} g_{\ast} \zeta(3)  \Tn^2 \log \left( \frac{4 \px \Tn}{\mu^2} \right) }.
	\end{equation} 
Again we compare with the reversal path length in the case of reflected radiated gauge bosons, Eq.~\eqref{eq:revcompfermsimple}, and now find it coincides with the dissipation path length. This is due to the $u$-channel singularity, which means hard scatterings dominate up to the IR cutoff in the effective particle mass. Physically this makes sense provided we treat the derived lengths as approximate up to $\mathcal{O}(1)$ factors. 
	
\paragraph{Compton scattering in scalar QED.}
As before, for Compton scattering in massless scalar QED we have
	\begin{equation}
	|\mathcal{M}|^2  \simeq  64 \pi^{2} \alphaD^{2},
	\end{equation}
and therefore 
	\begin{equation}
	\sigma_{\rm eff} \simeq \frac{4 \pi \alphaD^{2} }{ \hat{s} }.
    \label{eq:sigmaeffscalarqeddiss}
	\end{equation}
As in the fermion QED case, the scatterings of interest are dominantly hard, but there is now no $\hat{u} \to 0$ singularity which makes interpretation easier. Accordingly, the dissipation path length in the bath frame is
	\begin{equation}
	l_{\rm bath}  \simeq  \frac{ \pi \px }{ 2 \alphaD^2 g_{\ast} \zeta(3) \Tn^{2} }.
	\end{equation} 
Making the comparison to the reversal path length in the case of reflected radiated gauge bosons, Eq.~\eqref{eq:revcompscalsimple}, we find the path lengths coincide. Thus the scatterings which reverse the shell in the wall frame, also change its constituent particle momenta by $\mathcal{O}(1)$ factors in the scalar QED case.

\paragraph{Non-gauged PT.}
We again study this case by assuming interactions of the form in Eq.~\eqref{eq:nongaugePTmodel}. For the $\phi\phi \to \phi \phi$ scattering, using the approximate matrix element squared, Eq.~\eqref{eq:nongaugePTM2scalar}, we find 
    \begin{equation}
    \sigma_{\rm eff} \simeq \frac{1}{32 \pi \hat{s}} \left( \lambda^2 + \frac{ \lambda \mu_{3}^2 }{ \hat{s}} +  \frac{ \mu_{3}^4}{ \hat{s}^2} \right).
    \end{equation}
This gives a dissipation path length in the bath frame of
    \begin{equation}
   l_{\rm bath}  \simeq  \frac{ 64  \pi^3 \px }{ g_{\ast} \zeta(3) \Tn^2 \left( \lambda^2 + \frac{ \lambda \mu_{3}^2 }{ 4\Tn\px} +  \frac{ \mu_{3}^4}{ 16\Tn^2\px^2} \right)}
    \end{equation}
In the case of a hard scattering dominated process (the leading term given by $\lambda^2$), the dissipation and reversal path lengths coincide.  

Turning now to scattering with fermions,  $\phi f \to \phi f $, using Eq.~\eqref{eq:nongaugePTM2fermion}, we have 
    \begin{equation}
    \sigma_{\rm eff} \simeq \frac{1}{16 \pi \hat{s}} \left( y^{4} + \frac{ \mu_{3} y^{3} m_{f} }{ \hat{s} } + \frac{ \mu_{3}^2 y^{2} }{ \hat{s} } \right).
    \end{equation}
Note we do not integrate to small $-\hat{t}$, as we require large momentum shifts, and so do not find any logarithmic enhancements for the $t$-channel mediated process, as we did for the shell reversal).
The effective cross section gives a bath frame dissipation path length 
    \begin{equation}
   l_{\rm bath}  \simeq  \frac{ 32  \pi^3 \px }{ g_{\ast} \zeta(3) \Tn^2 \left( y^{4} + \frac{ \mu_{3} y^{3} m_{f} }{ 4\Tn\px } + \frac{ \mu_{3}^2 y^{2} }{ 4\Tn\px } \right)}
    \end{equation}
Again the dissipation and reversal path lengths coincide for the hard scattering dominated process (governed by the $y^4$ term). 
    
\subsubsection{Integral method}

For our $2 \to 2$ scattering, we see from Eq.~\eqref{eq:dissmomchange} that the momentum change of the shell particle in the bath frame is given by $\delta \px = -\hat{t}/4\Tn$. To take into account the possibility of a large number of soft scatterings adding up to give a momentum change of order $\px$, we can obtain an estimate of the path length by using an integral, as in Eq.~\eqref{eq:IBmomloss}. The only difference is that here we are working in the bath frame. Furthermore, we do not need to cut-off the change in momentum weighting, as hard scatterings correspond to $\delta \px \approx \px$, and not more. Accordingly, the momentum loss rate is given by
	\begin{equation}
	\frac{1}{\px} \frac{ d \px }{ dt }  \simeq  \frac{ n_{\rm bath} v_{\text{M\o{}l}} }{ \px } \int_{-4\pcm^{2}}^{-\mu^{2}} d\hat{t} \frac{ d\sigma}{d\hat{t}} \delta \px  \simeq  -\frac{ n_{\rm bath} v_{\text{M\o{}l}} }{ 4\px \Tn } \int_{-4\pcm^2}^{-\mu^{2}} d\hat{t} \frac{ d\sigma}{d\hat{t}} \hat{t}.
	\end{equation}

\paragraph{M\o{}ller scattering.}
We apply the above to $t-$channel gauge boson exchange, using Eq.~\eqref{eq:moldissipation}, and obtain
	\begin{equation}
	\frac{ d \log \px }{ dt } \simeq \frac{4  \alphaD^{2} \zeta(3) g_{\ast} \Tn^{2} }{ \pi \px } \log \left( \frac{4 \px T_n }{ \mu^2  } \right). 
	\end{equation}
Thus the dissipation path length in the bath frame is given by
	\begin{equation}
	l_{\rm bath}  \simeq  \frac{ \pi \px }{4  \alphaD^{2} \zeta(3) g_{\ast} \Tn^{2} \log \left( \frac{4 \px \Tn}{ \mu^2  } \right)},
	\end{equation}
which is logarithmically suppressed compared to the simple estimate, Eq.~\eqref{eq:Molsimplediss}, and replaces it as our preferred approximation. In the case of reflected radiated gauge boson shells, the ratio of dissipation to reversal path lengths remains the same when using the simple of integrated estimates, up to differences in the logarithmic factor. That is, up to logarithmic corrections, the dissipation path length is a factor of $\gamma_{\rm w} \Tn /(\sqrt{\alphaD} v_{\phi})$ longer than the reversal path length.

\paragraph{Compton scattering in scalar and fermion QED.}
As for the shell reversal, these dissipation path lengths are determined by hard scatterings, so the simple estimates are unchanged by use of the integral method. 

\paragraph{Non-gauged PT.}
In the $\phi \phi \to \phi \phi$ process we have a $1/\hat{t}^2$ term which receives a logarithmic correction. We find
    \begin{equation}
   l_{\rm bath} \simeq  \frac{64  \pi^3 \px }{ g_{\ast} \zeta(3) \Tn^2 \left[ \lambda^2 + \frac{ \lambda \mu_{3}^2 }{ 4\Tn\px} +  \frac{ \mu_{3}^4}{ 16\Tn^2\px^2} \log \left( \frac{ 4\Tn\px }{\mu^2} \right) \right]},
    \end{equation}
where $\mu^2 = m_{\phi,f}^2$ or $\mu^2 = m_{\phi,t}^2$ depending on whether the scattering is occurring in the symmetric or broken phases respectively. The path length for $\phi f \to \phi f$ does not receive any such correction.

\begin{figure}[p]
\centering
\vspace{-1.5cm}
{\makebox{\includegraphics[width=0.49\textwidth, scale=1]{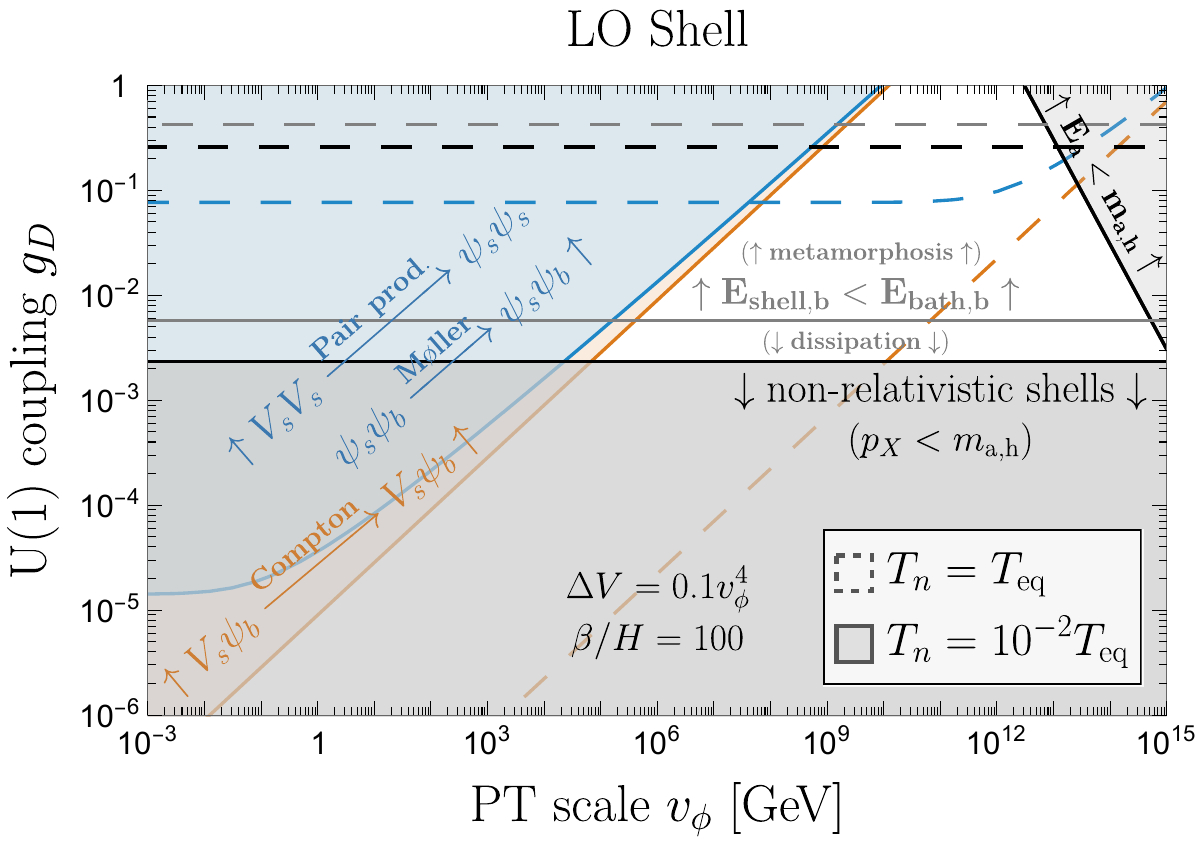}}}
{\makebox{\includegraphics[width=0.49\textwidth, scale=1]{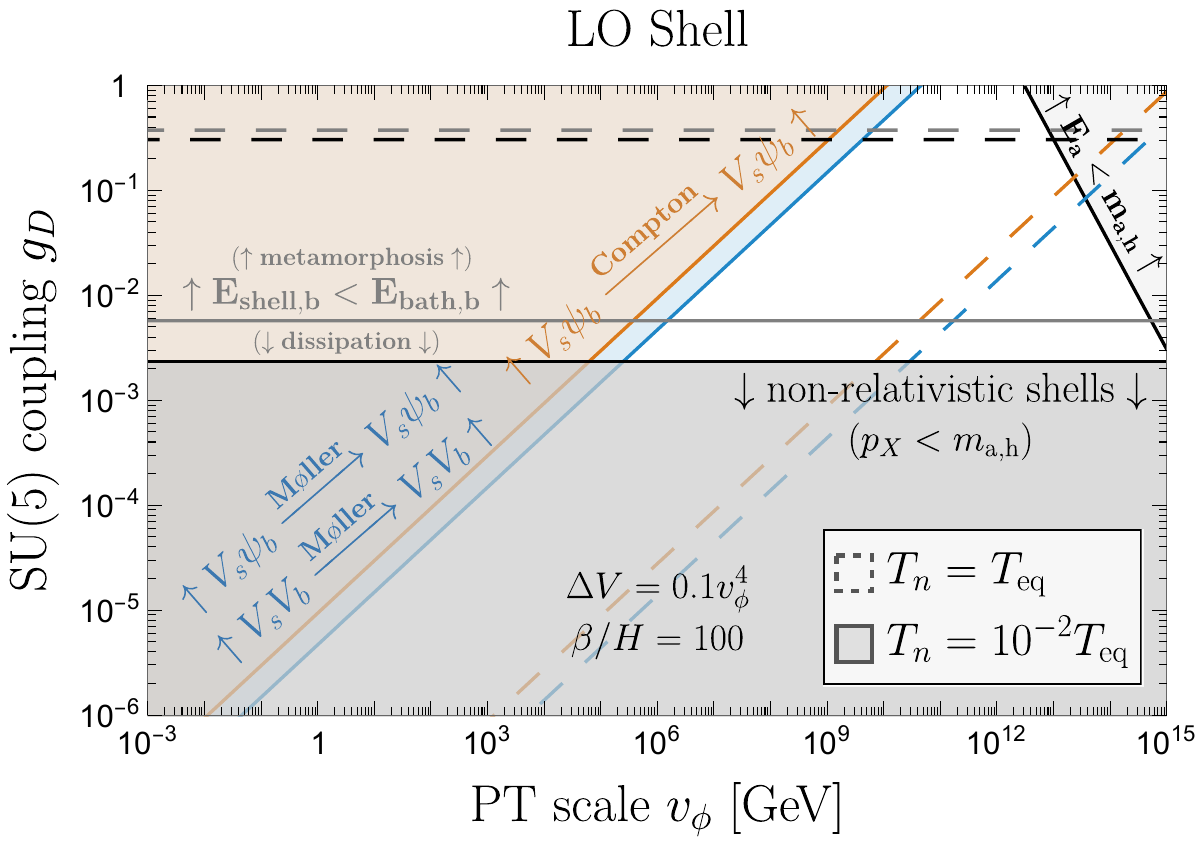}}}
{\makebox{\includegraphics[width=0.49\textwidth, scale=1]{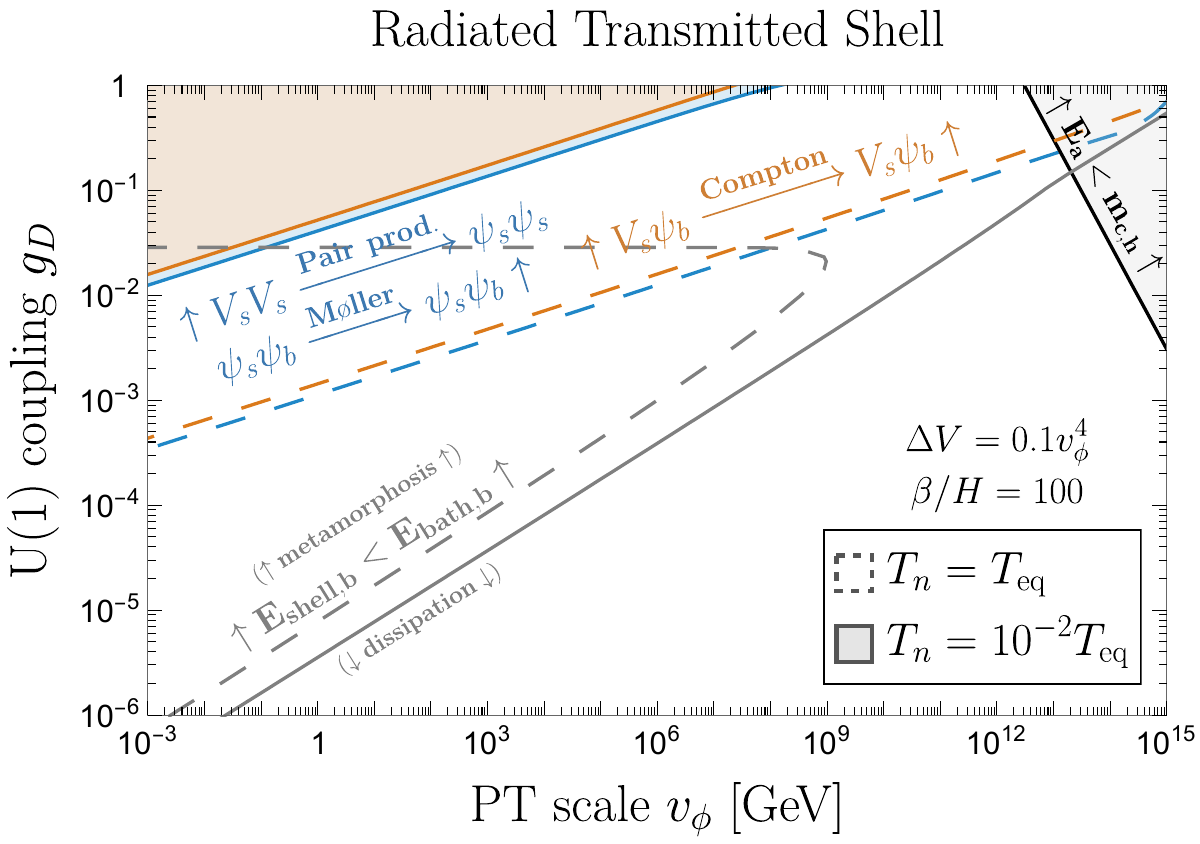}}}
{\makebox{\includegraphics[width=0.49\textwidth, scale=1]{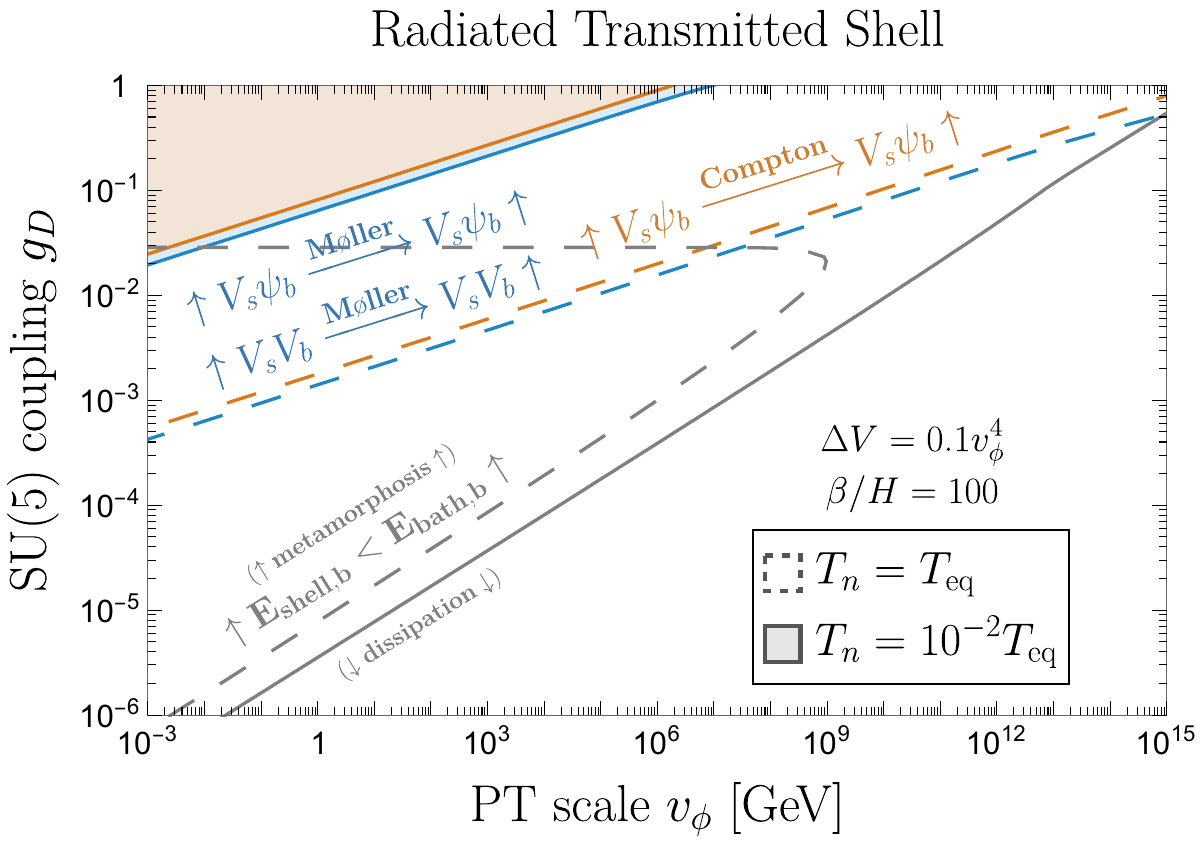}}}
{\makebox{\includegraphics[width=0.49\textwidth, scale=1]{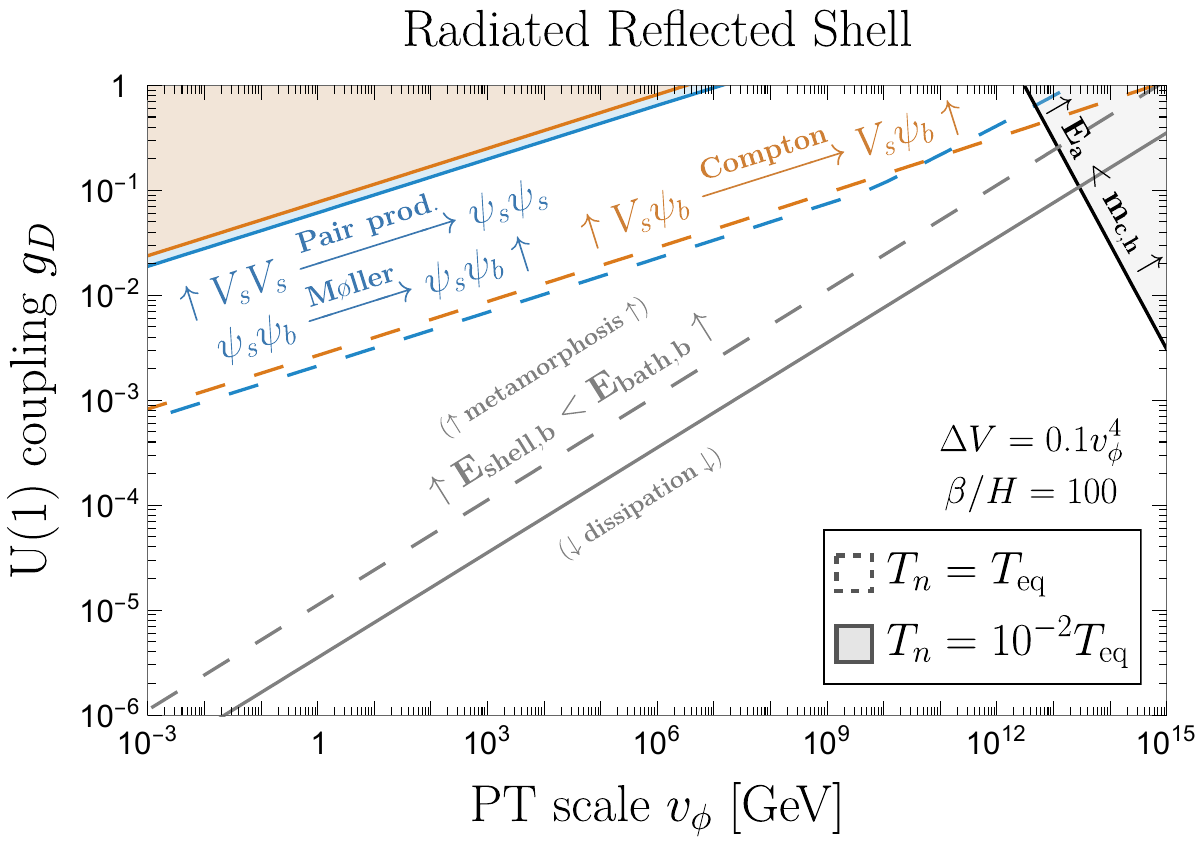}}}
{\makebox{\includegraphics[width=0.49\textwidth, scale=1]{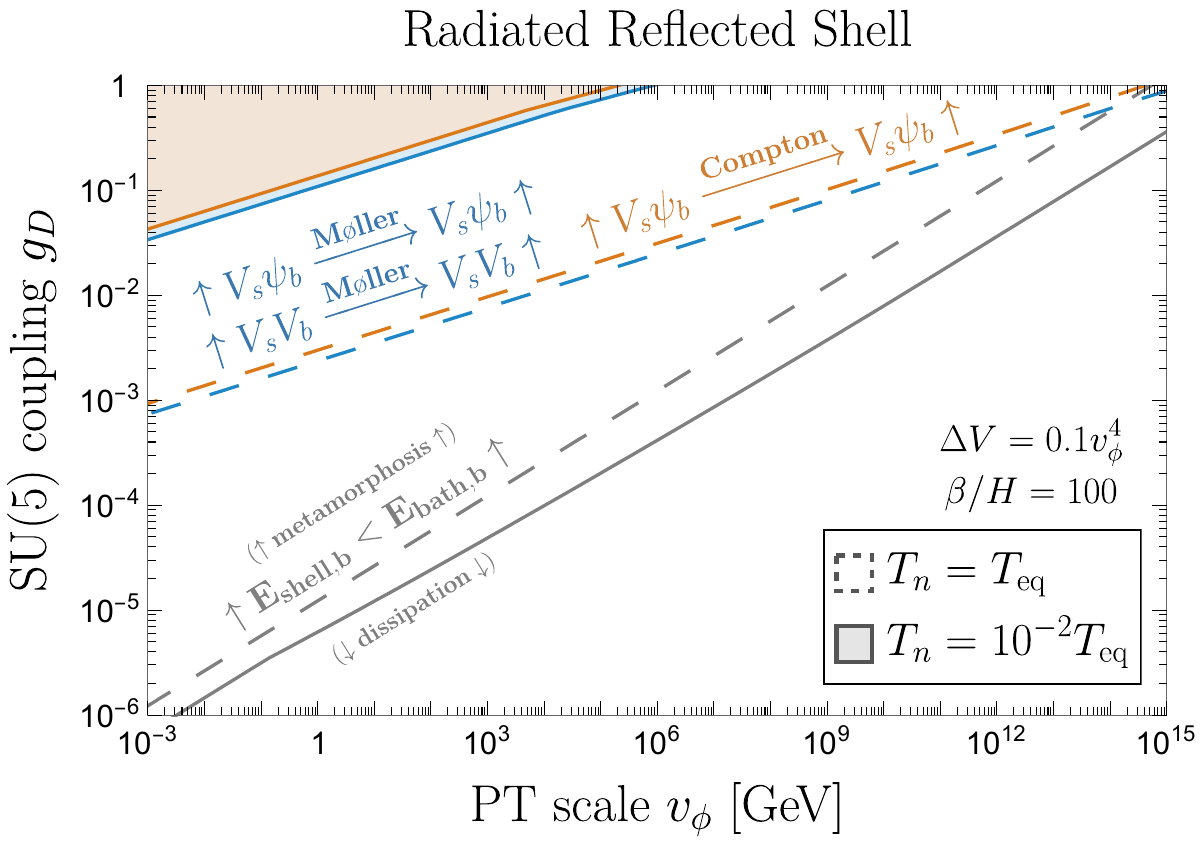}}}
\caption{\it \small   \textbf{Shell dissipation and metamorphosis:} in the  \textbf{blue} and \textbf{orange} regions, the shell particles undergo $\mathcal{O}(1)$ changes to their momentum in the bath frame, due to M\o{}ller (see Eq.~\eqref{eq:l_meta_moller}) and Compton (see Eq.~\eqref{eq:l_meta_compton}) scattering interactions with the bath, respectively. (The region for the scalar self interaction is very similar to the Compton scattering one for $\lambda = \gD^2$.) 
On one side of the \textbf{gray} lines, in the bath frame there is more energy in the shell than in the bath suggesting that it is energetically possible for the momentum of bath particles to be reversed in the bath frame (\textbf{shell metamorphosis}).  On the other side of the gray lines, the bath contains enough energy to dissipate the momentum of shell particles in the bath frame (\textbf{shell dissipation}). 
It is important to emphasize that the gray lines solely represent energetic considerations, and that they are relevant only when they fall within the blue and orange regions. Only within the shaded regions momentum loss is efficient, and momentum is lost via metamorphosis or dissipation depending on whether one is on one side or the other of the gray line.
The symbol $``V"$ denotes vector bosons while $``\psi"$ denotes fermions and scalars. The subscripts $``s"$ and $``b"$ denote particles from the shell and from the bath, respectively. In the Abelian case, M\o{}ller-type scattering can only happen if there are fermions or scalars in the shells, which itself depends on whether $2 \to 2$ pair production processes are active or not, see Eq.~\eqref{eq:2-to-2_scattering}.
We consider shell particles produced from $LO$ interaction (\textbf{top}), Bremsstrahlung radiation, either transmitted (\textbf{middle}) or reflected (\textbf{bottom}), assuming abelian (\textbf{left}) or non-abelian (\textbf{right}) gauge interaction, see  first two rows of Table~\ref{tab:production_mechanism}, assuming the amounts of supercooling $\Tn/T_{\rm eq} = 1$ (\textbf{dashed}) and $\Tn/T_{\rm eq} = 10^{-2}$ (\textbf{solid}).  We used $g_b = g_\SM + g_\text{emit}$ with $g_\SM = 106.75$ and $g_\text{emit} = 10 = N_F$. We fixed the PT completion rate $\beta/H=20$ to maximise the GW signal even though the results of this paper are only weakly dependent on this value.
\textbf{Grey} and \textbf{black} shaded areas as in Fig.~\ref{fig:free_stream_all_cases}.}
\label{fig:shell_meta}  
\end{figure}

\subsubsection{Summary of dissipation/metamorphosis lengths}
We now summarize the dissipation path lengths due to $2 \to 2$ interactions between shell and bath particles. When the shell interacts with the bath via $t-$channel gauge boson exchange the dissipation path length in the bath frame is approximately 
	\begin{equation}
 \label{eq:l_meta_moller}
	\boxed{ l_{\rm bath}  \simeq  \frac{ \pi \px }{4   \alphaD^{2} \zeta(3) g_{\ast} \Tn^{2} \log \left( \frac{4 \px \Tn}{ \mu^2  } \right)}, } \qquad
	\end{equation}
 as in M\o{}ller scattering. For Compton scattering with fermions we obtain
	\begin{equation}
  \label{eq:l_meta_compton}
	\boxed{ l_{\rm bath}  \simeq  \frac{  \pi \px  }{ \alphaD^{2} g_{\ast} \zeta(3)  \Tn^2 \log\left( \frac{4 \px \Tn}{\mu^2} \right) }, }
	\end{equation}
and for Compton scattering with scalars we obtain
	\begin{equation}
	\boxed{ l_{\rm bath}  \simeq  \frac{ \pi \px }{ 2  \alphaD^2 g_{\ast} \zeta(3) \Tn^{2} }. }
	\end{equation}
For non-gauged phase transitions, the leading effect will typically be given by
        \begin{equation}
   \boxed{ l_{\rm bath} \simeq  \frac{ 64  \pi^3 \px }{ g_{\ast} \zeta(3) \Tn^2  (\lambda^2 + 2 y^{4}) }, }
    \end{equation}
coming from the quartic scalar interactions or Yukawa interactions. Once a more precise field content and parameter space is defined, the techniques developed above can be of course used to check precisely the dominant contribution to the dissipation path length. 
We remind the reader that in these expressions, $\px$ is the shell particle momentum as measured in the bath frame, as found in Table~\ref{tab:production_mechanism}, and $\hat{s} \simeq 4 \px \Tn$.
Parameter space in which the shell is dissipated/metamorphosised for LO shells and radiated gauge bosons are shown in Fig.~\ref{fig:shell_meta}. Also shown is the energy condition, Eq.~\eqref{eq:Rshelldiss}, delineating the regions in which there is or is not sufficient energy in the bath for complete shell dissipation.

\section{Thermalization}
\label{sec:3to2}

While the thermal bath on its own is a system in thermal equilibrium, the particles in the shell are strongly boosted and strongly compressed. Interactions between the thermal bath and the shell, as well as interactions within the shell itself, give rise to out-of-equilibrium processes, which --- if happening sufficiently fast --- can lead to a new state of equilibrium, with the particles' energies and densities different with respect to the free-streaming case.

In Sec.~\ref{sec:2to2}, we will first consider inelastic $2 \to 2$ processes, or equivalently $2 \to 2$ annihilations, which change the type of particles present but not the overall number. This can occur via either shell-bath or internal shell-shell interactions. In Secs.~\ref{sec:3to2ratesa} -- \ref{sec:3to2ratesc}, we then go on to the more challenging problem and consider $2 \to 3$ and $3 \to 2$ type interactions, which also change the total number of shell particles.

\subsection{Processes changing individual but not total numbers}
\label{sec:2to2}

\subsubsection{Inelastic $2 \to 2$ processes between the shell and bath}
\label{sec:inelastic2to2}
Let us first consider $2\to 2$ annihilations where one initial particle belongs to the shell and the other one belongs to the bath. We anticipate that their inclusion does not result in stronger constraints on free-streaming, than the momentum-loss processes considered in Sec.~\ref{sec:mom_loss} (where we took into account elastic $2 \to 2$ processes between the shell and the bath). We give reasoning for this below.

The momentum loss could, in principle, proceed in two qualitatively different manners:
\begin{itemize}
\item[$\diamond$] First of all, the addition of inelastic scatterings between the shell and the bath simply increases the number and type of interactions and therefore reduces the free streaming length. This effect can be captured by an $\mathcal{O}(1)$ enhancement of the total effective cross section for momentum loss --- unless there is also an additional soft enhancement --- which could lead to qualitatively more efficient scattering (e.g.~giving  M\o{}ller type scattering $|\mathcal{M}|^2 \propto \hat{s}^{2}/\hat{t}^{2}$ when such a divergence was previously absent).    
\item[$\diamond$] Secondly, a two step process: efficient conversion via inelastic scattering to charged particles, followed by M\o{}ller type scattering. In particular, we saw in Sec.~\ref{sec:shell_reversal}, that for shell reversal, elastic scattering at $-\hat{t} \ll \hat{s}$ played an important role in the case of $t-$channel gauge boson exchange (due to the singular behaviour $|\mathcal{M}|^2 \propto \hat{s}^{2}/\hat{t}^{2}$). Thus, for shell reversal, there is a strong sensitivity to the composition of the shell, i.e.~whether it contains charged particles. We therefore need to check whether the consideration of inelastic processes between the shell and the bath (which could perhaps be enhanced at $-\hat{t} \ll \hat{s}$) could change the shell composition, and therefore also the type of scatterings leading to shell reversal, i.e.~M\o{}ller instead of Compton scattering.
\end{itemize}
We now check these two possibilities. In both cases, the interactions between the shell and bath occur at high $\sqrt{s}$ compared to broken phase particle masses, so we can work in the massless limit. Let us consider a shell of gauge bosons encountering gauge bosons of the bath. The polarization averaged matrix element in scalar QED for $V + V \to \phi + \phi^{\ast}$ is given by 
 	\begin{equation}
  \label{eq:scalar_QED_annihilating}
	|\mathcal{M}|^2  \simeq  32 \pi^{2} \alphaD^{2},
	\end{equation}
Resulting in a cross section 
    \begin{equation}
    \sigma v_{\text{M\o{}l}} =  \frac{|\mathcal{M}|^2}{8\pi \hat{s}}=  \frac{ 4\pi\alphaD^2 }{ \hat{s} },
    \end{equation}
where we used that $v_{\text{M\o{}l}}=2$.
The polarization averaged matrix element in fermion QED for $V + V \to f + \bar{f}$ is given by
        \begin{equation}
    \label{eq:fermion_QED_annihilating}
      |\mathcal{M}|^{2} = 64 \pi^{2} \alphaD^2 \left( \frac{ \hat{t} }{ \hat{u} } + \frac{ \hat{u} }{ \hat{t}} \right).
    \end{equation}
Giving an effective cross section 
    \begin{equation}
    \sigma v_{\text{M\o{}l}}  \simeq \frac{1}{8 \pi \hat{s}^2 }\int_{-\hat{s} }^{ - \mu^2 } d \hat{u} |\mathcal{M}|^2 = \frac{  16 \pi\alphaD^2 }{ \hat{s} } \log \left( \frac{ \hat{s} }{ \text{Max}[\mu^2, -\hat{t}_{\rm min}]} \right),
    \end{equation}
where $\mu^2$ is the IR cutoff from the effective plasma mass, and $-\hat{t}_{\rm min}$ is the minimum momentum exchange relevant for the process under consideration. We now comment on what enters the log factor depending on the process under consideration. We have $-\hat{t}_{\rm min} \sim \hat{s} \approx 4 T_{n} \pX $ for shell dissipation/metamorphosis from Eq.~\eqref{eq:dissmomchange}, $-\hat{t}_{\rm min} \sim 4\pxw^2 \ll \hat{s}$ for shell reversal from Eq.~\eqref{eq:t_reversal}, and $-\hat{t}_{\rm min} \sim \mu^2$ if we are simply interested in changing the composition of the shell.
All relevant cross sections for the $2\to2$ annihilation processes, computed in this Sec.~\ref{sec:2to2}, are reported synthetically in Table~\ref{tab:2-2_cross-section}.

Having found the cross sections, now let us consider the effect of the inelastic scatterings on shell momentum loss. For the scalar QED case, there is obviously no singular behaviour, and the inelastic cross section behaves the same way as the elastic one we previously saw in Eqs.~\eqref{eq:sigmaeffscalarqed} and \eqref{eq:sigmaeffscalarqeddiss}, and used in Sec.~\ref{sec:mom_loss}. Also the shell composition is only altered at the same rate as momentum loss, so there is no change, up to $\mathcal{O}(1)$ factors, due to the two-step process.

For the fermion QED case, we have $-\hat{t}, -\hat{u} \to 0$ singularities for $VV\to f\bar{f}$. But the same type of singularity has already been taken into account in Eqs.~\eqref{eq:sigmaefffermionqed} and \eqref{eq:shelldissfermqedsigmaeff} for elastic scattering in Sec.~\ref{sec:mom_loss}. Thus there is no enhanced momentum loss for shells of abelian gauge bosons in either the scalar or fermion QED cases. Moreover, all the constrained regions in Sec.~\ref{sec:mom_loss} remain unchanged, up to $\mathcal{O}(1)$ factors, when including inelastic processes between the shell and the bath.
We therefore do not display these regions in our plots, to avoid clutter.

The above discussion has focused on shells of abelian gauge bosons. Let us argue that the same conclusions hold for other type of shells.
First of all, for shells of non-abelian gauge bosons, charged fermions, or charged scalars --- which all scatter via gauge boson exchange --- we have already taken into account the soft enhancement in the matrix elements leading to momentum loss, $|\mathcal{M}|^2 \propto \hat{s}^{2}/\hat{t}^{2}$, in our previous discussion in Sec.~\ref{sec:mom_loss} (in general both M\o{}ller and Compton scatterings take place for such shells). Secondly, if we have a non-gauged PT, there exist divergences of the type $|\mathcal{M}|^2 \propto m^{4}/\hat{t}^2$ from Eqs.~\eqref{eq:nongaugePTM2scalar}, and $|\mathcal{M}|^2 \propto m^{2}/\hat{t}$ from \eqref{eq:nongaugePTM2fermion} ($m$ is some mass dimension one term), for which we have already accounted (amongst others, these are the most singular). But the addition of inelastic interactions will not give anything further enhanced, such as $|\mathcal{M}|^2 \propto \hat{s}^{2}/\hat{t}^{2}$ (where one has $\hat{s} \gg m^2$), simply because by definition no gauge bosons are involved.
Thus our conclusions of  Sec.~\ref{sec:mom_loss} remain also for other types of shells.

\subsubsection{Inelastic $2 \to 2$ processes within the shell}

Let us now consider inelastic $2\to2$ processes within the shell, i.e.~both initial annihilating particles belong to the same shell.
If efficient, these processes alter the identities of particles making up the shells. For example shells made of radiated reflected gauge bosons would become made of gauge bosons plus symmetry-breaking scalars plus, if they exist in the spectrum, light fermions charged under the gauge group, with relative abundances set by the respective numbers of internal degrees of freedom.

Even if efficient, however, $2\to 2$ processes starting with two shell particles do not alter the typical momenta of shells particles, $\pX$, nor their overall number densities, reported in Table~\ref{tab:production_mechanism}. Therefore, we will eventually include such $2\to 2$ equilibrated shells within our definition of free-streaming shells, so that we will not display the areas where $2\to 2$ processes are efficient in the summary Fig.~\ref{fig:free_stream_all_cases}. This is in contrast with the $3\to2$ processes studied below which alter both $\pX$ and the overall shell's number density. 

The caveat regarding the $2 \to 2$ processes and momentum loss discussed in Sec.~\ref{sec:inelastic2to2}, however, should be kept in mind. The two-step momentum loss process. In the case of radiated and reflected abelian gauge boson shells, we must also check whether $2\to 2$ processes within the shell allow for shell momentum-loss via: (i) conversion to charged scalars/fermions, (ii) which then undergo more efficient M\o{}ller scattering with incoming charged bath particles. 

Having made this specification, we calculate the rate of $2 \to 2$ processes within the shells, e.g.~annihilation of vector bosons into scalar bosons or fermions. If the probability that such processes occur before collision is larger than unity, then we should expect the particles composition of the shell (but not the shell's overall number density) to reach relative chemical equilibrium.
In the opposite case,
\begin{equation}
\label{eq:2-to-2_scattering}
\boxed{
    P_{2\to2} = n_{X,b} \, \sigma_{2\to 2}v_{\text{M\o{}l}} \, R_{c}~\ll ~1,
    }
\end{equation}
the composition of shell remains conserved during its propagation. The quantity $n_{X,b}$ is the shell number density given in Eq.~\eqref{eq:shelldensity1} and the bubble radius, $R_{c}$, is the effective distance over which the shell particles propagate before collision.
For definiteness, we consider the process of vector bosons annihilation into scalar bosons or fermions.
Its cross-section $\sigma$ multiplied by the M{\o}ller velocity $v_{\text{M\o{}l}} $, in the bath frame, can be approximated with naive-dimensional analysis as 
\begin{equation}
    \sigma_{2\to 2}v_{\text{M\o{}l}} 
    \simeq \frac{\gD^4}{16 \pi} \frac{1}{\pX^2}.
    \label{eq:sigma_2to2}
\end{equation}
For example, Eq.~(\ref{eq:sigma_2to2}) is indeed the result one obtains by considering the annihilation of vector bosons into a pair of scalars driving the PT.

In Fig.~\ref{fig:3to2}, we show in blue the regions where inelastic  $2 \to 2$  shell-shell (labelled `ss') interactions become efficient. As shells still free stream even if such processes are efficient, as we have argued in the beginning of this section, we leave their more complete study to future work. 

We can now also address the two-step shell momentum-loss process with the initial annihilation taking place within the shell. 
Let us first consider shell reversal. Comparing the `ss' vector boson annihilation region of Fig.~\ref{fig:3to2}, with the Compton scattering region of Fig.~\ref{fig:shell_reversal}, it turns out there are regions in which annihilation is efficient, but where reversal via Compton scattering is not. Therefore --- provided the annihilation into charged scalars or fermions is kinematically allowed --- there exist regions of parameter space in which radiated and reflected shells of gauge bosons can convert into charged particles, which are then efficiently reversed via M\o{}ller scattering back to the wall.
Thus, when such annihilations are efficient, then processes of shell reversal are important also for abelian reflected shells, and we shade the associated region both in Fig.~\ref{fig:shell_reversal} and in Fig.~\ref{fig:free_stream_all_cases}.
We obtain those regions by requiring that both the anihilation $VV \to \phi\phi^{\ast}$ (with the cross section given in Eq.~(\ref{eq:sigma_2to2})) and the momentum loss induced by $\phi \phi \to \phi \phi$ (where one initial $\phi$ belongs to the shell and one to the bath, and where we use $d\sigma/d\hat{t}$ from Eq.~(\ref{eq:Msq_Moller_fermion})) are efficient at the same time.
Coming finally to shell dissipation and metamorphosis, the one implied by the two-step process leads to no-free-streaming regions which overlap with those derived in Sec.~\ref{sec:shell_diss_meta}, as we show in Fig.~\ref{fig:shell_meta}. This is because the shell dissipation/metamorphosis requires significant momentum exchange, $-\hat{t}_{\rm min} \sim \hat{s}$, so differences between M\o{}ller and Compton scattering are negligible.


\subsection{Number changing processes: setup}
\label{sec:3to2ratesa}

In the rest of this section we consider those processes that involve number changing-interactions, see Fig.~\ref{fig:shelL_production}.
\begin{figure}[!ht]
\centering
\raisebox{0cm}{\makebox{\includegraphics[width=0.45\textwidth, scale=1]{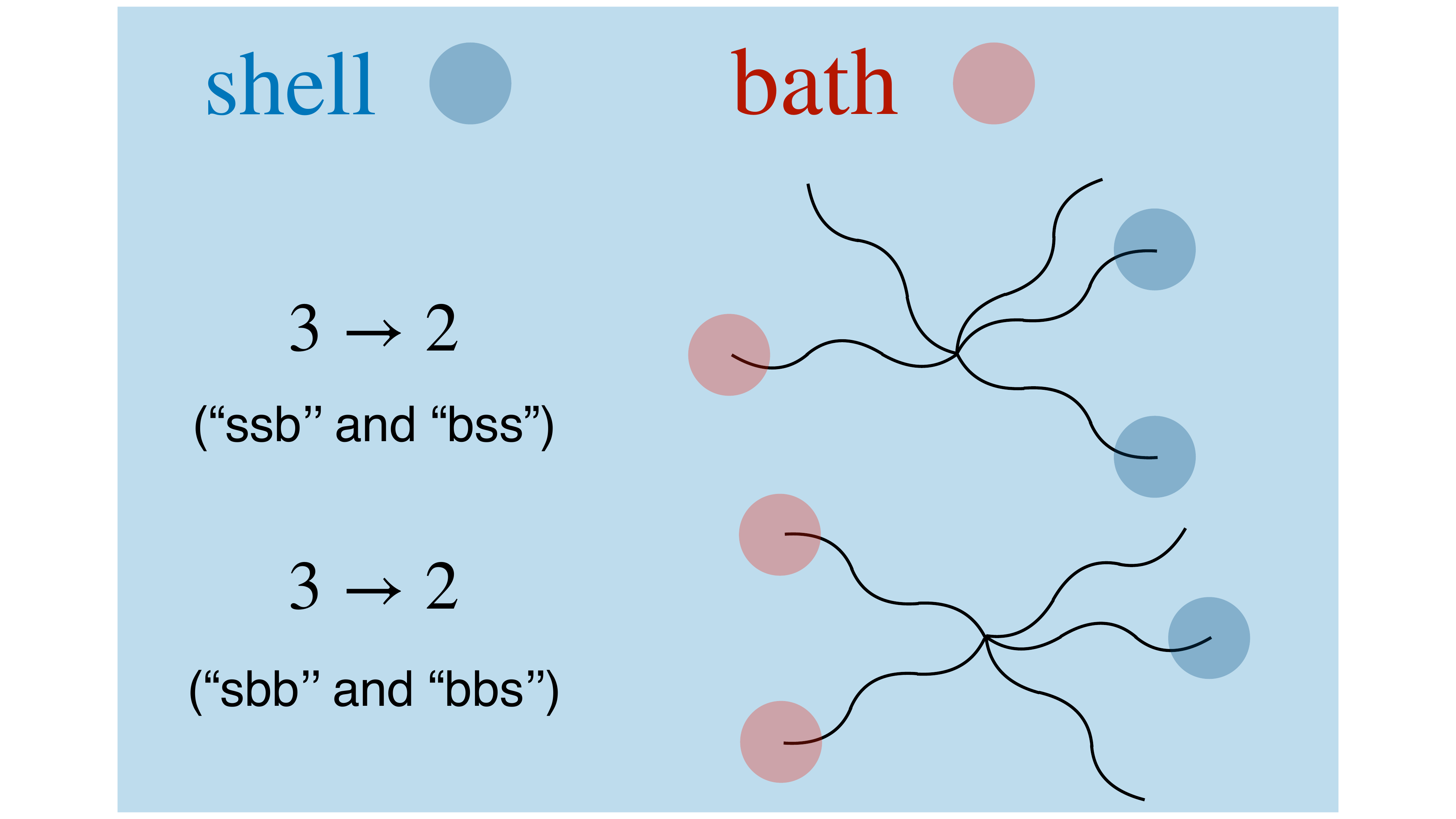}}}
{\makebox{\includegraphics[width=0.52\textwidth, scale=1]{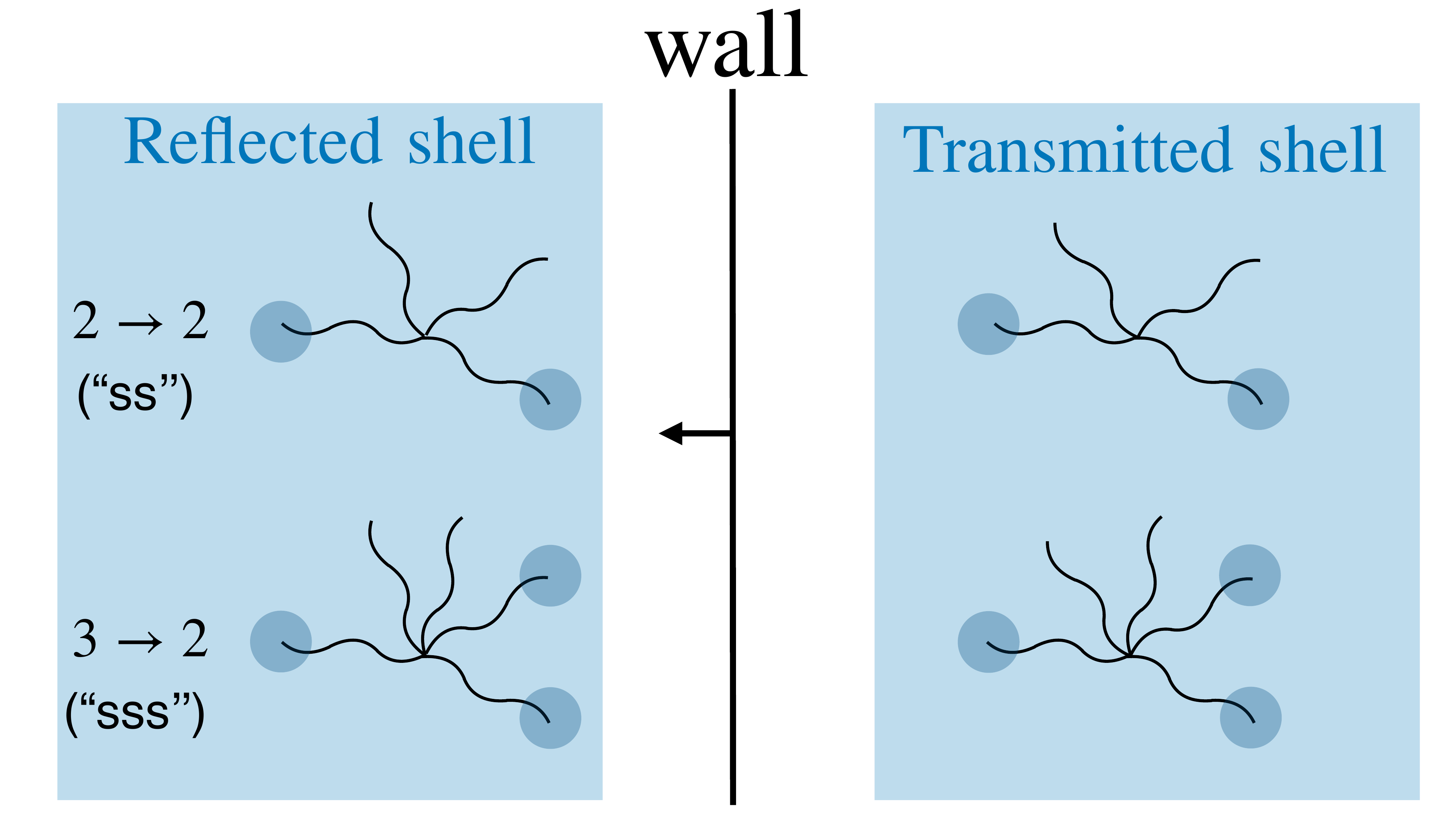}}}
\caption{\it \small  \textbf{Thermalization.} LEFT:  3-body interactions between particles of the bath and of the shells can be efficient enough to thermalize the shell with the bath. Red and blue circles represent incoming bath and shell particles respectively, while empty legs represent outgoing particles.
We denote the incoming particles originating from the shell and the bath by `s' and `b', respectively. 
The notations `ssb' and `bss' describe the same 3-body process -- two shell particles interacting with one bath particle -- given by the same cross-section. However, they are distinguished by their interaction probabilities: `ssb' signifies the interaction probability from the perspective of a shell particle, whereas `bss' reflects this probability from the standpoint of a bath particle.
The ratio of interaction probabilities is given by $\mathcal{P}_{ssb}/\mathcal{P}_{bss} \simeq n_b L_{\rm scat}^s/(n_s L_{\rm scat}^{\rm b})$, where $n_i$ denotes the number density, and $L_{\rm scat}^i$ represents the effective interaction distance before bubble collision, as detailed in Table~\ref{tab:L_E_n}. A similar distinction applies to the `sbb' and `bbs' configurations with $\mathcal{P}_{sbb}/\mathcal{P}_{bbs} \simeq n_b L_{\rm scat}^s/(n_s L_{\rm scat}^{\rm b})$. RIGHT: Inelastic 2-body and 3-body interactions between particles within the shells can be efficient enough to thermalize the shell.  See Sec.~\ref{sec:3to2} for the details.}
\label{fig:shelL_production} 
\end{figure}
To leading order in the couplings the relevant number-changing processes are $2 \to 3$ and $3 \to 2$.
At the initial stages of the evolution of a shell, it is enough to take into account only $3 \to 2$ processes, because $2 \to 3$ ones are suppressed by the final phase space. In the case of dense shells, the rate of $3 \to 2$ processes is further enhanced over the one for $2 \to 3$ processes due to an extra factor of the large shell's number densities. 
Therefore, for the purpose of determining the regions where shells free stream, it is enough to compute the effect of $3 \to 2$  interactions.
$2 \to 3$ interactions would have to be taken into account when shells approach some equilibration.

In case phase-space saturation is negligible (i.e. $1\pm f \simeq 1$ with $f$ the occupation number of the particle of interest, see Sec.~\ref{sec:phase_space_sat}), we can then write the effect of number changing interactions on the number density of particles $i$ as
\begin{equation}
    \frac{\mathrm{d} n_{i}}{\mathrm{d} t} \simeq \Gamma_{3 \to 2}^i \ n_{i} \,,
    \label{eq:dni3to2}
\end{equation}
where the interaction rate is given by
\begin{equation}
\label{eq:rate3to2}
    \Gamma_{3 \to 2}^i \simeq \frac{1}{256\pi^2} \frac{n_{j} n_{k}}{E_{1} E_{2} E_{5}} \times \int \mathrm{d} \Omega \lvert \mathcal{M} \rvert^2 \,,
\end{equation}
where $E_\ell$ is the energy of particle $\ell$ and $n_\ell$ is its number density, and where we have used the labeling of particles and momenta 
\begin{equation}
    1\;2\;5 \to 3\;4\,
\end{equation}
so $i,j,k \in \{1,2,5\}$ and $i \neq j \neq k \neq i$.
We further specify that we label with 1 and 2 the initial state particles which belong to the same population. More explicitly 1 and 2 can be either two shell particles, in which case 5 can be either shell or bath, or two bath particles, in which case 5 can only be a shell particle because usual number changing interactions within the bath alone are not the subject of our study.
For example, if we are interested in the evolution of bath particles due to scatterings with two shell particles, then in Eqs.~(\ref{eq:dni3to2}) and~(\ref{eq:rate3to2}) $i = 5, j = 1, k = 2$. 
$\int \mathrm{d} \Omega \lvert \mathcal{M} \rvert^2$ is the standard expression of the final two-particle phase-space integration of the spin-averaged squared matrix element $\lvert \mathcal{M} \rvert^2$ in the center-of-mass frame. 

The total probability of a particle $i$ undergoing  a $3 \to 2$ interaction before walls collide is
\begin{align}
	\boxed{
    \mathcal{P}_{3 \to 2}^i \simeq \Gamma_{3 \to 2}^i \times L_{\rm scat}^i \,,
    }
    \label{eq:P3to2}
\end{align}
where $\Gamma_{3 \to 2} $ is the interaction rate and $L_{\rm scat}^i$ is the effective distance (or equivalently time, because particles are ultra-relativistic) over which $3\to2$ interactions are possible for the particle $i$ under consideration.
The condition 
\begin{equation}
\label{eq:3-to-2_scattering}
    	\boxed{\mathcal{P}_{3 \to 2} \ll 1 },
\end{equation}
 will then imply that number changing interactions, between the bath and the shell or within the shell itself, do not affect the propagation of shells nor the evolution of the bath.

\subsection{Computation of $3\to2$ rates}
\label{sec:3to2ratesb}
\subsubsection{Energies, densities and scattering lengths}
A particle $i$ from the bath can only interact with shell particles when it traverses the shell. Thus, $L_{\rm scat}^i$ represents the effective thickness of the shell, corresponding to $L_{\rm b}$ listed in Table~\ref{tab:production_mechanism}. 
On the other hand, a particle $i$ originating from the shell has the potential to interact with either two other shell particles, one shell and one bath particle, or two bath particles throughout its entire journey. Hence, in this context, $L_{\rm scat}^i$ is equivalent to the bubble radius.

Having quantified $L_{\rm scat}^i$ in Eq.~(\ref{eq:P3to2}), we now turn to the quantities that enter $\Gamma_{3 \to 2}^i$ in Eq.~(\ref{eq:rate3to2}). The energy $E_i$ and the number density $n_i$ of initial bath particles are of course the thermal ones.
The energies and densities of initial shell particles can both be read off Table~\ref{tab:production_mechanism}, where the density is obtained by multiplying the bath one by $\mathcal{N}$.
A summary of these results is reported in Table~\ref{tab:L_E_n}.
We are left with the computation of $\int \mathrm{d} \Omega \lvert \mathcal{M} \rvert^2$, which we address next.

\begin{table}[htp!]
	\centering
	\begin{tabular}{c|c|c|c}
		$i$ &  $E_i$ & $n_i$ &$L_\text{scat}^i$\\
    \hline
    \hline
    & & & \vspace{-0.3 cm}\\
		bath particle
    & $3 \, \Tnuc$
    & $g_b \dfrac{\zeta(3)}{\pi^2} \Tnuc^3$ 
    & $L_{\rm b}$ \\
    & & & \vspace{-0.3 cm}\\
            \hline
    & & & \vspace{-0.3 cm}\\
		shell particle
    & $\pX$
    & $\mathcal{N} \dfrac{R_c}{3\,L_{\rm b}} \times g_\text{emit} \dfrac{\zeta(3)}{\pi^2} \Tnuc^3$
    & $R_c \simeq (\pi)^{\!\frac{1}{3}}\beta^{-1}$
    \\
	\end{tabular}
	\caption{ \it \small 
Typical values of a particle's energy, number density, and effective distance over which it undergoes $3\to 2$ interactions, depending on whether the particle belongs to a shell or to the bath. All quantities are given in the frame of the thermal bath. $g_b$ counts the bath degrees of freedom that participate in the scatterings of interest, $g_\text{emit}$ those that give rise to the shell of interest. See Table~\ref{tab:production_mechanism} for the values of $\pX$, $\mathcal{N}$ and $L_{\rm b}$ for the various shells. 
}
	\label{tab:L_E_n}
\end{table}

\subsubsection{Integrated Amplitudes}

We compute $\int \mathrm{d} \Omega \lvert \mathcal{M} \rvert^2$ 
for all possible $3\to 2$ processes involving in the initial state at least one gauge boson $V$, scattering with other $V$'s and/or fermions and/or scalars charged under the gauge group.
\\
We perform these computations for the cases of both an Abelian and a non-Abelian gauge symmetry, which for concreteness we take as $U(1)$ and $SU(N)$, because the self-interactions of the vectors eventually lead to important differences between the two.

We compute the matrix elements with the help of FeynRules \cite{Alloul_2014}, FeynArts \cite{Hahn_2001}, and FeynCalc \cite{Shtabovenko_2020,Shtabovenko_2016,Mertig:1990an}.
 We refer the reader to App.~\ref{app:3to2_phasespace} for a detailed explanation of our parametrisation and of our integration of the matrix elements over the final phase space phase, and to App.~\ref{app:more_on_3to2}
for detailed results of our computation of $\int \mathrm{d} \Omega \lvert \mathcal{M} \rvert^2$ in terms of scalar products of the 4-momenta in the center-of-mass frame $p_1+p_2+p_5 = p_3+p_4$, including for definiteness  only the terms that are leading order in $p_i \cdot p_j/\mu^2$.
Indeed, we anticipate from Sec.~\ref{sec:scalar_products} that one has the hierarchy $\mu^2 \ll p_{1} \cdot p_{2} \ll p_{1} \cdot p_{5} \simeq p_{2} \cdot p_{5}$ in case at least one initial particle belongs to the bath, while all scalar products are of the same order in case all initial particles belong to the same shell.
In the same App.~\ref{app:more_on_3to2}, we also report simple estimates of the upper limits one can expect on $\int \mathrm{d} \Omega \lvert \mathcal{M} \rvert^2$, as a check of our results and as an orientation to obtain analogous ones in the future.
For convenience of the reader, we report in Table~\ref{tab:Msq_parametrics} the parametric dependence, of the leading terms in $\int \mathrm{d} \Omega \lvert \mathcal{M} \rvert^2$, on the scalar products, the gauge coupling $\gD$ and, for $SU(N)$, on $N$.

\begin{table}[htp!]
	\centering
	\begin{tabular}{c|c|c||ccc}
		
		$\int \mathrm{d} \Omega \lvert \mathcal{M} \rvert^2_{125\to 34}$ &  $U(1)$ & $ SU(N)$ & & $p_1 \cdot p_2$ & $p_{1,2} \cdot p_5$\\
  \hline
  \hline
  & & & & &\vspace{-0.3 cm}\\
		125 = s\,s\,s
  & $4\pi \gD^6 \dfrac{1}{p_i\cdot p_j} \ln\frac{p_i\cdot p_j}{\mu^2}$
  &  $\dfrac{4 \pi \gD^6}{N^2} \dfrac{1}{p_i\cdot p_j} \ln\frac{p_i\cdot p_j}{\mu^2}$
  & & \multicolumn{2}{c}{$k_\perp^2 + \mu^2$}\\
   & & & & & \vspace{-0.3 cm}\\
            \hline
   & & & & & \vspace{-0.3 cm}\\
		s\,s\,b
  & $4\pi \gD^6 \dfrac{1}{p_1\cdot p_2} \ln\frac{p_1\cdot p_2}{\mu^2}$
  & $4\pi \gD^6 \dfrac{p_1\cdot p_5}{(p_1\cdot p_2)^2} \ln\frac{p_1\cdot p_2}{\mu^2}$
  & & $k_\perp^2 + \mu^2 $ & \multirow{3}{*}{$\Tnuc \pX + \mu^2$}\\
   & & & & & \vspace{-0.3 cm}\\
            b\,b\,s
  & $4\pi \gD^6 \dfrac{1}{p_1\cdot p_2} \ln\frac{p_1\cdot p_5}{\mu^2}$
  & $4\pi \gD^6 N^2 \dfrac{p_1\cdot p_5}{(p_1\cdot p_2)^2} \ln\frac{p_1\cdot p_2}{\mu^2}$
  & & $\Tnuc^2 + \mu^2$ &\\
	\end{tabular}
	\caption{ \it \small 
 \textbf{$\mathbf{U(1)}$ and $\mathbf{SU(N)}$ columns}. Parametric dependence of $\int \mathrm{d} \Omega \lvert \mathcal{M} \rvert^2$ in the scalar products, the gauge coupling $\gD$ and $N$. Only the leading terms in $\gD$, $N$ and the scalar products are kept, where
 in the ssb and bbs cases $\mu^2 \ll p_{1} \cdot p_{2} \ll p_{1} \cdot p_{5} \simeq p_{2} \cdot p_{5}$ (we remind we defined 1 and 2 to belong to the same population, i.e. to a shell `s' or to the bath `b'), while in the sss case there is no hierarchy among the different scalar products and $\mu^2 \ll p_{i} \cdot p_{j}$.
 The reported values assume that shells are made of vector bosons, and that they couple to fermions.
 $\int \mathrm{d} \Omega \lvert \mathcal{M} \rvert^2_{125\to 34}$ have the same or a more suppressed scaling in case i) there are no fermions charged under the gauge group, or ii) shells are made of fermions and/or scalars but not vectors (as e.g. reflected scalar shells in a global PT or shells from the Azatov-Vanvlasselaer mechanism).
 See App.~\ref{app:more_on_3to2} for results that keep O(1) factors, subleading terms and the identities of initial and final states.\\
 \textbf{$\mathbf{p_1 \cdot p_2}$ and $\mathbf{p_{1,2}\cdot p_5}$ columns}. Values of the scalar products of interest, where $k_\perp^2$ and $\pX$ depend on the shell of interest. $\pX$ can be read off Table~\ref{tab:production_mechanism}, while $k_\perp^2 = O(\gD^2 v_\phi^2+\Tnuc^2)$ for both reflected and transmitted radiated vectors, $k_\perp^2 = O(\Tnuc^2)$ for shells of particles getting a mass, see Eq.~(\ref{eq:kperpsq}) and text for more details.
 We add the IR cutoff $\mu^2$ to all scalar products.
 }
	\label{tab:Msq_parametrics}
\end{table}

\begin{figure}[!ht]
\centering
\raisebox{0cm}{\makebox{\includegraphics[width=0.49\textwidth, scale=1]{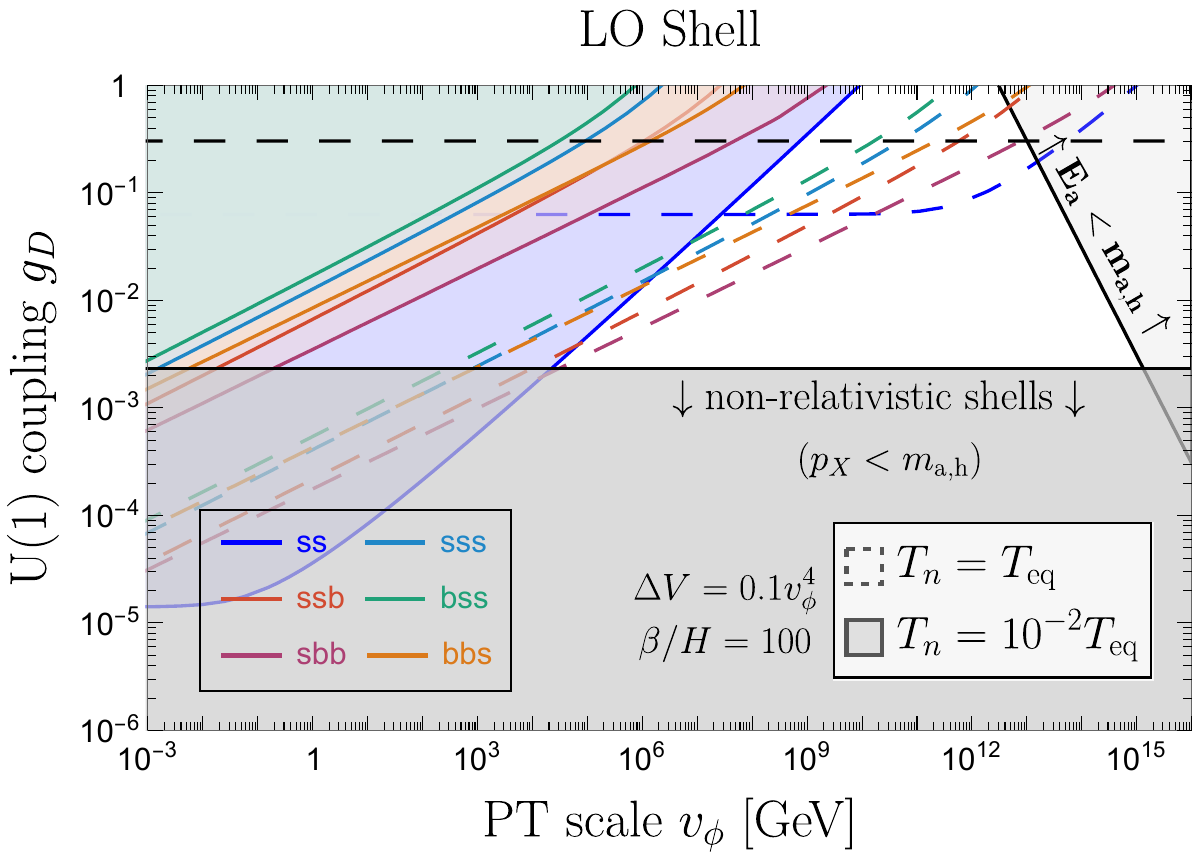}}}
{\makebox{\includegraphics[width=0.49\textwidth, scale=1]{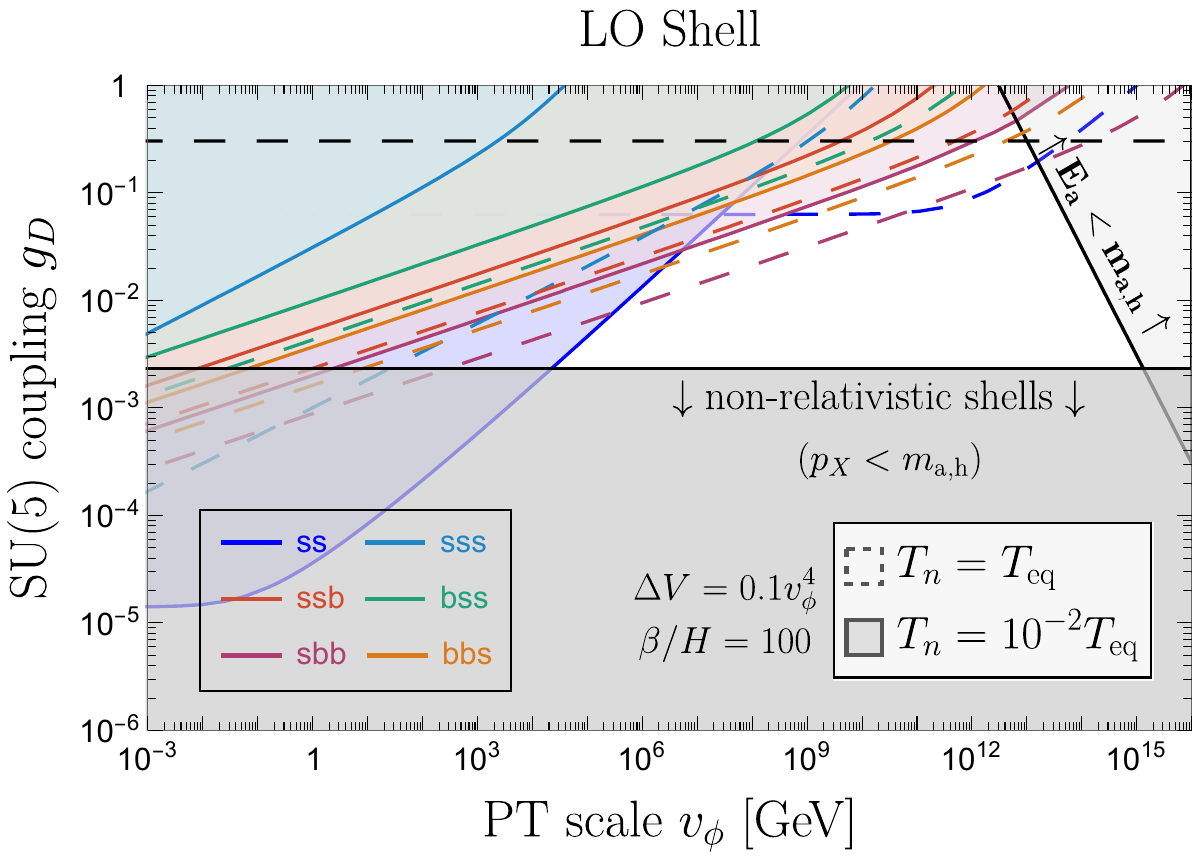}}}
{\makebox{\includegraphics[width=0.49\textwidth, scale=1]{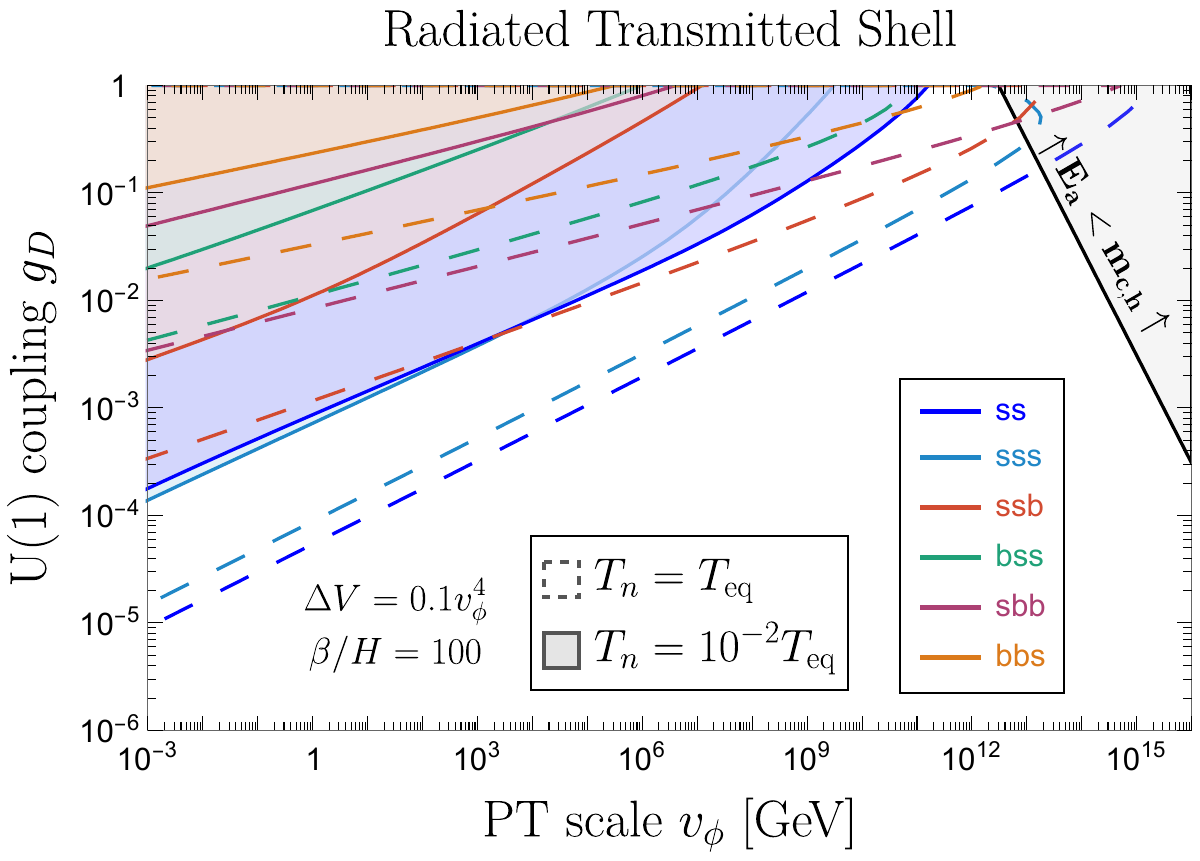}}}
{\makebox{\includegraphics[width=0.49\textwidth, scale=1]{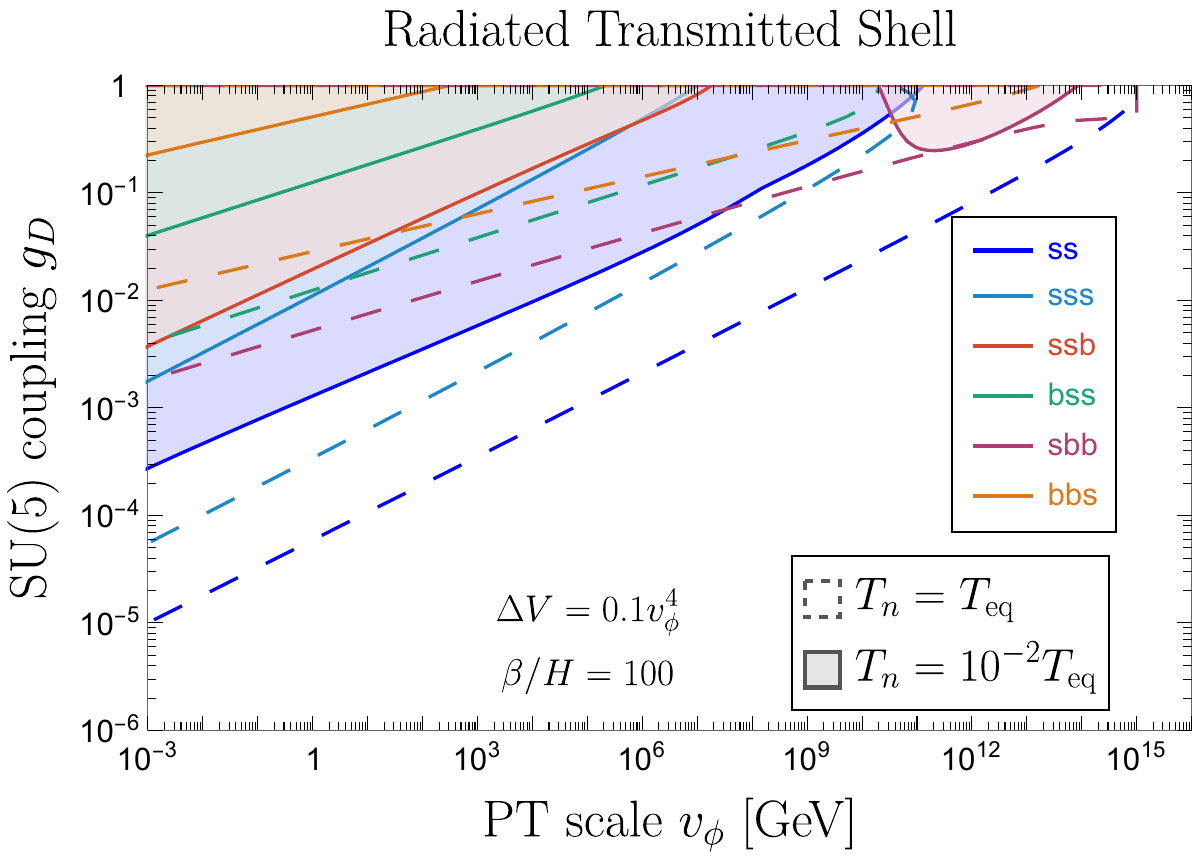}}}
{\makebox{\includegraphics[width=0.49\textwidth, scale=1]{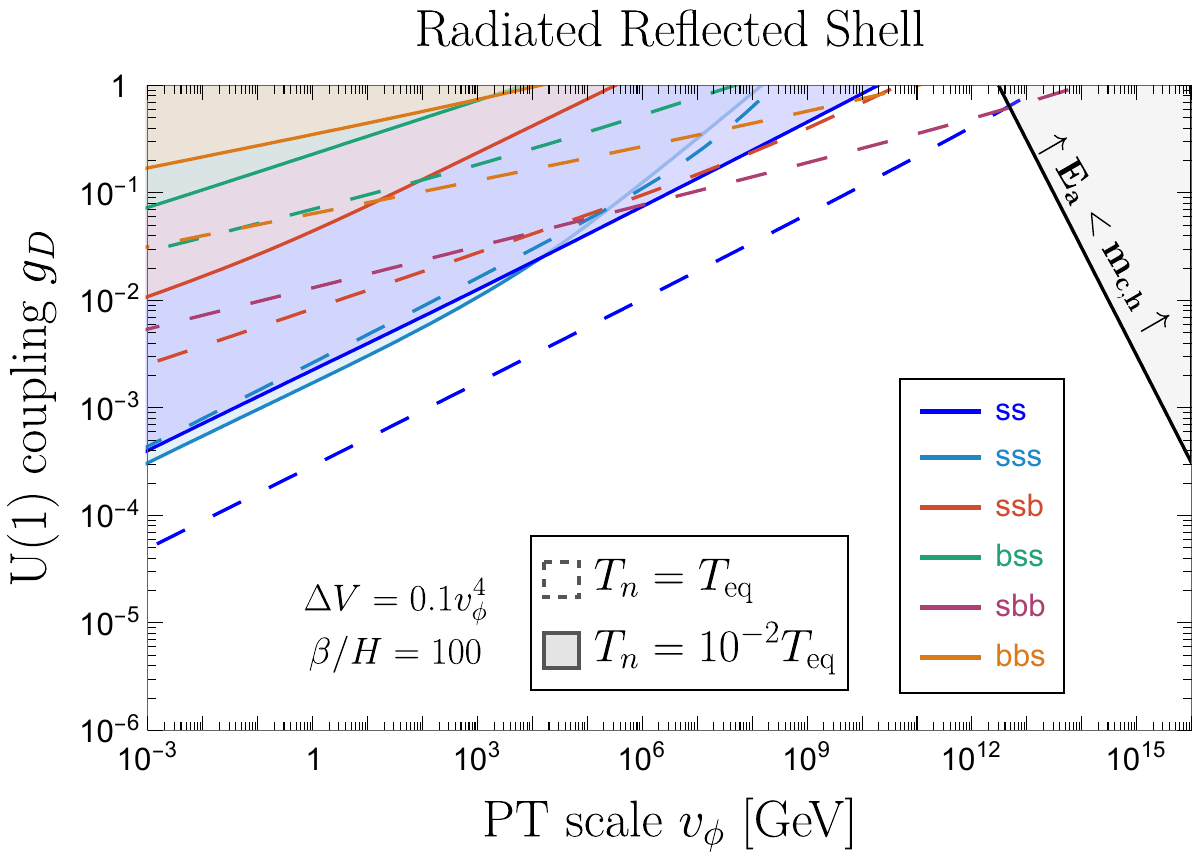}}}
{\makebox{\includegraphics[width=0.49\textwidth, scale=1]{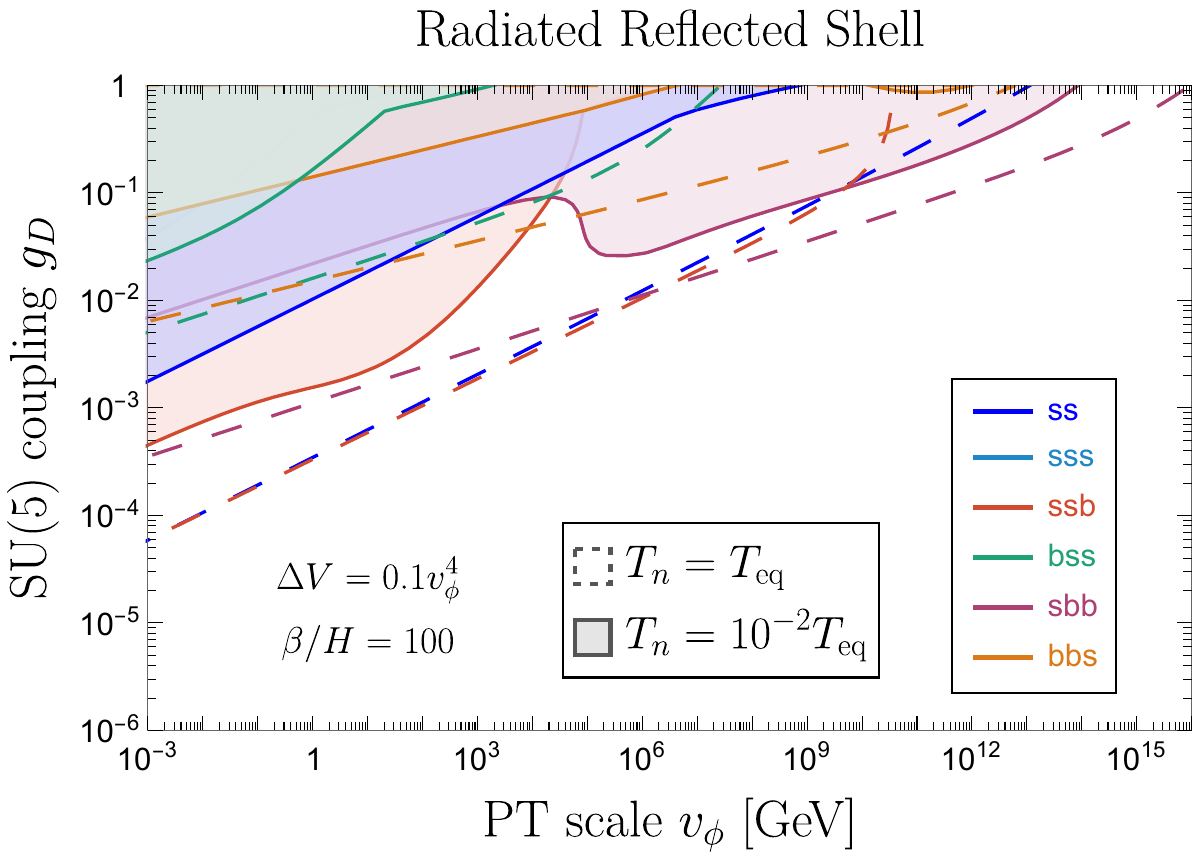}}}
\caption{\it \small \textbf{Shell thermalization:} In the coloured regions (on the left of the dashed lines) the $2\to2$ and $3\to2$ interactions specified in the insets (with ``s'' and ``b'' denoting incoming particles from shells and bath respectively), see Eqs.~\eqref{eq:2-to-2_scattering} and \eqref{eq:3-to-2_scattering}, become efficient and shells do not free stream, for $\Tnuc/\Teq = 1 (10^{-2})$.  We show shell particles produced from $LO$ interaction (\textbf{top}), Bremsstrahlung radiation, either transmitted (\textbf{middle}) or reflected (\textbf{bottom}), assuming abelian (\textbf{left}) or non-abelian (\textbf{right}) gauge interaction, see first two rows of Table~\ref{tab:production_mechanism}.
We used $g_b = g_\SM + g_\text{emit}$ with $g_\SM = 106.75$ and $g_\text{emit} = 10 = N_F$.
\textbf{Grey} and \textbf{black} shaded areas as in Fig.~\ref{fig:free_stream_all_cases}.}
\label{fig:3to2} 
\end{figure}

\subsubsection{Scalar Products}
\label{sec:scalar_products}

Depending on the identity of the initial-state scatterers, the scalar products have a different dependence on the parameters of the PT and of the theory, implying different results for the probability that the associated $3\to2$ interaction happens before collision, which is the question of our interest. We therefore now turn to derive expressions for the scalar products in terms of the parameters of the theory.

\paragraph{\bf Shell-shell-shell}
Let us start by the simple case where the 3 initial particles all belong to the shell.
Despite in the bath frame the energies of the three particles are very large, their scalar products do not track of course their frame-dependent energies, but rather the typical spread of their momenta, $p_i\cdot p_j \simeq k_\perp^2$. 
This spread depends on the shell of interest. We use
\begin{equation}
    k_\perp^2 =
    \begin{cases} 
    \Tnuc^2 
    & \text{particles~acquiring~mass}
    \\
    \langle k_{\perp,\mathcal{R}}^2 \rangle + \sigma_{k_{\perp,\mathcal{R}}}^2
    & \text{reflected~radiated~vectors}
    \\
    \langle k_{\perp,\mathcal{T}}^2 \rangle + \sigma_{k_{\perp,\mathcal{T}}}^2
    & \text{transmitted~radiated~vectors}
    \end{cases}
    \label{eq:kperpsq}
\end{equation}
where $\langle k_{\perp,\mathcal{R,T}} \rangle$ and $\sigma_{k_{\perp,\mathcal{R,T}}}$ are derived in App.~\ref{app:weak} and can be read off, respectively, in Eqs~(\ref{eq:k_perp_mean}) and (\ref{eq:k_perp_sigma}).

\paragraph{\bf Shell-shell-bath and bath-bath-shell}

Here $p_{1}$ and $p_{2}$ refer to particles that either both belong to the shell or both belong to the bath.
In particular for shell-shell-bath interactions we have $p_{1}$ and $p_{2}$ for the shell particles and $p_{5}$ for the bath particle; for bath-bath-shell interactions we have $p_{1}$ and $p_{2}$ for the bath particles and $p_{5}$ for the shell particle.
Again, in some frame $p_{1}$ and $p_{2}$ are nearly parallel (e.g. if they denote bath particles and one works in the wall frame), but their scalar product does not feel the large energies that they can have in that frame.
One then has the usual $p_{1} \cdot p_{2} \simeq \Tnuc^2$ for bath particles, and $p_{1} \cdot p_{2} \simeq k_\perp^2$ for shell particles, whose origin has already been explained in the shell-shell-shell case. 
One instead has $p_{1,2} \cdot p_{5} \simeq \pX \Tnuc$, where $\pX$ is the momentum of the given shell in the bath and can be read off Table~\ref{tab:production_mechanism}.
We report the resulting values for the various scalar products in Table~\ref{tab:Msq_parametrics}.

In this paper we are not interested in the impact of $3\to 2$ interactions on the evolution of the bath and the shell, but only in determining the region where they are relevant or not.
We still find it worth to note that, by a careful investigation (see App.~\ref{app:3to2_phasespace}) of just the scattering kinematics, it is straightforward to show that the two final state particles are of one type shell and one type bath, i.e. one final particle has the typical energy (up to $\mathcal{O}(1)$-factors) of a bath one and is traveling inside the bubble, and the other particle has the typical energy of a shell one and is traveling along with other shell particles.

\subsection{Results}
\label{sec:3to2ratesc}

Using Eq.~(\ref{eq:P3to2}) for the probability $\mathcal{P}_{3 \to 2}^i$ that a particle $i$ undergoes a $3\to 2$ interaction, with the rate from Eq.~(\ref{eq:rate3to2}) and the other inputs from Tables~\ref{tab:L_E_n}, \ref{tab:Msq_parametrics}, the condition $\mathcal{P}_{3 \to 2}^i < 1$ is not satisfied in the colored regions in Fig.~\ref{fig:3to2}, where we shade in different colors regions where (in parenthesis the momentum to assign to each particle, according to our definitions)
\begin{itemize}
\item[$\diamond$] $i (p_1), j (p_2), k (p_5)$ are all shell particles;
\item[$\diamond$] $i (p_1), j(p_2)$ are shell particles and $k (p_5)$ is a bath one;
\item[$\diamond$] $i (p_5)$ is a shell particles and $j (p_1), k (p_2)$ are bath ones;
\item[$\diamond$] $i (p_5)$ is a bath particles and $j (p_1), k (p_2)$ are shell ones;
\item[$\diamond$] $i (p_1), j (p_2)$ are bath particles and $k (p_5)$ is a shell one.
\end{itemize}
Among the 5 cases above, the one where the three initial particles all belong to the shell dominates the shaded regions for radiated shells (both transmitted and reflected) in the abelian case. This can be understood with the fact that, in those cases, $n_{\rm bath}/n_{\rm ej} \ L_{\rm eff}/ R_{c} \ll 1$.
The shaded regions for radiated shells, in the non-abelian case, instead all lie within the parameter space where our computation cannot be trusted because the plasma mass is larger than the vacuum one, see Sec.~\ref{sec:large_plasma_mass} and Fig.~\ref{fig:phase_space_sat}.

For completeness, in Fig.~\ref{fig:3to2} we also shade the regions where $2\to2$ processes within shells are efficient, discussed in Sec.~\ref{sec:2to2}: while they do not alter the total shell's number density and typical particle momentum and so do not make shells depart from free-streaming, they could affect the identities of the particles making up the shell.

Overall, thermalization processes are the strongest ones in determining the regions where radiated shells from an abelian gauge theory do not free stream, as well for LO shells from non-abelian theories at large $v_\phi$ values, see Fig.~\ref{fig:free_stream_all_cases}.
An example consequence of these findings is that the first processes to consider, in order to determine the evolution of radiated and transmitted shells at PTs associated with an abelian gauge group, would be $3\to2$ ones all involving shell particles in the initial state. We leave this interesting direction for future work.

\begin{figure}[!ht]
\centering
\raisebox{0cm}{\makebox{\includegraphics[width=0.85\textwidth, scale=1]{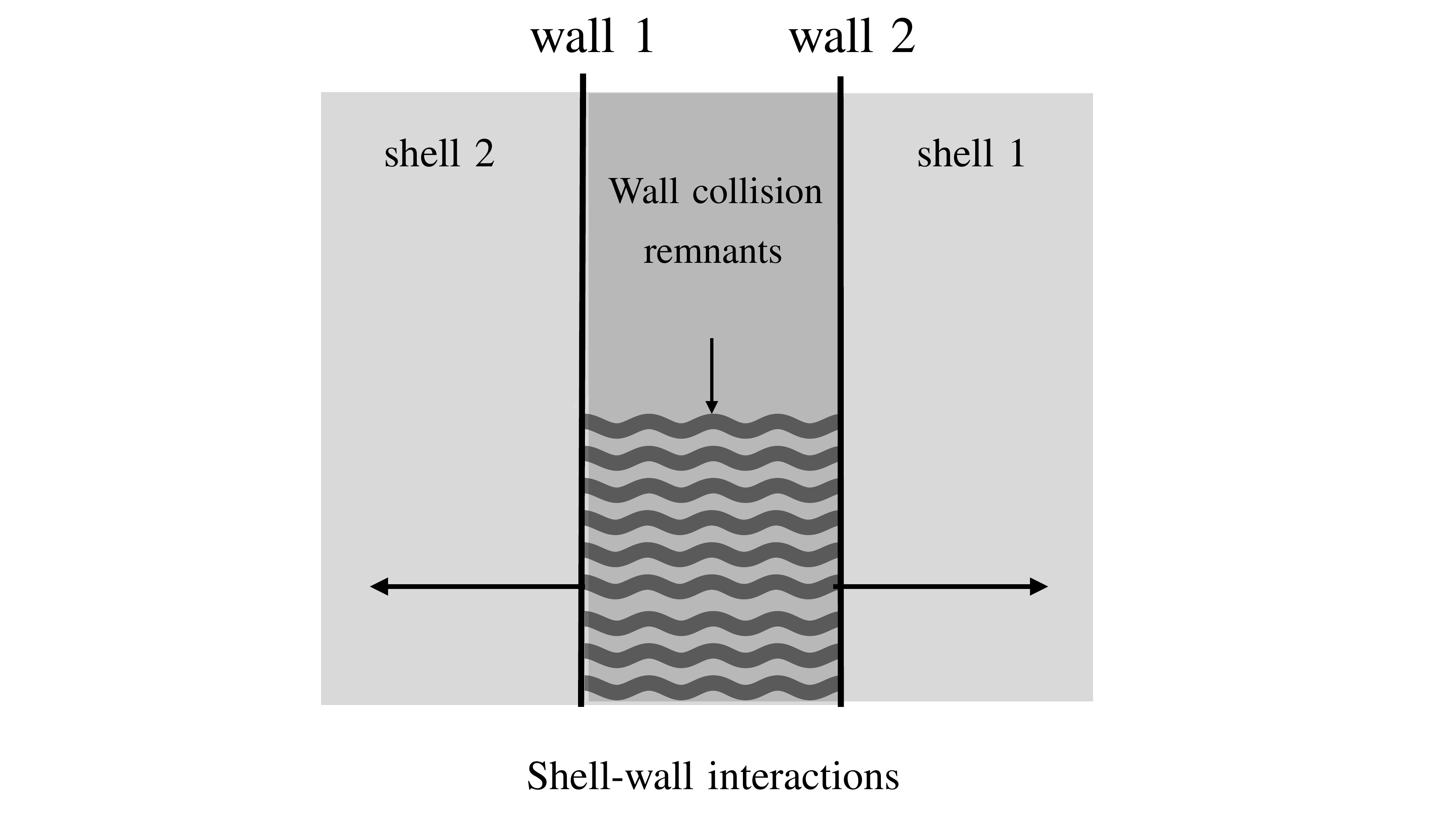}}}
\caption{\it \small \textbf{Shell-wall interactions:} After passing through one another, walls 1 and 2 progressively transform their gradient energy into an oscillating scalar field condensate (labelled as ``wall collision remnants'') that develops between them as they continue to move apart. Shell 1, associated with wall 1, can interact both with the gradient energy of wall 2 and with the wall collision remnants. Similarly, shell 2, accompanying wall 2, can interact both with the gradient energy of wall 1 and the wall collision remnants. See Sec.~\ref{sec:wall_int} for details of shell-wall interactions. Of course, shells 1 and 2 can also interact with each other, see \cite{Baldes:2023fsp}.}
\label{fig:shell_wall_collision} 
\end{figure}

\section{Interaction with bubble walls}
\label{sec:wall_int}

Field dynamics after wall-wall collision has been investigated by a number of authors, starting from~\cite{Hawking:1982ga}. The study of particle production from wall-wall collisions was first conducted in~\cite{Watkins:1991zt}, followed by subsequent applications and refinements in~\cite{Masiero:1992bv,Kolb:1996jr,Zhang:2010qg,Konstandin:2011ds,Falkowski:2012fb,Katz:2016adq,Jinno:2019bxw,Mansour:2023fwj,Shakya:2023kjf,Giudice:2024tcp}.
In Sec.~\ref{sec:wall_wall_collision} we characterize wall-wall collisions  following~\cite{Falkowski:2012fb} for the analytical expressions and~\cite{Jinno:2019bxw} for the numerical treatment. In Sec.~\ref{sec:wall_shell} we use these existing modelings to derive novel conclusions about the propagation of shells.

\subsection{Wall-wall collision}
\label{sec:wall_wall_collision}

\begin{figure*}[t!]
    \centering
\includegraphics[width=0.48\textwidth]{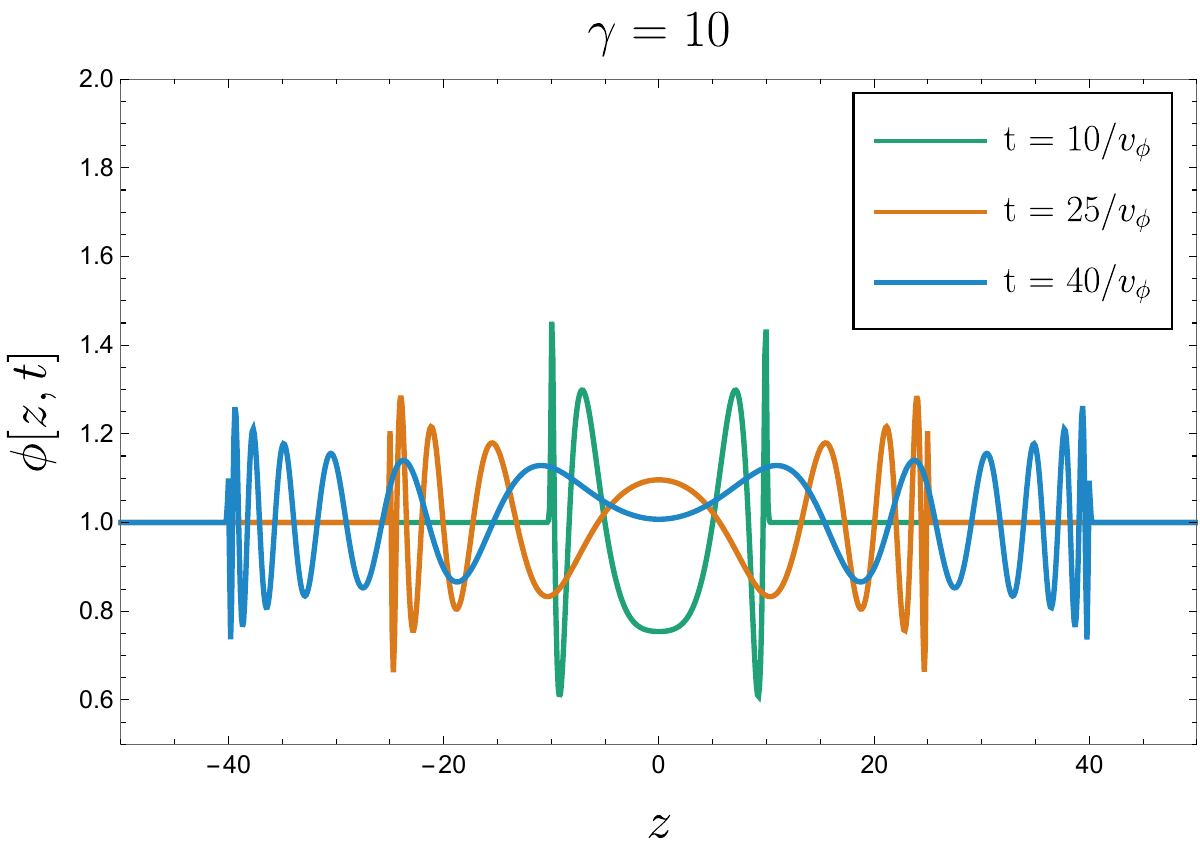}
\includegraphics[width=0.48\textwidth]{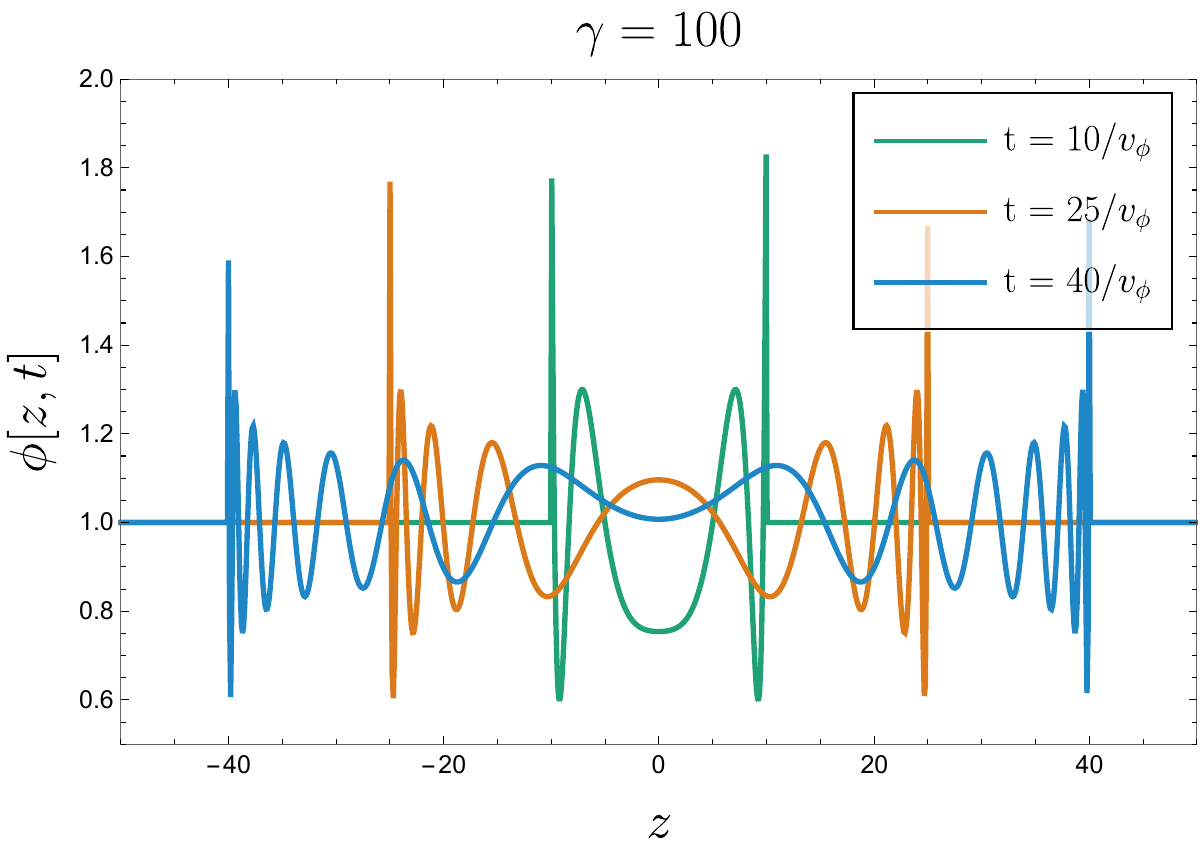}
\caption{\it \small  \label{fig:wallcollision} Evolution of the scalar field profile after wall collision with Lorentz factor $\gamma_{\rm w} = 10$ (\textbf{left}) and $\gamma_{\rm w} =100$ (\textbf{right}). We can distinguish the peaked energy distribution stored in the walls even after collision, and the more dilute energy fraction of the scalar field oscillating around the true vacuum. We used the analytical solution in Eq.~\eqref{eq:phi_profile_ana} of the scalar field equation of motion. This closely resembles the numerical solution shown in Fig.~\ref{fig:wallcollision_num}. }
\end{figure*}

\begin{figure*}[t!]
    \centering
\includegraphics[width=0.48\textwidth]{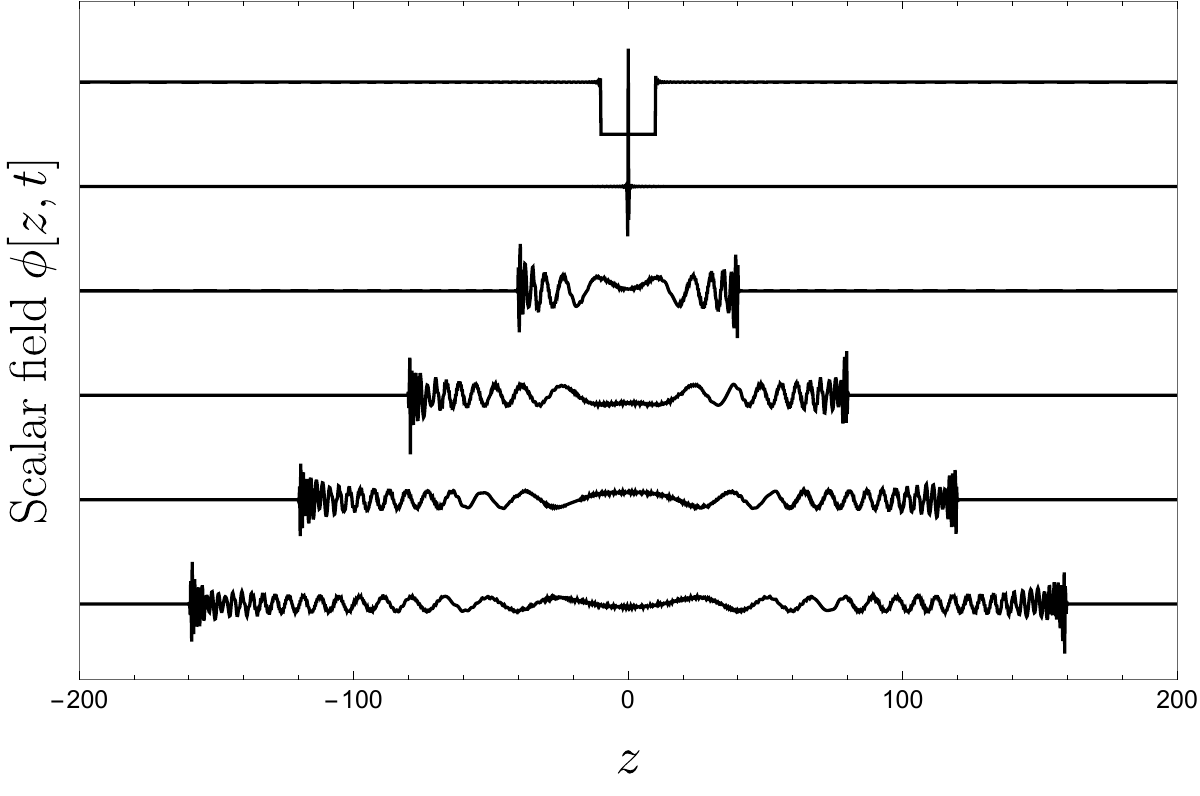}
\includegraphics[width=0.48\textwidth]{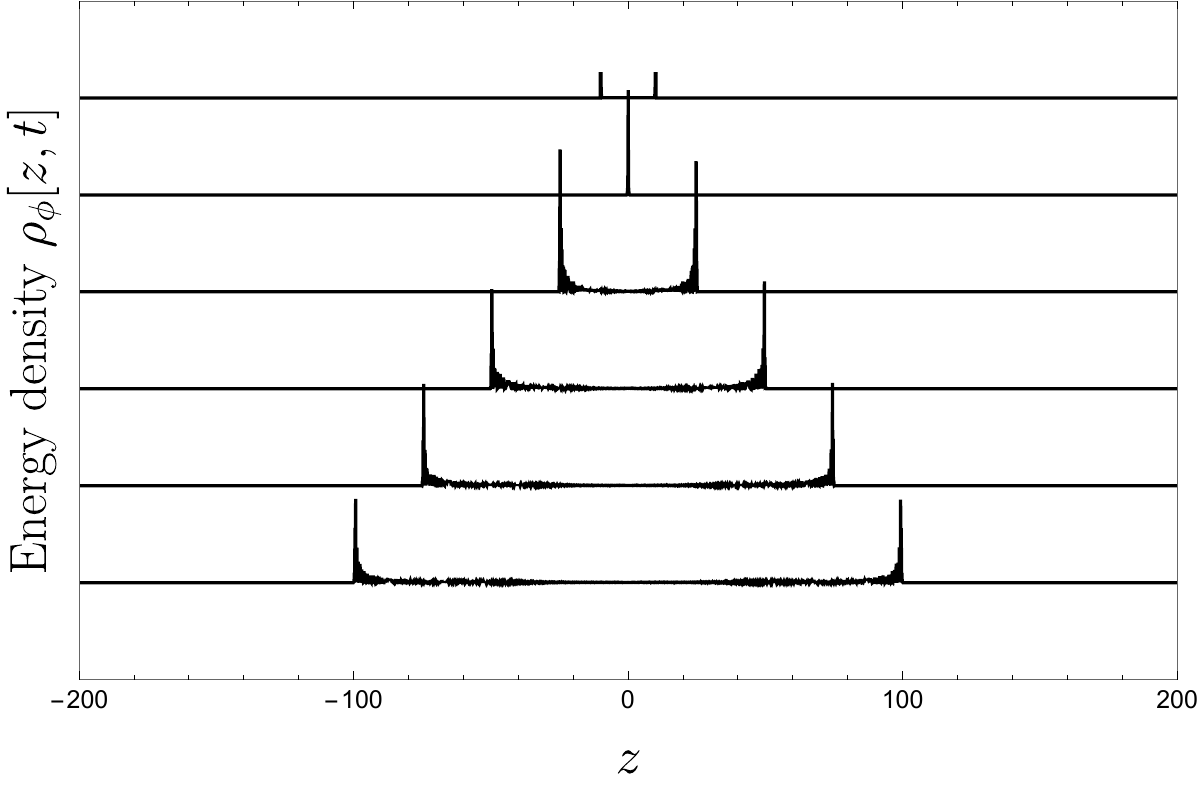}
\caption{\it \small  \label{fig:wallcollision_num}  Time evolution (from top to bottom) of the scalar field profile, its field value on the left and its energy density on the right, after wall collision. We numerically integrated the scalar field equation of motion in Eq.~\eqref{eq:eom_salar_field}. The wall Lorentz factor, measured as the ratio of the profile periodicity in the wall front to the periodicity in the oscillating condensate is $\gamma_{\rm w} \simeq 50$.}
\end{figure*}

We model two incoming bubble walls as two Heaviside functions propagating in opposing direction
\begin{equation}
\label{eq:phi_ini}
    \phi(z,t<0) = v_\phi +\frac{v_\phi}{2}\left[\textrm{tanh}\left( \gamma_w \frac{z+t}{L_{0}}\right) - \textrm{tanh}\left( \gamma_w \frac{z-t}{L_{0}}\right) \right],
\end{equation}
where $L_{0}$ is the wall thickness in the wall frame.
The evolution of the scalar field profile during bubble collision can be determined analytically by integrating the Klein-Gordon equation
\begin{equation}
\label{eq:eom_salar_field}
    (\partial_t^2 - \partial_z^2)\phi(z,t) + V'(\phi) = 0,
\end{equation}
with Eq.~\eqref{eq:phi_ini} as boundary condition. Taking $V(\phi) = m_\phi^2(\phi-v_\phi)^2/2$, one obtains \cite{Falkowski:2012fb}
\begin{equation}
\label{eq:phi_profile_ana}
    \phi(z,t>0) = v_\phi\left[1+\frac{L_{0}}{\gamma_w}\int_0^{\infty} dp_z \frac{p_z}{\sqrt{p_z^2+m_\phi^2}} \frac{\textrm{cos}(p_z z)}{\textrm{sinh}\left(\dfrac{\pi L_{0} p_z}{2\gamma_w}\right)}\textrm{sin}\left( \sqrt{p_z^2+m_\phi^2}t\right) \right].
\end{equation}
We show the analytical solution for the scalar profile in Fig.~\ref{fig:wallcollision}, which we find to be in good agreement with the profile obtained from numerically integrating the equation of motion in Eq.~\eqref{eq:eom_salar_field} following the recipe of \cite{Jinno:2019bxw}, and shown in Fig.~\ref{fig:wallcollision_num}.  

\begin{figure}[!ht]
\centering
\raisebox{0cm}{\makebox{\includegraphics[width=0.49\textwidth, scale=1]{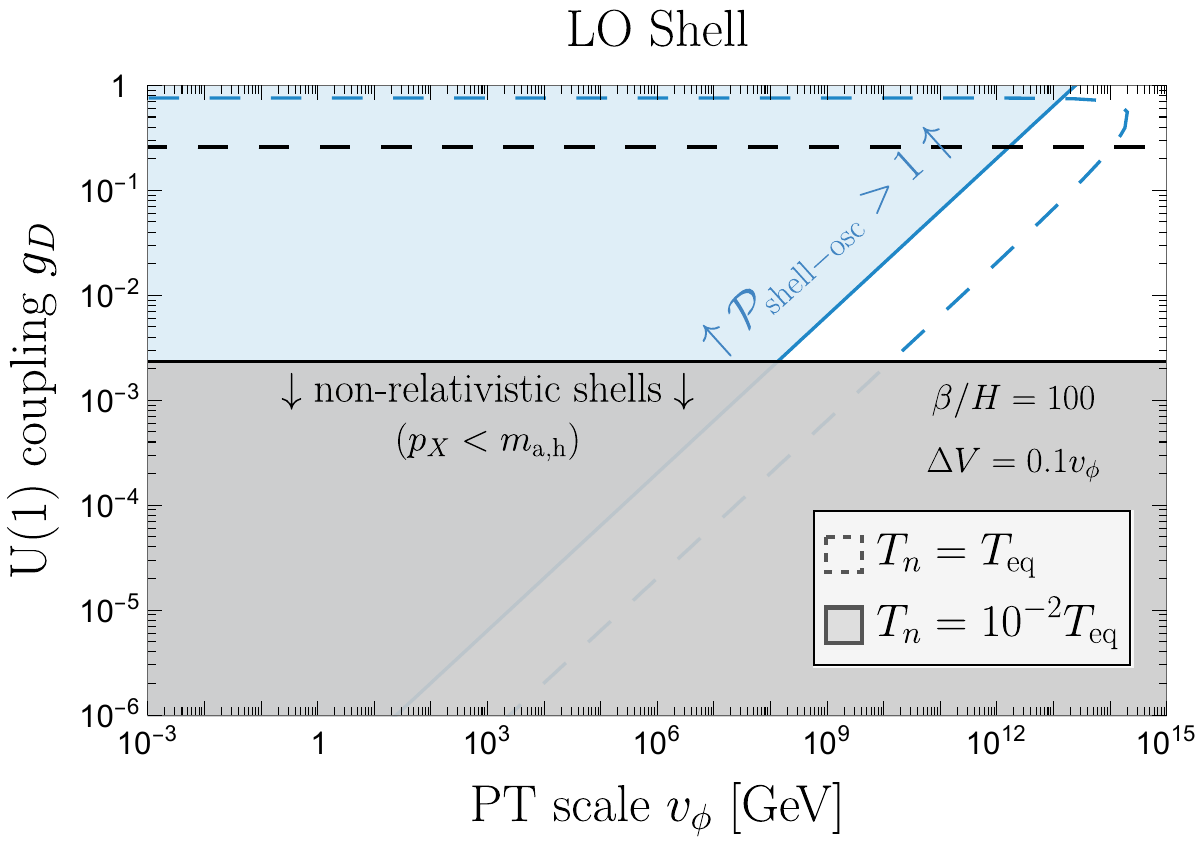}}}
{\makebox{\includegraphics[width=0.49\textwidth, scale=1]{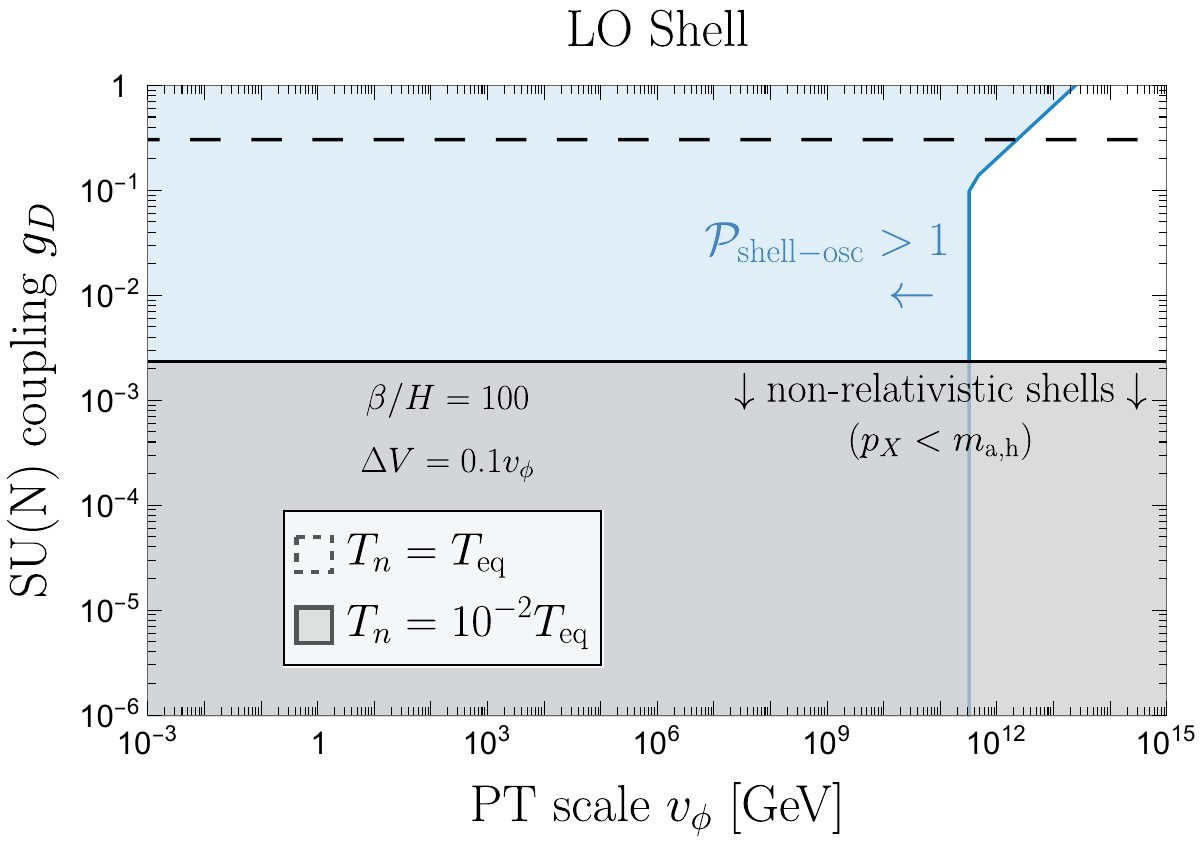}}}
{\makebox{\includegraphics[width=0.49\textwidth, scale=1]{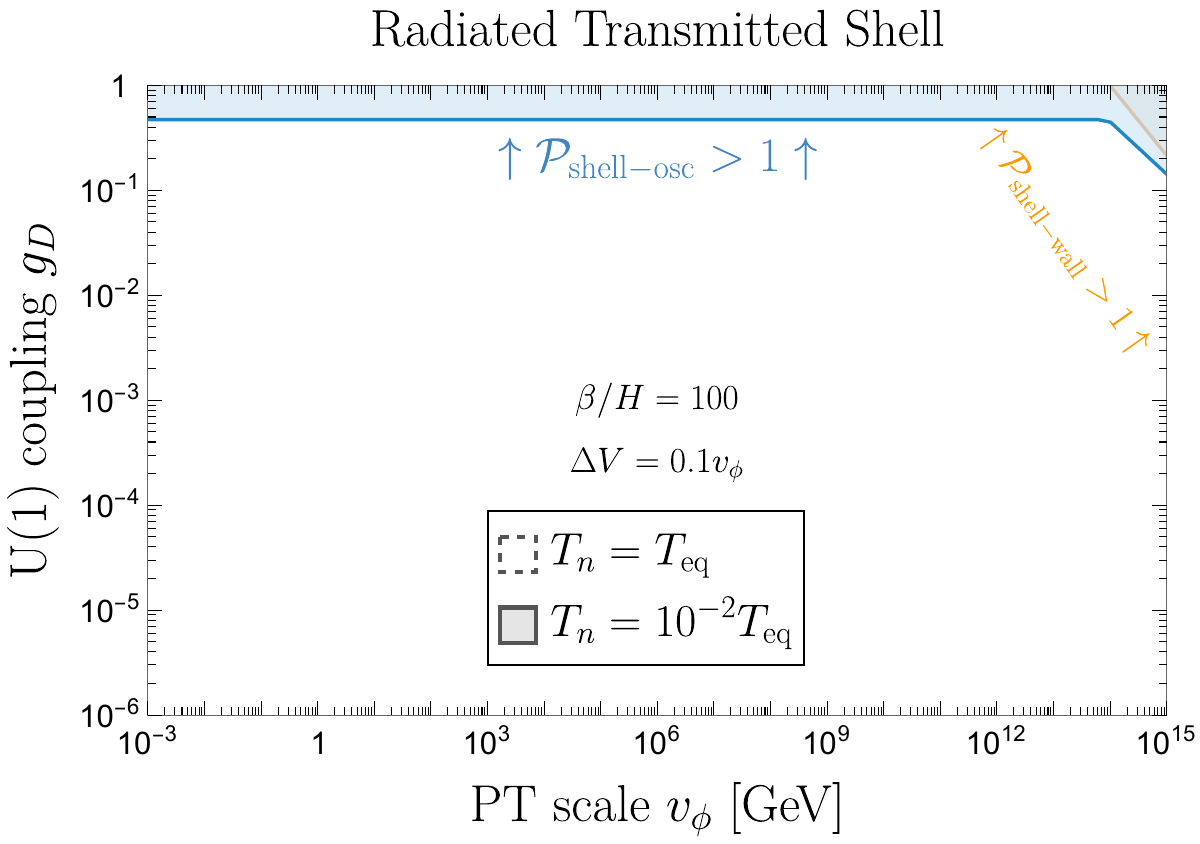}}}
{\makebox{\includegraphics[width=0.49\textwidth, scale=1]{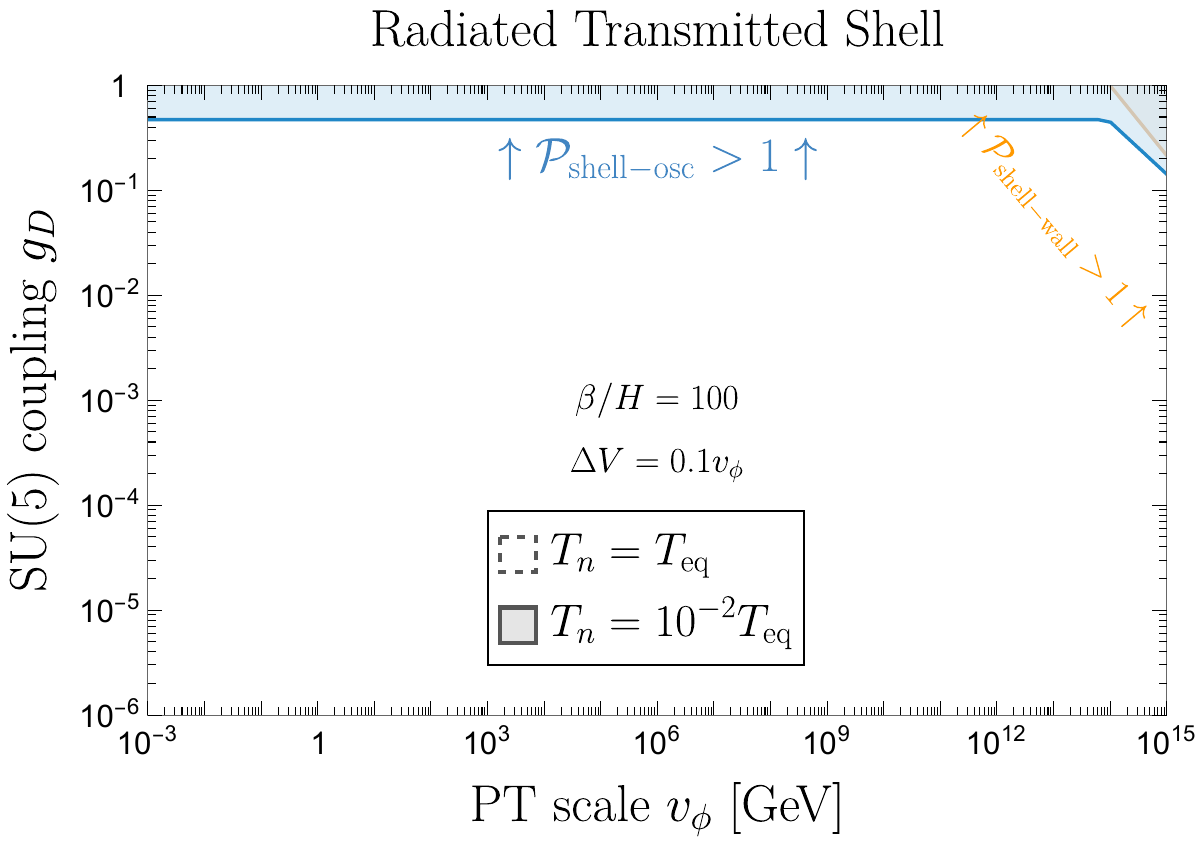}}}
{\makebox{\includegraphics[width=0.49\textwidth, scale=1]{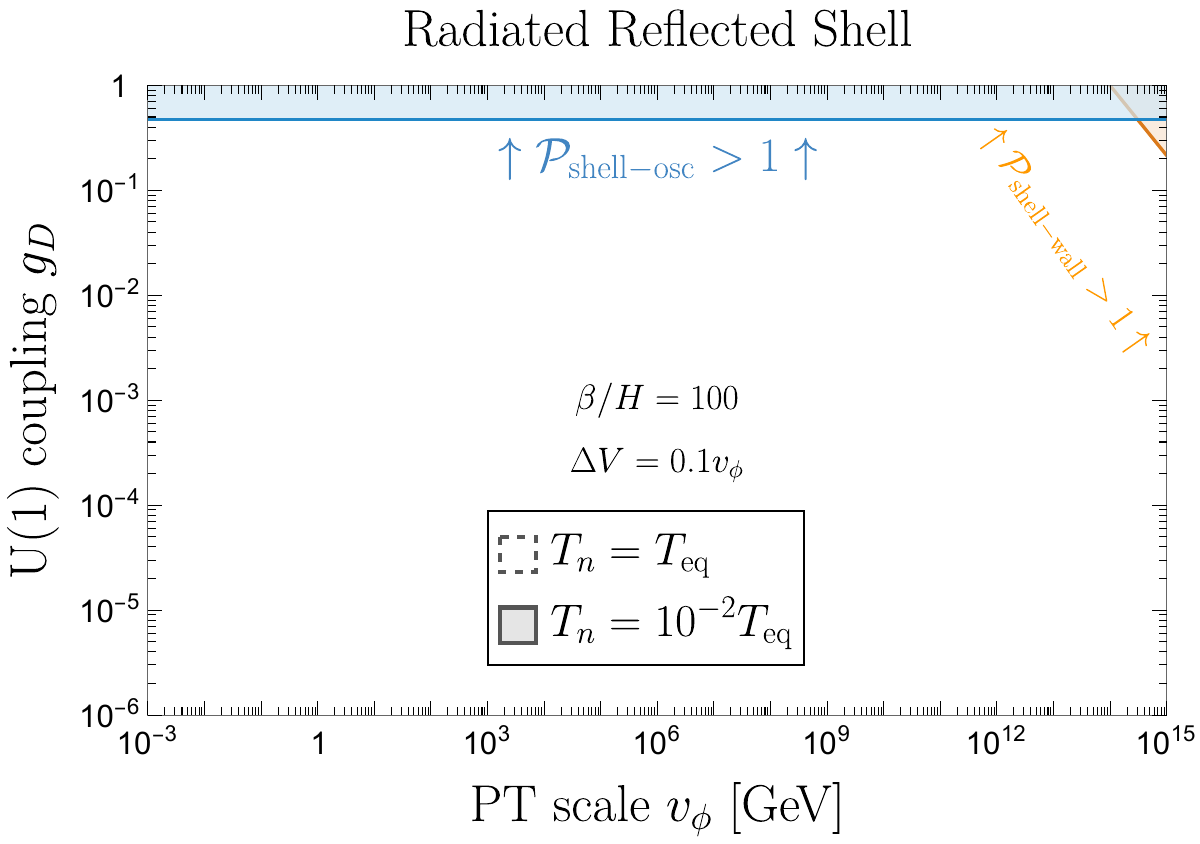}}}
{\makebox{\includegraphics[width=0.49\textwidth, scale=1]{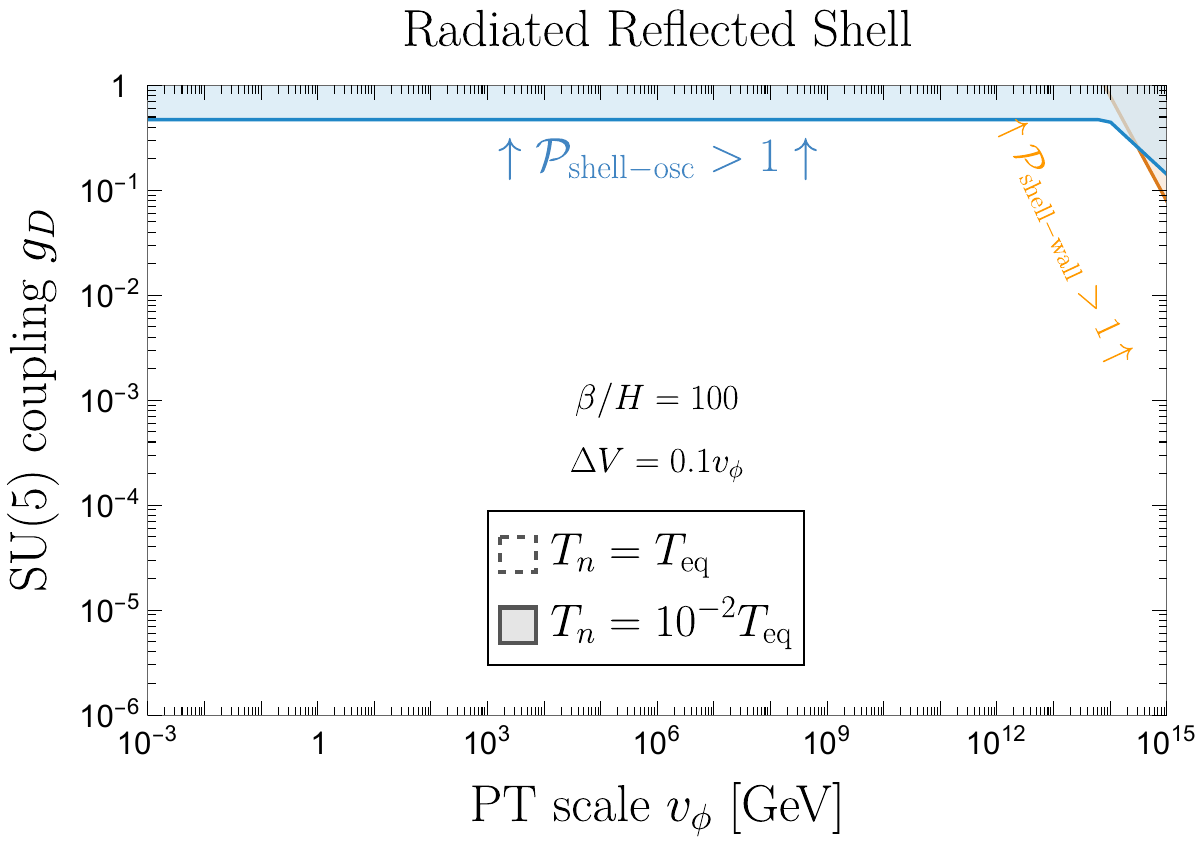}}}
\caption{\it \small \textbf{Wall-shell collision:} In the \textbf{blue} region, shells interact efficiently with the oscillating scalar condensate left behind collided walls. In the \textbf{orange} region, shells interact efficiently with the peak of scalar field gradient in bubble walls. We show shell particles produced from $LO$ interaction (\textbf{top}), Bremsstrahlung radiation, either transmitted (\textbf{middle}) or reflected (\textbf{bottom}), assuming abelian (\textbf{left}) or non-abelian (\textbf{right}) gauge interaction, see first two rows of Table~\ref{tab:production_mechanism}.
We used $g_b = g_\SM + g_\text{emit}$ with $g_\SM = 106.75$ and $g_\text{emit} = 10 = N_F$.
\textbf{Grey} and \textbf{black} shaded areas as in Fig.~\ref{fig:free_stream_all_cases}.}
\label{fig:wall_shell_collision} 
\end{figure}

After walls pass through each other, their energy gets released into oscillating modes of the scalar field as the walls continue to propagate. 
The energy stored in the bubble wall in the bath frame at the time of collision $t_c$ is
\begin{equation}
    \rho_{\rm wall}(t_c) \sim \gamma_{\rm w}^2 m_\phi^2 v_\phi^2,
    \label{eq:wallenergydensity}
\end{equation}
over a thickness of $L_{\rm wall}=L_{0}/\gamma_w \sim (\gamma_{\rm w} m_\phi)^{-1}$ in the bath frame in the case of a non-supercooled PT. With supercooling, $T_n \ll m_{\phi}, v_{\phi}$, the wall surface tension is larger and the above energy density instead exists over a thickness 
    \begin{equation}
    L_{\rm wall} \sim \frac{1}{\gamma_{\rm w} T_n},
    \end{equation}
in the bath frame, which one finds taking into account regions of energy density both from the leading edge of the bubble in which the potential term dominates and oscillations of the field about the true vacuum behind the wall~\cite{Baldes:2020kam,Gouttenoire:2021kjv}. (The temperature does not appear in the above modelling, Eqs.~\eqref{eq:phi_ini}-\eqref{eq:phi_profile_ana}, but we include such a correction to better match the supercooled case.)
The energy density of a scalar field oscillating in a quadratic potential is
\begin{equation}
    \rho_{\rm osc} \sim m_\phi^2 v_\phi^2/2.
\end{equation}
Conservation of the scalar field energy after collision, i.e. at $t > t_c$, gives
\begin{equation}
\label{eq:E_wall_E_osc}
    E_{\rm wall}(t)+E_{\rm osc}(t) = E_{\rm wall}(t_c),
\end{equation}
where $E_{\rm wall}(t)\simeq \rho_{\rm wall}L_{\rm wall}4\pi R^2$ and $E_{\rm osc}(t)\simeq \rho_{\rm osc}L_{\rm osc}4\pi R^2$ are the energies of the bubble wall and oscillating condensate, respectively, and $L_{\rm osc}=R-R_c$ is the growing distance between the wall and the collision point, with $R_c$ the bubble radius at collision. We obtain
\begin{equation}
\label{eq:rho_wall}
    \rho_{\rm wall}(t)\simeq \rho_{\rm wall}(t_c)\left(\frac{R_c}{R}\right)^2 - \rho_{\rm osc}\frac{R-R_c}{L_{\rm wall}}.
\end{equation}
where we have used that $L_{\rm wall}$ is constant in $t$ after collision.
 The wall energy is transferred into the oscillating condensate after a time $t_{\rm damp}$ when $\rho_{\rm wall}(t_{\rm damp})\simeq 0$ in Eq.~\eqref{eq:rho_wall}. We conclude that after collision, the walls continue propagating during a distance 
\begin{equation}
\label{eq:distance_d}
    L_{\rm osc}(t_{\rm damp}) \equiv R(t_{\rm damp})-R_c \simeq \frac{\rho_{\rm wall}(t_c)}{\rho_{\rm osc}}L_{\rm wall} \sim \gamma_{\rm w}^2 L_{\rm wall} \sim \frac{\gamma_{\rm w}}{\Tn},
\end{equation}
before being completely dissolved in the center-of-mass frame of the two colliding walls, which for simplicity we approximate to coincide with the bath frame in this work. 
In the second equality of Eq.~(\ref{eq:distance_d}) we have assumed that $R_c/R \simeq 1$, indicating that $L_{\rm osc}(t_{\rm damp}) \ll R_c$. This condition holds in the terminal velocity regime characterized by $\gamma_{\rm w}/\Tn \ll R_c$, and remains acceptable to an order of magnitude in the runaway regime where $\gamma_{\rm w}/\Tn \lesssim R_c$.

\subsection{Wall-shell collision}
\label{sec:wall_shell}

Shells of particles produced during the initial propagation of the walls can either interact with the collided wall (the peak of stress-energy tensor) or the oscillating left-over (the diluted part), see Fig.~\ref{fig:shell_wall_collision}. We model the two scalar field configurations as Bose-Einstein condensate with number densities 
    \begin{equation}
    n_{\rm wall} \sim \gamma_{\rm w} m_{\phi}v_\phi^2, \qquad  n_{\rm osc} \sim  m_{\phi}v_\phi^2
    \end{equation}
and energies 
    \begin{equation}E_{\rm wall} \sim \gamma_{\rm w} m_{\phi}, \qquad E_{\rm osc} \sim m_\phi,
    \end{equation}
respectively. The interaction probability of a particle $X$ with the two aforementioned scalar field configurations reads
    \begin{align}
    \mathcal{P}_{\rm shell-wall } = n_{\rm wall} L_{\rm wall} \sigma_{\rm eff}(\EX,E_{\rm wall})
    = \frac{ m_{\phi} v_{\phi}^2 }{ T_n } \sigma_{\rm eff}(\EX,E_{\rm wall}),  \\
    \mathcal{P}_{\rm shell-osc }  = n_{\rm osc} L_{\rm osc} \sigma_{\rm eff}(\EX,E_{\rm osc})
    = \frac{ \gamma_{\rm w} m_{\phi} v_{\phi}^2 }{ T_n } \sigma_{\rm eff}(\EX,E_{\rm osc}),
    \end{align}
where the effective cross section, $\sigma_{\rm eff}$, depends on the underlying interaction and shell particle energy $\EX \simeq \pX$.

As an example, consider a gauge interaction between the shell particle and quantum of the wall peak or oscillating condensate. Consider a change in momentum of order $\EX$. This may either be induced through t-channel gauge boson exchange, like M\o{}ller scattering, or Feynman diagrams without t-channel gauge bosons, like Compton scattering. The former features a soft enhancement, as we have seen previously in Sec.~\ref{sec:mom_loss}, leading to
\begin{equation}
\sigma_{\rm eff} \sim \frac{ \alphaD^2 }{ (-\hat{t}_{\rm min})} \qquad \text{(M\o{}ller scattering)}
\end{equation}
where $-\hat{t}_{\rm min} = \mathrm{Min}[\EX^2,\EX E_{\rm wall}]$, giving the required momentum exchange.\footnote{The real scalar quanta of the condensate couple off-diagonally with the imaginary component of the original complex field to the gauge bosons. However, the momentum exchange is large compared to the masses, so effective massless gauge boson exchange is valid to consider.} In the case of Compton scattering we instead just have 
    \begin{equation}
    \sigma_{\rm eff} \sim \frac{ \alphaD^2 }{ \EX E_{\rm wall} } \qquad \text{(Compton scattering),}
    \label{eq:comptonsigmaeff}
    \end{equation}
as the cross section is hard scattering dominated.

\subsubsection{Leading Order Shell}
Keeping in mind the above cross sections, we consider first a leading order shell, where $\EX = \Delta m_{X}^2/\Tn$, and assume all the mass is gained in the PT with $\Delta m_{X}^2 \simeq \alphaD v_{\phi}^{2}$. Then in the case of $t-$channel gauge boson exchange, such as  M\o{}ller scattering, we have\footnote{Note the first term is non-zero in the limit $\alphaD \to 0$, but this is not a problem, as we are calculating the scattering probability given the existence of a shell particle (which are not produced for $\alphaD = 0$), and the cancellation occurs because the required momentum squared exchange $-\hat{t} \approx \EX^2 \propto \alphaD^2$, cancels the $\alphaD^2$ factor in the numerator in the  $\alphaD \to 0$ limit.} 
    \begin{equation}
    \boxed{
    \mathcal{P}_{\rm shell-wall }  \approx \mathrm{Max}\left[ \frac{m_{\phi}T_n}{v_{\phi}^2}, \quad \frac{\alphaD}{\gamma_{\rm w}} \right],  \quad \textrm{and}\quad 
    \mathcal{P}_{\rm shell-osc } \approx \mathrm{Max}\left[ \frac{ \gamma_{\rm w} m_{\phi} T_n }{ v_{\phi}^2 }, \quad \gamma_{\rm w} \alphaD \right].}
    \end{equation}
This is applicable, e.g.~in the case of a leading order shell of gauge bosons in a non-abelian theory.
For interactions without $t-$channel gauge boson exchange, such as Compton scattering, we instead have
        \begin{equation}
    \boxed{
    \mathcal{P}_{\rm shell-wall }  \approx \frac{\alphaD}{\gamma_{\rm w}},  \quad \textrm{and}\quad 
    \mathcal{P}_{\rm shell-osc } \approx \gamma_{\rm w} \alphaD.}
    \end{equation}
This is applicable, e.g.~in the case of a leading order shell of gauge bosons in an abelian theory.

\subsubsection{Gauge boson shell $\alphaD \lesssim 1$}
Here we have $\EX \approx \gamma_w \sqrt{\alphaD} v_{\phi}$.  Then in the case of an interaction involving $t-$channel gauge boson exchange,  such as  M\o{}ller scattering, we have
    \begin{equation}
    \boxed{
    \mathcal{P}_{\rm shell-wall } \approx \mathrm{Max}\left[ \frac{ \alphaD m_{\phi} }{ \gamma_{\rm w}^2 T_n },  \quad  \frac{ \alphaD^{3/2} v_{\phi} }{ \gamma_{\rm w}^2 T_n } \right],  \quad \textrm{and}\quad
    \mathcal{P}_{\rm shell-osc }   \approx \mathrm{Max}\left[ \frac{ \alphaD m_{\phi} }{ \gamma_{\rm w} T_n },  \quad  \frac{ \alphaD^{3/2} v_{\phi} }{ T_n } \right].}
    \end{equation}
Again this is applicable for gauge bosons in a non-abelian theory. For Compton scattering, we instead have
    \begin{equation}
    \boxed{
    \mathcal{P}_{\rm shell-wall } \approx  \frac{ \alphaD^{3/2} v_{\phi} }{ \gamma_{\rm w}^2 T_n },  \quad \textrm{and}\quad
    \mathcal{P}_{\rm shell-osc }   \approx \frac{ \alphaD^{3/2} v_{\phi} }{ T_n } .}
    \end{equation}
Which is applicable for gauge bosons in an abelian theory. The results here does not depend on whether it is a reflected or transmitted gauge boson shell.

\subsubsection{Discussion}

For our representative examples, the regions where shells interact efficiently with either the gradient energy energy in bubble walls or the oscillating scalar field formed after the collision are shaded in Fig.~\ref{fig:wall_shell_collision}. 
For the abelian case, we have assumed shells of gauge bosons, which do not exchange $t-$channel gauge bosons in their interactions with the condensate, so the scattering is Compton. For the non-abelian case, we have assumed $t-$channel gauge boson exchange, so the scattering is M\o{}ller.

In the case of leading order shells, we see our estimates show that in much of the parameter space, following the bubble collisions, the shell particles will interact significantly with the oscillating condensate.
In the case of transmitted and reflected gauge bosons from bremsstrahlung, the free streaming regions after bubble collision are much larger. This is due to the higher energies of the shell particles. Note also, that in the case of radiated reflected particle shells, the shells coming from opposite bubbles always interact with each other first, before encountering the other bubble.
Therefore, if one is interested in particle production from shell collisions as in bubbletrons~\cite{Baldes:2023rqv}, one can use the free-streaming predictions even inside the shaded areas for radiated reflected shells, while one cannot inside the shaded areas for LO and radiated transmitted shells, because these shells find the walls before finding each other.
Other types of interactions and shells of Table~\ref{tab:production_mechanism} can be dealt with in a similar fashion as the examples we have showcased above.\footnote{
For example, we can apply the above method to the Azatov-Vanvlasselaer shells of free streaming non-cold DM produced in a PT studied in Ref.~\cite{Baldes:2022oev}. We find shell particle interactions with the oscillating condensate rule out the $\Tn < T_{\rm eq}$ case studied therein,
while allowed regions of parameter space survive in the $\Tn > T_{\rm eq}$ case, where DM with a mass of $10^9$~GeV produced from a weak scale PT could possibly leave an observable signal both in the matter power spectrum and in GW.}

Finally note that our results concerning wall-shells interactions have been derived following a rather simple approximation.
For example, our shaded areas would be affected by taking into account that the center-of-mass frame of two bubble walls is not still in the CMB frame, but will realistically move because the colliding walls do not have exactly the same velocity.
In addition, more complete treatment would require solving Boltzmann equations in the presence of a background field that varies both in time and in space, see e.g.~\cite{Ai:2023qnr} for a recent attempt in this direction, although in a completely different context and, as far as we understand, within approximations that do not straightforwardly apply to our case.
Refining our predictions, while interesting, goes beyond the purposes of this paper.

\section{Gravitational waves from free-streaming shells}
\label{sec:GW}

Another application of the present study deals with the GW spectrum resulting from a strong first-order PT.
In the free-streaming regime studied in this paper, the plasma dynamics can be reduced to the evolution of ultra-thin relativistic shells. As we argue in this section, this allows the application of established results \cite{Jinno:2017fby,Konstandin:2017sat} concerning the GW spectrum from infinitely-thin bubble walls. This extends the applicability of the ``bulk flow'' model \cite{Jinno:2017fby,Konstandin:2017sat}, usually restricted to the run-away regime, to the friction-dominated regime provided that shells free stream.

\paragraph{Energy budget.}
When a bubble expand until a radius $R$, the latent heat initially stored in the false vacuum phase $E_{\rm vac}=4\pi R^3\Delta V/3$  is converted into kinetic energy of the wall $E_{\rm wall} = 4\pi R^2\sigma \gamma_w$, where $\sigma$ is the wall surface tension.
The fraction $\kappa_{\phi}$ of the latent heat  energy stored in the scalar field gradient in bubble walls reads
\begin{equation}
\label{eq:kappa_phi}
    \kappa_{\phi} = \frac{E_{\rm wall}}{E_{\rm vac}} = \frac{3\gamma \sigma R_c}{\Delta V} = \frac{\gamma_{\rm coll}}{\gamma_{\rm run}},
\end{equation}
where $\gamma_{\rm coll}$ is the wall Lorentz factor at collision accounting for friction given by Eq.~\eqref{eq:gamma_final}, while $\gamma_{\rm run}$ is the same quantity neglecting
friction, given by Eq.~\eqref{eq:gamma_run}.
The remaining fraction of the latent heat
\begin{equation}
\label{eq:kappa_plasma}
    \kappa_{\rm plasma} = 1 - \kappa_{\phi},
\end{equation}
is converted into plasma excitations due to the presence of the friction pressure.

\paragraph{Bubble wall contribution.}
When $\kappa_{\phi} \simeq 1$ in Eq.~\eqref{eq:kappa_phi}, the latent heat is dominantly kept in the scalar field gradients stored in bubble walls. 
 GWs arising from scalar field gradients were initially calculated under the ``envelope'' approximation, where the assumption was that walls are infinitely thin and the parts that collided were disregarded \cite{Kamionkowski:1993fg,Caprini:2007xq, Huber:2008hg,Jinno:2016vai,Weir:2016tov}. Subsequently, the collided parts were included into the analysis through the ``bulk flow'' model both at the analytical \cite{Jinno:2017fby,Megevand:2021juo} and numerical levels \cite{Konstandin:2017sat,Lewicki:2020jiv,Lewicki:2020azd,Cutting:2020nla}.
 The bulk flow model relies on assuming that the collided parts can still be regarded as infinitely thin and expand freely \cite{Jinno:2017fby}. Using conservation of the total bubble wall energy $E_{\rm wall}$, this implies that the bubble wall stress-energy momentum tensor in the radial direction decreases as \cite{Jinno:2017fby}
 \begin{equation}
 \label{eq:T_rr_scaling}
     T_{\rm rr}
     \sim \rho_{\rm wall}
     \sim E_{\rm wall} / (R^2 \times L_{\rm wall}) \propto 1/R^2,
 \end{equation}
 until being dissipated, cf.~Eq.~\eqref{eq:rho_wall}. We have assumed the wall thickness $L_{\rm wall}$ in the bath frame to be constant after collision.  
The scaling in Eq.~\eqref{eq:T_rr_scaling} leads to an infrared enhancement of the GW spectrum, characterized by $\Omega_{\rm PT}\propto f^{1}$, in contrast to the previously assumed $\Omega_{\rm PT}\propto f^{3}$ \cite{Jinno:2017fby,Megevand:2021juo,Konstandin:2017sat,Lewicki:2020jiv,Lewicki:2020azd,Cutting:2020nla}. The  GW spectrum from the bulk flow model in the relativistic limit reads~\cite{Konstandin:2017sat}
	\begin{equation}
 	 \label{eq:Bulk_flow}
 \Omega_{\rm PT}h^2\simeq \frac{10^{-6}}{(g_*/100)^{1/3}} \left(\frac{H_*}{\beta} \right)^{\!2} \left( \frac{\alpha}{1+\alpha} \right)^{\!2}  S_{\rm PT}(f)S_{H}(f),
 \end{equation}
with the spectral shape $S_{\rm PT}(f)$ peaked on $f_\phi$
	\begin{equation}
	\label{eq:spectral_shape_scalar}
	S_{\rm PT}(f) = \frac{ 3(f/f_{\rm PT})^{0.9} }{2.1+0.9(f/f_{\rm PT})^{3}},\quad f_{\rm PT} = \left(\frac{a_*}{a_0}\right) 0.8 \left(\frac{\beta}{2\pi}\right),
	\end{equation}
and the redshift factor between percolation ``$*$'' and today ``$0$''
	\begin{equation}
	\label{eq:redshift_fac}
a_*/a_0 = 1.65 \times 10^{-2}~{\rm mHz}~\left(\frac{T_{\rm eq}}{100~\rm GeV}\right) \left( \frac{g_{\rm eff, \,reh}}{100} \right)^{1/6} H_{*}^{-1}.
	\end{equation}
We added the correction factor 
\begin{equation}
\label{eq:Hubble_expansion_fac}
S_{H}(f) = \frac{(f/f_{\ast})^{2.1}}{1+(f/f_{\ast})^{2.1}}, \quad f_{\ast} = c_*\left(\frac{a_*}{a_0}\right)\left(\frac{H_*}{2\pi}\right),
\end{equation}
with $c_* = \mathcal{O}(1)$ to impose an $f^{3}$ scaling for emitted frequencies smaller than the Hubble factor $ H_{\ast}/(2\pi)$ as required by causality~\cite{Durrer:2003ja,Caprini:2009fx,Cai:2019cdl,Hook:2020phx}.

\paragraph{Plasma contribution.}
Instead, if $\kappa_{\rm plasma}\simeq 1 $ in Eq.~\eqref{eq:kappa_plasma}, then the friction pressure in Eq.~\eqref{eq:P_LL} prevents bubble walls from accelerating further, see the shaded regions in Fig.~\ref{fig:runaway}, and the latent heat is dominantly dumped into the plasma.
Lattice calculations of the GW spectrum from fluid dynamics have been restricted to the regime of non-relativistic velocities $v_w\ll 1$, see e.g. \cite{Hindmarsh:2015qta,Jinno:2020eqg}. Apart from the very extreme case where bubbles expand in pure vacuum \cite{Cutting:2018tjt,Cutting:2020nla}, there is no lattice study of the
GW spectrum from first-order PTs in the relativistic limit $v_w\simeq 1$ due to three technical limitations. At first, hydrodynamic equations become highly non-linear once relativistic corrections are added \cite{Jinno:2019jhi}. Second, Lorentz contraction makes the fluid profile thinner than the lattice spacing. 
Third, the hydrodynamic regime is expected to break down as soon as the particle free-streaming length is larger than the fluid profile thickness, which applies to (but is not limited to) the case of free-streaming shells.

Instead, in the white region of Fig.~\ref{fig:free_stream_all_cases} the plasma excitations take the form of relativistic shells which propagate freely, retaining their ultra-thin profile. 
This suggests that the energy-conserving scaling in Eq.~\eqref{eq:T_rr_scaling} is applicable not only to bubble walls but also to free-streaming shells, through the substitutions $E_{\rm wall}\to E_{\rm shell}$ and $L_{\rm wall}\to L_{\rm shell}$.
From a gravitational perspective, an extremely spatially-walled momentum distribution within the plasma should be indistinguishable from an extremely spatially-walled momentum distribution within the scalar field, provided that the peak width is much smaller than the bubble radius. 
Therefore, we anticipate that the GW signal for relativistic bubble wall velocity in the combined friction-dominated and free-streaming regime should resembles the GW signal in the run-away regime, described by the bulk flow model in Eq.~\eqref{eq:Bulk_flow}. This is corroborated by the study~\cite{Jinno:2022fom} conducted in the moderately relativistic regime $\gamma_w \lesssim 10$ where the GW spectrum from free-streaming LO shells indeed resembles the one predicted in the bulk flow model.
Hence in all the white regions of this paper, the GW signal is expected to be well approximated by Eq.~\eqref{eq:Bulk_flow}.
Instead, in all the colored (or dashed) regions of this paper, the fate of those relativistic shells and the resulting GW spectrum, but also the friction pressure and bubble wall Lorentz factor, remain unknown.
The calculation of GW spectra along with the potential applicability of the bulk flow model $T_{\rm rr} \propto 1/R^2$, in regions where shells do not free stream, is left for future work.\footnote{
In Refs.~\cite{Lewicki:2020jiv,Lewicki:2020azd} and \cite{Lewicki:2022pdb} it is found that the stress-energy momentum tensor after collision of, respectively, the scalar field gradient and fluid profile in the hydrodynamical limit (i.e. when shells do not free-stream), decreases as $T_{\rm rr} \propto 1/R^3$. Such a result, if confirmed, suggests additional dissipation with respect to the naive expectation from energy conservation in Eq.~\eqref{eq:T_rr_scaling}, and would imply the need to improve the bulk-flow model.
}

\section{Conclusions}
\label{sec:conclusions}

First-order cosmological phase transitions (PT) with relativistic bubble walls are predicted in several motivated extensions of the standard model and allow for unique opportunities to solve some of its most pressing mysteries, like the nature of dark matter or the origin of the baryon asymmetry. Importantly, they are the only PT (as opposed to those with slower walls) that result in a gravitational wave (GW) background which one could hope to observe with current and foreseen GW telescopes.
A clear understanding of the dynamics associated with relativistic bubble walls has not yet been achieved by the community, possibly jeopardizing their predictions that attracted a lot of attention in recent literature, like those concerning dark matter, baryogenesis, and GW.

One key property of PTs with relativistic walls, that has so far received little attention, is that interactions between particles in the thermal bath and the background field that constitutes the wall necessarily generate secondary energetic particles~\cite{Bodeker:2017cim,Azatov:2020ufh,Baldes:2020kam,Gouttenoire:2021kjv,Baldes:2022oev,GarciaGarcia:2022yqb}. These form, and accumulate into, ultra-thin energetic shells, possibly very dense, on both sides of the walls.
In this paper, we have performed the first systematic study of such shells. In particular:
\begin{enumerate}
    \item
    In Sec.~\ref{sec:shells_general} we have listed all shells ever mentioned in the literature and identified new ones that, to our knowledge, had never been considered before. We have then provided, for the first time, a systematic quantitative characterisation of their properties, see Tab.~\ref{tab:production_mechanism} for a summary with pointers to more detailed derivations in the text. This constitutes a necessary ingredient for any subsequent study of shells.
    \item
    We have then determined the regions, of parameter space of first-order PTs, where these shells free stream (meaning they propagate freely through the universe) and therefore where one can trust many of the commonly used predictions from first-order PTs with relativistic walls. A summary of our results is reported in Fig.~\ref{fig:free_stream_all_cases}: we discovered that regions of parameter space widely considered in the literature lies in the region where shells do not free stream and where many properties of the PT (wall velocities, particle production, GWs, etc) will need to be re-evaluated.
    This is for example the case for the PTs explaining the recently observed GW at pulsar timing arrays, as well as for basically all PTs associated with the spontaneous breaking of a gauge group, unless the gauge coupling is orders of magnitude smaller than one or the breaking scale is much above a PeV.
\end{enumerate}
The processes that can prevent shells free-streaming, that we studied in this paper, are phase-space saturation effects (see Sec.~\ref{sec:phase_space_sat} and Fig.~\ref{fig:phase_space_sat}), interactions that affect the momentum of shell and/or bath particles, dominantly due to $2\to2$ processes (see Sec.~\ref{sec:mom_loss} and Figs.~\ref{fig:shell_reversal} and~\ref{fig:shell_meta}), interaction that change both the number and the momenta of shell and/or bath particles, dominantly due to $3\to2$ processes (see Sec.~\ref{sec:3to2} and Fig.~\ref{fig:3to2}) and shell interactions with the bubble walls (see Sec.~\ref{sec:wall_int} and Fig.~\ref{fig:wall_shell_collision}).

In exploring the evolution of shells, we obtained results that may be useful also in very different contexts.
They include
\begin{itemize}
\item[$\diamond$]
    The computation of the rate of momentum loss of particles through interaction with a surrounding thermal bath in Sec.~\ref{sec:mom_loss}, extending the methods of \cite{Baldes:2022oev}.
    \item[$\diamond$]
    An algorithm to simplify the phase-space integration of any 5-point amplitude, see App.~\ref{app:3to2_phasespace}.
    \item[$\diamond$]
    The computation of the $3\to2$ squared amplitudes and interaction rates for initial particles with very different momenta, which we report synthetically in Sec.~\ref{sec:3to2} and extensively in App.~\ref{app:more_on_3to2}, where we also provide tabulated results for all $3\to2$ squared amplitudes of any theory involving initial vectors, fermions or scalars.
\end{itemize}

One implication of our results, concerning gravitational waves from phase transitions with relativistic walls, is an extension of the applicability of their bulk-flow modeling~\cite{Jinno:2017fby,Konstandin:2017sat} to some of the cases where GWs are dominantly sourced by the particle shells.
GWs from shells are dominant over those from the background scalar field for friction-dominated walls, i.e. when walls do not run away but reach a terminal velocity.
The computation of GWs from shells lies outside the regime of validity of hydrodynamics modelings employed so far.
We have argued in Sec.~\ref{sec:GW} that the bulk-flow modeling should provide a good description of GWs sourced by the shells, both in the friction-dominated and runaway regimes, provided that shells free stream.

The study carried out in this paper opens several new avenues of investigation. It identifies the first processes that need to be studied, in case one would want to determine how shells evolve in the regions where they do not free-stream. For example, for shells made of radiated particles (either transmitted or reflected) at a PT associated with the spontaneous breaking of an abelian gauge symmetry, the most important process in inducing a departure from free-streaming is given by $3\to2$ interactions, see Fig.~\ref{fig:3to2}.
Determining the evolution of shells under these interactions would then allow to understand both the properties of those shells and the wall velocities, and in turn to obtain new predictions that depend on them like particle production via bubbletrons~\cite{Baldes:2023rqv}, as well as the GW spectrum predicted by the PT~\cite{Caprini:2015zlo,Caprini:2019egz}.

\section*{Acknowledgements}
We thank Aleksander Azatov, Jacopo Ghislieri, Ryusuke Jinno, Felix Kahlhoefer,  Henda Mansour, Guy Moore, Miguel Vanvlasselaer for very useful discussions. We also thank Miguel Vanvlasselaer for useful comments on the manuscript.

\paragraph*{Funding information.}
The work of IB was supported by the European Union’s Horizon 2020 research and innovation programme under grant agreement No 101002846, ERC CoG ``CosmoChart". YG is grateful to the Azrieli Foundation for the award of an Azrieli Fellowship. The work of FS is partly supported by the Italian INFN program on Theoretical Astroparticle Physics (TAsP), by the French CNRS grant IEA “DaCo: Dark Connections” and by COST (European Cooperation in Science and Technology) via the COST Action COSMIC WISPers CA21106.

\appendix

\section{Shell production mechanisms}
\subsection{Radiated vector bosons}
\label{app:weak}

\subsubsection{Number of radiated quanta}
Upon acquiring a mass inside bubble walls, a charged particle $a$, either a scalar a fermion or a vector boson, can radiate bremsstrahlung radiation $a\to bc$ \cite{Bodeker:2017cim,Azatov:2020ufh,Gouttenoire:2021kjv}. We denote by $m_{c,s}$ and $m_{c,h}$ the mass of the radiated vector boson in the symmetric and broken (``Higgs'') phase, by $x = E_c/E_a$ the parent-to-radiated energy ratio, and by $k_\perp$ the transverse momentum of the radiated particle. The parameter space allowed by kinematics can be approximated to \cite{Gouttenoire:2021kjv}
\begin{align}
\label{eq:range_x}
\sqrt{k_{\perp}^2 + m_c(z)^2}/E_{a}~ \leq ~ x ~ \leq ~1 ,\qquad 0 ~\leq ~ k_{\perp} ~ \leq~ E_{a}.
\end{align} 
The splitting probability reads \cite{Gouttenoire:2021kjv}
\begin{align}
dP_{a \to bc} =\zeta_a\frac{d k_{\perp}^2}{k_{\perp}^2}  \frac{d x}{x}~\Pi(k_\perp),
\label{eq:dP_abc_mcs}
\end{align}
where $\Pi(k_{\perp})$ contains the IR and UV suppression factors
\begin{equation}
\Pi(k_{\perp}) \equiv \left(\frac{k_{\perp}^2}{k_{\perp}^2+m_{c,s}^2}\right)^2 \left(\frac{m_{c,h}^2-m_{c,s}^2}{k_{\perp}^2+m_{c,h}^2}\right)^2, \label{eq:IR_UV_suppression_factor}
\end{equation}
and
\begin{equation}
\zeta_a \equiv \frac{\alphaD}{\pi} \sum_{b,c} C_{abc}, \qquad  \qquad \alphaD  \equiv \frac{\gD^2}{4\pi}, \label{eq:zeta_a_def}
\end{equation}
where $C_{abc}$ is the charge factor \cite{Hoche:2020ysm}.
In this work, we consider the mass in the broken and symmetric phase to be set by the Higgs mechanism and plasma effects respectively 
\begin{equation}
\label{eq:mch_mcs}
     m_{c,h}^2 \simeq 2\pi \alphaD v_\phi^2 + \mu^2,\qquad m_{c,s}^2 \simeq \mu^2,
\end{equation}
where $\mu$ is the IR cut-off sets by thermal corrections $\sim \gD T_n$ plus eventual finite-density corrections $\sim \gD \sqrt{n_c/p_c}$ in the non-abelian case, see Eq.~\eqref{eq:IRcutoff} for the precise expression.
\begin{figure}[!ht]
\centering
\includegraphics[width=0.49\textwidth, scale=1]{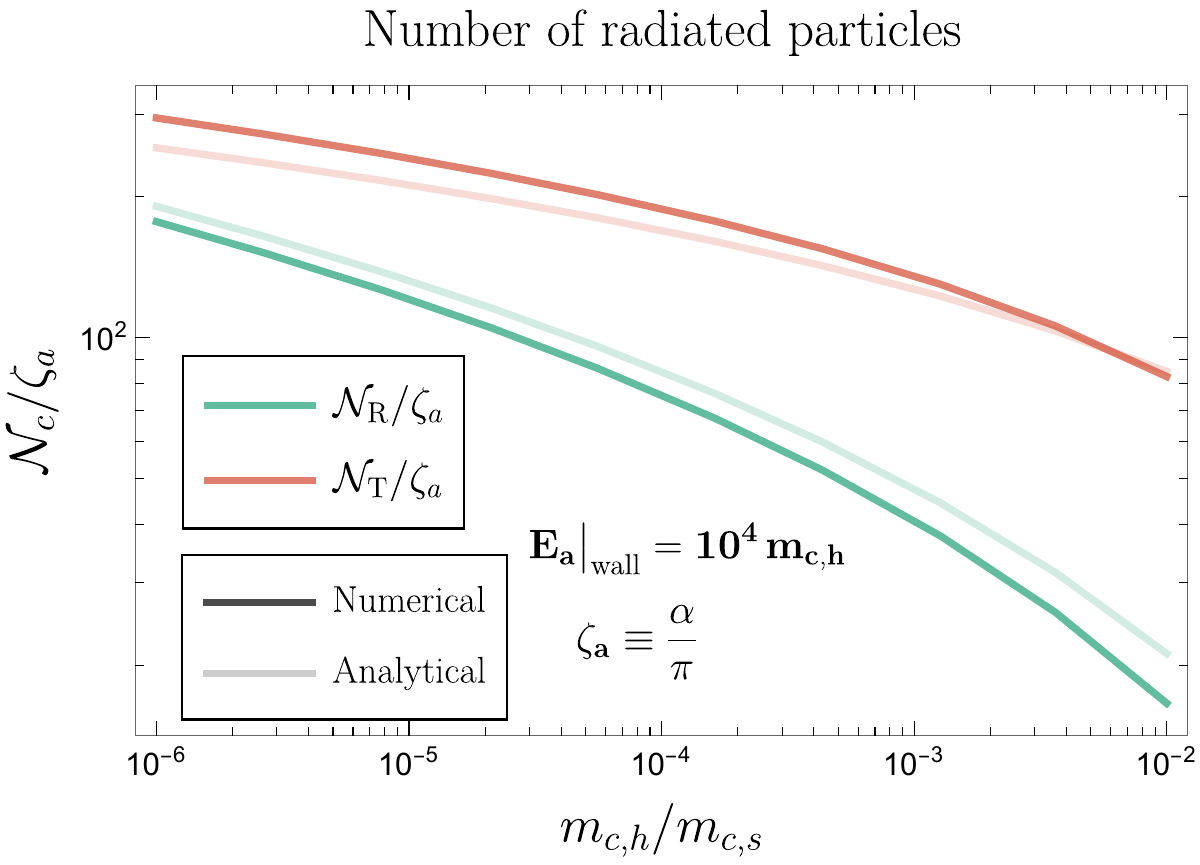}
\includegraphics[width=0.49\textwidth, scale=1]{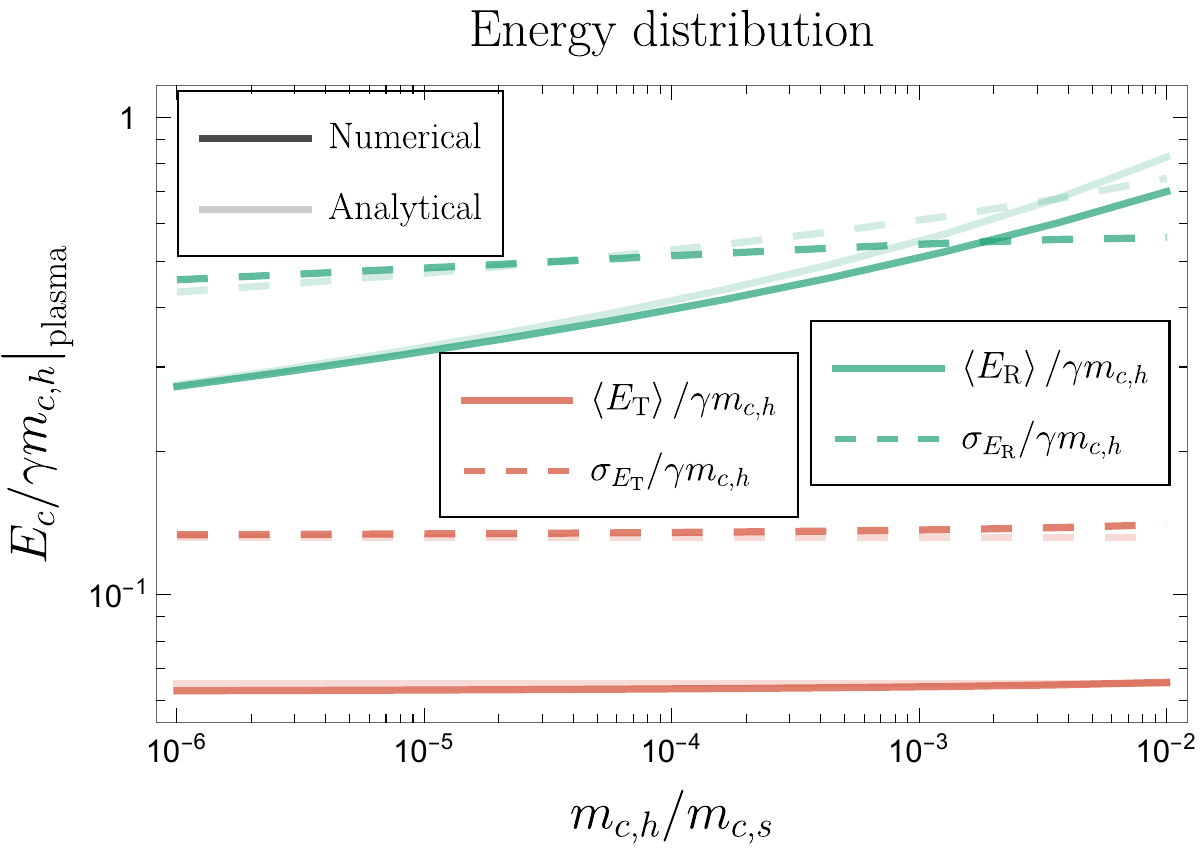}
\includegraphics[width=0.49\textwidth, scale=1]{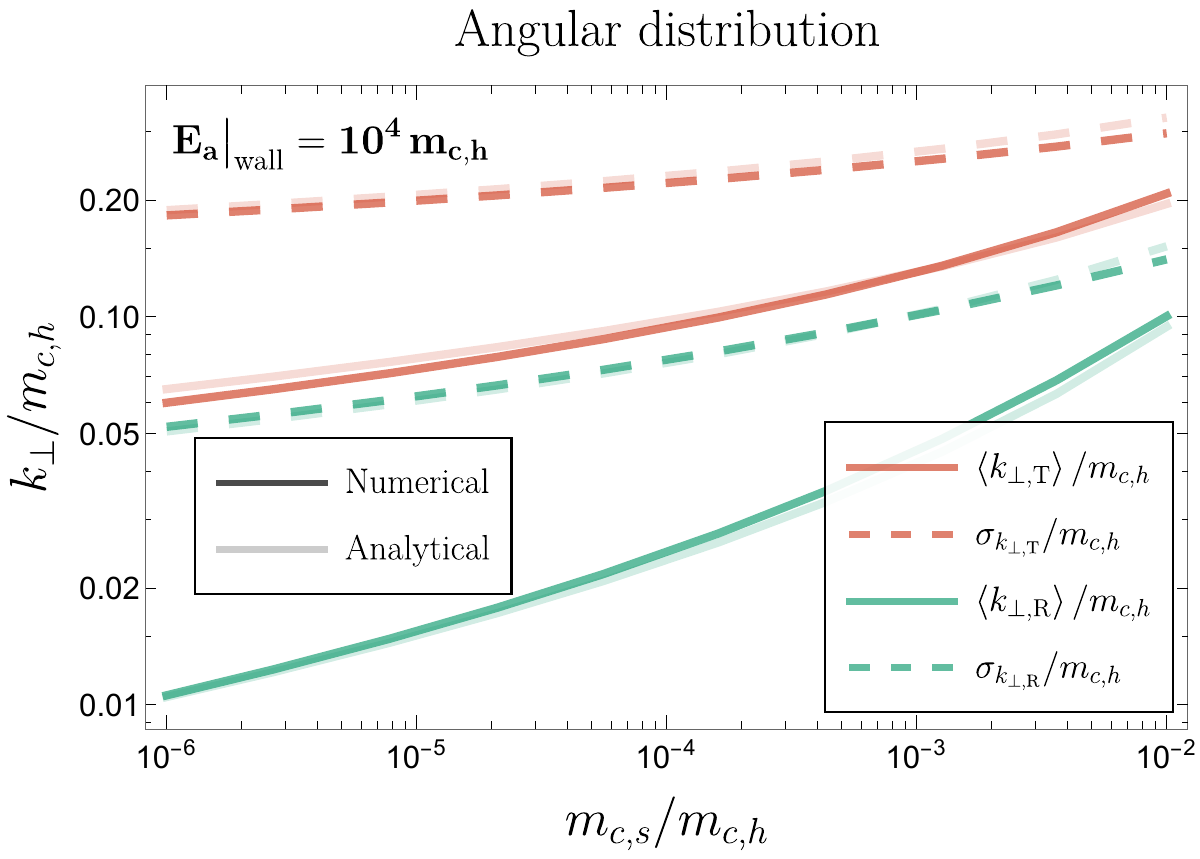}
\\[-1em]
\caption{\it \small We display the number of radiated particle 'c' (per single incoming particle 'a') and their associated energy and emission angle (mean and standard deviation). All quantities are expressed in the bath frame except the energy $E_a$ of incoming particles which is expressed in the wall frame.  }
\label{fig:number_energy_ang_dist}
\end{figure}
The number of reflected and transmitted vector bosons per incoming particle reads
\begin{equation}
\label{eq:rough_estimate}
\mathcal{N}_{a\to bc} = \mathcal{N}_{\rm R} +\mathcal{N}_{\rm T} , 
\end{equation}
where $\mathcal{N}_{\rm R}$ and $\mathcal{N}_{\rm T}$ are the reflected and transmitted population. We compute
\begin{align}
\label{eq:N_R}
 &\mathcal{N}_{\rm R} \simeq
\int_{0}^{E_a}dk^2_\perp \int_{\frac{\sqrt{k_{\perp}^2+m_{c,s}^2}}{E_{a}}}^{\frac{\sqrt{k_{\perp}^2+m_{c,h}^2}}{E_{a}}}dx\frac{dP_{a \to bc}}{dk^2_\perp dx}
\simeq \,\zeta_a \log^2\left(\frac{m_{c,h}}{m_{c,s}}\right)  , \\
\label{eq:N_T}
 &\mathcal{N}_{\rm T} \simeq 
\int_{0}^{E_{a}^2}dk^2_\perp \,
\int_{\frac{\sqrt{k_{\perp}^2+m_{c,h}^2}}{E_{a}}}^{1}dx\frac{dP_{a \to bc}}{dk^2_\perp dx}
\simeq 2\zeta_a\,\log\left(\frac{m_{c,h}}{m_{c,s}}\right)\text{log}\left(\dfrac{E_a}{m_{c,h}}\right) ,
\end{align}
where we have expanded the results of the integrals for $m_{c,s} \ll m_{c,h} \ll E_a$. Note that Eqs.~(\ref{eq:N_R}) and~(\ref{eq:N_T}) are consistent with Eq.~(59) in \cite{Gouttenoire:2021kjv}.

\subsubsection{Emitted energy and splitting angle}
The mean energies of reflected and transmitted radiated particles in the bath frame read 
\begin{align}
\label{eq:E_R}
&\left<E_{\mathrm{R}}\right> \simeq \left< \gamma_{\rm w} \beta xE_a +  \gamma_{\rm w} \sqrt{(xE_a)^2-k_{\perp}^2-m_{c,s}^2} \right> 
\simeq \frac{3.8\gamma_{\rm w} m_{c,h} }{\log\left( \frac{m_{c,h}}{m_{c,s}}\right)}  , \\
\label{eq:E_T}
&\left<E_{\mathrm{T}}\right> \simeq \left< \gamma_{\rm w} \beta xE_a - \gamma_{\rm w} \sqrt{(xE_a)^2-k_{\perp}^2-m_{c,h}^2} \right>\simeq   \frac{0.6\gamma_{\rm w} m_{c,h}}{\log\left( \frac{E_a}{m_{c,h}}\right) },
\end{align}
where the limit $\gamma_{\rm w} \to +\infty$ was taken. We also compute the standard deviation $\sigma_X\equiv \sqrt{\left<(X-\left<X\right>)^2\right>}$
\begin{align}
\label{eq:sigma_E}
\sigma_{E_{\mathrm{R}}} \simeq \frac{1.6\gamma_{\rm w} m_{c,h} }{\log^{\!1/2}\left( \frac{m_{c,h}}{m_{c,s}}\right)},\qquad
&\sigma_{E_{\mathrm{T}}} \simeq \frac{0.4\gamma_{\rm w} m_{c,h} }{\log^{1/2}\left( \frac{E_a}{m_{c,h}}\right)}.
\end{align}
The angular distributions read 
\begin{align}
\label{eq:k_perp_mean}
    \left< k_{\perp,\mathcal{R}}\right>^2
\simeq \left(\frac{2m_{_c,h}}{\log^2\left( \frac{m_{c,h}}{m_{c,s}}\right)}\right)^2+T_n^2, \qquad
\left< k_{\perp,\mathcal{T}}\right>^2
\simeq  \left(\frac{0.9m_{_c,h}}{ \log\left( \frac{m_{c,h}}{m_{c,s}}\right)}\right)^2+T_n^2.
\end{align}
and their standard deviation are
\begin{equation}
\label{eq:k_perp_sigma}
    \sigma_{k_{\perp,\mathrm{R}}}^2 \simeq  \left(\frac{0.7m_{_c,h}}{\log\left( \frac{m_{c,h}}{m_{c,s}}\right)}\right)^2+T_n^2,
    \qquad 
    \sigma_{k_{\perp,\mathcal{T}}}^2 \simeq  \left(\frac{0.7m_{_c,h}}{ \log^{1/2}\left( \frac{m_{c,h}}{m_{c,s}}\right)}\right)^2
    +T_n^2.
\end{equation}
In Eqs.~\eqref{eq:k_perp_mean} and \eqref{eq:k_perp_sigma}, we have included a second piece of order the nucleation temperature $T_n$ to account for the transverse momentum $p_{a,\perp} \simeq T_n$ in the thermal distribution of incoming quanta.
All the analytical expressions above are valid in the large log limits and the numerical coefficient are fitted on the numerical calculation. Fig.~\ref{fig:number_energy_ang_dist} shows a good agreement between the (semi-)analytical and numerical results. 
We conclude that vector bosons, which are radiated whenever the PT occurs in a gauge symmetry group, are produced with a rate controlled by the gauge coupling constant squared $\mathcal{N}\propto \gD^2$, with a large momentum in the bath frame $\pX \simeq \gamma_{w}m_{c,h}$ and highly collimated $k_\perp \ll \pX$. We refer to \cite{Gouttenoire:2021kjv} for a calculation of the resulting friction pressure on bubble walls.
In the main text, we account for effects from reflected and transmitted particles separately, with associated multiplicity factor $\mathcal{N}$ and momentum $\pX$ in the wall frame 
\begin{equation}
\label{eq:weak_K_N}
\mathcal{N}\simeq \left(\mathcal{N}_{\rm R},\mathcal{N}_{\rm T}\right),\qquad \pX \simeq  \left(\left<E_{\mathrm{R}}\right>,\left<E_{\mathrm{T}}\right> \right) .
\end{equation}
In all the figures presented in this paper, we incorporate the standard deviations of the kinematic quantities $X=\{E_{\mathrm{R}},E_{\mathrm{T}},k_{\perp,\mathcal{R}},k_{\perp,\mathcal{T}}\}$ by adding them directly to the means 
\begin{equation}
    X \to \sqrt{\left<X\right>^2 +\sigma_X^2}.
\end{equation}

We now discuss a complication which arises when the vector bosons receives a large mass from finite-density effects.

\subsection{Large IR cut-off limit}
\label{app:large_IR_cutoff}

The expression we have derived in the previous section, Sec.~\ref{app:weak}, are only valid in the limit $m_{c,h}\ll m_{c,s}$ (and also $E_a \gg m_{c,h}$). However as we discussed in Sec.~\ref{sec:IR_cutoff}, for non-abelian gauge group finite-density corrections act as an IR cut-off ($X\equiv c$)
\begin{equation}
\mu^2 \simeq 2\gD^2\frac{n_X}{\pX},
\end{equation}
where we neglected the thermal part, see Eq.~\eqref{eq:IRcutoff}.
In the ultra-relativistic limit, shells are dense and the IR cut-off is larger than the vector boson mass in the broken vacuum $m_{c,h}^{\infty}\simeq \gD v_{\phi}/\sqrt{2}$, far away from the wall. 
We suppose that the vector boson is massless in the symmetric vacuum, so that far away from the wall and up to thermal effects $m_{c,s}^{\infty} =0$.
Close to the wall within the shells, the vector boson mass receives finite density corrections 
\begin{equation}
m_{c,s}^2\simeq \mu^2 ,\qquad \textrm{and}\qquad m_{c,h}^2\simeq \mu^2 + \gD^2v_{\phi}^2/2.
\end{equation}
The numbers of reflected and transmitted radiated vector bosons -- valid both in the limit $m_{c,h}\gg m_{c,s} $ in which Eqs.~\eqref{eq:N_R} and \eqref{eq:N_T} were derived and when $\Delta m_c \equiv m_{c,h}-m_{c,s} \lesssim m_{c,h}$ -- are well approximated by the following fitting formula
 \begin{align}
\label{eq:N_R_22}
 &\mathcal{N}_{\rm R} \simeq
\int_{0}^{E_a}dk^2_\perp \int_{\frac{\sqrt{k_{\perp}^2+m_{c,s}^2}}{E_{a}}}^{\frac{\sqrt{k_{\perp}^2+m_{c,h}^2}}{E_{a}}}dx\frac{dP_{a \to bc}}{dk^2_\perp dx}
\simeq  \zeta_a \left( \frac{\Delta m}{m_{c,s}+\Delta m/L_m^{2/3}}\right)^3, \\
\label{eq:N__2T}
 &\mathcal{N}_{\rm T} \simeq 
\int_{0}^{E_{a}^2}dk^2_\perp \,
\int_{\frac{\sqrt{k_{\perp}^2+m_{c,h}^2}}{E_{a}}}^{1}dx\frac{dP_{a \to bc}}{dk^2_\perp dx}
\simeq 2\zeta_a\,\left( \frac{\Delta m}{m_{c,s}+\Delta m/\sqrt{L_mL_E}}\right)^2 ,
\end{align}
with $L_m=\log\left(\frac{m_{c,h}}{m_{c,s}}\right)$ and $L_E=\text{log}\left(\dfrac{E_a}{m_{c,h}}\right)$. Those formula are valid for all values of $\Delta m_c \equiv m_{c,h}-m_{c,s}$ .
The radiation pressure becomes suppressed with the mass difference $\Delta m_c$ as
\begin{align}
\label{eq:Delta_p_R}
\left< \Delta p_{\rm R} \right> &\simeq \zeta_a
\int_{0}^{E_a}\frac{dk_{\perp}^2}{k_{\perp}^2} \,
\int_{\frac{\sqrt{k_{\perp}^2+\mu^2}}{E_{a}}}^{\frac{\sqrt{k_{\perp}^2+m_{c,h}^2}}{E_{a}}}\frac{dx}{x}\,\frac{k_{\perp}^4}{(k_{\perp}^2+\mu^2)^2}\, \left(\frac{m_{c,h}^2-m_{c,s}^2}{k_{\perp}^2+m_{c,h}^2}\right)^2 \times 2xE_a  \\[0.2cm]
&\simeq 4 \,\zeta_a\,\Delta m_{c} \left(\frac{\Delta m_{c} }{m_{c,s}+\Delta m_{c}/L_m^{1/3}}\right)^3 ,
\label{eq:Delta_p_R_2}
\end{align}
and 
\begin{align}
\label{eq:Delta_p_T}
\left< \Delta p_{\rm T} \right> &\simeq \zeta_a
\int_{0}^{E_{a}^2}\frac{dk_{\perp}^2}{k_{\perp}^2} \,
\int_{\frac{\sqrt{k_{\perp}^2+m_{c,h}^2}}{E_{a}}}^{1}\frac{dx}{x}\,\frac{k_{\perp}^4}{(k_{\perp}^2+\mu^2)^2}\, \left(\frac{m_{c,h}^2-m_{c,s}^2}{k_{\perp}^2+m_{c,h}^2}\right)^2 \times \frac{k_{\perp}^2 + m_{c,h}^2}{2\,x\,E_{a}} \\[0.2cm]
&\simeq \zeta_a\,\Delta m_{c} \left(\frac{\Delta m_{c} }{m_{c,s}+\Delta m_c/L_m^{2/3}}\right)^{3/2},
\label{eq:Delta_p_T_2}
\end{align}
Hence both the number of particles produced and the friction pressure become suppressed in the limit $\Delta m_c \ll m_{c,s}$. We leave the implication of this results for the bubble wall velocity for future research.

\subsection{Radiated scalar bosons}
\label{app:scalar}
\subsubsection{Number of radiated quanta}
We now study the production of scalar bremsstrahlung radiation. We consider the vertex
\begin{equation}
\phi\to \phi \phi:\qquad \qquad V(z)=\lambda \phi(z).
\end{equation}
In the high-energy limit $p_c^z \gg m_{c,h}$ and thin-wall limit, we can approximate the mode function of $c$ by 
\begin{equation}
\chi_{c}(z) \simeq  {\rm exp} \left(i \int_0^z p_c^z(z') dz' \right) \simeq e^{i E_{c} z} \exp \left(  - \frac{i}{2E_{c}} \int_0^z(m_{c}^2(z')+k^2_\perp)~ dz' \right),
\label{eq:chi_z}
\end{equation}
and idem for $a$ and $b$. We deduce
\begin{equation}
\chi_a(z)\chi_b^*(z)\chi_{c}^*(z) = e^{i\Delta p_z z},
\end{equation}
with
\begin{equation}
\Delta p_z = \frac{1}{2E_a}\left( - m^2 +\frac{m^2 + k_{\perp}^2}{1-x}+\frac{m^2 + k_{\perp}^2}{x} \right) =\frac{1}{2E_a}\left( \frac{k_\perp^2+(1-x+x^2)m^2}{x(1-x)} \right) \simeq \frac{1}{2E_a}\left( \frac{k_\perp^2+m^2}{x(1-x)} \right).
\end{equation}
The matrix element becomes
\begin{equation}
\mathcal{M} = \int dz , V(z)\,\chi_a(z)\chi_b^*(z)\chi_{c}^*(z) \simeq i\left(  \frac{V_h}{\Delta p_{z,h}} - \frac{V_s}{\Delta p_{z,s}} \right) \simeq i \frac{V_h}{\Delta p_{z,h}},
\label{eq:M_matrix_2}
\end{equation}
with $V_s=0$ and $V_h=\lambda^2v_\phi^2$.
The matrix element squared becomes
\begin{equation}
|\mathcal{M}|^2 \simeq \frac{4E_a^2(1-x)^2x^2\lambda^2v_\phi^2}{(k_\perp^2+(1-x+x^2)m^2)^2}.
\end{equation}
The splitting probability reads
\begin{align}
dP_{\phi \to \phi\phi} &= \frac{d k_{\perp}^2}{64\pi^2 E_a^2}\frac{dx}{x} \left|M\right|^2 = \frac{\lambda^2v_\phi^2}{16\pi^2}\frac{d k_{\perp}^2}{(k_\perp^2+(1-x+x^2)m^2)^2}x (1-x)^2 dx .
\end{align}
Integrating over the phase space, we obtain the number of radiated scalar bosons (reflected and total)
\begin{align}
&\mathcal{N}_{R} = \int_{\mu^2}^{E_a^2} dk_\perp^2 \int_{\sqrt{\mu^2+k_\perp^2}/E_a}^{\sqrt{m^2+k_\perp^2}/E_a} dx\frac{dP_{a \to bc}}{dk_\perp^2 dx} \simeq  \frac{\lambda^2v_\phi^2}{32\pi^2 E_a^2}. \\
&\mathcal{N}_{\rm tot} = \int_{\mu^2}^{E_a^2} dk_\perp^2 \int_{\sqrt{\mu^2+k_\perp^2}/E_a}^1 dx\frac{dP_{a \to bc}}{dk_\perp^2 dx} \simeq \frac{\lambda^2v_\phi^2}{192\pi^2 m_{c,h}^2}, 
\end{align}

\subsubsection{Emitted energy and splitting angle}
The mean energies of reflected and transmitted radiated scalar in the bath frame can be expressed as
\begin{align}
\label{eq:E_R}
&\left<E_{\mathrm{R}}\right> \simeq \left< \gamma_{\rm w} \beta xE_a +  \gamma_{\rm w} \sqrt{(xE_a)^2-k_{\perp}^2-m_{c,s}^2} \right> 
\simeq 3\gamma_{\rm w} m_{c,h} , \\
\label{eq:E_T}
&\left<E_{\mathrm{T}}\right> \simeq \left< \gamma_{\rm w} \beta xE_a - \gamma_{\rm w} \sqrt{(xE_a)^2-k_{\perp}^2-m_{c,h}^2} \right>\simeq 2 \gamma_{\rm w} \frac{ m_{c,h}^2}{E_a}\log{\left( \frac{E_a}{m_{c,h}}\right)} ,
\end{align}
and the angular distributions are given by
\begin{align}
\label{eq:k_perp_mean_scalar}
    \left< k_{\perp,\mathcal{R}}\right>
\simeq \left< k_{\perp,\mathcal{T}}\right>
\simeq  m_{c,h}.
\end{align}
We conclude that in the case of scalar Bremsstrahlung, which would be the leading process if the phase transition occurs in a global (not gauged) symmetry group, scalar bosons are radiated with a large momentum in the bath frame only if they are reflected. However, their abundance relative to the transmitted ones is suppressed  $\mathcal{N}_{R}/\mathcal{N}_{\rm tot}\simeq 6(m_{c,h}/E_a)^2\ll 1$.

\section{$3 \to 2$ final phase space integration}
\label{app:3to2_phasespace}

In this appendix we describe an algorithm that allows one to perform the final phase space integral of $3 \to 2$ processes for any scattering amplitudes, where we define the momenta of incoming and outgoing particles as in the main text, i.e. $p_1 p_2 p_5 \to p_3 p_4$.
We anticipate that our result, although the integration will be performed in a specific frame, is Lorentz invariant, and can therefore be expressed in terms of three independent scalar products $p_{1} \cdot p_{2}$, $p_{1} \cdot p_{5}$, and $p_{2} \cdot p_{5}$.

The most complicated expressions of scalar products, in the denominators of squared amplitudes $\left| \mathcal{ M } \right|^2$, are those that come from interference terms. One $5$-point amplitude, consisting of only $3$-point or $4$-point interactions, contains at most two denominators from propagators, which are different from each other. In an interference term this would be multiplied by two more denominators, all possibly different.
Therefore we should be able to compute, and integrate over final phase space, terms with at least $4$ different denominators, e.g.
\begin{align}
	\left| \mathcal{ M } \right|^2 \supset \dfrac{1}{s_{13}^{a} s_{24}^{b}} \cdot \dfrac{1}{s_{23}^{c} s_{54}^{d}} \,,
\end{align}
where we have defined $s_{ij} = (p_i \pm p_j)^2$ with $-$ for one incoming and one outgoing momenta, i.e. for $i,j=3,5$, and $+$ for two incoming or outgoing momenta, i.e. for $i,j=1,5$ or $i,j=3,4$ (this corresponds to $s_{ij} = (p_i + p_j)^2$ if momenta were all defined as incoming).
One possible approach to recast those integrals is the method of Feynman parameters.
In loop-computations one usually completes the square to get rid of the linear terms, and then performs the loop integration. Here the approach would be to perform a $SO(3)$ rotation to get rid of any azimuthal angle in the denominator, such that the integral over the polar angle can be performed. After doing so one ends up with an integral over $3$ Feynman parameters. However, these integrals are not textbook integrals and evaluating them efficiently is not guaranteed.

In the following we will therefore describe an algorithm which is able to perform the final phase space integration for all possible $3\to2$ amplitudes.

\subsection{Partial fraction decomposition of scalar products}

We start from the following expression for the squared $3\to 2$ amplitude,
\begin{align}
	\left| \mathcal{ M } \right|^2 &= \sum_{n} \mathcal{C}_{n} \ \dfrac{\prod s_{ij}^{k}}{ s_{13}^{a} s_{14}^{b} s_{23}^{c} s_{24}^{d} s_{53}^{e} s_{54}^{f} } \,.
\end{align}
In addition, $\left| \mathcal{ M } \right|^2$ can also depend on combinations of $s_{12}$, $s_{15}$, $s_{25}$, $s_{34}$, but these do not depend on the final phase space integration. Therefore, we absorb them into the prefactors $\mathcal{C}_{n}$.

The next step is to reduce the numbers of different factors in the denominator. We achieve this by the method of partial fraction decomposition and by exploiting energy momentum conservation. This can be achieved systematically, in an easy way, by multiplying with one non-trivial representation of the identity.
We derive several non trivial representations of the identity by writing energy momentum conservation in the following way,
\begin{align}
	p_{k} \cdot \qty( p_{1} + p_{2} + p_{5} - p_{3} - p_{4} ) &= 0 \,,
\end{align}
which represents five independent equations for $k = 1,2,3,4,5$. We find, for example for $k=1$,
\begin{align}
	p_{1} \cdot p_{1} + p_{1} \cdot p_{2} + p_{1} \cdot p_{5} &= p_{1} \cdot p_{3} + p_{1} \cdot p_{4} \,,
\end{align}
or equivalently
\begin{align}
	s_{12} + s_{15} - 2m_{1}^2 - m_{2}^2 - m_{5}^2 - m_{3}^2 - m_{4}^2 &= - s_{13} - s_{14} \,.
\end{align}
Performing this manipulation for all $k$ we find several equations for the identity,
\begin{align}
	1 &= \dfrac{s_{13} + s_{14}}{\mathcal{C}_{1}} \,, \\
	1 &= \dfrac{s_{23} + s_{24}}{\mathcal{C}_{2}} \,, \\
	1 &= \dfrac{s_{53} + s_{54}}{\mathcal{C}_{5}} \,, \\
	1 &= \dfrac{s_{13} + s_{23} + s_{53}}{\mathcal{C}_{3}} \,, \\
	1 &= \dfrac{s_{14} + s_{24} + s_{54}}{\mathcal{C}_{4}} \,,
\end{align}
where the constants $\mathcal{C}_{k}$ do not dependent on the final phase space integration.

More relevant identities, which can be derived in similar fashion, are
\begin{align}
	1 &= \dfrac{s_{14} + s_{23} + s_{53}}{\mathcal{C}_{14}} \,, \\
	1 &= \dfrac{s_{13} + s_{24} + s_{54}}{\mathcal{C}_{13}} \,, \\
	1 &= \dfrac{s_{13} + s_{24} + s_{53}}{\mathcal{C}_{24}} \,, \\
	1 &= \dfrac{s_{14} + s_{23} + s_{54}}{\mathcal{C}_{23}} \,, \\
	1 &= \dfrac{s_{13} + s_{23} + s_{54}}{\mathcal{C}_{54}} \,, \\
	1 &= \dfrac{s_{14} + s_{24} + s_{53}}{\mathcal{C}_{53}} \,.
\end{align}
All $\mathcal{C}$'s are guaranteed to not introduce any additional singularities into the amplitude.

It is straightforward to see that by strategic multiplication with these identities one can step by step reduce the powers of particular denominators up to eliminate some (but not all) of them.
After the full reduction has been completed, we end up with the following basic integrands in $\left| \mathcal{ M } \right|^2$,
\begin{align}
	\mathcal{I}_{1} &\supset \dfrac{1}{s_{13}^{a}} \,,&\qquad \mathcal{I}_{2} &\supset \dfrac{1}{s_{13}^{a} s_{24}^{b}} \,,&\qquad \mathcal{I}_{3} &\supset \dfrac{s_{13}^{m}}{s_{53}^{a}} \,,
	\label{eq:3to2basisIntegrals}
\end{align}
where any $3$ can also be a $4$ and vice versa.

Note that any term in the numerator which depends on the final phase space integration was already present in the amplitude, before our reduction procedure. In other words, our reduction algorithm does not generate any additional term in the numerator.

\subsection{Parametrization of the integration region}
\label{app:scalar_products_parametrization}

It is most convenient to evaluate the two-particle final phase space integral in the center-of-mass frame, where after eliminating the delta-distribution ensuring energy momentum conservation we are left with an integral over two angles,
\begin{align}
	R_{2}( (p_{3} + p_{4})^2 ) &= \dfrac{\dd[3]{p_{3}}}{(2\pi)^3 2E_{3}} \dfrac{\dd[3]{p_{4}}}{(2\pi)^3 2E_{4}} \delta^{(4)} \qty( p_{1} + p_{2} + p_{5} - p_{3} - p_{4} ) = \dfrac{\sqrt{ \lambda( (p_{3} + p_{4})^2 , m_{3}^2 , m_{4}^2 ) }}{32 \pi^2 (p_{3} + p_{4})^2} \int \!\dd{\Omega} \,,
\end{align}
where $\lambda(x,y,z) = x^2 + y^2 + z^2 - 2xy - 2yz - 2zx$ is the K{\"a}ll{\'e}n function and $\int \dd{\Omega}$ is the standard integral over the $S^{2}$-sphere,
\begin{align}
	\int \!\dd{\Omega} &= \int_{0}^{\pi} \!\dd{\theta_{13}} \sin(\theta_{13}) \int_{0}^{2\pi} \!\dd{\varphi_{13}} = \int_{-1}^{+1} \!\dd{z_{13}} \int_{0}^{2\pi} \!\dd{\varphi_{13}}
\end{align}
It is worthy to point out that the final phase space integral is invariant under the exchange of the two momenta of the final states, $p_{3} \leftrightarrow p_{4}$. This is mostly useful for simplifying the amplitude in order to safe computational resources by not calculating the same integral twice, since $p_{3} \leftrightarrow p_{4}$ implies $z_{13} \leftrightarrow - z_{13}$. The integral is also invariant under $O(3)$ rotations on the $S^2$-sphere, which we will exploit heavily in order to being able to compute some of the integrals.

We now state our choice for the parameterization of the five momenta,
\begin{align}
	p_{1, {\rm com}} = 
	\begin{pmatrix}
		\sqrt{ m_{1}^2 + p_{I}^2 } \\ 0 \\ 0 \\ p_{I}
	\end{pmatrix}
	\,,\
	p_{2, {\rm com}} = 
	\begin{pmatrix}
		\sqrt{ m_{2}^2 + p_{J}^2 } \\ p_{J} \sin(\theta_{12}) \\ 0 \\ p_{J} \cos(\theta_{12})
	\end{pmatrix}
	\,,\
	p_{5, {\rm com}} = 
	\begin{pmatrix}
		\sqrt{ m_{5}^2 + p_{K}^2 } \\ -p_{J} \sin(\theta_{12}) \\ 0 \\ -p_{I} - p_{J} \cos(\theta_{12})
	\end{pmatrix}
	\,, \\
	p_{3, {\rm com}} = 
	\begin{pmatrix}
		\sqrt{ m_{3}^2 + p_{F}^2 } \\ p_{F} \sin(\theta_{13}) \cos(\varphi_{13}) \\ p_{F} \sin(\theta_{13}) \sin(\varphi_{13}) \\ p_{F}
	\end{pmatrix}
	\,,\
	p_{4, {\rm com}} = 
	\begin{pmatrix}
		\sqrt{ m_{4}^2 + p_{F}^2 } \\ - p_{F} \sin(\theta_{13}) \cos(\varphi_{13}) \\ - p_{F} \sin(\theta_{13}) \sin(\varphi_{13}) \\ - p_{F}
	\end{pmatrix}
	\,.
\end{align}

The on-shell condition $p_{5, {\rm com}}^2 = m_{5}^2$ fixes $p_{K} = \sqrt{ p_{I}^2 + 2 p_{I} p_{J} \cos(\theta_{12}) + p_{J}^2 }$. This allows us to replace the angle $\theta_{12}$ by the `length' $p_{K}$, which can be interpreted as describing a triangle either by two lengths $p_{I}$, $p_{J}$ and an angle $\theta_{12}$, or by three lenghts $p_{I}$, $p_{J}$, $p_{K}$. This constrains these parameters to the region $p_{I} \ge 0$, $p_{J} \ge 0 $, $p_{K} \ge 0$, $p_{K} \le p_{I} + p_{J}$, $p_{K} \ge \left|p_{I} - p_{J}\right|$.
Conservation of energy also fixes $p_{F}$ via the equation
\begin{align}
	\sqrt{ m_{1}^2 + p_{I}^2 } + \sqrt{ m_{2}^2 + p_{J}^2 } + \sqrt{ m_{5}^2 + p_{K}^2 } &= \sqrt{ m_{3}^2 + p_{F}^2 } + \sqrt{ m_{4}^2 + p_{F}^2 } \,.
\end{align}

After the integration the result depends only three variables $p_{I}$, $p_{J}$, $p_{K}$. However, we can exploit the Lorentz invariance of the amplitude, and re-express the result in terms of scalar products. This is because $p_{1} \cdot p_{2} = p_{1, {\rm com}} \cdot p_{2, {\rm com}}$, etc. defines a solvable system of three coupled algebraic equations which lets us perform the substitution
\begin{align}
	\qty{ p_{I} , p_{J} , p_{K} } &\Rightarrow \qty{ p_{1, {\rm com}} \cdot p_{2, {\rm com}} , p_{1, {\rm com}} \cdot p_{5, {\rm com}} , p_{2, {\rm com}} \cdot p_{5, {\rm com}} } \Rightarrow \qty{ p_{1} \cdot p_{2} , p_{1} \cdot p_{5} , p_{2} \cdot p_{5} } \,.
\end{align}
This trick avoids having to work out the explicit Lorentz transformation from the center-of-mass frame to your favorite frame, since one just computes the three scalar products in the preferred reference frame and subsitutes them into the result.

Note that because of $(p_{3} + p_{4})^2 = (p_{1} + p_{2} + p_{5})^2$ the scalar product $p_{3} \cdot p_{4}$ is not independent. This is consistent with the fact that integration over two integration variables should reduce the number of independent scalar products from five to three.

\subsection{Basis integrals}

Let us now proceed with the computation of the basis integrals of equation \eqref{eq:3to2basisIntegrals}. For a more transparent notation we are going to replace complicated expressions which do not depend on the final phase space angles with simple variables $c_{l}$. Also the substitution of integration variable $\cos(\theta_{13}) = z_{13}$ is going to be performed without extra comments.

The analytic integrals have been verified by comparing the results to those of numeric integration.

\subsubsection{$\mathcal{I}_{1}$}

Consider the integral
\begin{align}
	\int \dd{\Omega} \mathcal{I}_{1} &\supset \int \dd{\Omega} \ \dfrac{1}{ s_{13}^{a} } \\
	&= \int_{-1}^{+1} \dd{z_{13}} \int_{0}^{2\pi} \dd{\varphi_{13}} \ \dfrac{1}{ \qty( c_{0} + c_{1} \cdot z_{13} )^{a} } \,.
\end{align}
The azimuthal integration evaluates trivially to $2\pi$, and the integration over $z_{13}$ is a textbook integral which evalutes either to a rational function or a logarithm, depending on the value of $a$.

Now consider an integral of similar structure, but different momentum,
\begin{align}
	\int \dd{\Omega} \mathcal{I}_{1} &\supset \int \dd{\Omega} \ \dfrac{1}{ s_{23}^{a} } \\
	&= \int_{0}^{\pi} \dd{\theta_{13}} \sin(\theta_{13}) \int_{0}^{2\pi} \dd{\varphi_{13}} \ \dfrac{1}{ \qty[ c_{0} + c_{2} \cdot \sin(\theta_{13}) \cos(\varphi_{13}) + c_{1} \cdot \cos(\theta_{13}) ]^{a} } \,.
\end{align}
This integral can no longer be evaluated quickly. The solution is to exploit the invariance of the final phase space integral under $O(3)$-rotations, and rotate the vectors $p_{3, {\rm com}}$, $p_{4, {\rm com}}$ by an angle such that they are aligned with the direction of $p_{2, {\rm com}}$. Note that because of the linearity of the integral one has to apply this rotation only to this term, not to all the others. However, scalar products in the same term are invariant under rotations, since they are part of the Lorentz group, and therefore we do not need to perform any transformations an scalar products independent of $p_{3, {\rm com}}$ and $p_{4, {\rm com}}$.

Therefore, the method of integrating structures of type $\mathcal{I}_{1}$ is performing an appropriate $O(3)$ rotation, and then evaluating the integral over a rational function.

\subsubsection{$\mathcal{I}_{2}$}

Consider the integral
\begin{align}
	\int \dd{\Omega} \mathcal{I}_{2} &\supset \int \dd{\Omega} \ \dfrac{1}{s_{23}^{a} s_{54}^{b}} \,.
\end{align}
One possible way to proceed would be to perform a partial fraction decomposition with respect to the variable $z_{13}$, and then apply different $O(3)$ rotations to the different terms.

However, an easier way is Feynman parameterization, which is also easier to implement as a generally applicable algorithm. For two denominators we have
\begin{align}
	\dfrac{1}{A^{a} B^{b}} &= \dfrac{\Gamma(a+b)}{\Gamma(a) \Gamma(b)} \int_{0}^{1} \dd{x} \dfrac{x^{a-1} \bar{x}^{b-1}}{\qty( x A + \bar{x} B )^{a+b}} \,,\quad \bar{x} = 1-x \,.
\end{align}
Then for the above example we have
\begin{align}
	x A + \bar{x} B &= c_{0}^{\prime} - 2 \qty( x E_{2} E_{3} + \bar{x} E_{5} E_{4} ) + 2 \qty( x \ \vec{p}_{2} \cdot \vec{p}_{3} + \bar{x} \ \vec{p}_{5} \cdot \vec{p}_{4} ) \\
	&= c_{0}(x) + 2 \qty( x \vec{p}_{2} - \bar{x} \vec{p}_{5} ) \cdot \vec{p}_{3}
\end{align}
where $E_{k} = p_{k, {\rm com}}^{0}$ are the energies in the center-of-mass frame, and for readability we omitted the index ${\rm com}$ indicating the evaluation in the center-of-mass frame. In the last step we used that $p_{3, {\rm com}} = - p_{4, {\rm com}}$.

In order to evaluate the integral over the azimuthal and polar angle we make again use of the invariance under $O(3)$-rotations and rotate in such a way that the vector $\vec{p}_{3}$ is parallel to the vector $x \vec{p}_{2} + \bar{x} \vec{p}_{5}$. We can then write the scalar product as
\begin{align}
	2 \qty( x \vec{p}_{2} - \bar{x} \vec{p}_{5} ) \cdot \vec{p}_{3} &= 2 \abs{ x \vec{p}_{2} - \bar{x} \vec{p}_{5} } \abs{ \vec{p}_{3} } \cos(\theta_{13}) \,,
\end{align}
such that the integral over $\mathcal{I}_{2}$ reduces to
\begin{align}
	\int \dd{\Omega} \mathcal{I}_{2} &\supset \int \dd{\Omega} \ \dfrac{1}{s_{23}^{a} s_{54}^{b}} \\
	&= \dfrac{\Gamma(a+b)}{\Gamma(a) \Gamma(b)} \int_{0}^{1} \dd{x} \int_{-1}^{+1} \dd{z_{13}} \int_{0}^{2\pi} \dd{\varphi_{13}} \ \dfrac{x^{a-1} \bar{x}^{b-1}}{ \qty[ c_{0}(x) + c_{1}(x) z_{13}  ]^{a+b} } \,.
\end{align}
The integration over the azimuthal angle $\varphi_{13}$ evaluates trivially to $2\pi$.
Since this expression only appears for at least denominators, i.e. $a \ge 1$, $b \ge 1$, we have $a+b \ge 2$. Therefore, performing the integral over the polar angle $z_{13}$ is never going to yield a logarithm. It even turns out that, while the length of a vector contains a square root of the Feynman parameter $x$, the integral only depends on the length squared, and therefore we are left with an integral of the form
\begin{align}
	\int_{0}^{1} \dd{x} \ \dfrac{\sum_{i=0}^{k-1} a_{i} x^{i}}{\qty( b_{0} + b_{1} \cdot x + b_{2} \cdot x^2 )^k} \,.
\end{align}
This is a well known class of textbook integrals, and can be easily computed, for example with the method of partial fraction decomposition, or Mathematica.

Therefore, the method of integrating structures of type $\mathcal{I}_{2}$ is a combination of applying Feynman parameters and performing an appropriate $O(3)$ rotation.

\subsubsection{$\mathcal{I}_{3}$}

Consider the integral
\begin{align}
	\int \dd{\Omega} \mathcal{I}_{3} &\supset \int \dd{\Omega} \ \dfrac{s_{23}^{m}}{s_{53}^{a}} \,.
\end{align}
From the expressions before we have already seen that it is most convenient to eliminate any azimuthal angle $\varphi_{13}$ in the denominator. We achieve this as before by rotating $p_{3, {\rm com}}$ in a suitable way.

However, now the numerator is not independent of $p_{3, {\rm com}}$ and will be affected by the rotation.
This is easily dealt with by acting on the vector $\vec{p}_{3}$ with the standard rotation matrix $\mathcal{R}_{3}$. Since one can in general not choose the matrix $\mathcal{R}_{3}$ in such a way that any dependence on $\varphi_{13}$ disappears, we still have $\varphi_{13}$ in the numerator. Therefore, we have
\begin{align}
	\int \dd{\Omega} \mathcal{I}_{3} &\supset \int_{0}^{\pi} \dd{\theta_{13}} \sin(\theta_{13}) \int_{0}^{2\pi} \dd{\varphi_{13}} \ \dfrac{ \qty[ d_{0} + d_{2} \cdot \sin(\theta_{13}) \cos(\varphi_{13}) + d_{1} \cdot \cos(\theta_{13}) ]^{m} }{ \qty[ c_{0} + c_{1} \cdot \cos(\theta_{13}) ]^{a} }
\end{align}
We perform the integration over the azimuthal angle $\varphi_{13}$ first. When we expand the numerator, we get a series with terms proportional to
\begin{align}
	\int_{0}^{2\pi} \dd{\varphi_{13}} \ \qty[ \sin(\theta_{13}) \cos(\varphi_{13}) ]^{k} \,.
\end{align}
This integral vanishes for all odd powers of $k$. This is in so far convenient that keeping only the even powers eliminates all occurences of $\sin(\theta_{13}) = \sqrt{1 - z_{13}^2}$, which is not straightforward to integrate in general. Therefore after performing the azimuthal integration we are left with the integration over the polar angle, which takes the form
\begin{align}
	\int_{-1}^{+1} \dd{z_{13}} \ \dfrac{\sum_{m} a_{m} z_{13}^{m}}{ \qty[ c_{0} + c_{1} \cdot z_{13} ]^{a} } \,.
\end{align}
As in the previous subsection, this is a well known class of textbook integrals, and can be easily computed, for example with the method of partial fraction decomposition, or Mathematica.

Therefore, the method of integrating structures of type $\mathcal{I}_{3}$ is performing an appropriate $O(3)$ rotation and then proceeding with the integration over the angles as described above.

\section{More on the computation of $\int \mathrm{d} \Omega \lvert \mathcal{M} \rvert^2_{3\to 2}$}
\label{app:more_on_3to2}

\subsection{Detailed results of our computation}

In this appendix we report the detailed results of our computation of $\int \mathrm{d} \Omega \lvert \mathcal{M} \rvert^2$ for all possible $3\to 2$ processes involving in the initial state at least one gauge boson $V$, scattering with other $V$'s and/or fermions and/or scalars charged under the gauge group, where for completeness in the scalar case we included the self-coupling $\mathcal{L} \supset \lambda \lvert \phi \rvert^4$ in addition to the scalar gauge coupling to $V$.
We report $\int \mathrm{d} \Omega \lvert \mathcal{M} \rvert^2$ in terms of scalar products of the 4-momenta in the center-of-mass frame $p_1+p_2+p_5 = p_3+p_4$, including for definiteness  only the terms that are leading order in $p_i \cdot p_j/\mu^2$. 
In case at least one initial particle belongs to the bath, we remind that one has the hierarchy $\mu^2 \ll p_{1} \cdot p_{2} \ll p_{1} \cdot p_{5} \simeq p_{2} \cdot p_{5}$, leading to the expansions reported in Tables~\ref{tab:MsqAbelianFermion} (case of one fermion charged under the gauge group), \ref{tab:MsqAbelianScalar} (case of one scalar charged under the gauge group), \ref{tab:MsqAbelianFermionScalar} (case of both a fermion and a scalar charged under the gauge group).
In case all initial particles belong to the shell, then all scalar products are of the same order, leading to the leading-order results in Table~\ref{tab:MsqThermalization}.
The reader finds in the main text, for each case of interest, the values of the scalar products and the parametric dependence of the leading terms in $\int \mathrm{d} \Omega \lvert \mathcal{M} \rvert^2$ (Table~\ref{tab:Msq_parametrics}).

\begingroup
\def\arraystretch{1.5}
\begin{table}[htp!]
	\centering
	\begin{tabular}{c|c}
		$a(p_{1}) \ b(p_{2}) \ c(p_{5})$ & $\int \mathrm{d} \Omega \abs{ \mathcal{M} }^2$: $\mathcal{ G }_{V} = \mathrm{U}(1)$ \\
		$\to X Y$ & $\int \mathrm{d} \Omega \abs{ \mathcal{M} }^2$: $\mathcal{ G }_{V} = \mathrm{SU}(N)$ \\ \hline \hline
		$V V V \to f \bar{f}$ & $g^6 \ \frac{48\pi \Big( 2 \ln \Big( \frac{p_{1} \cdot p_{2}}{\mu^2} \Big) + 5 \ln 2 \Big)}{p_{1} \cdot p_{2}}$ \\
		 & $ g^6 \ \frac{\pi}{N^2(N^2-1)^2} \frac{1}{{ p_{1} \cdot p_{2} }} \Bigg[ \substack{9 N^2 (N^2-1) \ln \Big( \frac{p_{1} \cdot p_{5}}{\mu^2} \Big) + 2(N^4-7N^2+6) \ln \Big( \frac{p_{1} \cdot p_{2}}{\mu^2} \Big) \\ \qquad+ N^4(21\ln2 - 15) + N^2(9 -51\ln2) + 30\ln2 } \Bigg]$ \\ \hline
		$V V f \to V f$ & $g^6 \ \frac{24\pi \Big( 2 \ln \Big( \frac{p_{1} \cdot p_{2}}{\mu^2} \Big) + 5 \ln 2 \Big)}{p_{1} \cdot p_{2}}$ \\
		$V V \bar{f} \to V \bar{f}$ & \multirow{2}{*}{$g^6 \ \frac{N^2}{N^2-1} \frac{36\pi \ p_{1} \cdot p_{5} \Big( 5 \ln \Big( \frac{p_{1} \cdot p_{2}}{\mu^2} \Big) + 6 \ln 2 \Big)}{\big( p_{1} \cdot p_{2} \big)^2}$} \\ & \\ \hline
		$f \bar{f} V \to V V$ & $g^6 \ \frac{32\pi \Big( \ln \Big( \frac{p_{1} \cdot p_{2}}{\mu^2} \Big) + 3 \ln 2 \Big)}{p_{1} \cdot p_{5}}$ \\
		 & $g^6 \ \frac{8\pi \ p_{1} \cdot p_{5} \Big( (2N^2-3) \ln \Big( \frac{p_{1} \cdot p_{2}}{\mu^2} \Big) + 3 (N^2-1) \ln 2 \Big)}{\big( p_{1} \cdot p_{2} \big)^2}$ \\ & \\\hline
		$f \bar{f} V \to f \bar{f}$ & $g^6 \ \frac{4\pi \Big( \ln \Big(\frac{p_{1} \cdot p_{5}}{\mu^2} \Big) + 18 \ln \Big( \frac{p_{1} \cdot p_{2}}{\mu^2} \Big) - 1 + 29 \ln 2 \Big)}{p_{1} \cdot p_{2}}$ \\
		 & $g^6 \ \frac{\pi}{6N^2} \frac{1}{p_{1} \cdot p_{2}} \Bigg[ \substack{ 3N (N^2-1) \ln \Big( \frac{p_{1} \cdot p_{5}}{\mu^2} \Big) + 6(10N^3+N^2-10N-1) \ln \Big( \frac{p_{1} \cdot p_{2}}{\mu^2} \Big) \\ \qquad+ N^3(99\ln2-5) + 12N^2 \ln2 + N(3-99\ln2) - 12\ln2 } \Bigg]$ \\ & \\ \hline
		$f f V \to f f$ & $g^6 \ \frac{16\pi \Big( 14 \ln \Big( \frac{p_{1} \cdot p_{2}}{\mu^2} \Big) + 15 \ln 2 \Big)}{p_{1} \cdot p_{2}}$ \\
		$\bar{f} \bar{f} V \to \bar{f} \bar{f}$  & $g^6 \ \frac{N^2-1}{N^2} \frac{ \pi \Big( 4(5N-2) \ln \Big( \frac{p_{1} \cdot p_{2}}{\mu^2} \Big) + 30N \ln2 \Big) }{p_{1} \cdot p_{2}}$ \\ & \\ \hline
		$V f V \to V f$ & $g^6 \ \frac{\pi \Big( 10 \ln \Big( \frac{p_{1} \cdot p_{5}}{\mu^2} \Big) + 4 \ln \Big( \frac{p_{1} \cdot p_{2}}{\mu^2} \Big) + 5 + 34 \ln 2 \Big)}{p_{1} \cdot p_{2}}$ \\
		$V \bar{f} V \to V \bar{f}$ & $g^6 \ \frac{1}{N^2-1} \frac{10\pi \ p_{1} \cdot p_{5} \Big( 2(4N^2-1) \ln \Big( \frac{p_{1} \cdot p_{2}}{\mu^2} \Big) + 3(3N^2-1) \ln 2 \Big)}{\big( p_{1} \cdot p_{2} \big)^2}$ \\ & \\
	\end{tabular}
	\caption{\it \small Scatterings involving vectors and fermions, where $g=\gD$ is the gauge coupling and $\mu$ is the IR cut-off from Eq.~(\ref{eq:IRcutoff}).}
	\label{tab:MsqAbelianFermion}
\end{table}
\endgroup

\begingroup
\def\arraystretch{1.5}
\begin{table}[htp!]
	\centering
	\begin{tabular}{c|c}
		$a(p_{1}) \ b(p_{2}) \ c(p_{5})$ & $\int \mathrm{d} \Omega \abs{ \mathcal{M} }^2$: $\mathcal{ G }_{V} = \mathrm{U}(1)$ \\
		$\to X Y$ & $\int \mathrm{d} \Omega \abs{ \mathcal{M} }^2$: $\mathcal{ G }_{V} = \mathrm{SU}(N)$ \\ \hline \hline
		$V V V \to \phi \bar{\phi}$ & $g^6 \ \frac{8\pi \Big( 9 \ln \Big( \frac{p_{1} \cdot p_{5}}{\mu^2} \Big) + 11 \ln 2 \Big)}{p_{1} \cdot p_{5}}$ \\
		 & $g^6 \ \frac{2N^2-3}{2(N^2-1)^2} \frac{3\pi}{p_{1} \cdot p_{2}}$ \\  & \\ \hline
		$V V \phi \to V \phi$ & $g^6 \ \frac{4\pi \Big( 9 \ln \Big( \frac{p_{1} \cdot p_{5}}{\mu^2} \Big) - 9 + 11 \ln 2 \Big)}{p_{1} \cdot p_{5}}$ \\
		$V V \bar{\phi} \to V \bar{\phi}$ & $g^6 \ \frac{\pi \ p_{1} \cdot p_{5}}{N^2-1} \frac{ 8(4N^2-1) \ln \Big( \frac{ p_{1} \cdot p_{2} }{\mu^2} \Big) + 12(3N^2-1) \ln 2 }{\big( p_{1} \cdot p_{2} \big)^2}$ \\ & \\ \hline
		$\phi \bar{\phi} V \to V V$ & $g^6 \ \frac{16\pi \Big( \ln \Big( \frac{p_{1} \cdot p_{2}}{\mu^2} \Big) + 3 \ln 2 \Big)}{p_{1} \cdot p_{5}}$ \\
		 & $g^6 \ \pi \ p_{1} \cdot p_{5} \ \frac{ 4(2N^2-3) \ln \Big( \frac{ p_{1} \cdot p_{2} }{\mu^2} \Big) + 12(N^2-1) \ln 2 }{\big( p_{1} \cdot p_{2} \big)^2}$ \\ & \\ \hline
		$\phi \bar{\phi} V \to \phi \bar{\phi}$ & $g^6 \ \frac{\pi}{p_{1} \cdot p_{2}}$ \\
		 & $g^6 \ \frac{2N^2-3}{24N} \frac{\pi}{p_{1} \cdot p_{2}}$ \\ & \\ \hline
		$\phi \phi V \to \phi \phi$ & $g^6 \ \frac{96\pi \Big( \ln \Big( \frac{p_{1} \cdot p_{2}}{\mu^2} \Big) + \ln 2 \Big)}{p_{1} \cdot p_{5}}$ \\ 
		$\bar{\phi} \bar{\phi} V \to \bar{\phi} \bar{\phi}$ & $\frac{ \pi }{4N^2} \frac{1}{p_{1} \cdot p_{5}} \left[\rule{0cm}{1.2cm}\right. \substack{ (N+1) (4 \lambda g^4 N + g^6 (N-1) ) \ln \Big( \frac{ p_{1} \cdot p_{5} }{\mu^2} \Big)   -    8 (N^2-1) ( 4\lambda g^4 N - g^6 (3N^2-1) ) \ln \Big( \frac{ p_{1} \cdot p_{2} }{\mu^2} \Big)   \\ \qquad+ 16 \lambda^2 \gD^2 \Big( N^3 (3\ln2 - 1) + N^2 (3\ln2 -2) \Big) -8 \lambda g^4 \Big( N^3(5\ln2 + 1) + N^2 - N (5\ln2 + 2) \Big) \\ \qquad+ g^6 \Big( N^3 (35\ln2 - 1) + 13N^2 \ln 2 - N(35\ln2 - 3) - 13\ln2 - 2 \Big) } \left.\rule{0cm}{1.2cm}\right]$ \\ & \\ \hline
		$V \phi V \to V \phi$ & $g^6 \ \frac{8\pi}{p_{1} \cdot p_{2}}$ \\
		$V \bar{\phi} V \to V \bar{\phi}$ & $g^6 \ \frac{\pi \ p_{1} \cdot p_{5}}{N^2-1} \frac{ 8 (4N^2-1) \ln \Big( \frac{ p_{1} \cdot p_{2} }{\mu^2} \Big) + 12(3N^2-1) \ln 2 }{\big( p_{1} \cdot p_{2} \big)^2}$ \\ & \\
	\end{tabular}
	\caption{\it \small Scatterings involving vectors and scalars, where $g=\gD$ is the gauge coupling and $\mu$ is the IR cut-off from Eq.~(\ref{eq:IRcutoff}).}
	\label{tab:MsqAbelianScalar}
\end{table}
\endgroup

\begingroup
\def\arraystretch{1.5}
\begin{table}[htp!]
	\centering
	\begin{tabular}{c|c}
		$a(p_{1}) \ b(p_{2}) \ c(p_{5})$ & $\int \mathrm{d} \Omega \abs{ \mathcal{M} }^2$: $\mathcal{ G }_{V} = \mathrm{U}(1)$ \\
		$\to X Y$ & $\int \mathrm{d} \Omega \abs{ \mathcal{M} }^2$: $\mathcal{ G }_{V} = \mathrm{SU}(N)$ \\ \hline \hline
		$\phi \bar{\phi} V \to f \bar{f}$ & $g^6 \ \frac{2\pi \Big( \ln \Big( \frac{p_{1} \cdot p_{5}}{\mu^2} \Big) + 3 \ln 2 -1 \Big)}{p_{1} \cdot p_{2}}$ \\
		 & $g^6 \ \frac{\pi}{12N} \frac{ 3 (N^2-1) \ln \Big( \frac{p_{1} \cdot p_{5}}{\mu^2} \Big) + 9 (N^2-1) \ln 2 - 5N^2 + 3  }{p_{1} \cdot p_{2}}$  \\ & \\ \hline
		$f \bar{f} V \to \phi \bar{\phi}$ & $g^6 \ \frac{2\pi}{p_{1} \cdot p_{2}}$ \\
		 & $g^6 \ \frac{2N^2-3}{12 N} \frac{\pi}{p_{1} \cdot p_{2}}$ \\ & \\ \hline
		$\phi f V \to \phi f$ & $g^6 \ \frac{8\pi \Big( 2 \ln \Big( \frac{p_{1} \cdot p_{2}}{\mu^2} \Big) + 3 \ln 2 \Big)}{p_{1} \cdot p_{2}}$ \\
		$\bar{\phi} \bar{f} V \to \bar{\phi} \bar{f}$ & $g^6 \ \frac{N^2-1}{N} \frac{\pi \Big( 2 \ln \Big( \frac{p_{1} \cdot p_{2}}{\mu^2} \Big) + 3 \ln 2 \Big)}{p_{1} \cdot p_{2}}$ \\  & \\ \hline
		$\phi \bar{f} V \to \phi \bar{f}$ & $g^6 \ \frac{8\pi \Big( 2 \ln \Big( \frac{p_{1} \cdot p_{2}}{\mu^2} \Big) + 3 \ln 2 \Big)}{p_{1} \cdot p_{2}}$ \\
		$\bar{\phi} f V \to \bar{\phi} f$ & $g^6 \ \frac{N^2-1}{N} \frac{\pi \Big( 2 \ln \Big( \frac{p_{1} \cdot p_{2}}{\mu^2} \Big) + 3 \ln 2 \Big)}{p_{1} \cdot p_{2}}$ \\ & \\ \hline
		$V V V \to V V$ & $0$ \\
		 & $g^6 \ \frac{N^3}{(N^2-1)^2} \frac{ 144\pi \ p_{1} \cdot p_{5} \Big( 5 \ln \Big( \frac{p_{1} \cdot p_{2}}{\mu^2} \Big) + 6 \ln 2 \Big)  }{\big( p_{1} \cdot p_{2} \big)^2}$ \\ & \\
	\end{tabular}
	\caption{\it \small Scatterings involving vectors, fermions, and scalars, where $g=\gD$ is the gauge coupling and $\mu$ is the IR cut-off from Eq.~(\ref{eq:IRcutoff}).}
	\label{tab:MsqAbelianFermionScalar}
\end{table}
\endgroup

\begingroup
\def\arraystretch{1.5}
\begin{table}[htp!]
	\centering
	\begin{tabular}{c|c}
		$a(p_{i}) \ b(p_{j}) \ c(p_{k})$ & $\int \mathrm{d} \Omega \abs{ \mathcal{M} }^2$: $\mathcal{ G }_{V} = \mathrm{U}(1)$ \\
		$\to X Y$ & $\int \mathrm{d} \Omega \abs{ \mathcal{M} }^2$: $\mathcal{ G }_{V} = \mathrm{SU}(N)$ \\ \hline \hline
		$V V V \to f \bar{f}$ & $g^6 \ \frac{24\pi \Big( 5 \ln \Big( \frac{p_{i} \cdot p_{j}}{\mu^2} \Big) + 5\ln3 - \ln2 \Big)}{p_{i} \cdot p_{j}}$ \\
		 & $g^6 \ \frac{\pi}{2N^2 (N^2-1)^2} \frac{1}{p_{i} \cdot p_{j}} \Bigg[ \substack{ 15 (N^4-3N^2+2) \ln \Big( \frac{p_{i} \cdot p_{j}}{\mu^2} \Big) \\ \qquad+ N^4(29 \ln2 -3) +N^2(-9 + 105\ln2 - 90\ln3) +24\ln3 - 6\ln2 } \Bigg] $  \\ \hline
		$V V V \to \phi \bar{\phi}$ & $g^6 \ \frac{36\pi \Big( \ln \Big( \frac{p_{i} \cdot p_{j}}{\mu^2} \Big) + 3\ln2 - \ln3 \Big)}{p_{i} \cdot p_{j}}$ \\
		 & $ g^6 \ \frac{\pi}{4N^2 (N^2-1)^2} \frac{1}{p_{i} \cdot p_{j}} \Bigg[ \substack{  9 (N^4-3N^2+2) \ln \Big( \frac{p_{i} \cdot p_{j}}{\mu^2} \Big) \\ \qquad+ N^4(3 + 11\ln2) + N^2 (9 - 129 \ln2 + 54\ln3) - 18\ln3 + 54\ln2 } \Bigg] $ \\ \hline
		$V V V \to V V$ & $0$ \\
		 & $g^6 \ \frac{N^3}{(N^2-1)^2} \frac{ 2\pi \Big( 1323 \ln \Big( \frac{p_{i} \cdot p_{j}}{\mu^2} \Big) - 3 - 396 \ln 3 + 1987 \ln 2 \Big) }{p_{i} \cdot p_{j}}$
	\end{tabular}
	\caption{\it \small Self-thermalization within the shell, where $g=\gD$ is the gauge coupling and $\mu$ is the IR cut-off from Eq.~(\ref{eq:IRcutoff}).}
	\label{tab:MsqThermalization}
\end{table}
\endgroup

\subsection{Simple upper estimates of the integrated amplitudes}
\label{app:estimate32}

In this Appendix we give useful formulae to estimate an upper limit on the integrated spin-averaged squared matrix element, without actually having to compute the amplitude and performing the final phase space integration.
In general, these estimates are very conservative in the sense that they overestimate the amplitudes. These estimates can then serve as a guidance on the computation of new amplitudes.

\subsubsection{Equilibration between bath and shell}

Assume that we have a sizeable spread of particles, i.e. $p_{1} \cdot p_{2} \gtrsim \mu^2$.

We take the following basis
\begin{align}
	p_{1,com} = 
	\begin{pmatrix}
		p_{I} \\ 0 \\ 0 \\ p_{I}
	\end{pmatrix}
	\,,\ p_{2,com} =
	\begin{pmatrix}
		p_{J} \\ p_{J} \sin(\theta_{12}) \\ 0 \\ p_{J} \cos(\theta_{12})
	\end{pmatrix}
	\,,\ p_{3,com} =
	\begin{pmatrix}
		\sqrt{ p_{I}^2 + 2 p_{I} p_{J} \cos(\theta_{12}) + p_{J}^2 } \\ -p_{J} \sin(\theta_{1,2}) \\ 0 \\ -p_{I} - p_{J} \cos(\theta_{12})
	\end{pmatrix}
	\,,\\ p_{4,com} =
	\begin{pmatrix}
		p_{F} \\ p_{F} \sin(\theta_{13}) \cos(\varphi_{13}) \\ p_{F} \sin(\theta_{13}) \sin(\varphi_{13}) \\ p_{F} \cos(\theta_{13})
	\end{pmatrix}
	\,,\ p_{5,com} =
	\begin{pmatrix}
		p_{F} \\ - p_{F} \sin(\theta_{13}) \cos(\varphi_{13}) \\ - p_{F} \sin(\theta_{13}) \sin(\varphi_{13}) \\ - p_{F} \cos(\theta_{13}) \,,
	\end{pmatrix}
\end{align}
which is the one from appendix \ref{app:scalar_products_parametrization}, with all particles massless. It is helpful to define
\begin{align}
	p_{IJ} &= \sqrt{ p_{I}^2 + 2 p_{I} p_{J} \cos(\theta_{12}) + p_{J}^2 } \ge 0 \,,
\end{align}
sucht that energy conservation implies
\begin{align}
	2 p_{F} &= p_{I} + p_{J} + p_{IJ} \,.
\end{align}
We find these scalar products,
\begin{align}
	p_{1} \cdot p_{5} &= \dfrac{1}{2} \qty[ \qty( p_{I} + p_{IJ} )^2 - p_{J}^2 ] \\
	p_{2} \cdot p_{5} &= \dfrac{1}{2} \qty[ \qty( p_{J} + p_{IJ} )^2 - p_{I}^2 ] \\
	p_{1} \cdot p_{2} &= \dfrac{1}{2} \qty[ \qty( p_{I} + p_{J} )^2 - p_{IJ}^2 ]
\end{align}
Note, that two scalar products are the same if the two associated energies are the same, i.e. $p_{1} \cdot p_{5} = p_{2} \cdot p_{5}$ if $p_{I} = p_{J}$.

We do not look at interference terms, since they cannot be more divergent than a single diagram squared. We have the principal form of the spin-averaged squared amplitude,	
\begin{align}
	\abs{ \mathcal{ M } }^2 &= \alpha^3 \sum \dfrac{\prod s_{mn}}{ \qty( s_{ij} - \mu )^{a} \qty( s_{kl} - \mu )^{b} } \,.
\end{align}
One case is given by
\begin{align}
	\abs{ \mathcal{ M } }^2 \supset \dfrac{1}{\qty( s_{13} - \mu^2 )^{a} \qty( s_{45} - \mu^2 )^{b}} \,,
\end{align}
where if $p_{1} \parallel p_{2}$, then also $p_{1} \parallel p_{5}$, and both denominators become divergent at the same time. Since they are thermally distributed, this is not exactly the case, so they diverge at different integration regions. Therefore we find the suppression by the IR regulator (Debye mass) to be at maximum
\begin{align}
	\qty( \mu^2 )^{-2+1} \,.
\end{align}

We also have
\begin{align}
	\abs{ \mathcal{ M } }^2 \supset \dfrac{1}{\qty( s_{12} - \mu^2 )^{a} \qty( s_{45} - \mu^2 )^{b}} \,,
\end{align}
which after integrating scales at worst like
\begin{align}
	\dfrac{1}{ \qty( p_{1} \cdot p_{2} )^{2} } \qty( \mu^2 )^{-2+1} \,.
\end{align}

Therefore we find
\begin{align}
	\int \dd{\Omega} \abs{ \mathcal{ M } }^2 &\simeq \alpha^3 \dfrac{1}{p_{1} \cdot p_{5}} \qty( \dfrac{p_{1} \cdot p_{5}}{p_{1} \cdot p_{2}} )^{k} \qty( \dfrac{p_{1} \cdot p_{5}}{\mu^2} )^{l} \,,\qquad k \le 2 \,,\ l \le 1 \,.
 \label{eq:Msq_estimate}
\end{align}
Here we inserted the only leftover scale $p_{1} \cdot p_{5}$ to fix the correct mass dimension, which is conservative from the point of view of finding an upper limit on the amplitude squared. For simplicity we have also neglected any appearances of logarithms, since they are not much larger than $O(1)$ factors.

Comparing Eq.~(\ref{eq:Msq_estimate}) with the calculated processes in Tables \ref{tab:MsqAbelianFermion}, \ref{tab:MsqAbelianScalar}, \ref{tab:MsqAbelianFermionScalar}, we find that we do not always have maximal divergence with $k=2$, $l=1$. This is due to cancellations at amplitude level depending on the specific physical process.

\subsubsection{Equilibration within the shell}

For equilibration within the shell itself consider the following momentum basis,
\begin{align}
	p_{1,com} &\simeq p_{2,com} \simeq p_{5,com} \simeq 
	\begin{pmatrix}
		\mu \\ 0 \\ 0 \\ 0
	\end{pmatrix}
	\,, \\ \quad p_{3,com} &= 
	\begin{pmatrix}
		p_{f} \\ p_{f} \sin(\theta_{13}) \\ 0 \\ p_{f} \cos(\theta_{13})
	\end{pmatrix}
	\,, \quad p_{4,com} = 
	\begin{pmatrix}
		p_{f} \\ - p_{f} \sin(\theta_{13}) \\ 0 \\ - p_{f} \cos(\theta_{13}) 
	\end{pmatrix} \,,
\end{align}
sucht that $p_{f} = \tfrac{3}{2} \mu$. This holds under the assumption that the spread is small, which implies that in the COM-frame each initial particle is approximately at rest with sub-leading momentum component.

Then we find
\begin{align}
	p_{1} \cdot p_{2} \simeq p_{1} \cdot p_{5} \simeq p_{2} \cdot p_{5} &\simeq m_{c,h}^2 \\
	p_{1,2,5} \cdot p_{3} &\simeq \dfrac{3}{2} \mu^2 \qty( 1 - z_{13} ) \\
	p_{1,2,5} \cdot p_{4} &\simeq \dfrac{3}{2} \mu^2 \qty( 1 + z_{13} ) \\
	p_{3} \cdot p_{4} &\simeq 2 p_{f}^2 = \dfrac{9}{2} \mu^2
\end{align}
An important observation is that $m_{c,h} \gg \mu$, i.e. having spread gives us a fictitious suppression, compared to the maximal possible upper limit. The maximal amplitude is therefore given diagrams scaling like
\begin{align}
	\abs{ \mathcal{M} }^2 \propto \dfrac{1}{\qty( p_{1} \cdot p_{3} )^2} \dfrac{1}{\qty( p_{2} \cdot p_{4} )^2}
\end{align}
which, since they only contain one energy scale, gives us immediately the scaling
\begin{align}
	\int \dd{\Omega} \abs{ \mathcal{M} }^2 &\sim \dfrac{1}{\mu^2} \,.
 \label{eq:Msq_sss_estimate}
\end{align}
Comparing Eq.~(\ref{eq:Msq_sss_estimate}) to the calculated processes in Table~\ref{tab:MsqThermalization}, we find that we not always have maximal divergence. As before, this is due to cancellations at amplitude level depending on the specific physical process.

While the estimates in Eq. (\ref{eq:Msq_estimate}) and Eq. (\ref{eq:Msq_sss_estimate}) serve as a good guideline and a valuable sanity check for the full computation, we find that they are in most cases too conservative. Due to less divergent pole structures of specific amplitudes one expects to find a result which is smaller by orders of magnitudes by performing a full computation of the specific process. We refer again to the results for a few selected processes in Tables \ref{tab:MsqAbelianFermion}, \ref{tab:MsqAbelianScalar}, \ref{tab:MsqAbelianFermionScalar}, \ref{tab:MsqThermalization}, for which we have performed the full calculation.

\medskip
\small
\FloatBarrier

\bibliographystyle{JHEP}
\bibliography{biblio}
\end{document}